\documentclass[review]{elsarticle}
\usepackage{graphicx}
\usepackage{tabularx}
\usepackage{pdflscape}
\usepackage{rotating}
\usepackage{multicol}
\usepackage[utf8]{inputenc}
\usepackage{threeparttable}
\usepackage{rotating}
\usepackage{adjustbox}
\usepackage{lineno,hyperref}
\usepackage{float}
\usepackage{color}
\modulolinenumbers[5]

\journal{Journal of \LaTeX\ Templates}

\def\be{\begin{equation}} 
\def\ee{\end{equation}}

\def\msun{{\Msun}}

\def\HH{${\rm {H_2}}\,\,$}

\def\HI{\hbox{H~$\scriptstyle\rm I\ $}} 
\def\HII{\hbox{H~$\scriptstyle\rm II\, $}}

\def\HII{{\rm HII\,}}

\def\gsim{\lower.5ex\hbox{\gtsima}} 
\def\lsim{\lower.5ex\hbox{\ltsima}} \def\gtsima{$\; \buildrel > \over \sim \;$} \def\ltsima{$\; \buildrel < \over \sim \;$} \def\prosima{$\; 
\buildrel \propto \over \sim \;$} \def\gsim{\lower.5ex\hbox{\gtsima}} 
\def\lsim{\lower.5ex\hbox{\ltsima}} 
\def\simgt{\lower.5ex\hbox{\gtsima}} 
\def\simlt{\lower.5ex\hbox{\ltsima}} 
\def\simpr{\lower.5ex\hbox{\prosima}}   
  
 \def\gtsima{$\; \buildrel > \over \sim \;$} 
\def\ltsima{$\; \buildrel < \over \sim \;$} 
\def\gsim{\lower.5ex\hbox{\gtsima}} 
\def\lsim{\lower.5ex\hbox{\ltsima}} 
\def\simgt{\lower.5ex\hbox{\gtsima}} 
\def\simlt{\lower.5ex\hbox{\ltsima}} 
\def\simpr{\lower.5ex\hbox{\prosima}}

\def\msun{\,{\rm \Msun}}

\def\E3{{\cal E}_{\rm g}^{III}}

\def\Msun{\rm M_\odot}
\def\Zsun{\rm Z_\odot}

\def\Msun{\rm M_\odot}
\def\Zsun{\rm Z_\odot}
\def\M*{M_*}
\def\Z*{Z_*}
\def\L*{L_*}
\def\muv{\rm M_{UV}}

\def\c4{\rm CIV}
\def \muv{\rm M_{UV}}

\begin{document}

\begin{frontmatter}

\title{Early galaxy formation and its large-scale effects}

\author{Pratika Dayal\fnref{myfootnote1}}
\address{Kapteyn Astronomical Institute, Rijksuniversiteit Groningen, Landleven 12, Groningen, 9717 AD, The Netherlands}
\fntext[myfootnote1]{p.dayal@rug.nl}

\author{Andrea Ferrara\fnref{myfootnote2}}
\address{Scuola Normale Superiore, Piazza dei Cavalieri 7, Pisa, 56126, Italy}
\fntext[myfootnote2]{andrea.ferrara@sns.it}

\begin{abstract}
Galaxy formation is at the heart of our understanding of cosmic evolution. Although there is a consensus that galaxies emerged from the expanding matter background by gravitational instability of primordial fluctuations, a number of additional physical processes must be understood and implemented in theoretical models before these can be reliably used to interpret observations. In parallel, the astonishing recent progresses made in detecting galaxies that formed only a few hundreds of million years after the Big Bang is pushing the quest for more sophisticated and detailed studies of early structures.  In this review, we combine the information gleaned from different theoretical models/studies to build a coherent picture of the Universe in its early stages which includes the physics of galaxy formation along with the impact that early structures had on large-scale processes as cosmic reionization and metal enrichment of the intergalactic medium. 
\end{abstract}


\end{frontmatter}

\tableofcontents

\newpage
\newpage
\section*{List of acronyms}
 \begin{center}
    \begin{tabular}[h]{cccccc}\hline
      {\bf Acronym} & & & & & {\bf Extended name} \\
      \hline
      \hline

{AGB} & & & & & Asymptotic Giant Branch \\      
{AGN} & & & & & Active Galactic Nuclei \\
{ALMA} & & & & & Atacama Large Millimeter Array \\
{AMR} & & & & & Adaptive Mesh Refinement \\

{BH} & & & & & Black Holes \\
{BBN} & & & & & Big Bang Nucelosynthesis \\

{($\Lambda$)CDM} & & & & & (Lambda) Cold Dark Matter \\
{CGM} & & & & & Circum-galactic medium \\
{CMB} & & & & & Cosmic Microwave Background \\
{COBE} & & & & & Cosmic Background Explorer\\

{DLAs} & & & & & Damped Lyman Alpha systems \\
{DM} & & & & & Dark Matter \\

{E-ELT} & & & & & European Extremely Large Telescope \\
{EoR} & & & & & Epoch of Reionization \\

{GMCs} & & & & & Giant molecular Clouds \\
{GRBs} & & & & & Gamma Ray Bursts \\

{HERA} & & & & & Hydrogen Epoch of Reionization Array \\
{HMF} & & & & & Halo Mass Function \\
{HST} & & & & & Hubble Space Telescope \\

{IGM} & & & & & Inter-galactic medium \\
{IMF} & & & & & Initial Mass Function \\
{ISM} & & & & & Inter-stellar medium \\

{JWST} & & & & & James Webb Space Telescope \\

{LABs} & & & & & Lyman Alpha Blobs \\
{LAEs} & & & & & Lyman Alpha Emitters \\
{LBGs} & & & & & Lyman Break Galaxies \\
{LCGs} & & & & & Lyman continuum emitting galaxies \\
{LyC} & & & & & Lyman continuum  \\
{LF} & & & & & Luminosity function \\
{LLS} & & & & & Lyman limit systems \\
{Lofar} & & & & & Low Frequency Array \\
{LW} & & & & & Lyman Werner \\

{MW} & & & & & Milky Way \\

\end{tabular}
  \end{center}

\newpage 

 \begin{center}
    \begin{tabular}[h]{cccccc}\hline
      {\bf acronym} & & & & & {\bf extended name} \\
      \hline
      \hline
      
{PISN} & & & & & Pair Instability Supernovae \\
{PopIII} stars & & & & & Population III (metal-free) stars \\

{QSO} & & & & & Quasi-stellar object \\
      
{r.m.s.} & & & & & Root mean square \\

{SAM} & & & & & Semi-analytic models \\
{SFRD} & & & & & Star formation rate density \\
{SKA} & & & & & Square Kilometre Array \\
{SMBH} & & & & & Super massive black holes \\
{SMD} & & & & & Stellar mass density \\
{SN} & & & & & Supernova \\
{SNII} & & & & & TypeII Supernova \\
{SPH} & & & & & Smoothed-particle hydrodynamics \\

{UV} & & & & & Ultra-violet \\
{UVB} & & & & & Ultra-violet background \\
{UV LF} & & & & & Ultra-violet luminosity function\\

{VLT} & & & & & Very Large Telescope\\

{WDM} & & & & & Warm Dark Matter \\
{WMAP} & & & & & Wilkinson Microwave Anisotropy Probe

\end{tabular}
  \end{center}

\section*{List of symbols}
 \begin{center}
    \begin{tabular}[h]{ccc}\hline
      {\bf symbol} &  {\bf Definition } \\
      \hline
      \hline

$a(t)\,  {\rm or} \,a(z)$ &  Scale factor of the Universe at time $t$ or redshift $z$ \\  
$c$ &  Speed of light  ($3 \times 10^{10} \, {\rm cm \, s^{-1}}$) \\
$\chi_{HI}$ &  Neutral hydrogen fraction   \\

$f_{esc}$ &  Escape fraction of \HI ionizing photons from the galaxy \\
$G $ &  Gravitational constant ($6.67 \times 10^{-8}\, {\rm cm^3 \, gm^{-1} \, s^{-2}}$) \\
${\rm h} $ &  Planck's constant ($6.67 \times 10^{-27}\, {\rm cm^2 \, gm \, s^{-1}}$) \\
$k_B $ &  Boltzmann's constant  ($1.38 \times 10^{-16}\, {\rm cm^2 \, gm \, s^{-2}}\, K^{-1}$) \\
$H_0 $ &  Hubble constant ($100 h\, {\rm km\, s^{-1} \, Mpc^{-1}}$) \\
$m_p$ &  Proton mass ($1.6726 \times 10^{-24}\, {\rm gm}$)  \\
$m_H$ &  Hydrogen mass ($1.6737 \times 10^{-24}\, {\rm gm}$)  \\
$\mu$ &  Mean molecular weight \\
$\dot N_s$ & Production rate of LyC photons \\
$\Omega_b$ &  Density parameter for baryons \\
$\Omega_c$ &  Density parameter cold dark matter \\
$\Omega_m$ &  Density parameter for total matter \\
$\Omega_\Lambda$ &  Density parameter for dark energy \\
$\rho_{crit}$ &  Critical density of the Universe \\
$R_{vir}$ &  Virial radius of halo of mass $M_h$\\
$\sigma_8 $ &  r.m.s. fluctuations on scales of $8 h^{-1}$Mpc\\
$\tau$ &  CMB electron scattering optical depth \\
$T_{vir}$ &  Virial temperature of halo of mass $M_h$\\
$T_{\gamma}$ & CMB temperature\\
$V_{vir}$ &  Virial temperature of halo of mass $M_h$\\
$z_{re}$ &  Reionization redshift \\

 \end{tabular}
  \end{center}


\newpage


\section{Galaxy formation: a brief historical perspective}
The modern history of galaxy formation theory began immediately after the second World War. In fact, the mere term ``galaxy'' came into play only after the pivotal discovery, by Hubble in 1925, that the objects until then known as ``nebulae'', since their original discovery by Messier towards the end of the ${\rm 18^{th}}$ century, were indeed of extra-galactic origin. It was then quickly realised that these systems were germane to understanding our own galaxy, the Milky Way (MW). The years immediately after blossomed with key cosmological discoveries: in 1929 Hubble completed his study to prove cosmic expansion and, soon after, in 1933, Zwicky suggested that the dominant fraction of mass constituting clusters of galaxies was unseen, laying the ground for the concept of ``Dark Matter'' (DM). As early as 1934, one of the first modelling attempts \citep{stromberg1934} proposed galaxies to have formed out of primordial gas whose condensation process could be followed by applying viscous hydrodynamic equations. Similar pioneering attempts were made by \citet{Weizscker1951} who proposed that galaxies fragmented out of turbulent, expanding primordial gas, although this suggestion was unaware of the later finding that vorticity modes in the early Universe decay rapidly in the linear regime. In addition, \citet{Hoyle1951} had already pointed out that the spin of a proto-galaxy could arise from the tidal field of its neighbouring structures. A few years later, \citet{Hoyle1953} produced the first results emphasising the role of gas radiative cooling, fragmentation and the Jeans length (the scale just stable against gravitational collapse) during the collapse of a primordial cloud. These were used to explain the observed masses of galaxies and the presence of galaxy clusters. The limitation of these early models, that attempted to discuss the formation of galaxies independently of a cosmological framework, however, was that a number of ad-hoc assumptions had to be made. 

Soon after, \citet{Sciama1955} reiterated the point already made by \citet{Gamow1953}, while working on his proposal of an initially hot and dense initial phase of the Universe (``Hot Big Bang'') leading to primordial nucleosynthesis, that the galaxy formation problem must be tied to the cosmological one. As a result of the initially hot state of the Universe, Gamow also predicted the existence of a background of thermal radiation (now known as the Cosmic Microwave Background; CMB) with a temperature of a few degrees Kelvin. As at that time the steady-state model proposed by Bondi, Gold \& Hoyle (for an overview see, e.g. \citet{Bondi1952}), in which the continual creation of matter enabled the large-scale structure of the Universe to be independent of time despite its expansion, was still under serious consideration, Sciama applied the previous ideas of galaxy formation in that context. His idea was that the gravitational fields of pre-existing galaxies could produce density concentrations in the intergalactic gas resulting in the formation of new galaxies due to gravitational instability. Other ideas, that were sometimes subsequently rejuvenated, were also circulating. The most seminal of these was the ``explosive galaxy formation'' scenario put forward by \citet{Vororitsov1961} in which galaxies could be torn apart by powerful ``repulsive'' forces and broken into sub-units. A more modern version of this statement would use the energy deposition by massive stars and, possibly, massive black holes (via processes collectively referred to as ``feedback'') to trigger a propagating wave of galaxy formation (see e.g. \citet{Ostriker1981}).   

The situation changed dramatically with the experimental discovery of the CMB by Penzias \& Wilson in 1965 \citep{penzias-wilson1965}. \citet{Peebles1965} immediately realised that the Jeans scale after the matter-radiation decoupling (i.e. when the hydrogen became neutral), in a Universe filled with black-body radiation, would be close to the mass of present-day globular clusters ($\approx 10^6 M_\odot$). This  remarkable result inextricably tied cosmic evolution to galaxy formation. However, \citet{Harrison1967} emphasised a problem related to the growth of structure, namely that if galaxies have to emerge out of amorphous initial conditions via a gravitational or thermal instability process, the rate of growth of the perturbations should be short compared to the cosmic age. However, due to Hubble expansion, the growth rate is only a power-law rather than an exponential as in classic theory. Harrison then incorrectly concluded, partly due to the then-sketchy knowledge of the cosmological parameters and the role of dark matter, that this condition was not met. Thus, he favoured the ``primordial structure hypothesis'' which presupposes that structural differentiation originates with the Universe. In modern language this is equivalent to requiring a very large amplitude of the primordial fluctuation field, later clearly disfavoured by the measurements of CMB anisotropies. A different prediction was made by \citet{Silk1968} who instead suggested a much larger ($\approx 10^{11} M_\odot$) critical mass of the first galaxies. This argument was based on his discovery of the ``Silk damping'' effect that predicts fluctuations below a critical mass to be optically thick and damped out by radiative diffusion on a short time scale as compared to the Hubble expansion time. This proposal, and the contrasting one by Peebles, set the stage for the two scenarios of galaxy formation that survived until the end of the last century. These are the ``top-down'' scenario, in which small galaxies are formed by fragmentation of larger units, and the ``bottom-up'' scenario, where large galaxies are hierarchically assembled from smaller systems.

Enormous theoretical efforts were undertaken in the 1970's to discriminate between these two contrasting paradigms of galaxy formation. The idea that galaxies emerged out of fluctuations of the primordial density field was gradually becoming more accepted (notwithstanding the unknown origin and amplitude of the perturbations) after the seminal works by \citet{Harrison1970} and \citet{Zeldovich1972} who showed that only a specific, scale-invariant form of the fluctuation power spectrum would be compatible with the development of such fluctuations into proto-galaxies. The next urgent step was then to clarify the details of the non-linear stages of the growth of these fluctuations. \citet{Zeldovich1970} obtained an approximate solution, valid for large scale perturbations and pressure-less matter. In this solution, matter would collapse into a disk-like structure (popularly known as a ``Zeldovich pancake'') before developing a complex shock structure and eventually fragmenting into lower-mass structures. This large-scale approximation seemed appropriate given the Silk damping of small scale perturbations. Moving in a similar direction, \citet{Gunn1972} laid down the basic formalism describing the nonlinear collapse of a perturbation of arbitrary scale embedded in the expanding cosmic flow. They showed that after an initial expansion phase, the radius of the (spherical) perturbation would ``turn around", reverting the motion into a contracting one. These results ultimately clarified the key importance of mass accretion onto the initial seed perturbations to form fully-fledged galaxies. This study was nicely complemented by the work of \citet{press-sch1974} who developed a simple, yet powerful, method to compute the mass distribution function of collapsed objects from an initial density field made up of random gaussian fluctuations. This paper resulted in strong support for the bottom-up scenario by successfully postulating that once the condensation process has proceeded through several scales, the mass spectrum of condensations becomes ``self-similar" and independent of the initially assumed spectrum. Hence, larger mass objects form from the nonlinear interaction of smaller masses. This ``Press-Schechter formalism" has become the underlying paradigm of galaxy formation theory and we will devote a later section (Sec. \ref{dm_assembly}) of this review to highlight its main features. Since this result was derived by analysing the statistical behaviour of a system of purely self-gravitating particles, that in modern terms we would classify as dark matter, the theory fell short of the description of baryonic physics. The latter represents a key ingredient in predicting dissipative processes, star formation and ultimately the actual appearance of observed galaxies. This was pointed out by \citet{Larson1974} who worked out a model representing the collapse of an initially gaseous proto-galaxy and the subsequent transformation of gas into stars. In this study, Larson also pioneered the use of numerical simulations that he applied to the solution of the relevant fluid-dynamic equations describing the collapse.  

The time was then ripe for the production of the milestone paper by \citet{White1978}, built on arguments proposed by a number of other works including \citet{Gott1976, Silk1977, binney1977} and  \citet{rees1977}. The basic idea was to combine the Press-Schechter theory of gravitational clustering with gas dynamics following the earlier proposals by, e.g., \citet{Larson1974}. Dark matter, whose existence in clusters, such as Coma, was becoming accepted, was finally self-consistently included in galaxy formation theories. In practice, the \citet{White1978} model is a two-staged one: dark matter first condenses into halos via pure gravitational collapse after which baryons collapse into the pre-existing potential wells, dissipating their energy via gas-dynamical processes. As a result, the final galaxy properties are partly determined by the assembly of their parent dark matter halo. Once simple, semi-empirical, prescriptions for star formation are implemented one can simultaneously compute statistical properties, like the co-moving number density of galaxies as a function of halo mass/luminosity/stellar mass and redshift, along with their structural/morphological properties (including the gas/stellar mass/star formation rate and sizes). Although admittedly simplified, the model by \citet{White1978} provided a first prediction of the luminosity function (the number density of galaxies in a given luminosity bin) that was roughly consistent with the scarce data available at that time. This class of two-staged galaxy formation models, originating from the above cited original works, has become known as ``semi-analytical models'' (SAMs). Due to their simplicity and flexibility, although present-day descendants have acquired a remarkable degree of complexity, SAMs are a standard tool in the galaxy formation field. Their more technical features are briefly discussed in Sec. \ref{theo_tools}.   

At approximately the same time, the idea that galaxies are embedded in a dynamically-dominant dark matter halo received decisive support from the observations led by Vera Rubin, starting with the fundamental study presented in \citet{Rubin1978}. Measuring the rotation curves of 10 spiral galaxies using the Balmer-alpha (or H$\alpha$) line at 6562 \AA\, in the galaxy rest-frame Rubin and her collaborators reached the surprising conclusion that all
rotation curves were approximately flat out to distances as large as $r = 50$ kpc rather than showing the expected Keplerian decline.  They also noted that the maximal velocity was not correlated with the galaxy luminosity. Rather, it was a measure of the total mass and radius, clearly indicating the need for a dark matter halo to explain the observed rotation curves.  

However, one crucial element, necessary for completing a coherent framework for galaxy formation, was still missing. This was a precise knowledge of the primordial density fluctuation power spectrum. Indeed, the powerful two-stage approach by \citet{White1978} was still plagued by this limitation. In their study, these authors assumed a simple power-law dependence of the root mean square (r.m.s.) amplitude of the perturbation on mass as $\sigma \propto M^{-\alpha}$, with $\alpha$ essentially being an unknown parameter. Fortunately, soon after \citet{Guth1981} and \citet{Sato1981} independently proposed the basic ideas of inflation \citep[for a review see, e.g.,][]{mazumdar2010}. They noted that, despite the (assumed) highly homogeneous state of the Universe immediately after the Big Bang, regions separated by more than 1.8 degrees on the sky today should never have been causally connected (the horizon problem). In addition, the initial value of the Hubble constant must be fine-tuned to an extraordinary accuracy to produce a Universe as flat as the one we see today (flatness problem). According to the  original proposal (also known as ``old inflation'') de-Sitter inflation occurred as a first-order transition to true vacuum. However, a key flaw in this model was that the Universe would become inhomogeneous by bubble collisions soon after the end of inflation. To overcome the problem, \citet{Linde1982} and \citet{Albrecht1982}, instead, proposed the concept of slow-roll inflation with a second-order transition to true vacuum. Unfortunately, this scenario also suffers from the fine-tuning problem of the time required in a false vacuum to lead to a sufficient amount of inflation. \citet{Linde1983} later considered a variant version of the slow-roll inflation called chaotic inflation, in which the initial conditions of scalar fields are chaotic. Chaotic inflation has the advantage of not requiring an initial thermal equilibrium state. Rather, it can start out close to the Planck density, thereby solving the problem of the initial conditions faced by other inflationary models.  

Most importantly, in the context of galaxy formation, inflation offered precise predictions for the origin of the primordial fluctuations and their power spectrum. \citet{Bardeen1983} showed that fluctuations in the scalar field driving inflation (the ``inflaton'') are created on quantum scales and expanded to large scales, eventually giving rise to an almost scale-free spectrum of gaussian adiabatic density perturbations. This is the Harrison-Zeldovich spectrum where the initial density perturbations can be expressed as $P(k) \propto k^{n_s}$ with the spectral index $n_s \approx 1$.  Quantum fluctuations are typically frozen by the accelerating expansion when the scales of fluctuations leave the Hubble radius. Long after inflation ends, the scales cross and enter the Hubble radius again. Indeed, the perturbations imprinted on the Hubble patch during inflation are thought to be the origin of the large-scale structure in the Universe. In fact temperature anisotropies, first observed by the Cosmic Background Explorer ({\it COBE}) satellite in 1992, exhibit a nearly scale-invariant spectrum as predicted by the inflationary paradigm - these have been confirmed to a high degree of precision by the subsequent Wilkinson Microwave Anisotropy Probe ({\it WMAP}) and {\it Planck} experiments. As $\sigma \propto M^{-(3+n_s)/6}$ for the inflationary spectrum, the value of $\alpha$ in the \citet{White1978} theory could now be uniquely determined.

The final step was to compute the linear evolution of the primordial density field up to the recombination epoch. The function describing the modification of the initial spectrum, due to the differential growth of perturbations on different scales and at different epochs, is called the ``transfer function'' and was computed for dark matter models made by massive (GeV-TeV) particles, collectively known as Cold Dark Matter (CDM) models, by \citet{Peebles1982, Bardeen1986} and later improved by the addition of baryons by \citet{Sugiyama1995}.

Based on the above theoretical advances and exploiting the early availability of computers, it was possible to envision, for the first time, the possibility of simulating, {\it ab-initio}, the process of structure formation and evolution through cosmic time. This second class of galaxy formation models is referred to as ``cosmological simulations'' which is another major avenue of research in the field today. More technical details will be given in a following dedicated section (Sec. \ref{theo_tools}). Here, it suffices to stress how these theoretical and technological advances have revolutionised our view of how galaxies form. In the first attempt, \citet{Navarro1991} simulated the dynamical evolution of both collisionless dark matter particles and a dissipative baryonic component in a flat universe in order to investigate the formation process of the luminous components of galaxies at the center of galactic dark halos. They assumed gaussian initial density fluctuations with a power spectrum of the form $P(k) \propto k^{-1}$ meant to reproduce the slope of the power spectrum in the range of scales relevant for galaxies. Although the available computing power for their smoothed-particle hydrodynamic (SPH scheme; for details see Sec. \ref{theo_tools})  allowed only the outrageously small number of about 7000 particles (compared to the tens of billions used in modern computations) the emergence of the ``cosmic web'', made up of 
filaments, knots and voids, was evident. At the same time, their study represented a change of paradigm marking a transition from the early idealised, loosely informed spherical models to a physically-motivated realization of cosmic structure. A similar effort was also carried out by \citet{Katz1991}, who introduced key technical improvements, such as TREESPH, in which gravitational forces are computed using a hierarchical tree algorithm. The adaptive properties of TREESPH were ideal to solve collapse-type problems. 

The amazing progress made in the past century has paved the road for the current research in cosmic structure and galaxy formation whose most modern incarnation we aim at presenting in this review. Although the general scenario has now been established, many fundamental problems remain open, in particular those concerning the formation of the first galaxies. This area represents the current frontier of our knowledge. As such it will be central for the next decade with the advent of many observational facilities that will allow us to peer into these most remote epochs of the Universe.

\newpage

\section{Cosmic evolution in a nutshell}
\label{ch2}

Our understanding of the Universe has improved remarkably over the last two decades. In addition to strengthening and refining the foundations of the Big Bang standard model, including the Hubble expansion, the CMB and the abundance of light elements, cosmology has given us several genuine, irrefutable surprises. The most prominent among these are that {\it(i)} we live in a flat Universe, corroborating predictions of inflationary theories; {\it (ii)} roughly 85.3\% of cosmic matter is constituted by some kind of, as yet unknown, dark matter particles; {\it (iii)} the Hubble expansion is accelerating, possibly due to a non-clustering, negative-pressure fluid called dark energy; {\it (iv)} black holes a billion times the mass of the Sun were already in place 1 billion years after the Big Bang; and {\it (v)} distant galaxies, and products of their stellar activity, including Gamma-Ray Bursts (GRBs), have recently been detected out to redshifts $z >11$, corresponding to a cosmic age shorter than 0.42 Gyr. Do we have a complete, exact theory to explain these puzzling experimental evidences? Unfortunately, not yet. Gaps remain in the basic cosmological scenario and admittedly some of them are rather large.

It is now widely accepted that the Universe underwent an inflationary phase early on that sourced the nearly scale-invariant primordial density perturbations that have resulted in the large-scale structure we observe today. However, inflation requires non-standard physics and, as of now, there is no consensus on the mechanism that made the Universe inflate; moreover, observations allow only few constraints on the numerous inflationary models  available. This inflationary expansion forced matter to cool and pass through a number of symmetry-breaking phase transitions (grand unification theory, electroweak, hadrosynthesis, nucleosynthesis) until recombination when electrons and protons finally found it energetically favourable to combine into hydrogen (or helium) atoms. Due to the large value of the cosmic photon-to-baryon ratio, $\eta\simeq 10^9$, this process was delayed until the temperature dropped to 0.29 eV, about 50 times lower than the binding energy of hydrogen. Cosmic recombination proceeded far out of equilibrium because of a ``bottleneck" at the $n = 2$ level of hydrogen since atoms can only reach the ground state via slow processes including the two-photon decay or Lyman-alpha resonance escape. As the recombination rate rapidly became smaller than the expansion rate, the reaction could not reach completion and a small relic abundance of free electrons was left behind at $z < 1078$.
 
\begin{figure*}[h!]
\center{\includegraphics[scale=0.6]{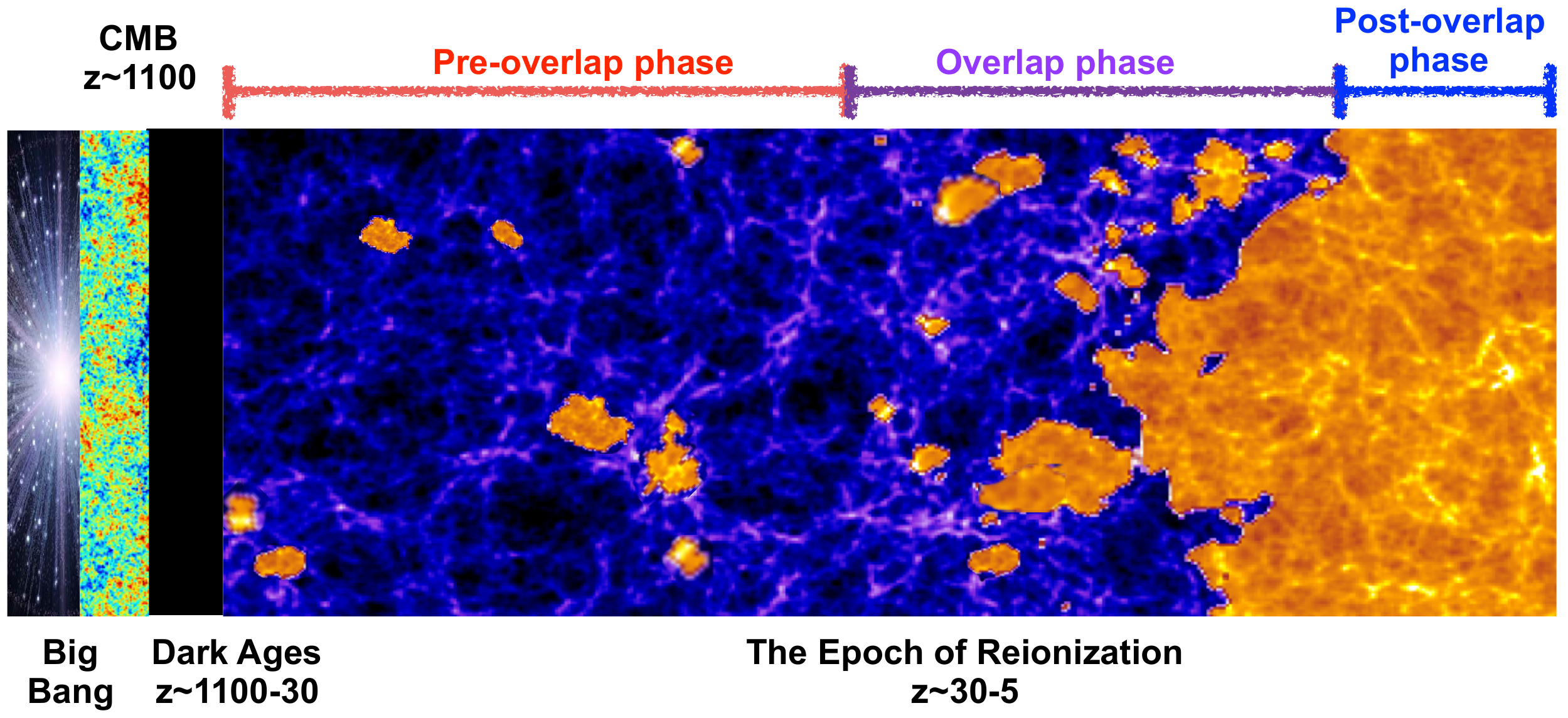}}
\vspace{-0.75cm}
\caption{A timeline of the first billion years of the Universe. According to our current understanding, immediately after its inception in the Big Bang, the Universe underwent a period of accelerated expansion (``inflation") after which it cooled adiabatically. At a redshift $z \sim 1100$, the temperature dropped to about 0.29 eV at which point matter and radiation decoupled (``decoupling") giving rise to the CMB and electrons and protons recombined to form hydrogen and helium (``recombination"). This was followed by the cosmic ``Dark Ages" when no significant radiation sources existed. These cosmic dark ages ended with the formation of the first stars (at $z \lsim 30$). These first stars started producing the first photons that could reionize hydrogen into electrons and protons, starting the ``Epoch of cosmic Reionization" which had three main stages: the ``pre-overlap phase" where each source produced an ionized region around itself, the ``overlap  phase" when nearby ionized regions started overlapping and the ``post-overlap phase" when the IGM was effectively completely ionized. (Reionization simulation credit: Dr. Anne Hutter).}
\label{fig_timeline} 
\end{figure*}

Immediately after the recombination epoch, the Universe entered a phase called the {\it Dark Ages}, where no significant radiation sources existed, as shown in Fig. \ref{fig_timeline}. The hydrogen remained largely neutral at this stage. The small inhomogeneities in the dark matter density field present during the recombination epoch started growing via gravitational instability giving rise to highly nonlinear structures, i.e., collapsed haloes (Sec. \ref{ch3}). It should, however, be kept in mind that most of the baryons at high redshifts do not reside within these haloes - they are rather found as diffuse gas in the Intergalactic Medium (IGM).

These collapsed haloes formed potential wells, whose depth depends on their mass and, within which baryons could fall. If the mass of the halo was high enough (i.e. the potential well was deep enough), the gas would be able to dissipate its energy, cool via atomic or molecular transitions and fragment within the halo (Sec. \ref{ch4}). This produced conditions appropriate for the condensation of gas and the formation of stars in galaxies. Once these luminous objects started forming, the Dark Ages were over. The first population of luminous stars and galaxies generated ultraviolet (UV) radiation through nuclear reactions. In addition to galaxies, perhaps an early population of accreting black holes (quasars) and the decay or annihilation of dark matter particles also generated some amount of UV light. The UV photons with energies $> 13.6$ eV were then able to ionize hydrogen atoms in the surrounding IGM, a process known as ``cosmic reionization" (Sec. \ref{ch7}). Reionization is thus the second major change in the ionization state of hydrogen (and helium) in the Universe (the first being recombination).

Cosmic reionization, like an Ariadne's thread, connects many of these gaps. While its evolution is shaped by the matter-energy content and geometry of the Universe, it reflects the way in which galaxies and black holes formed, affects the visibility of distant objects and modifies the properties of the CMB, to mention a few aspects. Disentangling the complicated physical interplay amongst these factors holds the key to filling some of the aforementioned gaps remaining in the standard cosmological model. 

According to our current understanding, reionization started around the time when the first structures formed which is currently believed to be around $z\approx 30$. In the simplest picture, shown in Fig \ref{fig_timeline}, each source first produced an ionized region around itself (the ``pre-overlap" phase) which overlapped and percolated into the IGM during the ``overlap" phase. The process of overlapping seems to have been completed around $z = 6-8$ at which point the neutral hydrogen fraction, expressed as $\chi_{HI}(z) = n_{HI}(z)/n_{H}(z)$ where $n_{H}$ and $n_{HI}$ are the densities of hydrogen and neutral hydrogen respectively, fell to values of $\chi_{HI}< 10^{-4}$. Following that, a never-ending ``post-reionization" (or ``post-overlap") phase started which implies that the Universe is largely ionized at the present epoch. Reionization by UV radiation was also accompanied by heating: electrons released by photo-ionization deposited the photon energy in excess of 13.6 eV into the IGM. This IGM reheating could expel the gas and/or suppress cooling in low mass haloes possibly affecting the cosmic star formation during and after reionization. In addition, the nuclear reactions within the stellar sources potentially altered the chemical composition of the medium if the star exploded as a supernova (Sec. \ref{ch6}), significantly changing the star formation mode at later stages.

The Epoch of reionization (EoR) is of immense importance in the study of structure formation since, on the one hand, it is a direct consequence of the formation of first structures and luminous sources while, on the other, it affects subsequent structure formation. Observationally, the reionization era represents a phase of the Universe which is only starting to be probed. While the earlier phases are probed by the CMB, the post-reionization phase ($z < 6$) has been mapped out by a number of probes ranging from quasars to galaxies and other sources. In addition to the importance outlined above, the study of Dark Ages and cosmic reionization has acquired increasing significance over the last few years because of the enormous repository of data that is slowly being built-up (Sec. \ref{gal_prop}): over the last few years observations have increasingly pushed into the EoR with the number of high-redshift galaxies and quasars having increased dramatically. This has been made possible by a combination of state-of-the-art facilities such as the Hubble Space Telescope ({\it HST}), Subaru and Very Large telescopes ({\it VLT}) and refined selection techniques. In the latter category, the Lyman break technique, pioneered by \citet{steidel1996}, has been successfully employed to look for Lyman Break Galaxies (LBGs), that are up to three orders of magnitude fainter than the Milky Way, at $z \simeq 7$ \citep{ellis2012, atek2015}. Using the power afforded by lensing, such techniques have now detected viable galaxy candidates at redshifts as high as $z \simeq 11$, corresponding to only half a billion years after the Big Bang. Further, the narrow-band Lyman Alpha technique has been successfully used to look for Lyman Alpha Emitters (LAEs) and recently, the broad-band colours are being used to identify galaxies by means of their nebular emission \citep[e.g.][]{smit2014, zitrin2015}. Finally, some of the most distant spectroscopically confirmed cosmic objects are GRBs that establish star formation was already well under way at those early epochs, further encouraging deeper galaxy searches. These data sets are already being supplemented by those from facilities including the Atacama Large Millimeter Array ({\it ALMA}), the Low Frequency Array ({\it Lofar}) and the Hydrogen Epoch of Reionization Array ({\it HERA}). In the coming decade, cutting-edge facilities including the James Webb Space Telescope ({\it JWST}), the Square Kilometre Array ({\it SKA}) and the European Extremely Large Telescope ({\it E-ELT}) are expected to further provide unprecedented glimpses of the early Universe. This observational progress has naturally given rise to a plethora of theoretical models, ranging from analytic calculations to semi-analytic models to numerical simulations, often yielding conflicting results. In the following Sections we will try to summarise these discoveries and progresses in a useful manner. As conventional, we cite all magnitudes in the AB system \citep{oke-gunn1983} throughout this review.


\newpage

\section{The cosmic scaffolding: dark matter halos}
\label{ch3}

In this Section, we start by discussing the standard Lambda Cold Dark Matter ($\Lambda$CDM) cosmological model, which forms the backdrop for galaxy formation, in more detail. We then discuss the collapse of halos in the context of linear perturbation theory before progressing to the non-linear collapse and assembly of the first dark matter halos via merger events and accretion. 

\subsection{The standard cosmological model}
\label{cosm_model}
As discussed in Sec. \ref{ch2}, the Hot Big Bang scenario, the prevailing paradigm describing the Universe from its earliest stages to the present day, envisages it to have started in a hot, dense phase with the energy density largely dominated by radiation down to redshift $z \simeq 10^4$. At $z \simeq 1100$, for the first time, the temperature of the Universe fell to $T \simeq 3000$ K, making it energetically favourable for electrons and protons to recombine into neutral hydrogen (\HI) atoms. This Epoch of Recombination was imminently followed by the decoupling of the matter and radiation fluids, allowing photons to free-stream to us from the last scattering surface. Such photons are currently observed as the CMB. Indeed, along with the CMB measurements, the homogeneous \citep[e.g.][]{maartens2011} and isotropic \citep[e.g.][]{saadeh2016} expansion of the Universe and the confirmation of the abundance of light elements predicted by Big Bang Nucleosynthesis \citep[BBN;][]{alpher1948} form the three pillars of the Hot Big Bang model. This homogeneous, isotropic and expanding Universe is described by the maximally-symmetric Robertson-Walker (RW) metric: 
\begin{equation}
ds^2 = c^2dt^2 - a^2(t) \bigg\{ \frac{dR^2}{1-k R^2} + R^2 d\theta^2 + R^2 sin^2 \theta d\phi^2 \bigg\},
\end{equation} 
where $c$ is the speed of light, $R, r, \theta$ and $\phi$ are co-moving co-ordinates and $a(t)$ is the cosmic scale factor. Finally, $k$ represents the curvature of the Universe which is positive, zero and negative for a closed, flat and  open Universe, respectively.

\begin{figure*}[h!]
\center{\includegraphics[scale=0.82]{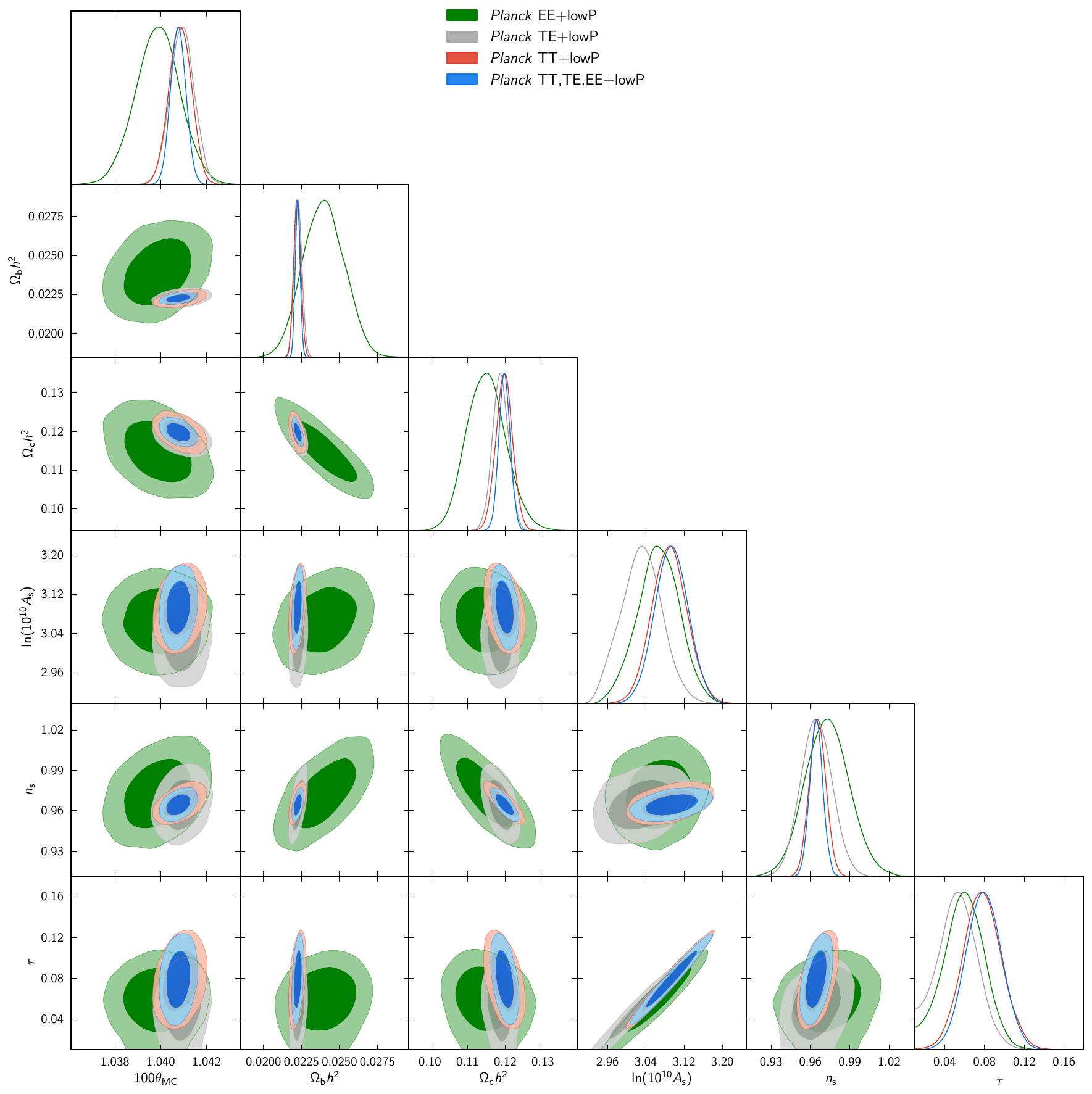}}
\caption{The base [$\Omega_b h^2$, $\Omega_c h^2$, $100 \theta_{MC}$, $n_s$, $\tau$ and ${\rm ln}(10^{10} A_s)$] and derived [$h$, $\sigma_8$, $\Omega_m$ and $\Omega_\Lambda$] $\Lambda$CDM parameter constraints from {\it Planck} data \citep{planck2015}. As marked, the green, gray, red and blue contours show parameter values derived using EE and low $l$ polarization data, TE and low $l$ polarization data, TT and low $l$ polarization data and combining TT, TE, EE and low $l$ polarization data, respectively, with low $l$ referring to modes with $l \leq 29$; the dark and shaded regions represent the $1$ and $2-\sigma$ contours, respectively. See text in Sec. \ref{cosm_model} for details.}
\label{ch1_fig_planck} 
\end{figure*}

\begin{figure*}
\center{\includegraphics[scale=1.75]{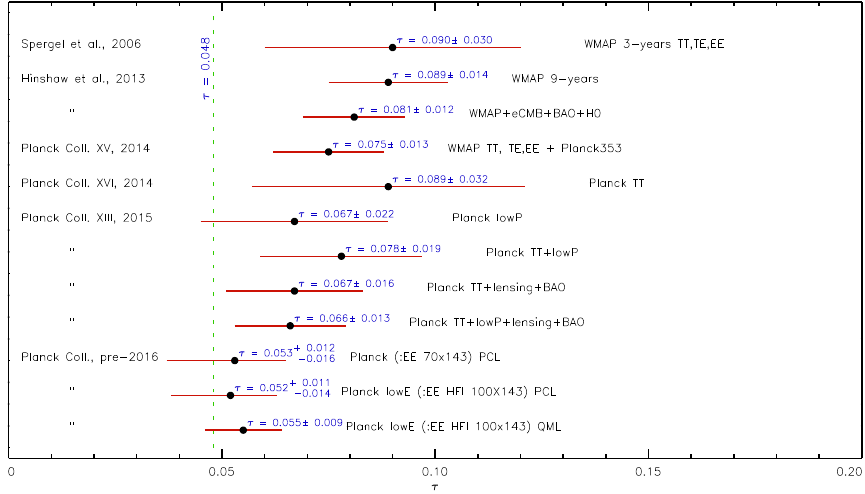}}
\vspace{-0.2cm}
\caption{The evolution of the electron scattering optical depth, $\tau$, over the past 12 years \citep{planck2016}, inferred from CMB data from the {\it WMAP} and {\it Planck} results as marked. Although not plotted here, the initial {\it WMAP} 1-year results indicated a very high value of $\tau = 0.17\pm 0.04$ resulting in a reionization redshift of $z_{re} = 20^{+10}_{-9}$ \citep{bennett2003}. Since then, the value of 
$\tau$ has shown a consistent decrease with the (much lower) final value being $0.055\pm 0.009$ yielding a reionization redshift of $z_{re} \sim 7.8-8.8$. }
\label{ch1_fig_tau_year} 
\end{figure*}

Since its accidental discovery, resulting in the award of the 1978 Nobel prize, by Penzias and Wilson in 1965 \citep{penzias-wilson1965}, the CMB has been studied in exquisite detail by a number of observatories including {\it BOOMERanG} (Balloon Observations of Millimetric Extragalactic Radiation and Geomagnetics), {\it MAXIMA} (Millimeter Anisotropy eXperiment IMaging Array), DASI (Degree Angular Scale Interferometer), {\it WMAP} and, most recently, the {\it Planck} satellite. These advances in observational cosmology have led to the establishment of a ``concordance" $\Lambda$CDM model that specifies the three key components of the Universe: {\it (i)} Dark Energy ($\Lambda$); {\it (ii)} Cold Dark Matter (CDM) that is now thought to provide the cosmic scaffolding for the entire cosmic web; and {\it (iii)} ordinary ``baryonic'' matter. This cosmological model is fully defined once the 6 basic parameters are derived from CMB data, as shown in Fig. \ref{ch1_fig_planck}. These include: $\Omega_b h^2$ specifying the baryonic density, $\Omega_c h^2$ specifying the CDM density, $100 \theta_{MC}$ measuring the sound horizon at last scattering, $\tau$ measuring the electron scattering optical depth and $n_s$ and ${\rm ln}(10^{10} A_s)$ specifying the spectral index and amplitude of the initial density perturbations, respectively; $n_s=1$ implies the scale-invariant Harrison-Z'eldovich spectrum as noted. This model also includes 4 additional derived parameters: the Hubble parameter ($h$), the root mean square mass fluctuations on scales of $8 h^{-1}\, {\rm Mpc}$ ($\sigma_8$), the total matter density ($\Omega_m$) and the adimensional density of Dark Energy ($\Omega_\Lambda$). Combining the temperature-temperature (TT) spectra, temperature-E mode polarization (TE) spectra and the E mode-E mode polarization (EE) spectra, the best fit values of these parameters are found to be $\Omega_b h^2 = 0.02225 \pm 0.00016$, $\Omega_c h^2 = 0.1198 \pm 0.0015$,  $100 \theta_{MC} = 1.04077 \pm 0.00032$, $\tau = 0.079 \pm 0.017$, $n_s=0.9645 \pm 0.0049$, ${\rm ln}(10^{10} A_s) = 3.094 \pm 0.034$, $h = 0.6727 \pm 0.0066$, $\sigma_8 = 0.831 \pm 0.013$, $\Omega_m = 0.3156 \pm 0.0091$ and $\Omega_\Lambda = 0.6844 \pm 0.0091$ \citep{planck2015}. Combining temperature data and low multipole polarization data, the latest {\it Planck} results \citep{planck2016} yield a lower lower value of $\tau = 0.055 \pm 0.009$, which has important implications for reionization and its sources as discussed in Sec. \ref{sources_reio}. Given its import for reionization, we also show the evolution of $\tau$, over 12 years, from {\it WMAP} data that implied a reionization redshift of $z_{re} \sim 9-14$ till the latest {\it Planck} 2016 results implying a much lower value of $z_{re} \sim 7.8-8.8$ in Fig. \ref{ch1_fig_tau_year}. 

The $\Lambda$CDM model has been remarkably successful in predicting the large scale structure of the Universe, the temperature anisotropies measured by the CMB and the Lyman Alpha forest statistics \citep[e.g.][]{peebles1971, blumenthal1984, bond-szalay1983, cole2005, hinshaw2013, planck2014, slosar2013}. However, as recently reviewed by \citet{weinberg2013}, CDM exhibits a number of small scales problems including: {\it (i)} the observed lack of, both, theoretically predicted low-mass \citep[``the missing satellite problem'';] []{moore1999, klypin1999} and high-mass \citep[``too big to fail problem'';][]{boylan2011,boylan2012} satellites of the Milky Way; {\it (ii)} predicting dark matter halos that are too dense (cuspy) as compared to the observationally preferred constant density cores \citep[``the core-cusp problem'';][]{moore1999b, navarro1997, subramanian2000}, and {\it (iii)} facing difficulties in producing typical disks due to ongoing mergers down to $z \simeq 1$ \citep{wyse2001}. The limited success of baryonic feedback in solving these small scale problems \citep[e.g.][]{boylan2012, teyssier2013} has prompted questions regarding the nature of dark matter itself. One such alternative candidate is provided by Warm Dark Matter (WDM) with particle masses $m_x \sim \mathcal{O}$(keV) \citep[e.g.][]{blumenthal1984,bode2001}. In addition to its particle-physics motivated nature, the WDM model has been lent support by the observations of a 3.5 keV line from the Perseus cluster that might arise from the annihilation of light ($\sim 7$ keV) sterile neutrinos into photons \citep{bulbul2014,boyarsky2014,cappelluti2017}. However, other works \citep{maccio2012,schneider_wdm2014} caution that the power-suppression arising from WDM makes it incompatible with observations, leaving the field open to alternative models including fuzzy CDM consisting of ultra light $\mathcal{O}$(10$^{-22}$eV) boson or scalar particles \citep{hui2017,du2017}, self-interacting (1 MeV - 10 GeV) dark matter \citep{spergel2000, rocha2013,vogelsberger2014_wdm}, decaying dark matter \citep{wang2014} and interactive CDM where small-scale perturbations are suppressed due to dark matter and photon/neutrino interactions in the early Universe \citep[e.g.][]{boehm2001, dvorkin2014}. We, however, limit our discussion to the standard $\Lambda$CDM model for the purposes of this review.

\subsection{The start: linear perturbation theory} 
The growth of small perturbations under the action of gravity was first undertaken by Jeans \citep{jeans1902} and Lifshitz \citep{lifshitz1946}, with comprehensive derivations in the books by \citet{peebles1993} and \citet{mo2010}. In brief, the aim of linear perturbation theory is to understand the growth of small perturbations of a density contrast $\delta({\bf x},t) = \rho({\bf x},t)/\bar \rho(t) \ll 1$ at a given space and time, where $\bar \rho(t)$ represents the average background density at time $t$. Such growth is described by the linearized form of the hydrodynamical equations (mass and momentum conservation plus the Poisson equation) for a fluid in a gravitational field. These perturbations grow against the background matter distribution represented by an unperturbed medium of constant density and pressure with a null velocity field. Further making reasonable assumptions of homogeneity, isotropy and adiabatic perturbations, these equations can be combined to yield the time-evolution of the density contrast, in comoving co-ordinates, as

\begin{equation}
\frac{d^2 \delta({\bf x},t)}{dt^2} + 2 \frac{\dot a(t)}{a(t)} \frac{d \delta ({\bf x},t)}{dt} = \frac{c_s^2}{a(t)^2} \nabla^2 \delta({\bf x},t) + 4 \pi G \bar \rho(t) \delta({\bf x},t) ,
\label{pert_growth}
\end{equation}
where $c_s$ is the fluid sound speed and $G$ is the Gravitational constant. The physical interpretation of Eqn. \ref{pert_growth} is straightforward: the gravitationally-driven perturbation growth (second term on the RHS) is opposed by the pressure term (first term on RHS) as well as cosmological expansion (second term on LHS). Writing the density contrast in terms of a Fourier series, $\delta= \sum \delta_{\bf k}\, {\rm exp}\, i{\bf k}\cdot{\bf x}$, where ${\bf k}$ is the wave-vector, Eqn. \ref{pert_growth} can be written as
\begin{equation}
\frac{d^2 \delta_{\bf k}}{dt^2} + 2 H(t) \frac{d \delta_{\bf k}}{dt} = \delta_{\bf k} \bigg( 4 \pi G \bar \rho(t) - \frac{k^2 c_s^2}{a(t)^2}  \bigg).
\label{pert_growth2}
\end{equation}
The source term (RHS) vanishes at a scale where the pressure gradient balances gravity resulting in the {\it Jeans wavelength}, the scale that is just stable against collapse, expressed as
\begin{equation}
\lambda_J = \frac{2 \pi a(t)}{k_J} = c_s \bigg(\frac{\pi}{G \bar \rho}\bigg)^{1/2}.
\end{equation}
The mass enclosed within a radius $\lambda_J/2$ is referred to as the {\it Jeans mass} which can be expressed as
\begin{equation}
M_J = \frac{4\pi}{3} \rho \bigg(\frac{\lambda_J}{2}\bigg)^3.
\end{equation}

On the other hand, at $\lambda \simgt \lambda_J$, the response time for pressure support is longer than the perturbation 
growth time,
\begin{equation}
t = \frac{1}{(4 \pi G \rho)^{1/2}},
\end{equation}
and collapse occurs. The power of the Jeans scale lies in the fact that it can be applied to bound structures ranging from galaxy clusters (with physical scale of a few Mpc) to star-forming molecular clouds (on scales of 10-100pc) in the interstellar media (ISM) of galaxies. 

These arguments can be extended to a time immediately after matter-radiation decoupling, at $z \approx 1100$, in order to obtain the Jeans length and mass. At this time, the sound speed can be approximated assuming a non-relativistic monoatomic gas such that
\begin{equation}
c_s^2 = \bigg( \frac{5 k_B T_{\gamma}}{3m_p} \bigg),
\end{equation}
where $k_B$ is the Boltzmann constant, $m_p$ is the proton mass and $T_{\gamma}$ is the CMB temperature. This results in a co-moving Jeans length \citep{mo2010}:
\begin{equation}
\lambda_J \approx 0.01 (\Omega_{b} h^2)^{-1/2} \, {\rm Mpc},
\end{equation}
where $\Omega_{b}$ is the baryonic density parameter at $z=0$, yielding a constant Jeans mass 
\begin{equation}
M_J \approx 1.5 \times 10^5 (\Omega_{b} h^2)^{-1/2} \, M_\odot,
\end{equation}
a value comparable to that of present-day globular clusters. However, we caution that the Jeans mass is not qualitatively accurate in the era of the first galaxies - for these the relevant mass scale is instead provided by the ``filtering mass" discussed in Sec. \ref{uvb_fb}.

Inflation predicts a that primordial perturbations have a scale invariant power spectrum $P(k) \propto k^{n_s}$ where the spectral index has been measured to have a value $n_s = 0.9645\pm0.0049$ \citep{planck2015}. The gravitational growth of perturbations modifies the primordial spectrum by depressing its amplitude on scales smaller than the horizon at the matter-radiation equality. The result is that, on small scales, $P(k) \propto k^{n_s-4}$, while the largest scales retain the original quasi-linear spectrum $\propto k^{n_s}$ \citep[e.g.][]{Bardeen1986}. Given that most of the power in the standard CDM model is concentrated on small scales, these are the first to go nonlinear, resulting in the formation of bound halos as now explained.

\subsection{Nonlinear collapse and hierarchical assembly of dark matter halos}
\label{dm_assembly}
Once density perturbations grow beyond the linear regime, i.e. $\delta({\bf x},t)\sim1$, the full non-linear collapse must be followed. The dynamical collapse of a dark matter halo can be solved exactly in case of specific symmetries, the simplest of which is the collapse of a ``top-hat" spherically symmetric density perturbation. The radius of a mass shell in a spherically symmetric perturbation evolves as \citep{mo2010}
\begin{equation}
\frac{d^2r}{dt^2}  = - \frac{GM}{r^2} + \frac{\Lambda}{3} r,
\end{equation}
with the non-zero cosmological constant contributing to the gravitational acceleration. Integrating this equation yields
\begin{equation}
\frac{1}{2}\bigg(\frac{dr}{dt}\bigg)^2 - \frac{GM}{r} - \frac{\Lambda c^2}{6} r^2 = {\cal E},
\end{equation}
where ${\cal E}$ is the (constant) specific energy of the mass shell. Let us assume that at its maximum expansion, before the perturbation detaches from the expanding Hubble flow (``turn-around'') and starts to collapse, the shell has a radius $r_{max}$ at time $t_{max}$ where 
\begin{equation}
t_{max} = \frac{1}{H_0} \bigg ( \frac{\zeta}{\Omega_\Lambda}\bigg)^{1/2} \int_0^1 dx\bigg[ \frac{1}{x}-1+\zeta (x^2-1)\bigg], 
\end{equation}
where $\zeta = (\Lambda c^2 r_{max}^3/6GM)$. Further the initial radius $r_i$ can be linked to $r_{max}$ as
\begin{equation}
\frac{r_i}{r_{max}} \approx \bigg( \frac{w_i}{\zeta} \bigg)^{1/3} \bigg[ 1-\frac{1}{5}(1+\zeta)\bigg(\frac{w_i}{\zeta}\bigg)^{1/3}\bigg],
\end{equation}
where $w_i = \Omega_\Lambda(t_i)/\Omega_m(t_i) = (\Omega_\Lambda/\Omega_m)(1+z_i)^{-3}$. Combining this with the fact that $M = (1+\delta_i) \Omega_{m,i} \bar \rho(t_i) (4\pi r_i^3/3)$ yields
\begin{equation}
\delta_i = \frac{3}{5} (1+\zeta) \bigg(\frac{w_i}{\zeta}\bigg)^{1/3}.
\end{equation}
This can be linearly evolved till the present time to obtain $\delta_0$ as
\begin{equation}
\delta_0 = \frac{a_0 g_0}{a_i g_i} \delta_i =  \frac{3}{5} g_0 (1+\zeta) \bigg(\frac{w_i}{\zeta}\bigg)^{1/3},
\end{equation}
where $w_i = w_0/(1+z_i)^3$ and $g$ is the linear growth factor. Assuming that the shell collapses at $t_{col} = 2 t_{max}$, we can derive the linear overdensity at the collapse time as
\begin{equation}  
\delta_c (t_{col}) = \frac{3}{5} g(t_{col}) (1+\zeta) \bigg(\frac{w(t_{col})}{\zeta}\bigg)^{1/3} \approx 1.686[\Omega_m(t_{col})]^{0.0055},
\end{equation}
i.e. the linear over-density at collapse time $t_{col}$ is weakly dependent on the density parameter with $\delta_{crit} \simeq 1.686$ for all realistic cosmologies \citep{peebles1980, barkana-loeb2001, mo2010}.

Given that the scale factor evolves as $a \propto t^{2/3}$ in a matter-dominated Universe, the redshift of maximum expansion of the perturbation, $z_{max}$, can be related to the collapse redshift, $z_c$, as
\begin{equation}
\frac{1+z_{max}}{1+z_c} \propto \bigg(\frac{t_{col}}{t_{max}}\bigg)^{2/3} = 2^{2/3}, 
\end{equation}
implying a very swift collapse after turn-around. In reality, however, particles in the mass shell considered will cross the mass shells inside it. Indeed, by $t = 2 t_{max}$ the mass shells enclosed by the collapsed shell will have crossed each other many times resulting in a dark matter halo in virial equilibrium i.e. where the gravitational energy has been converted into the kinetic energy of the particles involved in the collapse. Subsequently, baryons, being the sub-dominant matter component, almost passively start to fall into these dark matter potential wells.

A halo of mass $M_h$ collapsing at redshift $z$ has a physical virial radius ($R_{vir}$), circular velocity ($V_{vir}$) and virial temperature ($T_{vir}$) given by \citep{barkana-loeb2001, ciardi-ferrara2005}
\begin{eqnarray}
R_{vir} & = & 0.784 \bigg(\frac{M_h}{10^8 h^{-1} M_\odot}\bigg)^{1/3} \bigg[ \frac{\Omega_{m} \Delta_c}{\Omega_m^z 18 \pi ^2} \bigg] \bigg( \frac{1+z}{10}\bigg)^{-1} \,\, h^{-1} {\rm kpc}. \\
V_{vir} & = & 23.4 \bigg(\frac{M_h}{10^8 h^{-1} M_\odot}\bigg)^{1/3} \bigg[ \frac{\Omega_{m} \Delta_c}{\Omega_m^z 18 \pi ^2} \bigg]^{1/6} \bigg( \frac{1+z}{10}\bigg)^{1/2}  {\rm km \, s^{-1}}.\\
\label{eqn_rvt_circ}
T_{vir} & = & 1.98 \times 10^4 \bigg( \frac{\mu}{0.6}\bigg) \bigg(\frac{M_h}{10^8 h^{-1} M_\odot}\bigg)^{2/3} \bigg[ \frac{\Omega_{m} \Delta_c}{\Omega_m^z 18 \pi ^2} \bigg]^{1/3} \bigg( \frac{1+z}{10}\bigg) \,\, {\rm K} ,
\label{eqn_tvir}
 \end{eqnarray}
where \citep{bryan1998}
\begin{eqnarray}
\Delta_c & = & 18\pi^2 + 82 (\Omega_m^z-1) - 39 (\Omega_m^z-1)^2 \, ,\\
\Omega_m^z & = & \frac{\Omega_{m}(1+z)^3}{\Omega_{m}(1+z)^3 + \Omega_{\Lambda} },
\end{eqnarray}
and $\mu$ is the mean molecular weight. 


Once dark matter collapses to form halos, according to the hierarchical CDM structure formation model, these small-scale bound halos merge through time to form successively larger structures. Further, the properties of dark matter halos, such as the density profile, naturally have a critical impact on the properties of the baryons bound in halos. We refer readers to \citet{taylor2011} for a detailed review on dark matter halos, and only focus on two key aspects- relating to their density profiles and the evolution of their number density through time- in what follows. 

\citet{press-sch1974} were the first to provide an elegant analytic formalism to track the evolution of the number density (per comoving volume) of dark matter halos, the Halo Mass Function (HMF), through time using their ``peak-formalism". Essentially, this formalism identifies peaks above a certain density threshold after smoothing an initial gaussian random density field with a filter. The probability that the density, at a given spatial position and time, exceeds the critical density ($\delta_{crit} \simeq 1.686$) collapsing into a bound object, is given by
\begin{equation}
p [>\delta_{crit}]= \frac{1}{\sqrt{2 \pi} \sigma(M)} \int_{\delta_{crit}}^\infty e^{-\frac{\delta_s^2}{2 \sigma^2(M)}} d\delta_s = \frac{1}{2} {\rm erfc} \, [\delta_{crit}/\sqrt{2} \sigma(M)],
\end{equation}
where $\delta_s$ represents the smoothed density field and $\sigma(M)$ represents its mass variance, obtained by convolving the initial power spectrum $P(k)$ with a window function. Press and Schechter soon realised that only half of the (initially over-dense) dark matter would be able to form collapsed objects in this scenario. However, initially under-dense regions can be enclosed inside larger over-dense regions, leading to a finite probability of their resulting in bound objects. They therefore argued that material in initially under-dense regions would be accreted onto collapsed objects, doubling their mass whilst leaving the shape of the mass function unchanged. Including this factor of 2 the Press-Schechter HMF, expressing the number density of halos between mass $M_h$ and $M_h+dM_h$ at redshift $z$, can be written as
\begin{equation}
n(M_h,z) dM_h = \sqrt{\frac{2}{\pi}} \frac{\bar\rho}{M_h^2}  \nu e^{-\nu^2/2}\bigg| \frac{d \ln \nu}{d \ln M_h}\bigg| dM_h,
\label{ps_hmf}
\end{equation} 
\begin{figure*}
\center{\includegraphics[scale=0.59]{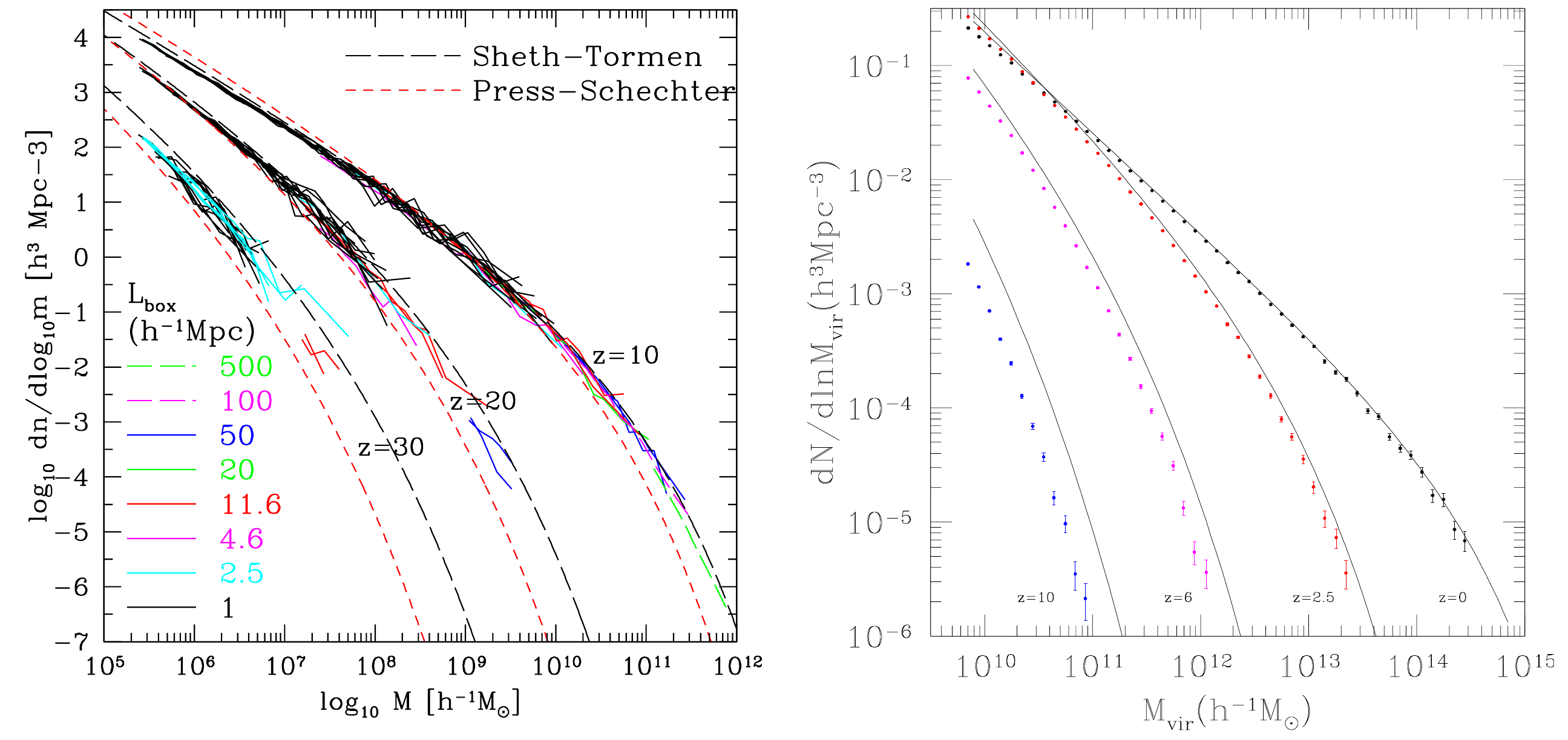}}
\caption{A comparison of the Sheth-Tormen HMF with those derived from N-body simulations by \citet{reed2007} ({\it left panel}) and \citet{klypin2011} ({\it right panel}). As seen, while the Sheth-Tormen HMF is in agreement with simulation results at $z=0$ (right panel), it successively over-predicts the number density of halos of all masses with increasing $z$ in both panels. These results are independent of the halo finder used: while the left panel has used the friends-of-friends halo finder, the right panel has used the Bound-Density-Maxima algorithm.}
\label{fig_hmf} 
\end{figure*}
where $\nu = \delta_{crit}(z)/\sigma(M)$. Further, $\delta_{crit}(z)$ is related to its value at $z=0$ such that $\delta_{crit}(z) = \delta_{crit}(0) D(z)$, where $D(z)$ is the linear growth rate normalised to the present day value so that $D(z=0)=1$ and $D(z)$ can be written as \citep{carroll1992}:
\begin{equation}
D(z) = \frac{g(z)}{g(0)(1+z)},
\end{equation}
where
\begin{equation}
g(z) = 2.5 \Omega_m [\Omega_m^{4/7}-\Omega_\Lambda + (1+\Omega_m/2)(1+\Omega_\Lambda/70)]^{-1}.
\end{equation}
This implies that the Press-Schechter HMF is fully defined once the initial conditions ($P(k)$, $\sigma(M)$) are specified. Eqn. \ref{ps_hmf} shows that the HMF has a characteristic mass, $M_{h*}(z)$, below which halos follow a power-law mass-number density relation, with the number density falling off exponentially above this mass. This simple spherical collapse model was extended to the more realistic ellipsoidal collapse scenario by \citet{sheth-tormen1999, sheth-tormen2002} resulting in the Sheth-Tormen HMF, whose free parameters are fit by comparing to observations. A number of works have shown that while the Sheth-Tormen HMF faithfully reproduces the number density of low-$z$ systems with mass ranging from dwarfs to clusters, it  progressively over-predicts the number density of high-mass halos with increasing $z$ \citep[][and references therein]{reed2007, klypin2011, watson2013} as shown in Fig. \ref{fig_hmf}. This inconsistency between the HMFs derived from N-body simulations and the analytical HMF holds irrespective of the (halo finding) technique used to identify particles bound to a given potential well. As shown in Fig. \ref{fig_hmf}, while \citet{reed2007} have used the friends-of-friends halo finder \citep{davis1985}, that links all particles within a chosen linking length into a single group, \citet{klypin2011} have used the Bound-Density-Maxima algorithm \citep{klypin1997} that locates maxima of density and removes unbound particles to identify bound groups. Motivated by this problem, a number of works \citep[e.g.][]{klypin2011, trac2015} have suggested correction factors to bring the Sheth-Tormen HMF into better agreement with simulations. For example, \citet{klypin2011} suggest the following correction factor, $F(z)$, to ensure better than 10\% agreement between analytic and numerical HMFs at all $z$:
\begin{equation}
F(z) = \frac{5.5 D(z)^4}{1+ [5.5 D(z)]^4}.
\end{equation}
Encoding the abundance of dark matter halos as a function of mass and redshift, the HMF is a powerful probe of cosmology: while the amplitude of the HMF on the scale of galaxy clustering at $z=0$ has been used to jointly derive limits on $\sigma_8$ and the matter density parameter $\Omega_m$, the redshift evolution of the amplitude has been used to characterise the dark energy equation of state \citep{allen2011} and even confirm the concordance model of cosmology \citep{harrison2012}.

As for the dark matter halo density profile, the simplest method of calculating it is to consider a halo to be made up of collapsing concentric shells of different radii and densities. This yields a radial density profile described by a steep power-law with a constant slope \citep[e.g.][]{bertschinger1985}. The real breakthrough in this field was made when \citet{navarro1996, navarro1997} ran high-resolution CDM simulations to show that the internal density profile of dark matter halos is well described by the Navarro, Frenk and White (NFW) profile that is shallower (stepper) than $r^{-2}$ at small (large) radii such that
\begin{equation}
\rho(r) = \frac{\rho_{crit} \delta_c} {(r/r_s)(1+r/r_s)^2},
\end{equation}
where $\rho_{crit}$ is the critical density, $r_s$ represents the halo-dependent scale radius and $\delta_c$ represents the characteristic over-density defined by
\begin{equation}
\delta_c = \frac{200}{3} \frac{c_h^3}{[{\rm ln}(1+c_h) - c_h/(1+c_h)]},
\end{equation}
where $c_h = R_{vir}/r_s$ represents the concentration parameter that reflects the critical density of the Universe at the ``collapse" redshift \footnote{While the original NFW paper, and most works following it, denote the concentration parameter by $c$, we use the notation $c_h$ since $c$ denotes the speed of light through-out this work }. While the NFW profile works extremely well over a wide range of halo masses, ranging from galactic to cluster scales \citep[e.g.][]{ludlow2017}, a number of high resolution simulations \citep{springel2008, stadel2009, dutton2014} find the profile in the inner-most regions ($r\lsim 0.01 R_{vir}$) to be better fit by the Einasto profile \citep{einasto1965}. On the other hand, \citet{burkert1995} have showed that the halo density profiles of dwarfs galaxies are better fit by a Burkert profile of the form 
\begin{equation}
\rho(r) = \frac{\rho_{0} r_0^3} {(r+r_s)(r^2+r_s^2)},
\end{equation}
where $\rho_0$ and $r_s$ are the two free parameters representing the central density and scale radius respectively. Resulting in an isothermal core of fixed density in the inner-most regions, this profile has now been shown to also fit the halo profiles for spirals that are a hundred times more massive \citep{salucci2000}. 

\begin{figure*}
\center{\includegraphics[scale=1.25]{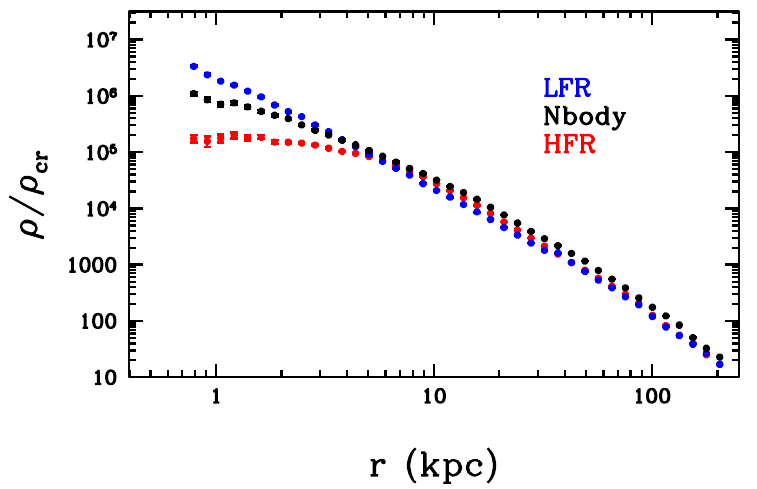}}
\caption{The dark matter density profile (compared to the critical density) of a MW-mass ($10^{12} M_\odot$) halo in three different realisations \citep{maccio2012}. The black dotted line shows results using an N-body only simulation which predicts an NFW profile. The blue and red dotted lines show results from SPH simulations with low (Kroupa IMF \citep{kroupa1993} and $4 \times 10^{50}$ erg of energy per SN) and high (Chabrier IMF \citep{chabrier2003} and $10^{51}$ erg of energy per SN) feedback, respectively. As shown, SN feedback in the latter case has a significant impact, resulting in a density profile that is much flatter (and cored in the central 8 kpc) compared to the NFW. }
\label{fig_dm_den} 
\end{figure*}

Finally, we note that the concentration parameter depends on the complex assembly history of a given halo: with a longer phase of relatively quiescent growth, massive rare halos are found to be more concentrated with the concentration parameter varying weakly with $z$. On the other hand, low mass halos, that have experienced a recent major merger, are found to be less concentrated with $c_h$ varying rapidly as a function of $z$ \citep[e.g.][]{gao2008, zhao2009, ludlow2014}. A growing body of work has focused on using high-resolution SPH simulations, that simultaneously track the assembly of dark matter halos and the baryonic component, to study the impact of supernova (SN) feedback affecting the dark matter density profile. As in Fig. \ref{fig_dm_den}, it has now been shown that the high star formation (and hence SN) efficiency associated with $M_h \gsim 10^{10} M_\odot$ halos is powerful enough to perturb the dark matter profile, resulting in a flat, cored, density profile in the central few kpc \citep{maccio2012, chan2015,governato2010, governato2012, dicintio2014, pontzen2014}.


\newpage

\newpage
\newpage
\newpage

\section{Basic physics of galaxy formation}
\label{ch4}

We are now ideally placed to discuss the key physical processes (including gas cooling, star formation, radiation fields and SN feedback) involved in galaxy formation (Sec. \ref{building_gal}). We then discuss the key theoretical approaches used to model early galaxies (Sec. \ref{theo_tools}) before ending by detailing the key physics implemented in a number of (semi-analytic and numerical) models, over a range of physical scales, in Table \ref{table31}.  
  
\subsection{Physical ingredients}
\label{building_gal}
Although the physics of gravitational instability, connecting the primordial density fluctuation field to nonlinear dark matter structures, offers a solid starting point for galaxy formation theories, the most difficult challenge is to obtain a sound description of the fate of baryons. This is a particularly important point as most of the information obtained from observations is carried by electromagnetic signals from this component. It is therefore critical that any valuable model predicts both the global (such as luminosity functions, star formation rates, stellar and gas masses) and structural (sizes, morphology, accretion, outflows, thermal state of the ISM, turbulence) properties of galaxies. These predictions are far from trivial as they involve a large number of microphysical and hydrodynamical processes interacting on a multi-scale level in a complex feedback network. However, because some of the key processes cannot be described purely from a fundamental physics perspective (a classical example is the way in which star formation is implemented in {\it any} type of model), one is forced to take a more heuristic approach in which theory must include some {\it Ansatz} based on empirical evidence. For this reason, the output of any model must be carefully confronted against observations to test whether the assumptions made hold against a wider application range. 
In the following we try to succinctly describe the key relevant processes for the formation of a galaxy. As it will become clear in the following Sections, though, details of their specific implementation might noticeably change the conclusions reached by different studies. With these caveats, that will be better elucidated later on, we proceed with a first reconnaissance of the key processes involved in galaxy formation. 

\subsubsection{Gas cooling}
\label{Cool}
Once the baryons have been shock-heated to the virial temperature of their host dark matter halo, further evolution can occur only if the gas is able to radiatively dissipate its internal energy. The associated loss of pressure causes the gas to collapse towards the center, likely forming a rotationally supported disk-like structure. Cooling processes largely involve collisionally-excited line emission from atoms and continuum radiation (e.g. bremsstrahlung, recombination and collisional ionizations) with heavy elements (including C, N, O and Fe) being the most efficient radiators. However, the first galaxies must have formed in an (almost) primordial environment where the abundances of these elements, as compared to hydrogen (referred to as the ``metallicity"), was extremely low. Lacking the key coolants, radiative dissipation had to rely essentially on two species: (a) molecular hydrogen and (b) collisional excitation of the 1s-2p transition of the hydrogen atom i.e. the Lyman-alpha (Ly$\alpha$) transition. Both processes present some criticalities: while molecular hydrogen, lacking a dipole moment, is an inefficient radiator, Ly$\alpha$ requires temperatures $\simgt 10^4$ K for electrons to be sufficiently energetic to excite the transition. Thus, while H$_2$ supplies the necessary cooling in relatively small halos with $T_{vir} \simlt 10^4 $ K, Ly$\alpha$ dominates the process for larger halos. Once heavy elements are formed the cooling rates increase, essentially linearly with metallicity ($Z$), and the gas can collapse and form a disk on the cooling timescale, $t_{cool}$, that is typically shorter than the free-fall time of the system, $t_{ff}$. For a detailed review on the subject see e.g. \citet{Ferrara2008}.   

\subsubsection{Star formation}
\label{Star}
Once gas has cooled into a central, baryon dominated structure, which often resembles a rotating disk, clumps form as a result of the ongoing process of gravitational instability. Large gas agglomerates, known as Giant Molecular Clouds (GMC), form under the action of self-gravity. Their typical masses are, to a first approximation, set by the Jeans length which is equal to the product $\lambda_J \approx c_s t_{ff}$, where $c_s$ is the gas sound speed in the disk, and $t_{ff}$ the local free-fall time.  Turbulent motions and magnetic fields, if present in the early Universe, can contribute extra pressure in addition to the thermal one, playing an important role in supporting the cloud against collapse. These GMCs are highly structured systems, composed of a dense core and filaments, embedded in a more diffuse (interclump) gas component. Arising from the combination of turbulence and gravitational effects, the gas density distribution spans a wide range of values, ranging from $\approx 10^2 {\rm cm}^{-3}$ in the diffuse, interclump medium to $> 10^7 {\rm cm}^{-3}$ in the densest cores. As a result, the gas in GMCs, as suggested by the name, is predominantly molecular (including H$_2$, CO and other more complex molecules). How these dense cores turn into stars is a difficult question that has been challenging experts since the 1960's. Although many details are still unclear, a general scenario that emerges is one in which cores, undergoing gravitational collapse, become centrally concentrated, with
a density profile that scales approximately as $\rho(r) \propto r^{-2}$. In the innermost regions of the collapsing core, the opacity eventually becomes large enough that the gas switches from approximately isothermal to adiabatic behaviour. Once the gas is hot enough to dissociate the molecular hydrogen, a second collapse ensues and a protostar is formed. These complex processes cannot be followed in detailed by even the most advanced galaxy formation simulations. Therefore, one usually resorts to empirical prescriptions to determine the fraction of the gas in the core that is turned into stars. This efficiency, $\epsilon_{core}$, is relatively high, and observationally inferred to be $\approx 0.3-0.5$. Perhaps more useful in large scale approaches is the analogous overall efficiency per free-fall time for the entire GMC. This value is estimated to be $\epsilon_{ff}\approx 0.01$ and it is encoded in the empirical Kennicutt-Schmidt relation that links the gas surface density to the star formation rate. Another important quantity to be specified in any galaxy model is the Initial Mass Function (IMF) of the stars, i.e. their mass distribution at birth. Whether the IMF is best described through the Salpeter function as observed in the Galaxy (where the number of stars of a given mass $N(M_s) \propto M_s^{-2.35}$) or follows alternative forms such as the Kroupa \citep{kroupa2001} or Chabrier \citep{chabrier2003}, as well as its its dependence on other physical quantities (e.g. metallicity, environment etc.) remains highly uncertain. Excellent reviews on star formation physics are given in \citet{McKee2007} and \citet{Krumholz2015}.  

\subsubsection{Radiation fields}
\label{Rad}
Radiation from (massive) stars plays an important role in the thermal and dynamical evolution of galaxies. This is particularly true for the earliest systems, which are characterized by virial temperatures not greatly  exceeding those resulting from photo-heating (typically $\sim 10^4$ K) by UV photons. In addition, radiative emission is fundamental in order to detect and study these systems. The power (i.e. ``luminosity", $L_s$) and the spectrum of the radiation field emitted by a star depends sensitively on its mass ($M_s$). Massive stars tend to be more luminous, approximately following the relation $L_s\propto M_s^{7/2}$ for most of their lifetime, spent on the so-called main sequence. They are also hotter: assuming that they emit as black bodies, and given the main-sequence mass dependence of the stellar radius, $R_s \propto {\sqrt M_s}$, one obtains a temperature that scales as $T_s \propto M_s^{5/8}$. Thus the spectrum of massive stars contains a larger fraction of photons with energy  h$\nu > 1 \, {\rm Ryd} = 13.6$ eV capable of ionizing hydrogen atoms and, for the extreme case of metal-free (Pop III) stars, also helium. This process creates ionized bubbles through which photons can freely escape from galaxies. {The first stars also produce Lyman-Werner (LW) photons ($11.2-13.6$eV) establishing a LW background that can impact further star formation (see Sec. \ref{first_sf}). Non-ionizing UV photons (h$\nu < 1$ Ryd) are equally important given that they also heat the gas via photoelectric effect mediated by dust grains; such energy deposition allows the creation of the multiphase ISM structures observed in galaxies. The background created by UV radiation (the UV background; UVB) from stars can also  photoevaporate, and ultimately disperse, GMCs hampering further star formation. All these processes are thoroughly discussed in the classical book by \citet{Spitzer1978}.   

\subsubsection{Supernova explosions}
\label{Sup}
Massive stars have short lifetimes: as the available energy from nuclear fusion $\propto M_s$, the lifetime can be roughly estimated to be $M_s c^2/L_s \propto M_s^{-5/2}$. Thus a massive star with $M_s \simgt 8 M_\odot$ only lives for $\simlt 30$ Myr, a short timescale compared to the typical dynamical time of a galaxy which is approximately equal to the rotational time $t_{dyn}=2\pi R/V$, where $R$ and $V$ are the disk radius and circular velocity, respectively. As an example, the dynamical timescale for the Milky Way is estimated to be $t_{dyn}\approx 300$ Myr. The core collapse, marking the death of massive stars, ignites a powerful explosion (``supernova") followed by shock waves propagating into the surrounding gas, and carving hot ($T\approx 10^6$ K) bubbles. The explosion leaves behind a remnant which could be a neutron star or a black hole (BH)\footnote{In the particular case of pair instability supernovae, occurring in the narrow range $140 < M_s/\Msun < 260$ there is no compact remnant and the star is totally disrupted.}. Approximately 99\% of the total supernova energy (typically corresponding to 1 Bethe = 10$^{51}$ erg almost independent of $M_s$) is carried away by neutrinos. The ejecta initially expand freely at velocities exceeding 1000 km s$^{-1}$, but once the inertia of the swept up gas becomes important, the expansion enters an adiabatic Sedov-Taylor phase in which the radius of the bubble increases with time only as $ t^{3/5}$ rather than linearly. Finally, radiative energy losses in the post-shock gas become important, causing a drop in the internal energy, resulting in an additional deceleration, until the cold shell simply coasts along conserving momentum. The role of SN energy injection can be hardly overlooked as it might have dramatic effects on the evolution of a galaxy. The production of hot gas quenches star formation (a process often refereed to as ``mechanical feedback"), shocks inject bulk motions, vorticity and turbulence in the ISM and explosions disperse heavy elements and dust enhancing the cooling ability of the gas. Finally, the hot gas is buoyant and tends to leave the galaxy in the form of a galactic wind (see below). Modelling these processes is extremely difficult and no clear consensus has been reached on how to self-consistently include them in theoretical models and in simulations that cannot properly resolve the relevant physical scales. The interested reader can find a detailed theoretical description of SN explosions in \citet{Ostriker1988}. 

\subsubsection{Mass and energy exchange}
\label{Exc}
Galaxies can hardly be modelled as closed systems. Cosmological simulations show that galaxies form at the intersection of the filaments constituting a complex network, known as the ``cosmic web". Matter flows into galaxies along these ``pipes" feeding the system with dark matter and baryons (presumably in a ratio close to the cosmological $\Omega_c/\Omega_b \approx 5.8$). The accreted intergalactic gas represents new fuel for star formation. At the same time, given its more pristine composition, intergalactic gas acts to dilute the heavy element content of the interstellar medium. The cycle is closed by the fact that supernovae associated with newly born stars might eject a sizeable fraction of their gas (and metal) mass into the IGM again \citep[for a review see][]{Veilleux2005}. Note that, under some conditions, part of the accelerated wind gas might rain back onto the galaxy in a so-called ``galactic fountain". It is also possible that the above dynamical process reaches a sort of steady state in which the rate of gas accretion balances the sum of star formation and outflow rate (this model is informally known as the ``bath-tub" model). Accretion is not the only way in which galaxies gain their mass. An alternative mechanism relies on galaxy mergers occurring after close encounters. This is probably the dominant mode of growth at early cosmic times as the merger rate for typical galaxies is a steep function of redshift, scaling as $\propto (1+z)^\gamma$ with $\gamma={2.2-2.5}$. Mergers in hierarchical formation models are particularly important as, in addition to triggering bursts of star formation driven by the increased density of the compressed gas, they could drive dramatic changes in galaxy morphology and dynamics. For example, elliptical galaxies are thought to result from the merger of two disk galaxies. These issues are reviewed in detail by \citet{Conselice2014}.          

\subsection{Theoretical tools}
\label{theo_tools}
The vast range of scales involved in modelling galaxies in a cosmological context, ranging from the sub-pc scales of star formation and its associated feedback to the super-Mpc scales associated with reionization, have spawned a variety of theoretical techniques. Here, we briefly discuss the three main ``classes" of theoretical (semi-analytical, semi-numerical and hydrodynamic) models adopted for modelling early galaxies and interested readers are referred to the review by \citet{somerville2015} for more details on the modelling techniques discussed here. The key aims, parameters (volumes modelled and dark matter particle resolution mass) and physics implemented in a number of theoretical models, ranging from small to large scales, is summarised in Table \ref{table31}.  A fourth, empirical method based on {\it abundance-matching} attempts to map the properties of the underlying dark matter halos onto observable galaxy properties \citep[e.g.][]{behroozi2015, mashian2016}. However, given that such models do not explicitly implement physical processes from first principles, we do not discuss them further in this Section. 

We start with the assembly of dark matter halos that provide the backbone for galaxy formation. Dark matter merger trees, describing the (merger- and accretion-driven) growth of halos can be constructed in two ways. The first approach, colloquially refereed to as building ``binary merger trees" uses the extended Press-Schechter formalism to successively fragment a halo into its (at most two) higher-$z$ progenitors; progenitors below the chosen resolution mass limit are then regarded as ``smooth-accretion" of dark matter from the IGM. Alternatively, dark matter only (``N-body simulations") can be run where the motion and collapse of dark matter particles, initially distributed according to the cosmological initial conditions, are followed by calculating the forces from all other particles using Poisson's equations. These calculations can either be carried out using particle-based ``tree" methods (where forces from families of distant particles are computed via multiple moments), using particle mesh codes (particles move along potential gradients computed via Fourier transform of the density field) or a combination thereof. 

Once these codes are run, a key challenge is to identify a bound ``halo" and the constituent sub-structures (sub-halos) arising from distinct halos that formed and merged into the central halo at earlier times. Simulations rely on three main categories of halo finding techniques \citep[for a detailed comparison see][]{knebe2011}: {\it (i) Friends-of-friends (FoFs) method:} that identifies bound groups by linking together all particles separated by a chosen ``linking length" which is usually a fraction of the average particle separation in the simulation; {\it (ii) Spherical over-density (SO) method:} that identifies peaks in the matter density field around which spherical shells are grown out to a point where the density falls below a certain pre-defined value; and {\it (iii) 6D phase-space methods:} an extension of the FoF method, these identify bound halos by including an additional proximity condition in velocity space. We are now in a position to discuss how these dark matter halos can be combined with baryonic calculations to yield galaxy populations.

\subsubsection{Semi-analytic models (SAMs)}
SAMs are based on using sets of equations to implement, some or all of, the physical processes discussed in Sec. \ref{building_gal} into dark matter merger trees, jointly tracking the assembly of baryons and their host dark matter halos through cosmic time - a typical SAM is shown in Fig. \ref{fig_sam}. Depending on the aim, the key processes (gas accretion, star formation, chemical/mechanical/radiative feedback and metal enrichment) can be included with varying degrees of complexity: e.g. gas and metals can be distributed following the dark matter density profile, gas can be split into cold/hot phases, outflows can preferentially carry away metal enriched gas and reionization feedback can be implemented, to name a few. As a result of their very nature of treating baryons through equations, lacking ``sub-grid" resolution, these models generally yield average baryonic properties over a given halo. That being said, the appeal of SAM lies in their (relative) simplicity of implementation, the flexibility of incorporating physical processes and the large physical/mass dynamic ranges and physical parameter space that can be explored over reasonably short timescales with relatively low computing requirements. 

\begin{figure*}[h!]
\center{\includegraphics[scale=0.4]{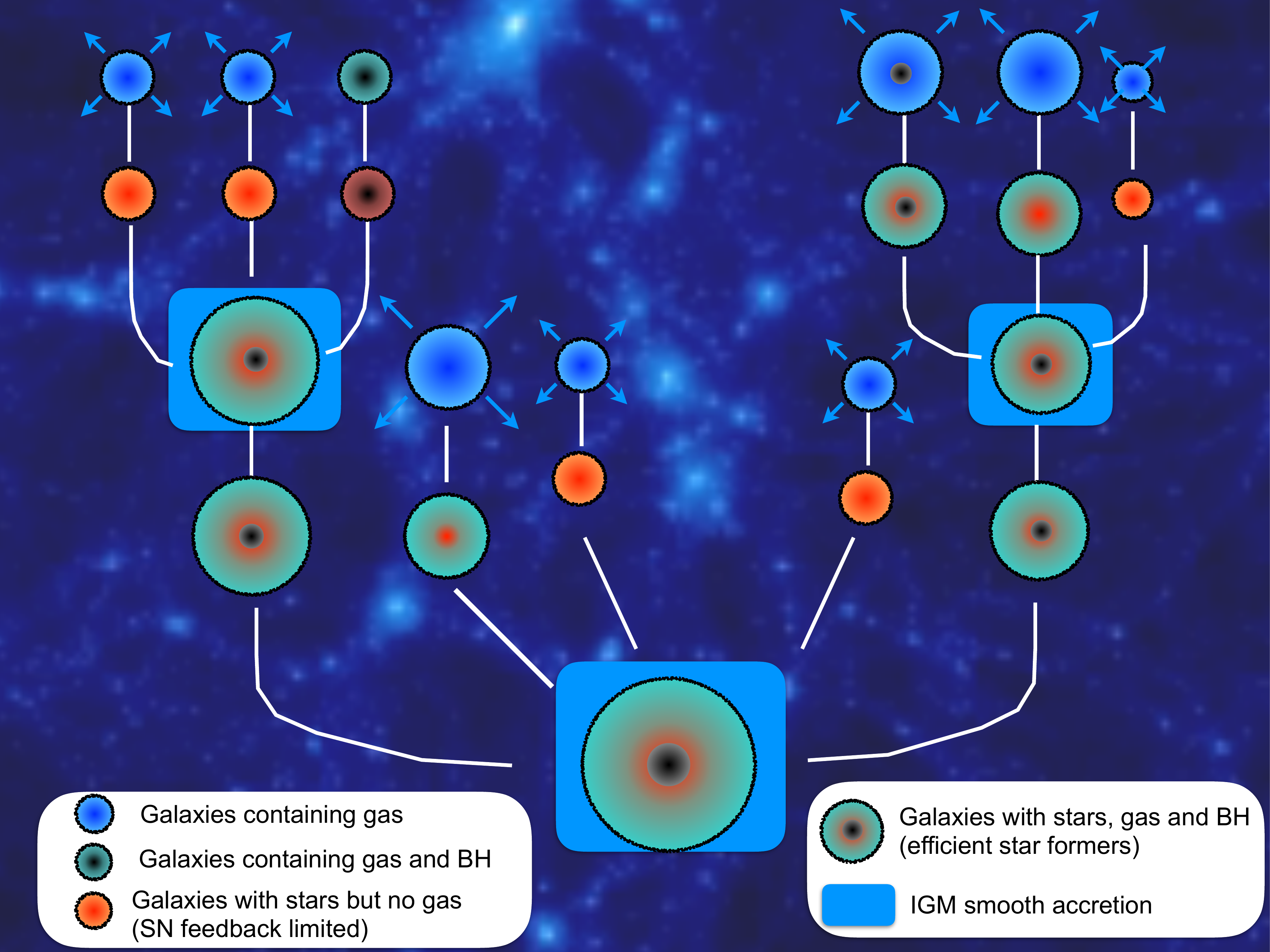}}
\caption{A SAM jointly tracking the (merger and accretion driven) buildup of dark matter halos as well as their baryonic component. SAMs present a powerful tool for capturing the key physics involved in galaxy formation including: the stellar mass growth due to star formation, the merger and accretion driven gas mass growth, the role of SN (and possibly black holes) in ejecting gas from low-mass halos and tracking the resulting impact on the subsequent growth of more massive systems via halo mergers and gas accretion. As shown, galaxies assemble as a result of both wet mergers of ``efficient star formers" that do not lose all their gas mass after star formation/black hole accretion and dry mergers of supernova/black hole ``feedback limited" systems that lose all their gas resulting in dry mergers. This naturally leads to a variety of galaxy assembly histories and properties for a given halo mass. }
\label{fig_sam} 
\end{figure*}

\subsubsection{Hydrodynamic simulations}
The most complex approach to galaxy formation involves hydrodynamic simulations. In this approach, one simultaneously solves the relevant equations of gravity, hydrodynamics and thermodynamics over particles or grid cells that represent dark matter, gas and stars. This is an extremely powerful, albeit computationally expensive, method that can track and predict the interplay between dark matter and baryons and the resulting galaxy/large scale structure properties (depending on the simulation size) once appropriate sub-grid (scales below the resolution limit) recipes for gas cooling, star formation, and feedback are implemented. 

Hydrodynamic simulations can be divided into two main categories: (i) {\it Lagrangian or SPH}: where particles carry the reference frame, and particle properties  - including mass, temperature, metallicity - are ``smoothed" by weighing over all neighbours closer than a chosen separation (called the linking length). An example of a typical SPH simulation is shown in Fig. \ref{fig_sph}; and (ii) {\it Eulerian}:  these are based on a cartesian reference frame where fluid particles are discretised into cells and physical properties are calculated across fixed cell boundaries. Given the dynamic range involved in galaxy formation, the latter method has been extended to Adaptive Mesh Refinement (AMR) techniques where cells satisfying a given criterion - for example in density or temperature - are split into sub-cells leading to a higher resolution in such regions. Issues with these techniques \citep[see e.g.][]{springel2010}, such as the inability of SPH to resolve shocks and the sensitivity of AMR results to bulk velocities, have led to the development of a new class of ``unstructured" mesh models, lying between Lagrangian and Eulerian methods, that use Voronoi tesselation to sub-divide the space around particles. Based on polyhedral cells, this mesh continually de-forms and re-forms as particles move. Despite its obvious advantage in simultaneously tracking both dark matter and gas particles, the enormous computational effort required for hydrodynamic simulations naturally places a limit on the physical volume that can be simulated and the physical parameter space that can be explored for a given mass resolution. 

\begin{figure*}
\center{\includegraphics[scale=0.2]{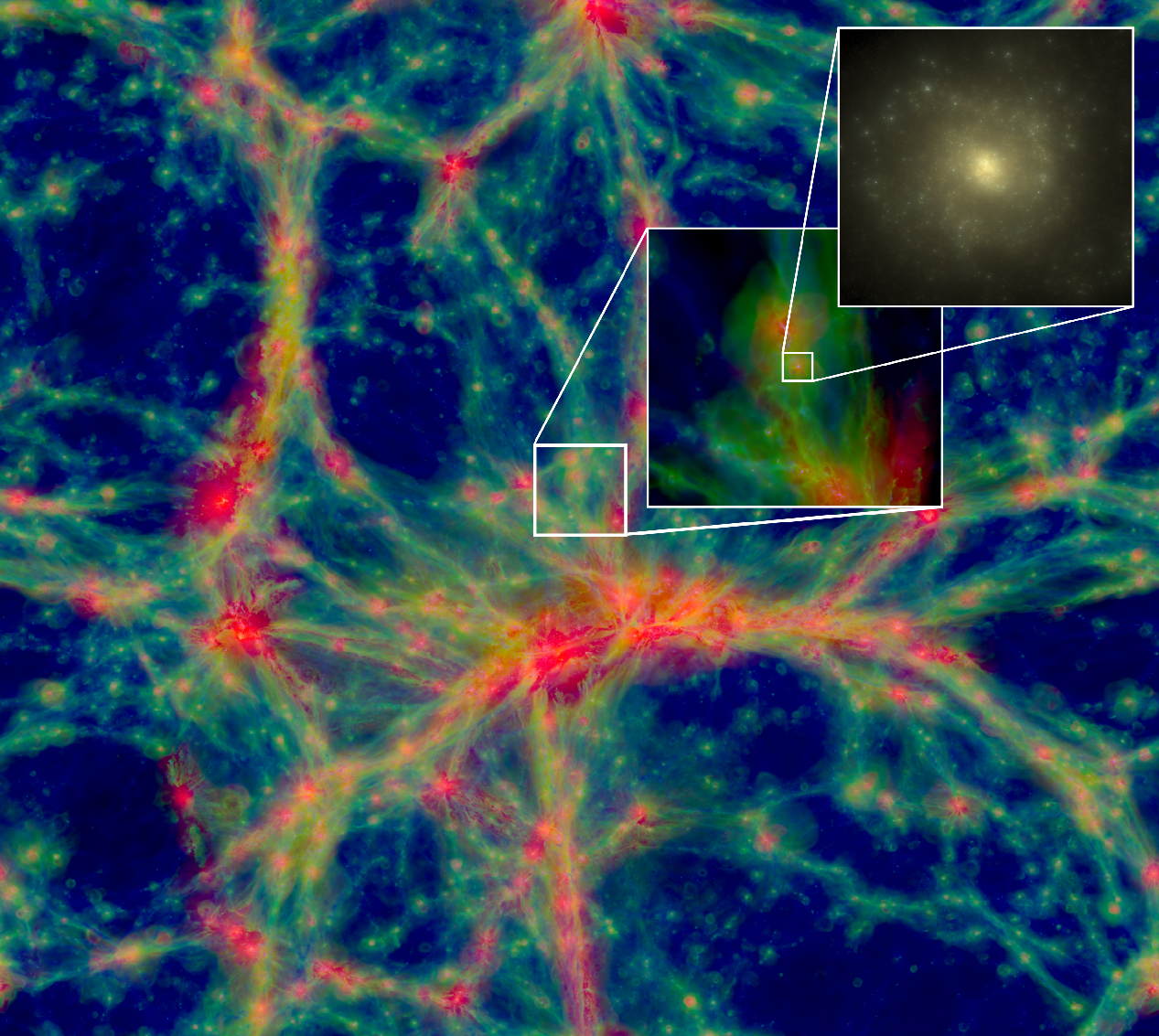}}
\caption{A $100\times 100 \times 20$ comoving Mpc (cMpc) slice through the EAGLE simulation, illustrating the dynamic range 
attainable with state-of-the-art SPH simulations. The intensity represents the gas
density while the color indicates the gas temperatures (blue - green - red from cooler to
hotter). The inset shows a region 10 cMpc and 60 ckpc on a side. The zoom in to an individual galaxy
with stellar mass $3 \times 10^{10} \Msun$ shows the optical band stellar light \citep{schaye2015}.}
\label{fig_sph} 
\end{figure*}

\subsubsection{Semi-numerical models}
As detailed in Sec. \ref{uvb_fb}, the past years have seen an increasing realisation of the necessity of coupling galaxy formation - on kiloparsecs scales - with the impact of the radiative feedback generated during reionization - on tens of Megaparsec scales. Indeed, \citet{iliev2014} have shown that, while $(100 h^{-1}\, {\rm cMpc})^3$ boxes are sufficiently large for deriving convergent reionization histories, the morphology of the ionized bubbles remains poorly described for box sizes smaller than $(200 h^{-1} \rm {cMpc})^3$. The rise of 21cm cosmology, and associated statistics including the r.m.s, skewness and kurtosis of the differential 21cm brightness temperature (the temperature of the redshifted 21cm emission with respect to the CMB), have therefore led to the development of a new class of ``semi-numerical" models \citep[for a review see][]{trac2011}. The key idea is to couple semi-analytic models of galaxy formation, run on large $(\gsim 100{\rm cMpc})^3$ N-body simulations, with analytic/numerical models of radiative transfer. This is the only computationally tractable approach of consistently describing the complexity of the galaxy formation-reionization interplay as will be discussed in more detail in Sec. \ref{ch7}.

\begin{table}
\caption{An illustrative list of models for the first billion years. For each model we list the key aim (column 2), the simulation technique (column 3), box size (column 4), DM mass resolution (column 5), the key physics implemented with the footnote explaining the letters used (column 6) and the model name and reference (column 7). The explanations of the letters used in columns 1 and 6 follows on the next page.} 
\vspace{-0.9cm}
\begin{center} 
\begin{turn}{90}
\begin{tabular}{cccccccc}
\hline
No. & Main aim & Technique & box size [cMpc] & $M_{DM} [\Msun]$ & Key Physics & Code [reference]\\
\hline
\hline
& & & & Small-scale models  &  & \\
\hline
\hline
1 & SF in GMC & Resimulated & $1-10 R_{vir}$ & $\sim 10^2 - 10^6$ & AIKO & FIRE \citep{hopkins2014} \\
2 & SF, GF, EoR & AMR & 1 & 1840 & DIKO  & \citep{wise2012b}\\
3 & GF, EoR & AMR+RT &  4 $h^{-1}$ & $4 \times 10^6$ & EO & EMMA \citep{aubert2015} \\
4 & UV fb & SPH+RT & 5 & $2.5 \times 10^5$ & EIO  & \citep{hasegawa2013}\\
5 & UV fb, GF & SPH+RT &  3-6 $h^{-1}$ & $0.18-1.4 \times 10^6$ & GIKO & \citep{finlator2011}\\
6 & ISM,CGM & AMR & 9.7$h^{-1}$ kpc & $9.5 \times 10^4$ & AIKL & \citep{pallottini2017} \\
7 & UV fb, GF & SPH+RT &  10  $h^{-1}$ & $4.3\times 10^7 $ & GIO  & \citep{petkova2011} \\
8 & GF, EoR & Eulerian+RT &  20 & $4.8 \times 10^5$ & EIKO & \citep{norman2015} \\
9 & GF, EoR & AMR & 40 & $3 \times 10^4$ & diko & Renaissance \citep{oshea2015, xu2016} \\
10 & GF, EoR & AMR+RT &  20-40 $h^{-1}$ & $7 \times 10^6$ & AIO & CROC \citep{gnedin2014} \\
\hline
\hline
& & & & Intermediate-scale models  &  & \\
\hline
\hline
1 & GF, EoR & N-body+semi-numerical RT & 100 & $3.9 \times 10^6$ & DIKO & DRAGONS \citep{mutch2016} \\
2 & GF & SAM & - & $M_{min}=10^8$ & CIP & DELPHI \citep{dayal2014a} \\
3 & EoR (LG) & Eulerian+RT & $91$ & $3.5 \times 10^5$ & EIO & CoDA \citep{ocvirk2016} \\
4 & GF, EoR & SPH+ RT &  12.5-100 $h^{-1}$ & $10^6-8 \times10^7$ & EIKO & Aurora \citep{pawlik2017} \\
5 & $f_{esc}$ & SPH & $10-100 h^{-1}$ & $6\times10^6 - 9 \times 10^8$ & GIKLM &  \citep{yajima2011} \\
6 & GF & SPH & $25-100 h^{-1} $ & $1.2-9.7 \times 10^6$ & FIJKM & EAGLE \citep{schaye2015} \\
7 & GF & Unstructured mesh & $106$ & $6.2 \times 10^6$ &  GIJK  & Illustrus \citep{vogelsberger2014} \\

\hline
\hline
& & & & Large-scale models   &  & \\
\hline
\hline
1 & EoR & N-body+RT & 114-425 $h^-1$ & $0.55-5 \times 10^7$ & HO & \citep{iliev2014}  \\
2 & GF & SPH & $400 h^{-1}$ & $1.7 \times 10^7$ & GIJKM & BlueTides \citep{feng2016} \\
3 &  GF & SAM &  500$h^{-1}$ & $1.3 \times 10^9$ & BIJKP & GALFORM \citep{lacey2016} \\
\hline
\end{tabular}
\end{turn}
\label{table31} 
\end{center}
\end{table}

\newpage

The explanations of the letters used in column 6 of Table \ref{table31}.
\begin{itemize}
{
 \item[] A: SFR $\propto \rho_{\rm H_2}$
  \item[] B: SFR $\propto mass_{\rm H_2}$
  \item[]  C: SFR $\propto mass_{\rm gas}$ 
  \item[]  D: SFR $\propto mass_{\rm cold\, gas}$ 
   \item[] E: SFR $\propto \rho_{gas}$
   \item[] F: SFR $\propto Pressure_{gas}$
  \item[] G: Multi-phase SF using \citet{springel2003} prescription
  \item[] H: SFR $\propto M_h$
   \item[]  I: SN feedback 
  \item[] J: SMBH feedback
  \item[] K: Metal enrichment
  \item[]  L: Dust enrichment
 \item[]  M: Feedback from homogeneous UVB \citep[e.g.][]{haardt2001, faucher2009}
  \item[]  N: Feedback from self-consistently calculated homogeneous UV fields
   \item[]  O: Feedback from self-consistently calculated inhomogeneous UVB
\item []P: SFR =0 (for $M_h<M_{cr}$ or $V_{vir}<V_{cr}$ halos at $z <z_{re}$)
    }
 \end{itemize}

Further, in Table \ref{table31}, GF stands for galaxy formation, UV fb stands for UV feedback and LG stands for Local group.

\newpage
\newpage
\newpage

\section{The birth of the first galaxies}
\label{ch5}

We are now in a position to discuss how early galaxies built-up their gas mass and formed the first stars. As noted in Sec. \ref{ch4}, the gas in the first galaxies would have been metal-free, being composed of a primordial mixture of ($\sim 75\%$) hydrogen and ($\sim 25\%$) helium. Continuing the classification where solar metallicity stars are termed Population I, the stars forming out of such metal-free gas are refereed to as {\it Population III} stars \citep[PopIII; e.g.][]{bond1981}: this classification has been refined to include PopIII.1 (PopIII.2; formerly known as PopII.5) stars that formed out of primordial gas (that was photo-ionized) prior to its gravitational collapse into halos \citep[e.g.][]{yoshida2007}. While we limit ourselves to an overview of the first stars, we refer interested readers to excellent reviews \citep[e.g.][]{bromm-larson2004, glover2005, bromm2009, bromm-yoshida2011, wise2012} for more details. 

\subsection{Cold accretion and gas assembly of the first galaxies}
\label{cold_acc}
Once a dark matter halo forms a potential well, it accretes gas from the intergalactic medium to form its interstellar medium and circum-galactic medium (CGM). The standard paradigm of galaxy formation \citep[e.g.][]{rees1977, Silk1977, fall1980} sketches a picture of ``hot-mode accretion" where gas falling into a dark matter potential well, with circular velocity $V_{vir}$, is shock heated to the halo virial temperature $T_{vir}(z)= 10^6 (V_{vir}(z)/167 \, {\rm km\, s^{-1}})^2$ K. The gas inside a ``cooling radius", within which the cooling time is shorter than the free-fall time, can cool radiatively and settle into a centrifugally supported disk where star formation can take place. Mergers can scatter stars off their orbits, producing spheroidal systems that might re-assemble into a disk after subsequent gas accretion. However, as early as 1977 it was pointed out that, for plausible physical conditions, only a small fraction ($\simeq 10\%$) of gas would be heated to $T_{vir}$, with most never heated to temperatures above $2-3 \times 10^4$ K \citep[e.g.][]{binney1977, Katz1991,Fardal2001}. This has evolved into the current paradigm of ``cold-mode accretion" that suggests most gas is accreted at temperatures much lower than $T_{vir}$. Indeed, given that cooling is extremely efficient at $T \simeq 1-2.5 \times 10^5$ K, this provides a natural temperature threshold separating hot and cold-mode accretion \citep{birnboim2003, keres2005, ocvirk2008}. 

\begin{figure*}[h!]
\center{\includegraphics[scale=0.45]{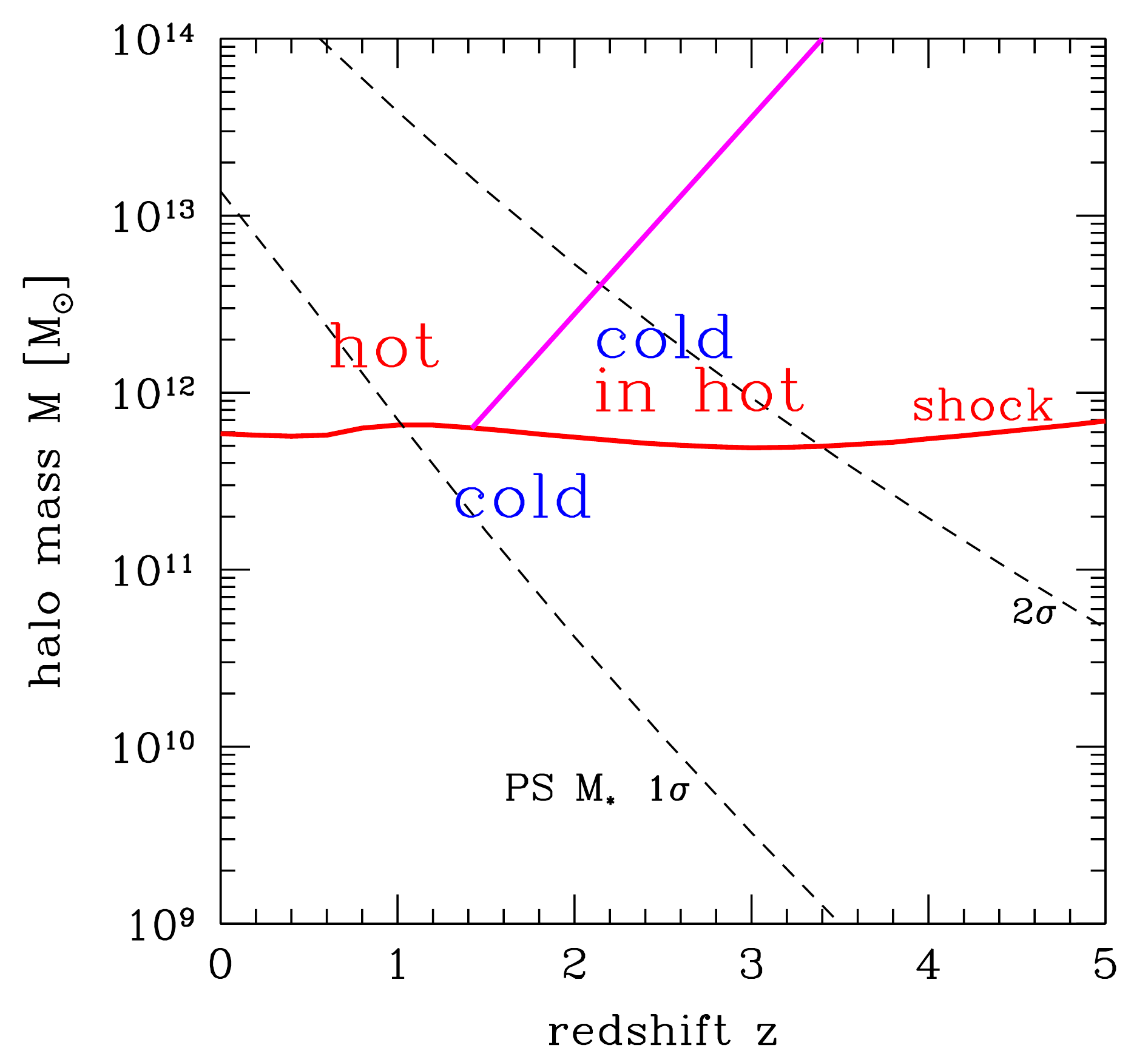}}
\caption{Cold streams and shock-heated medium as a function of halo mass and redshift \citep{dekel2006}. The (nearly) horizontal red curve shows the typical mass above (below) which gas flows into galaxies are predominantly hot (cold). The dashed lines show the halo mass corresponding to 1 and 2-$\sigma$ fluctuations as a function of redshift. The inclined solid curve shows the upper mass limit (roughly corresponding to the rare, massive 2-$\sigma$ fluctuations at $z \gsim2$) where the hot interstellar medium can host cold streams at $z \simeq 1-2$ allowing disc growth and star formation. On the other hand, similar mass halos at lower $z$ are all dominated by hot-mode accretion that shuts-off gas supply and star formation. }
\label{fig_mh_coldacc} 
\end{figure*}

A number of works \citep[e.g.][]{dekel2006, dekel2009} have shown that pressure can build-up, allowing a stable virial shock to develop, when the compression rate, required to restore pressure, is larger than the cooling rate of the post-shocked gas. However, there are two regimes in which shock heating of gas can be avoided: firstly, low mass galaxies are unable to support stable shocks at any redshift. Secondly, even in the presence of shocks, cold gas can be accreted along filaments that penetrate deep into the halo, feeding the galaxy with gas brought in from much larger distances than in the standard picture of hot-mode accretion \citep{keres2005, dekel2006, ocvirk2008,dekel2009, keres2009, vandevoort2011, vandevoort2012}.  This naturally implies a critical {\it shock-heating} mass, $M_{shock}$, below which galaxies pre-dominantly assemble through filamentary cold flows, with quasi-spherical hot-accretion dominating above this scale. As shown in Fig. \ref{fig_mh_coldacc}, this critical mass has a value of $M_{shock} \simeq 10^{11.5-12}\Msun$, essentially independent of redshift, and represents an equipartition scale where halos accrete their gas equally via hot and cold mode accretion \citep{keres2005, ocvirk2008}. $M_h \lsim M_{shock}$ halos effectively accrete via the cold-mode at all $z$. However, the situation is more complicated for larger mass halos: while cold streams, penetrating through the hot ISM, can contribute to the gas mass of the rare and massive $M_h \gsim M_{shock}$ galaxies at high-$z$ ($z \gsim 2$), at lower redshifts such halos are purely dominated by hot-mode accretion \citep{dekel2006, dekel2009}.

Simulations show that the rate of cold-gas accretion, which is both a function of halo mass and redshift, can be approximated as \citep{dekel2013}
\begin{equation}
\dot M_b \simeq 7.36 \bigg(\frac{M_h}{10^{12} \Msun} \bigg)^{1.15} (1+z)^{2.25} f_{0.184} \,\,\,\, [\Msun {\rm yr}^{-1}],
\end{equation}
where $f_{0.184}=0.184$ is the baryonic fraction using the latest cosmological values \citep{planck2016}. As shown in Fig. \ref{fig_acc_keres}, this increase in the accretion rate, combined with the HMF shifting to lower masses at earlier times, naturally results in cold-mode accretion becoming increasingly important at high-$z$ \citep{keres2005}; we however caution the reader that such studies are limited to $z \lsim 5$ and can therefore not be extrapolated to higher redshifts. Indeed, most ($\simeq 60\%$) gas is accreted cold, with $T \simeq 10^4$ K at $z \simeq 5$. As the cold-flow rate drops with decreasing redshift a bimodal temperature distribution emerges, at $z \simeq 2.5$, with roughly half the gas being accreted cold (at $10^{4.5}$ K) and the other half hot (at $10^{6.5}$ K) culminating with most gas being accreted via hot-mode accretion, with $T \simeq 10^{5-7}$ K, by $z \simeq 0$.

\begin{figure*}
\center{\includegraphics[scale=0.8]{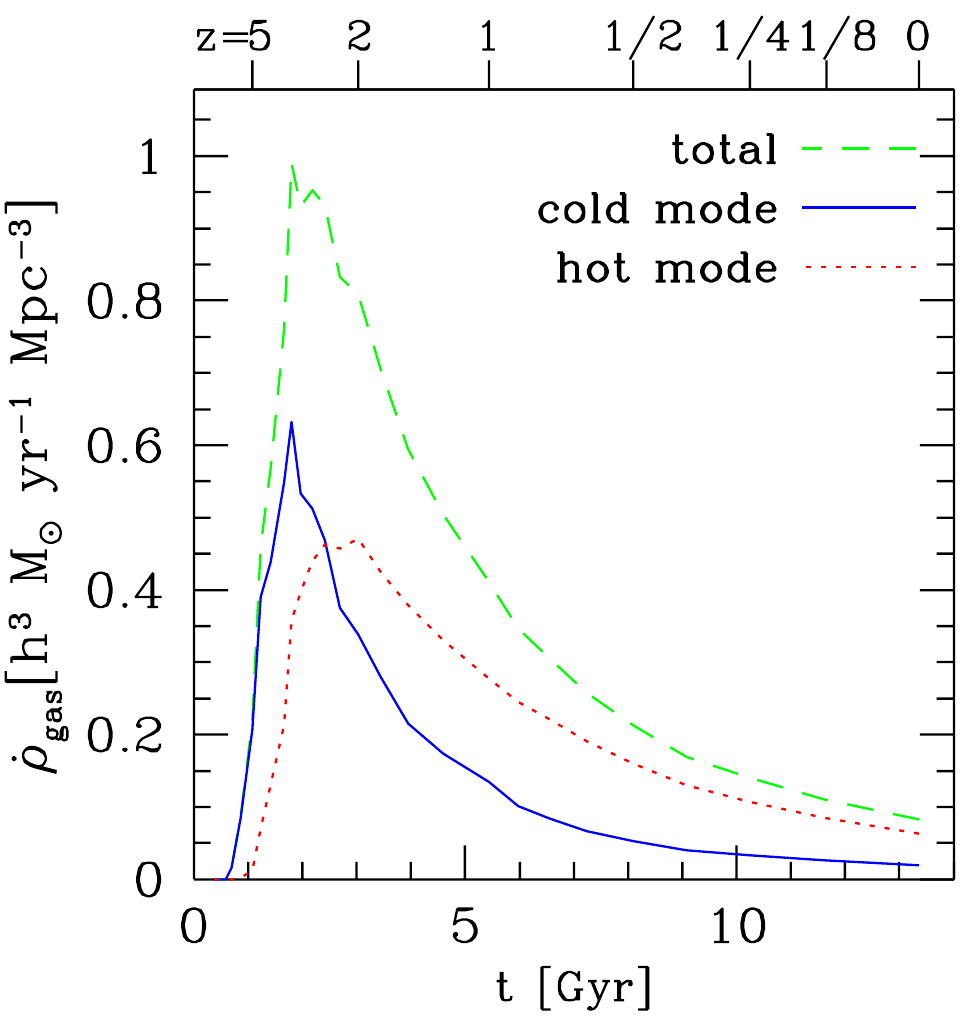}}
\caption{Redshift history of total smooth gas accretion (dashed green line) with the deconstructed contributions from cold- (solid blue line) and hot-mode (red dotted line) accretion \citep{keres2005}. These lines show results from all galaxies in a 22.2 $h^{-1}$ cMpc box with $128^3$ dark matter particles and initially the same number of gas particles. As shown, given the HMF shifting to lower masses with increasing redshift, cold-mode accretion provides the dominant mechanism for building the gas content of galaxies at $z \gsim 2$, shifting to hot-mode accretion at lower-$z$.}
\label{fig_acc_keres} 
\end{figure*}

\begin{figure*}[h!]
\center{\includegraphics[scale=1.0]{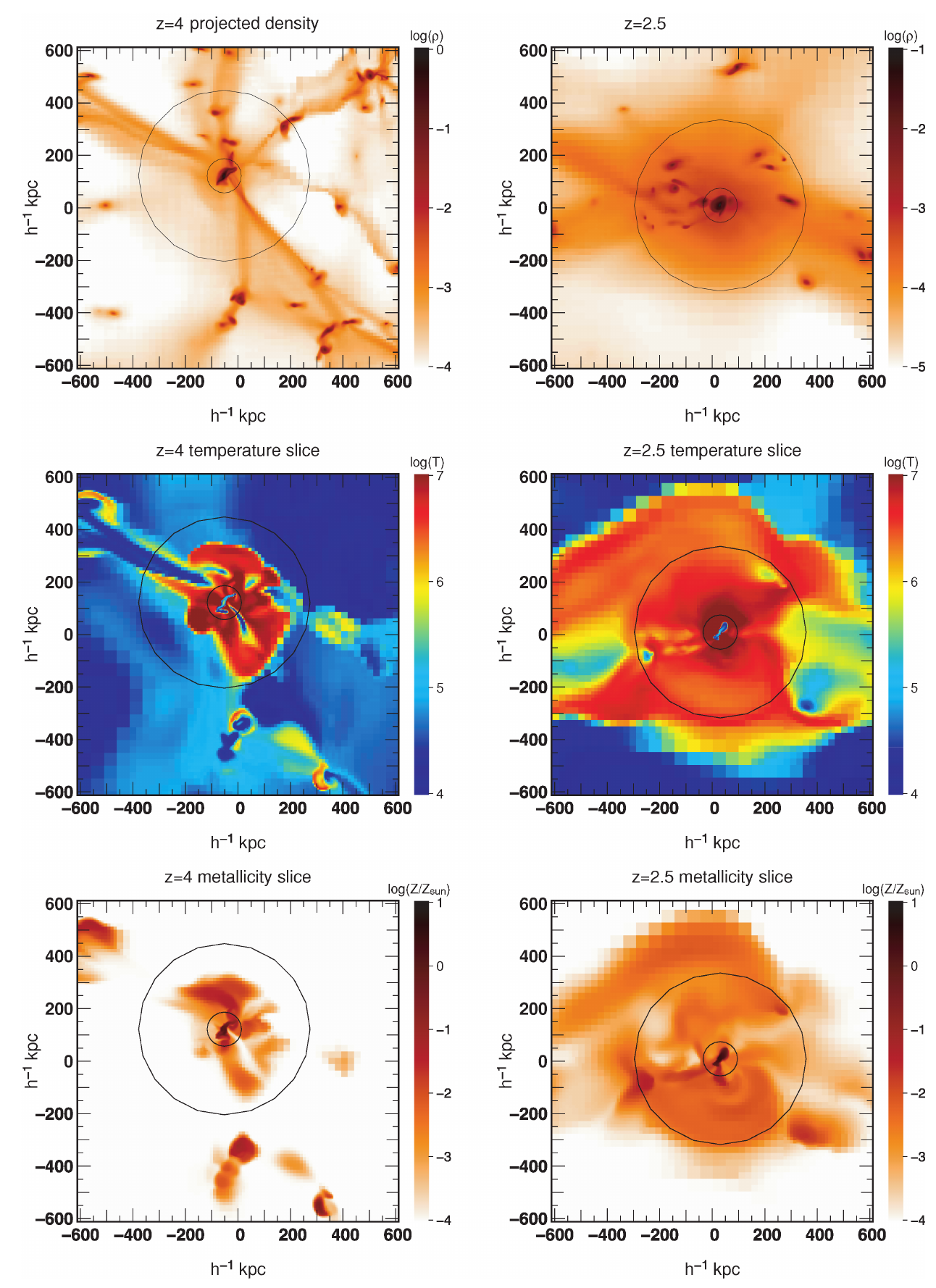}}
\caption{Simulation maps \citep{ocvirk2008} showing cold accretion onto a typical galaxy with $M_h = 2\times 10^{12}\Msun$ at $z=4$ ({\it left column}) and $z=2.5$ ({\it right column}). The rows show the projected density ({\it top panels}), temperatures ({\it middle panels}) and metallicity ({\it bottom panels}). In each panel, the large and small circles show the location of $R_{vir}$ and $0.2 R_{vir}$, respectively. While this galaxy would predominantly accrete via cold, high-density filaments showing a wide range of metallicities at $z=4$, it would accrete from a hot quasi-spherical, higher metallicity medium at a lower redshift of $z=2.5$.}
\label{fig_acc_prop} 
\end{figure*}

Running high-resolution AMR simulations \citet{ocvirk2008} have confirmed such results for a  MW-mass ($M_h = 2 \times 10^{12}\Msun$) galaxy as shown in Fig. \ref{fig_acc_prop}. As can be seen, such a galaxy, at $z=4$ is predominantly fed by well-defined cold streams ($T \lsim 10^{5.5}$ K) at $R_{vir}$ with the gas reaching $T_{vir}$ only after penetrating as deep as $0.5 R_{vir}$ \citep[see also][]{keres2005}. On the other hand, a similar-mass galaxy at $z=2.5$ accretes from a hot quasi-spherical medium with $T \gsim 10^{6.5}$ K at all $r\lsim R_{vir}$. Critically, this model highlighted a crucial missing ingredient: the metallicity of the accreted gas that, together with its density, determines the cooling rate. As seen from Fig. \ref{fig_acc_prop}, gas accreted via the cold mode shows a wide range of metallicities ranging from metal-poor ($Z \sim 10^{-4}\Zsun$) filamentary gas tunnelling in from $\sim 2 R_{vir}$ to the central galaxy to metal-rich ($Z \sim 0.1\Zsun$) satellites. In comparison, gas accreted via the hot-mode has a much higher metallicity, $Z \gsim 0.01\Zsun$, at all radii. 

However, the importance of cold-accretion sensitively depends on the feedback strength implemented in simulations \citep{benson2011}. Firstly, SN feedback can heat a significant fraction of the ISM gas, diluting incoming cold flows, reducing the importance of cold-flow accretion. Secondly, photo-evaporation or heating of gas from low mass halos due to the ionizing UVB (see Sec. \ref{uvb_fb}) has a bearing on the importance of cold-flows in two ways: (i) the lack or decrease of gas in low-mass halos results in a lower amount of ``merged gas", with the rest being reclassified as ``cold accretion"; and (ii) the UVB establishes an IGM temperature floor such that only halos with $T_{vir}>T_{IGM}$ can accrete in the cold mode, thereby affecting the minimum value of $M_{shock}$ \citep{brooks2009}.  


By providing ready-to-use gas, cold flows are critical to the process of galaxy formation: firstly, they are responsible for assembling most of the fuel for star formation in high-$z$ galaxies, with mergers only playing an important role in the assembly of groups and clusters
\citep{brooks2009, vandevoort2011, dayal2014a}. Secondly, they fuel the extremely high star formation rates ($\simeq 100-200\Msun \, {\rm yr^{-1}}$) observed in galaxies at $z \simeq 2-3$ whose extended clumpy disks are incompatible with the compact, perturbed kinematics expected from mergers \citep[see e.g.][]{dekel2009}. Thirdly, they are directly responsible for most of the disk growth at high-$z$ where filamentary accretion can keep disks intact \citep{dekel2009, brooks2009}. Finally, they are probably responsible for the observed bimodality in galaxy populations separating massive, red ellipticals from low mass, blue star forming disks \citep[e.g.][]{kauffmann2003b}.

Despite their importance in driving galaxy formation, observations of cold accretion have taken a long time to build up. 
A key reason is the small solid angle covered by cold-flow accretion \citep[e.g.][]{dekel2009, faucher2011, goerdt2012}. 
Forming at the intersections of filaments, rare, massive halos are typically fed by a few filaments with an opening angle of $20^{\circ}-30^{\circ}$ or a few percent of the sphere ($\sim 0.4\, {\rm rad}^2$). On the other hand, forming deep inside filaments, lower-mass halos accrete from a wider opening angle \citep{dekel2009}. Using Lyman Limit Systems (LLS) and Damped Lyman Alpha systems (DLAs) as probes of cold-mode accretion, \citet{faucher2011} find a covering fraction of $< 15\%$ for a $3 \times 10^{11}\Msun$ halo at $z=2$ that increases to $\sim 30\%$ for a lower mass ($7 \times 10^{10}\Msun$) halo at $z=4$. This low-covering fraction, compared to the order unity covering fraction for galactic winds, leads to signatures of cold streams being swamped by those of outflows in stacked spectra \citep{faucher2011}. Additionally, signatures of inflows can easily be masked by interstellar absorption at the systemic redshift \citep{rubin2012} and when selecting absorbers via metal-lines, ambiguity remains in separating a galactic scale outflow falling back onto the galaxy (``galactic fountain") from material being accreted from the IGM for the first time. As a result, observers have mostly had to resort to hunting for indirect signatures of cold flows although there are a few hints of direct detections, as summarised below. 

(i) {\it Direct evidences of cold flow accretion}: Using a sample of about 680 UV-selected galaxies at $z \simeq 2.4$, \citet{rakic2012} find the optical depth of Ly$\alpha$ absorbers to be stronger in the transverse direction than along the line-of-sight, which has been interpreted as resulting from cold accretion on scales of $1.4-2$ pMpc. Further, \citet{rauch2011, rauch2013} have detected extended, filamentary Ly$\alpha$ emission near faint LAEs, possibly powered by cold-gas accretion, which might be lower-mass analogues of the Ly$\alpha$ blobs (LABs) discussed next. Finally, \citet{martin2015, martin2016} claim the Ly$\alpha$ emitting filament detected near a quasi-stellar object (QSO) at $z =2.8$ and the associated proto-disk (of mass $4 \times 10^{12}\Msun$) could only have formed as a result of cold-accretion from the cosmic web.

(ii) {\it Observing cold flows via Ly$\alpha$ blobs}: First detected at $z \simeq 3$ by \citet{steidel2000}, LABs are extremely luminous ($L_{Ly\alpha} \approx 10^{44} \, {\rm erg \, s^{-1}}$) and extended ($\sim 100$ kpc) regions of Ly$\alpha$ emission. A number of theoretical studies have shown that LABs might result from the dissipation of gravitational energy through collisional excitations as cold gas flows into the potential well of a massive ($M_h \gsim 10^{12}\Msun$) dark matter halo \citep[e.g.][]{dijkstra2006, dijkstra2009, goerdt2010, rosdahl2012}. Indeed, ruling out alternative sources, namely Ly$\alpha$ emission originating from a dust-enshrouded highly star busting galaxy with large-scale superwinds or being powered by QSO activity, a number of works have claimed the detection of LABs as unambiguous evidence of cold-accretion at $z \simeq 2-3$ \citep{nilsson2006, smith2007, smith2008, erb2011} although see \citep{cantalupo2017} for a summary to the contrary.

(iii) {\it Observing cold flows via Lyman Limit (LLS) and Damped Ly$\alpha$ (DLA) systems}: Simulations show that cold-accretion is a dominant contributor to the cross-section of optically thick LLS ($10^{17} < N_{{\rm HI}} <
10^{20.3}\, {\rm cm}^{-2}$) and low-column density DLAs ($10^{20.3} < N_{{\rm HI}} <10^{21}\, {\rm cm}^{-2}$), with most of this gas never having experienced a virial shock \citep{faucher2011, fumagalli2011, vandevoort2012}. Indeed it has been argued that, despite comprising $\lsim 20\%$ of the LLS population at $z \simeq 2.5-3.5$ \citep{fumagalli2016b, lehner2016}, low-metallicity LLS in the proximity of metal-enriched galaxies present a compelling case for metal-poor cold accretion \citep[e.g.][]{ribaudo2011, crighton2013, fumagalli2016a}. 

(iv) {\it Observing cold flows via absorption line systems}: While individual low-metallicity metal line absorbers, e.g. MgII \citep{giavalisco2011, rubin2012, martin2012}, have been suggested as evidences of cold accretion, it is difficult to make strong statements from such claims given \HI near galaxies does not necessarily need to be inflowing despite its low-metallicity. 

\subsection{The formation of the first stars}
\label{first_sf}
Once the first galaxies gain and cool their gas, the first stars can finally form a few hundred million years after the Big Bang, ending the ``cosmic dark ages" of the Universe. Forming out of gas of primordial composition these PopIII stars bridge the gap between the primordial ($Z \sim 5 \times 10^{-9} \Zsun$) gas predicted by BBN and the lowest-metallicity Population II stars ($Z \sim 0.005-0.05\Zsun$) observed. In addition, these stars could help explain the constant Lithium abundance of metal-poor halo stars (which is lower than BBN predictions) in addition to providing the seeds for super-massive black holes \citep[see e.g.][]{ciardi-ferrara2005}. 

The first stars are predicted to form in low mass ($M_h \simeq 10^6 \Msun$) halos (the so-called minihalos) at $z \simeq 20-30$ 
that, corresponding to $3-4\sigma$ peaks in the gaussian random field of primordial density perturbations \citep[e.g.][]{tegmark1997}, can host dense gas (with hydrogen number density $n_H \gsim 500/{\rm cm^3}$) which can cool on timescales shorter than the free-fall time. Assuming a top-hat collapse and idealised virialized halos, these structures can attain a molecular hydrogen fraction that scales with the virial temperature as $f_{H_2} \propto T_{vir}^{1.5}$ \citep{tegmark1997}. Given that the virial temperature of these halos is $\sim 10^3$ K, i.e. well below the atomic cooling threshold temperature of $10^4$ K, gas cooling in these halos predominantly relies on \HH \citep{saslaw1967}. As early as in 1961, it was realised that \HH can form in the gas phase through the following channel \citep{mcdowell1961}:
\begin{eqnarray}
\label{eqnhminus}
{\rm H}+e^- & \rightarrow & {\rm H}^- + h\nu \\
{\rm H}^- + {\rm H} & \rightarrow & {\rm H_2} + e^-,
\end{eqnarray} 
where the free electrons are either remnants from the epoch of recombination \citep{seager2000} or are produced by collisional ionizations in accretion shocks during galaxy assembly \citep{maclow1986}. \HH can also form through the alternative channel:
\begin{eqnarray}
\label{eqnh2plus}
{\rm H}+{\rm H^+} & \rightarrow & {\rm H_2^+} + h\nu \\
{\rm H_2^+}  + {\rm H} & \rightarrow & {\rm H_2} + {\rm H^+}.
\end{eqnarray} 
However, as a result of ${\rm H}^-$ (Eqn. \ref{eqnhminus}) forming more efficiently as compared to ${\rm H_2^+}$ (Eqn. \ref{eqnh2plus}), the former channel dominates in most circumstances \citep[e.g.][]{tegmark1997}.

One of the most critical quantities pertaining to PopIII stars is their ``characteristic" mass indicating the scale, below which the IMF flattens or declines and, that contains most of the stellar mass depending on the chemical, dynamical and thermal properties of the primordial gas. In the early 2000's, two main numerical approaches were adopted to model PopIII(.1) stars. The first, using AMR, allowed simulating the formation of PopIII stars ranging from cosmological ($\sim 10$ Kpc) to proto-stellar ($\sim 10^{-3}$pc) scales \citep{abel2000, abel2002}. On the other hand, although possessing a smaller dynamical range, the SPH approach allowed modelling sink particles and a wider range of initial conditions \citep{bromm1999, bromm2002}. Irrespective of the initial conditions and methodologies, however, both approaches found PopIII stars to be very massive, with masses ranging between $M_s=30-300\Msun$. They also show a ``preferred" state of the primordial gas, solely arising as a result of the micro-physics of \HH cooling, corresponding to a critical temperature $T_c \sim 200$ K and a critical density $n_c \sim 10^4 \, \mathrm{cm}^{-3}$. This critical temperature arises as a result of the fact that collisional excitations and radiative decays of rotational transitions are responsible for cooling gas at $T\lsim 10^3$ K, with the two lowest \HH rotational energy levels being separated by about 512 K. Although collisions with H atoms populating the high-energy tail of the Maxwell-Boltzmann distribution can lead to slightly lower temperatures, \HH cooling cannot proceed below $T \lsim 100$ K. On the other hand, the critical density arises as a result of the competition between radiative decays (that cool the gas) and collision de-excitations (that do not). Indeed, at $n<n_c$, the cooling rate is $\propto n^2$ , decreasing to $\propto n$ for higher densities. Once these characteristic properties are reached, \HH cooling is rendered inefficient and the gas undergoes a period of quasi-hydrostatic contraction, cooling and collapsing once the Jeans \citep[or Bonnor-Ebert;][]{ebert1955, bonnor1956} mass is reached at
\begin{equation}
M_J \simeq 500 \bigg( \frac{T}{200K} \bigg)^{3/2} \bigg( \frac{n}{10^4 \, \mathrm{cm}^{-3}} \bigg)^{-1/2} \Msun,
\label{pop3_mj}
\end{equation} 
where $T$ and $n$ represent the gas temperature and particle number density, respectively. This gas eventually reaches a density of $n \sim 10^{8-10}\, \mathrm{cm}^{-3}$ at which point three-body \HH formation ($3{\rm H} \rightarrow {\rm H_2+H}$) becomes effective \citep{palla1983}. Continued collapse leads to the gas becoming optically thick to \HH at $n \sim 10^{10}\, \mathrm{cm}^{-3}$ \citep{yoshida2006} before it finally becomes optically thick also in the continuum at $n \sim 10^{16} \, \mathrm{cm}^{-3}$ \citep{yoshida2008}. The cooling then transitions from \HH dissociation cooling to being fully adiabatic once all the \HH is destroyed. We briefly note that PopIII.2 star formation is intrinsically different in the initial stages because a higher ionization fraction in the recombining gas enables enhanced \HH formation. Allowing the gas to cool to a lower temperature, this increases the effectiveness of cooling via HD (deuterated hydrogen molecule) that can cool the gas to the CMB temperature \cite[e.g.][]{mckee2008}. This lower temperature and higher critical density of HD produce lower mass ($\lsim 100 \Msun$) fragments (see Eqn. \ref{pop3_mj}) after which the collapse proceeds as described for PopIII.1 stars. 

\begin{figure}
\center{\includegraphics[scale=0.9]{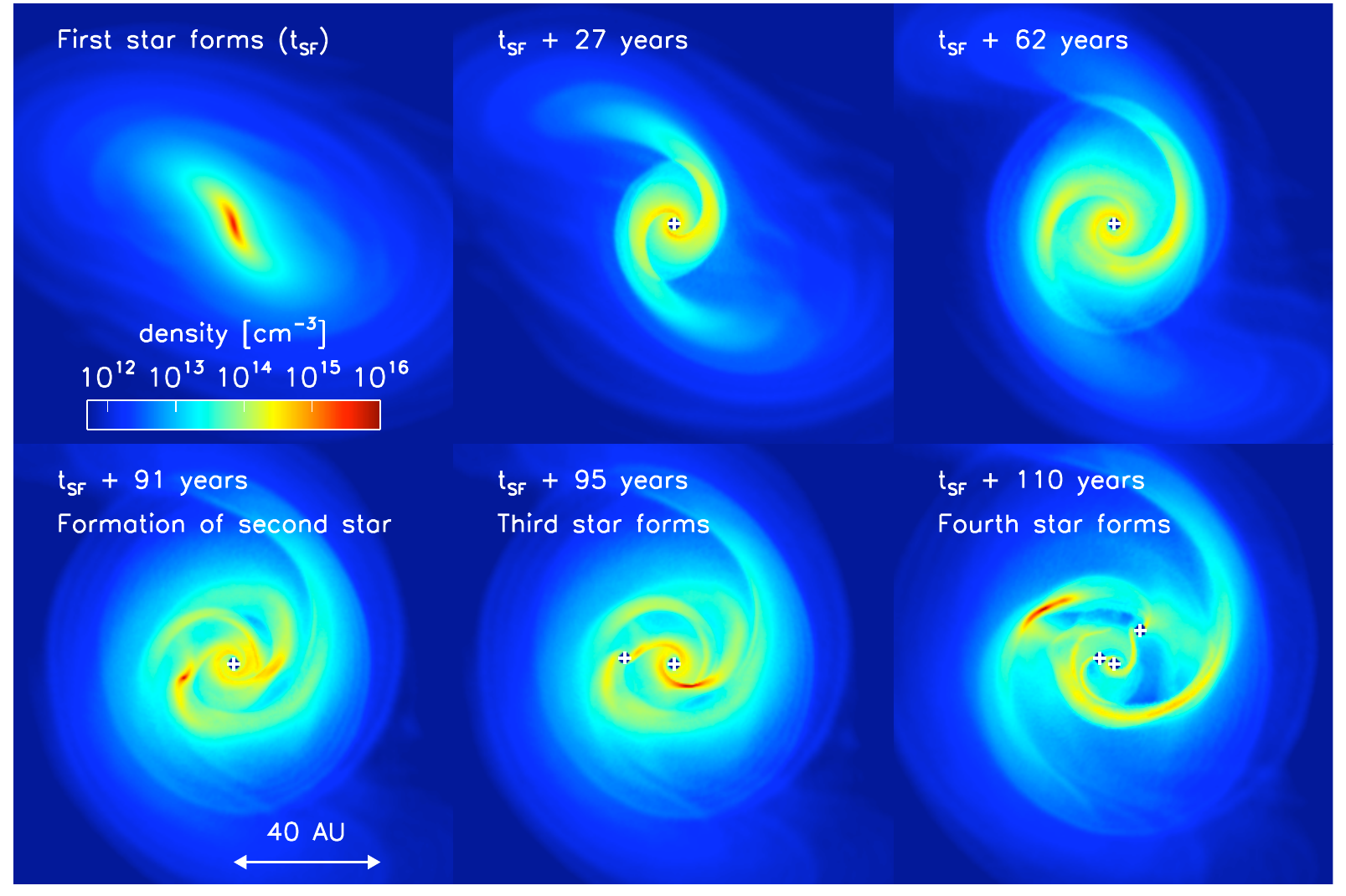}}
\caption{The density evolution in a 120 AU region around the first protostar, showing the build-up
of the protostellar disk and its eventual fragmentation on the order of a few tens of Myrs \citep{clark2011b}; wakes in the low-density
regions are produced by the previous passage of the spiral arms. These results have been obtained by re-simulating a collapsing over-dense region of $1000\Msun$ taken from a 200 kpc SPH simulation implemented with the idea of sink particles, allowing the authors to follow the fragmentation of the proto-stellar disk.}
\label{fig_pop3clark} 
\end{figure}

Initially, it was though that each of these primordial protostellar clouds would form a single star, since no gas fragmentation was seen in simulations \citep[e.g.][]{abel2002, bromm2004b, yoshida2008, mckee2008}, yielding a final PopIII stellar mass of the order of $30-300 \Msun$. However, a paradigm shift occurred in year 2010 due to the growth in brute-force computational power that allowed simulations to use the concept of  ``sink particles'' \citep{bate1995} and follow analyses beyond the formation of a protostar. The main idea was to replace gravitationally-bound gas that had collapsed below the resolution scale by a sink particle of the same mass that would only accrete gas gravitationally. Incorporating sink particles, both SPH \citep{stacy2010, clark2011, clark2011b} and AMR \citep{turk2009, machida2009, greif2011, prieto2011,xu2013} simulations started showing gas settling into a protostellar disk around a hydrostatic core. This accretion disk is highly susceptible to gravitational fragmentation, as shown in Fig. \ref{fig_pop3clark}, even for velocity dispersions as low as 20\% of the sound speed, with a more active and widespread fragmentation in PopIII.1 halos \citep{clark2011}. The multiplicity (fragmentation into binary or multiple-stellar systems) and normalisation of the resulting mass function depends on details including the sink implementation \citep{greif2011}, gas metallicity - with lower metallicity gas being more prone to fragmentation \citep{machida2009} - and the simulation resolution \citep{turk2012}. Such multiplicity, however, could possibly be reduced by the late-time viscous merging of fragments in the disk as shown by recent works \citep[e.g.][]{hirano2017}.

Interestingly the mass of the initial PopIII hydrostatic core corresponding to $M_{core} \sim 5 \times 10^{-3}\Msun$, forming in the centre of the dense ($n_H \sim 10^{22} \, \mathrm{cm}^{-3}$) collapsing cloud, is almost the same as that inferred for PopII star formation \citep{omukai1998} since it essentially corresponds to the Jeans mass within the accreting protostellar disk. The final mass, however, depends on the accretion rate, $\dot M_{acc} \propto c_s^3/G \propto T^{3/2}$, a result that has now been confirmed using three-dimensional numerical simulations \citep{oshea2007, wise2008}. A comparison of typical gas temperatures in the birth clouds of PopIII ($T \sim 200$ K) and PopII ($T \sim 10$ K) stars clearly shows that the former accrete at a rate that is about two orders of magnitude higher than PopII star formation. The accretion rate is nevertheless limited by the Eddington luminosity, $L_{edd}$, such that 
\begin{equation}
G \frac {M_p}{R_p} \dot M_{acc} \simeq L_{edd} \rightarrow \dot M_{acc} \simeq \frac{L_{edd}}{G} \frac{R_p}{M_p}, 
\label{macc_prot}
\end{equation} 
where $M_p$ and $R_p$ are the mass and radius of the central protostar. Using a value of $R_p = 5 R_\odot$, typical for PopIII stars, yields $\dot M_{acc} \simeq 5 \times 10^{-3} \Msun \, \mathrm{yr}^{-1}$, a result that has also been confirmed numerically \citep{abel2002, omukai2003, yoshida2006, oshea2007}. With accretion, the mass of the protostellar core grows as $M_p \sim \int\dot M_{acc} dt \propto t^{0.45}$ which naturally implies that objects forming at lower redshifts have larger accretion rates as also shown by simulations \citep{oshea2007}. However, it must be noted that disk fragmentation produces strong gravitational torques that cause the accretion rate to become extremely variable \citep{clark2011b}. If the accretion rate becomes sub-critical early on, the dramatic expansion of $R_p$, that can stop accretion (see Eqn. \ref{macc_prot}), can be avoided resulting in the star reaching a final mass as high as $700 \Msun$ at an age of about 3Myr \citep{bromm-larson2004}. Interestingly, two-dimensional simulations, that do not follow disc fragmentation or non-asymmetric three-dimensional effects, obtain similar results with a final stellar mass of the order of $1-1000\Msun$ \citep{hirano2014}. 

Whether or not such a high mass can be achieved, however, depends on a number of heavily debated feedback effects, such as \HH photodissociation and radiation pressure from Ly$\alpha$ photons, that can terminate accretion. On the one hand, \citet{mckee2008} have shown that feedback effects do not significantly reduce the accretion rate until the ionized hydrogen (\HII) region extends beyond the gravitational escape radius which generally occurs once the star has already reached $50-100\Msun$, with a similar range of PopIII masses ($\sim 1-100\Msun$) being found by more recent hydrodynamic simulations \citep{susa2014}. However, hydrodynamic simulations also show scenarios where (i) most of the PopIII ionizing radiation escapes along the poles \citep{hosokawa2011}, photo-evaporating the disk and halting accretion at about 0.7 Myr, resulting in a final PopIII stellar mass of $43 \Msun$ \citep{yoshida2008}; or (ii) accretion rates drop, although gas can still fragment forming binaries, once the region of gas photo-heated by LW reaches the disk size, limiting PopIII stars to having masses $\sim 0.05-20\Msun$ \citep{stacy2012, stacy2016}. These authors find a wide range of PopIII masses resulting in a broad, logarithmically-flat(ish) top-heavy IMF \citep[see also][]{turk2009, stacy2010, clark2011} as shown in Fig. \ref{fig_imf_pop3}.

\begin{figure}[h]
\center{\includegraphics[scale=0.9]{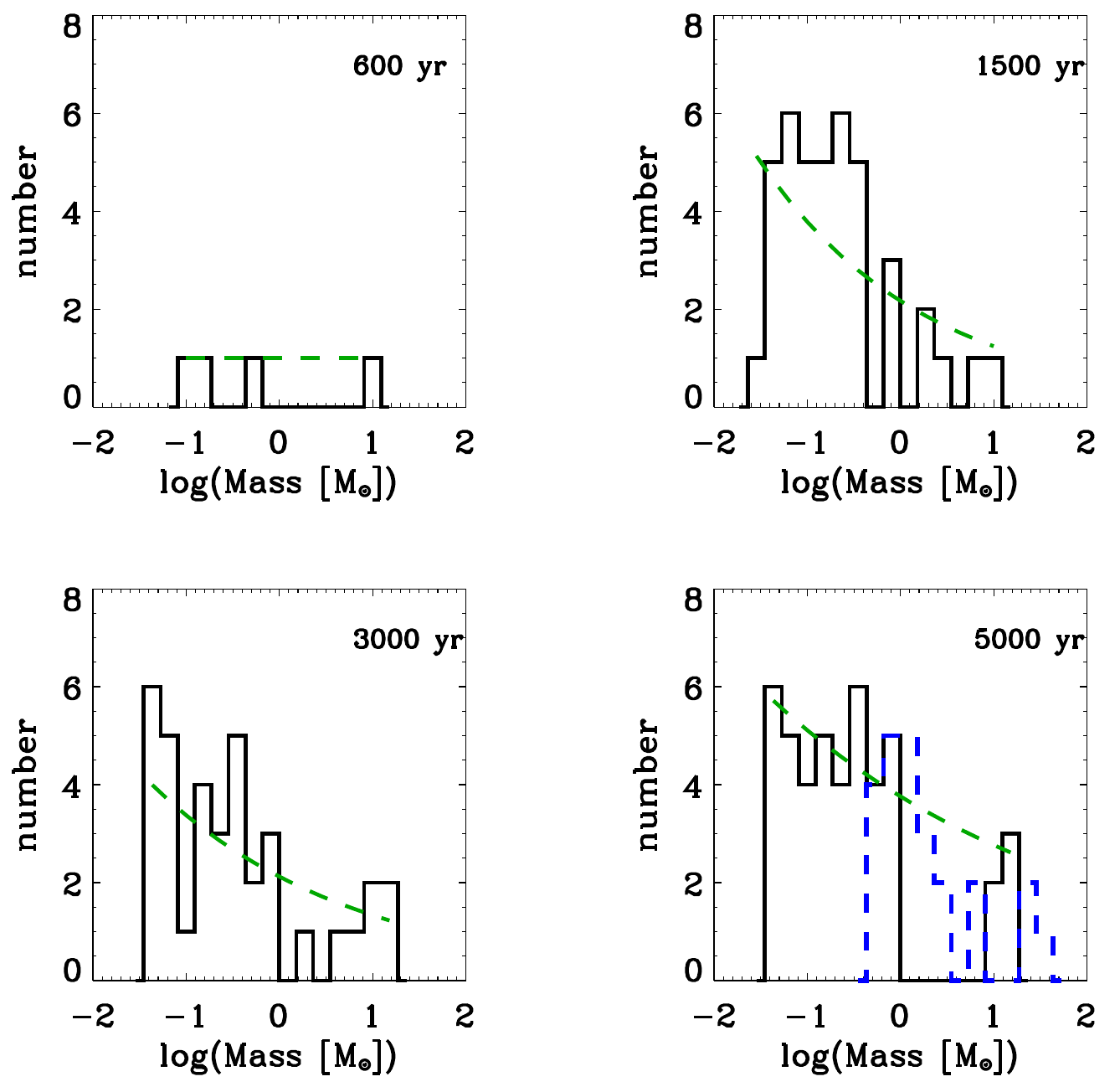}}
\caption{Simulating PopIII stellar systems starting from cosmological initial conditions and a resolving scales as small as 1AU, \citep{stacy2016} show the time evolution - from 600-5000 yr - of the the sink mass function (black histograms) that clearly shows the fragmentation of metal-free gas. The dashed green line in each panel shows the least-square fit to the function $dN/dm \propto M_p^{-x}$ where $x=(0,1.24, 1.2, 1.13)$ at $t=(600,1500,3000, 5000)$ yrs as shown. The dashed blue line in the last panel shows results from a lower resolution simulation without feedback.}
\label{fig_imf_pop3} 
\end{figure}

Given the hierarchical assembly of galaxies in the $\Lambda$CDM model the feedback from these PopIII stars, significantly impacting the formation and evolution of subsequent galaxy generations \citep[e.g.][]{wise2008b, greif2010}, can be broadly divided into 3 main categories \citep{ciardi-ferrara2005}: 

(i) {\it Chemical feedback}: PopIII stars in the mass range of $140-260\Msun$ explode as Pair Instability Supernovae (PISN), with the range being extended down to $85 \Msun$ for a rapidly rotating progenitor \citep{chatz2012}, completely disrupting the progenitor \citep{fryer2001}; PopIII stars outside this range end up as black holes. These PISN have extremely large metal yields, on average converting roughly half their mass into metals, with iron making up the largest portion of the metal mass for the most massive PISN \citep{heger2002}. It has been numerically shown that, with PISN explosion energies of the order of $10^{53} \, {\rm ergs}$, roughly 90\% of these metals can escape their host minihalos and enrich the neighbouring $\sim 1 $kpc to a metallicity $Z \gsim 10^{-2}\Zsun$ about 4 Myrs after the explosion \citep{bromm2003, chiaki2018}. Simulations of SN blastwaves \citep{greif2007, wise2008} show that the early enrichment from PopIII stars is very inhomogeneous with low-density voids being enriched earlier than higher density filaments and virialised halos \citep{cen2008}, extending the era of PopIII star formation. It is therefore clear that, enriching the Universe, albeit patchily, with its first metals, these PopIII stars mark the start of the end of metal-free star formation. The fine structure line cooling from singly ionized C (CII) and neutral oxygen (OI) enables gas to cool, to lower temperatures \citep{bromm2003}, and fragment leading to a shift from a top-heavy PopIII to a normal bottom-heavy metal-enriched PopII/I star formation once the critical metallicity threshold of $Z_{crit} \simeq 10^{-3.5} - 10^{-5\pm1}$ is exceeded \citep{bromm2001, schneider2002, schneider2006, santoro2006, smith_zcrit2007}. Further fragmentation can be achieved via dust continuum cooling if the dust-to-gas ratio exceeds a value of $2.6-6.3 \times 10^{-9}$ \citep{schneider2012, chiaki2014}. The exact value of $Z_{crit}$ is rendered unimportant by the fact that even a few SN explosions can enrich a $10^8\Msun$ halo above $Z_{crit}$ \citep{wise2008b, karlsson2008, greif2010, maio2011, jeon2015}. However, even in the presence of metals, the minimum temperature floor is set by the CMB, $T_{\gamma} = 2.732 (1+z)$, resulting in a $z \simeq 15$ characteristic stellar mass of 
\begin{equation}
M_s \sim 20 \frac{n_f}{10^4 \mathrm{cm}^{-3}} \Msun,
\end{equation}
where $n_f$ is the density at which gas opacity prevents further fragmentation, giving rise to the idea of PopIII.2 star formation modulating the gap between PopIII and PopII stars \citep[e.g.][]{mackey2003}.  

(ii) {\it Mechanical feedback}: PISN have explosion energies $E_{PISN} \sim 10^{51-53} \, {\rm erg}$ \citep{heger2002}, which is at least an order of magnitude larger than that produced by ordinary metal-enriched Type II SN \citep[SNII;][]{woosley1986}. Such high explosion energies are larger than the binding energy of low-mass halos at early cosmic epochs (e.g. $\sim 3\times 10^{50}\, {\rm erg}$ for a $10^6\Msun$ halo at $z \sim 20$), leading to a complete disruption of gas mass in the host halo \citep{greif2007, wise2008}, suppressing further star formation for a Hubble time $\sim 300\, {\rm Myrs}$ at $z\sim 10$ \citep{greif2007, jeon2015}. However, \citet{muratov2013} show that the effects of such outflows are temporary, with cosmological inflows of gas restoring the baryon fraction to its universal value. Further, these authors show that, both, metal production and stellar feedback from PopII stars overtake those from PopIII stars in a few tens to hundreds of Myrs, depending on the galaxy mass. 
However, it must be noted that SN effects, although being debilitating for the host halo, could have exerted a positive mechanical feedback on neighbouring minihalos by shock-compressing the gas in their cores \citep{greif2007}. 

(iii) {\it Radiative feedback}: PopIII stars have a very hard spectrum, measured as the ratio of singly ionized He and hydrogen ionizing  photons. The ionizing rate ratio is $Q(He^+)/Q(H) \sim -1.4$ for PopIII stars as compared to $\sim -2.6$ for stars with $Z=0.25\Zsun$. PopIII stars also produce about 3 times more hydrogen ionizing photons compared to stars with $Z \simeq \Zsun$ for the same underlying IMF \citep{schaerer2003}. As a result, PopIII stars initiate both the epoch of hydrogen and helium reionization. Following the growth of \HII regions from PopIII stars using radiation hydrodynamic simulations, a number of works have shown that a $100\Msun$ star can create an \HII\, region of about $3 \, {\rm kpc}$ \citep{kitayama2004, whalen2004,alvarez2006}. The extremely high IGM temperatures of about $T \sim 20,000$ K in such ionized regions can result in the photo-evaporation the gas from low-mass halos as discussed in Sec. \ref{uvb_fb} that follows. Indeed, while the threshold virial mass for PopIII star formation is effectively independent of redshift in the absence of a heating UVB, varying between $1.5-7 \times 10^5\Msun$ at $z \sim 19-33$ \citep{oshea2007}, it increases with decreasing redshift as more of the IGM is ionized. The \HII\, ionization front can also drive shock waves (with a speed approximately equal to $\sim 30$ km/s) that can expel about 90\% of the gas from minihalos (escape velocity $\sim 3$ km/s), leaving behind a warm ($T \sim 3 \times 10^4$ K), diffuse ($0.1 \, \mathrm{cm}^{-3}$) medium, delaying subsequent star formation by as much as 100 Myrs \citep{johnson2007}. 

The UV radiation from the first stars would have also established a cosmological LW background radiation field \citep[e.g.][]{ciardi2000}. \HH can be easily dissociated by LW photons through the Solomon process (${\rm H_2+ \gamma \rightarrow H_2^* \rightarrow 2H}$), where ${\rm H_2^*}$ is an excited electronic state. Although galaxies more massive than the earliest $\sim 10^6 \Msun$ minihalos could self-shield, the LW background could photo-dissociate their \HH at a rate $\propto J_{21} \mathrm{min}[1,(N_{H_2}/10^{14} \mathrm{cm}^{-2})^{-0.75}]$ (with the first and second terms representing the LW background intensity and the shielding factor) in a static medium \citep[e.g.][]{draine1996, bromm-larson2004}. In the absence of \HH, cooling can proceed via atomic transitions in $T_{vir}>10^4$ K halos where gas at the virial temperature can collapse into massive ($\sim 10^6\Msun$) compact clouds in the centre of the halo, providing the seeds for the Super Massive Black Holes (SMBH; $M_{BH} \sim 10^9\Msun$) \citep{regan2009, volonteri2012} observed by the Sloan Digital Sky Survey (SDSS) at $z \sim 6$ \citep{fan2006}. However, as mentioned above, since \HH formation is catalysed by the presence of free electrons, processes such as X-ray production from SN remnants \citep{oh2001}, accretion onto black holes \citep{machacek2003}, the presence of \HII regions \citep{oh2003} and collisional ionizations driven by structure formation \citep{maclow1986} can also have a positive effect on \HH formation. We end by noting that if PopIII stars had only reached masses up to a few tens of $\Msun$ \citep[as per the calculations of][]{hosokawa2011, stacy2012, susa2014}, their disruptive effects would be less important than those discussed here.

We now briefly discuss observations of metal-free stars. We start with the field of ``stellar archaeology" which focuses on hunting for metal-poor stars in the local Universe and we refer readers to a recent review \citep{frebel2015} for more details. In brief, the idea is to use the fact that the metallicity of a star reflects the chemical conditions of its birth clouds - the chemical abundance of a star can therefore be used as a proxy for its age. Combining this with the fact 
that the lifetime of a star (of mass $M_s$) scales roughly as $10(\Msun / M_s)$ Gyr, observations aim to hunt for $M \lsim 0.8\Msun$ stars, with a lifetime $\sim 19$ Gyr, in the Galactic halo and in satellite dwarfs, both, to find metal-free stars and to get hints on the (metal) enrichment of the Universe in its first few hundred million years. Moving farther to high-$z$, as recognised by \citet{partridge1967a}, the earliest galaxies can be confirmed through nebular recombination lines (see Sec. \ref{obs_gal}) including Ly$\alpha$ (1216 \AA) and HeII (1640 \AA). Starting with Ly$\alpha$, \citet{kashikawa2012} have detected a LAE at $z=6.538$ with a Ly$\alpha$ equivalent width of $EW = 872^{844}_{-298}$\AA\, that either implies an age $\lsim 4$ Myr and a metallicity of $Z \lsim 10^{-5}\Zsun$ or a clumpy dust distribution. Additionally, as a result of their extremely hot atmospheres \citep[e.g.][]{schaerer2003}, PopIII stars produce copious amounts of HeII ionizing photons that can result in a HeII recombination line with EW of $\gsim 100$ \AA \citep[e.g.][]{johnson2009}. A number of works have now shown that the {\it JWST} will be able to detect the signatures of PopIII star formation in the high-$z$ Universe. This could be achieved by detecting the HeII emission line from $z=10$ galaxies containing a minimum PopIII stellar mass of $10^5\Msun$. Alternatively,  using an integration time of 100 hours \citep{zackrisson2011a} the {\it JWST} could detect both the H$\alpha$ and HeII lines in $10^9\Msun$ halos if the PopIII IMF is top-heavy \citep{pawlik2011}. Finally, it is possible to identify Pop III galaxies at $z=8-10$ from  their peculiar colors resulting from the presence of strong (Ly$\alpha$ and H$\alpha$) nebular emission lines and the absence of any metal emission lines (e.g. [OIII]$\lambda5007$ \AA). This method could be sensitive to Pop III mass fractions as low as $\lsim 1\%$ of the total stellar mass \citep{zackrisson2011b}. These observations could also be rendered more sensitive using gravitational lensing to boost the luminosity of faint background sources \citep[e.g.][]{windhorst2018}. In addition to line emission, the extremely luminous PISN expected from PopIII stars ($\sim 130-250\Msun$) could be detected by the hundreds by the {\it JWST} \citep{mesinger2006}. The challenge in this technique lies in building a statistical sample across multiple fields and separated by a long enough cadence to study the slowly evolving light-curves. Finally, the technique of intensity mapping - that focuses on observing the cumulative large scale fluctuations in intensity from faint, undetectable sources - could be employed by a future 2m space-based ultraviolet observatory to hunt for PopIII star formation at $ z=10-20$ using multiple lines (e.g. HeII-CO or HeII-21cm) in order to spatially correlate fluctuations and eliminate contaminants \citep{visbal2015}.

\newpage

\section{Heavy elements: production, transport and ejection}
\label{ch6}
BBN is a key component of the Hot Big Bang model that explains the synthesis of light elements in the early Universe \citep[for a detailed explanation see e.g.][]{kolb1990}. Taking place at an epoch when temperatures ranged between $\sim 10-0.1$ MeV, BBN predicts that a substantial mass fraction of $^4 {\rm He} = 0.2485 \pm 0.0002$ and trace amounts (of the order of $10^{-5}$) of Deuterium/H and $^3{\rm He/H}$ formed about 3 minutes after the Big Bang, using the latest {\it Planck} cosmological parameters \citep{planck2014}. As the expansion of the Universe cooled matter, thermonuclear fusion could not proceed efficiently beyond $^7{\rm Li}$ resulting in an extremely low abundance of $^7{\rm Li/H} \sim 10^{-9}$; BBN is inefficient in generating $^6{\rm Li}, ^9{\rm Be}$ and $^{10,11}{\rm B}$. Moving further along the atomic table, Carbon is produced along inefficient paths involving $^{11}{\rm B}$ with radiative capture of $^{12}{\rm C}$ leading to production of trace amounts of Nitrogen and Oxygen with abundances of $(^{12}{\rm C}, ^{14}{\rm N}, ^{16}{\rm O})/{\rm H} = (4.4 \times 10^{-16}, 2.6 \times 10^{-17}, 1.8 \times 10^{-20})$ \citep[e.g][]{iocco2007}. Therefore metals heavier than $^4{\rm He}$ were only formed in substantial quantities after the first stars started the ``metal-age" of the Universe at a (median) redshift of $z=65.8$, roughly 30 Myrs after the Big Bang, at which there is a 50\% probability of finding the first star \citep[e.g.][]{naoz2006}. Producing the first heavy metals after BBN, these stars paved the way for the transition from PopIII to metal-rich PopII star formation as explained in what follows.

\subsection{Metal production and mixing in galaxies}
\label{metal_prod}
Before discussing the metal production in (PopIII and PopII) stars, we introduce a few key concepts. Firstly, the metallicity of gas is defined as the ratio of the total metal mass (of elements heavier than $^4{\rm He}$) to the total gas mass such that $Z = \sum_{i>He} M_i/M_{gas}$. This can be normalised to the solar metallicity that has a value $\Zsun = 0.0122$ \citep{asplund2005}. The abundance of a specific element (say, Iron) relative to the Sun can then be expressed as $$[{\rm Fe/H}]= {\rm Log}\bigg(\frac{n_{He}}{n_H} \bigg) - {\rm Log} \bigg( \frac{n_{Fe}}{n_H}\bigg)_\odot.$$ The amount of metals produced by a given stellar population depends on a number of key parameters including the stellar mass, the initial stellar metallicity, the SN explosion energy and the IMF. 

As noted in Sec. \ref{first_sf} the first stars formed out of gas of primordial composition with $Z<Z_{crit} \sim 10^{-3.5}-10^{-5\pm1} \Zsun$ \citep[e.g.][]{schneider2006, smith_zcrit2007}, had masses of up to $300 \Msun$ and a top-heavy IMF. PopIII stars in the range of $140-260\Msun$ explode as PISN, releasing roughly half their mass into metals \citep{heger2002}. We can carry out a simple order-of-magnitude estimate of the volume that can be metal-polluted by such PISN following the calculations in \citet{ferrara2016}: let us assume that the total amount of metals produced by a PISN is $M_z \approx 2 E_{51}\Msun$ where the SN explosion energy is expressed in terms of $E_{51} = E/10^{51} {\rm erg}$. Assuming a fraction $f_w$ of the explosion energy to be converted into kinetic energy, these metals will be dispersed over a volume $V = f_w E / \rho c_s^2$ where $\rho$ and $c_s$ are the density and sound speed in the surrounding medium, respectively. This can be used to obtain an estimate of the corresponding shell mass enriched by metals as $M_m = \rho V = f_w E/c_s^2$. Using typical values, of $c_s = 10$ km/s and $f_w=0.1$ appropriate for high-$z$ galaxies \citep[see e.g.][]{dayal2014a}, yields a shell metallicity of $\langle Z\rangle = M_Z/M_m = 3.2 \times 10^{-3} \Zsun \gg Z_{crit}$. Although the shell mass used here is an upper limit, given that radiative losses can decelerate the progress of the shock wave and we have assumed perfect mixing, a single PopIII star with an explosion energy of $E = 10 E_{51}$ can enrich the surrounding $5 \times 10^5 \Msun$ of gas above the critical metallicity, quenching all further PopIII star formation. 

Continuing along these lines, we can also calculate the volume density of PISN required to enrich all of the gas in the Universe at $z=15$ (at critical density) to $Z_{crit} = 10^{-5}\Zsun$, assuming each PISN to have a mass of $100 \Msun$, 45\% of which is converted into metals, as:
\begin{equation}
\frac{1}{V} = \frac{\rho(z=15) Z_{crit}}{M_Z} \sim 10 \, {\rm Mpc^{-3}}.
\end{equation}
This implies that 1 PISN in every $\approx 10^8 \Msun$ halo would be sufficient to enrich the entire IGM to $Z_{crit}$, modulo the fact that we have neglected metal enrichment from PopII stars and assumed all metals from PopIII to be homogeneously distributed throughout the IGM. In reality, however, the clustering of $3-4\sigma$ density fluctuations, that host the first stars, results in the densest regions being patchily metal-enriched first as shown by simulations and discussed in Sec. \ref{met_igm} that follows. However, a number of simulations have shown that, given that PopIII stars contribute less than $1-10\%$ to the total star formation at redshifts as early as $z \simeq 15$ \citep{maio2010}, they have a negligible contribution to the total metal budget of the Universe \citep{tornatore2007, oppenheimer2009, xu2013, pallottini2014}. We now discuss metal production from PopII stars which are regarded as the key cosmic metal factories. 

The IMF-averaged PopII SN yield of a given heavy element, $y_i$, can be calculated as
\begin{equation}
y_i = \frac{\int_{M_l}^{M_u} M_i \phi(M) dM}{\int_{M_l}^{M_u} \phi(M) dM},
\end{equation}
where $\phi(M)$ expresses the IMF which has a slope of -2.35 for the commonly used Salpeter IMF \citep{salpeter1955}. Further, $M_l$ and $M_u$ express the lower and upper mass ranges of stars that can explode as SNII and contribute to the metal content. However given that stars with $M_s = 50 \pm 10 \Msun$ form black holes without contributing to the metal content \citep{tsujimoto1995} and the uncertain pre-SN evolution of $8-11\Msun$ stars, SN metal yields have only been tabulated between $12-40\Msun$ \citep{woosley1995}. For this specific case, the number of SNII per unit stellar mass is found to be $\nu = [280 \Msun]^{-1}$. We tabulate the yields of some common elements in Table \ref{table_yields} - we note that while metal yields are almost independent of metallicity for $10^{-4} \lsim Z \lsim 10^{-2}$, they vary by about 10\% above this value \citep{woosley1995} as shown in the table.

\begin{table}
\caption{SN metal yields [Oxygen (column 5), Silicon (column 6), Sulphur (column 7), Iron (column 8) and total (column 9)] for a range of SN models (column 1) using a Salpeter IMF with the lower and upper SN mass limits shown in columns 2 and 3, respectively, and the initial metallicity values shown in column 4. The IMF limits for the entire stellar population range between 
$0.1-100\Msun$ for SN \citep{woosley1995} and between $0.07-100 \Msun$ for the SNII+HN (hypernova) case \citep{nomoto2006}, yielding a SNII rate of $\nu = [280 \Msun]^{-1}$ SNII. In this specific table, solar metallicity corresponds to $\Zsun = 0.02$. }
\begin{tabular}{|c|c|c|c|c|c|c|c|c|c|}
\hline
SN & $M_l$ & $M_u$ & $Z$ & $y_O $ & $y_{Si} $ & $y_{S} $ & $y_{Fe}$ & $y_Z $ & Reference \\
& $[\Msun]$ & $[\Msun]$ & $ [\Zsun]$ & $ [\Msun]$ & $[\Msun]$ & $ [\Msun]$ & $ [\Msun]$ & $[\Msun]$ &  \\
\hline
SNII & 12 & 40 & $10^{-2}$ & 1.430 & 0.133 & 0.064 & 0.136 & 2.03 & \citep{woosley1995} \\
SNII & 12 & 40 & $1$ & 1.560 & 0.165 & 0.078 & 0.115 & 2.23 & \citep{woosley1995}\\
SNII+HN & 13 & 40 & 1 & 1.806 & 0.144 & 0.061 & 0.108 & 3.11 & \citep{nomoto2006}\\
SNIa &  &  & $$ & 0.148 & 0.158 & 0.086 & 0.744 & 1.23 & \citep{gibson1997}\\
\hline
\end{tabular}
\label{table_yields} 
\end{table}

Using the same calculations as carried out for PopIII star formation above, and using the yield and $\nu$ values from Table. \ref{table_yields}, we can also calculate the average metallicity of a galaxy after an epoch of PopII star formation resulting in a total stellar mass $M_*$ as
\begin{equation}
\langle Z\rangle = y_Z \nu M_* \times \frac{c_s^2}{f_w E_{51}}.
\end{equation}
Plugging in a reasonable estimate for the star formation efficiency corresponding to 0.3\% \citep{dayal2014a} (which yields $M_* = 1.4 \times 10^6$), again using $f_w=0.1$, $c_s = 10$ km/s, and an initial stellar metallicity of $10^{-2}\Zsun$, we find $\langle Z \rangle \sim 0.16\Zsun$ for SNII metals only. This simple estimate assumes that most of the metals mix in the galaxy and do not escape from it via outflows (i.e. a closed-box solution), so it is perhaps an upper limit to $Z$ for a given efficiency/yield as discussed in Sec. \ref{met_igm} that follows. 

However, the details of how metals produced in stars are distributed and mixed into the surrounding gas and on increasingly large scales remains an outstanding question. Historically, this process has represented a major factor of uncertainty in building galaxy evolution models, and particularly so when numerical simulations are involved. This is because the mixing process involves a number of complex phenomena, such as conductive interfaces, dynamical instabilities, diffusion and turbulence cascades, acting on an extremely wide range of scales in a multi-phase ISM. As a side note, the term mixing is often used synonymously with dispersal in the literature; the two terms, however, involve physically distinct processes affecting different spatial scales. While dispersal strictly refers to the transport of bulk enriched material on (super-)galactic scales independently of its degree of homogeneity, mixing is produced by the above mentioned microscopic processes enforcing a homogeneous chemical composition. 

Metals initially contained in dense supernova ejecta are separated from the ambient medium by a contact discontinuity. The first step is then to break such separating surfaces. When the cooling time in the cavity becomes shorter than its age, a cooled shell will form that rapidly becomes unstable \cite{Gull73, madau2001} to Rayleigh-Taylor (RT) and Kelvin-Helmoltz (KH) instabilities. At this stage metals are distributed in fragments whose size depends on the energetics of the SN explosion and the thermodynamical properties of the ambient medium.The RT fingers produced by RT instabilities penetrate deep into the high metallicity gas and may be eroded by KH instabilities due to the passage of rapidly moving hot gas, i.e., a mixing layer \cite{Slavin93}. The shear flows produced in the mixing layer efficiently mix the freshly produced metals with more pristine gas.  These fragments are then processed by thermal conduction (as they are still immersed into the hot cavity gas) and shocks. Both processes heat up the metal enriched gas until a sort of equilibrium temperature, close to the geometric mean of the fragments and cavity gas, is attained. This gas will then cool by emitting characteristic optical, infrared and ultraviolet lines. Cooling often involves non-equilibrium ionization and self-photoionization. 

The transfer of heat eventually balances the temperature of the processed ejecta and ambient gas. However mixing requires also transfer of matter. This process results from diffusion \cite{Tenorio96} eventually resulting in a perfect mixing of the new heavy elements with those already present in the gas.  While a complete theory of mixing is lacking, a simple dimensional analysis can provide some insight. The characteristic diffusion time over a length $L$ is $t_D = L^2/D$, where $D$ is the diffusion coefficient which is usually expressed as the product of the particle mean free path, $\ell$, and the r.m.s. velocity, $\sigma = (3 k T/\mu m_H)^{1/2}$. Using standard values for the density and temperature of the hot halo and the \HI disk, on finds that the diffusion time scale is much shorter in the halo ($t_D \approx 4\times 10^5$ yr) than in the disk ($t_D \approx 6 \times 10^{11}$ yr). The fact that the mixing timescale in the disk is even larger than the present Hubble time raisses a thorny problem. Thus, it seems that efficient mixing can be guaranteed only if enriched matter cycles from the star forming regions into the halo and then rains back in a sort of galactic fountain \cite{Pezzulli16}. 

Alternatively one can envisage the presence of strong turbulent motions, initially suggested by \citet{Roy95} whose natural mixing timescale would be the eddy turnover time, $L/\sigma_T$, where $\sigma_T$ is the r.m.s. velocity of the turbulent fluid. In practice, this process is akin to that at work in the mixing layers discussed above, and can uniformly mix metals on scales down to a pc. Along these lines, \citet{deAvillez02} used simulations to compute the timescale for the variance decay of concentration fields with no continuing sources in SN-driven interstellar turbulence. However, they did not give a detailed study or discussion of the central physical issues in turbulent mixing, such as the production of small-scale scalar fluctuations and their effect on the mixing timescales. Similarly, while \citet{Klessen03} considered the dispersal of tracer particles in supersonic turbulence, they did not study mixing in the sense of homogenization of concentration fluctuations.

Detailed modelling of metal mixing is not particularly abundant in the literature. Recent hydrodynamics simulations have employed parametric models to account for
unresolved sub-grid metal mixing. One of the most successful examples is given by the approach developed by \citet{Pan10}, who have carried out the first systematic numerical study of passive scalar mixing in isothermal supersonic turbulence. Other studies have, instead, concentrated on the effects of mixing on the global chemical properties of galaxies  (e.g. \cite{Feng14, Armillotta18}, and in the enrichment process and chemical signatures of the first stars, e.g. \cite{Cen08, Jeon15, Smith15, Sarmento18}). More sophisticated hydrodynamics 
simulations incorporating a realistic multi-phase ISM, and improved microphysics and turbulence treatment are required to understand the details of metal mixing in galaxies.

\subsection{Galactic winds and the enrichment of the IGM}
\label{met_igm}
Radiative cooling is very efficient in dense low-mass halos at high-$z$. Left unchecked, this leads to an over-production of stars and too many baryons being locked up in condensed halos (as compared to observations), a problem canonically termed  ``overcooling'' \citep{benson2003, springel-hernquist2003}. This problem can be alleviated by introducing SN feedback that can reduce the star-formation efficiency of small halos by ejecting their gas and quenching further star formation \citep[e.g.][]{maclow1999, springel2003,greif2007}. We start by discussing the maximum amount of star formation that can be supported by a halo, purely based on equating the halo binding energy to the SN ejection energy \citep{dayal2014a}. The formation of an amount $M_*(z)$ of stars at redshift $z$ can impart the ISM with a total SNII energy ($E_{SN}$) given by
\begin{equation}
E_{SN} = f_w E_{51} \nu M_*(z) \equiv f_w V_s^2 M_*(z),
\end{equation}
where each SNII is assumed to impart an (instantaneous) explosion energy of $E_{51}$ to the ISM. Further, assuming all stars with mass larger than $8 \Msun$ to explode as SNII yields $\nu = [134 \, {\rm \Msun}]^{-1}$ for a Salpeter IMF between 
$0.1-100 \Msun$. The values of $E_{51}$ and $\nu$ yield $V_s= 611$ km s$^{-1}$. Finally, $f_w$ is the fraction of the SN explosion energy that is converted into kinetic energy and drives winds. 

For any given halo, the energy $E_{ej}$ required to unbind and eject all the ISM gas can be expressed as
\begin{equation}
E_{ej} = \frac{1}{2} [M_{g,i}(z)-M_*(z)] V_e^2,
\end{equation}
where $M_{g,i}(z)$ is the gas mass in the galaxy at epoch $z$; the term $M_{g,i}(z)-M_*(z)$ 
accounts for the fact that SN explosions have to eject the part of the initial gas mass not converted into stars. 
Further, the escape velocity $V_e$ can be expressed in terms of the halo rotational velocity, $V_{vir}$, as $V_e = \sqrt 2 V_{vir}$.

\begin{figure}[h]
\center{\includegraphics[scale=0.48]{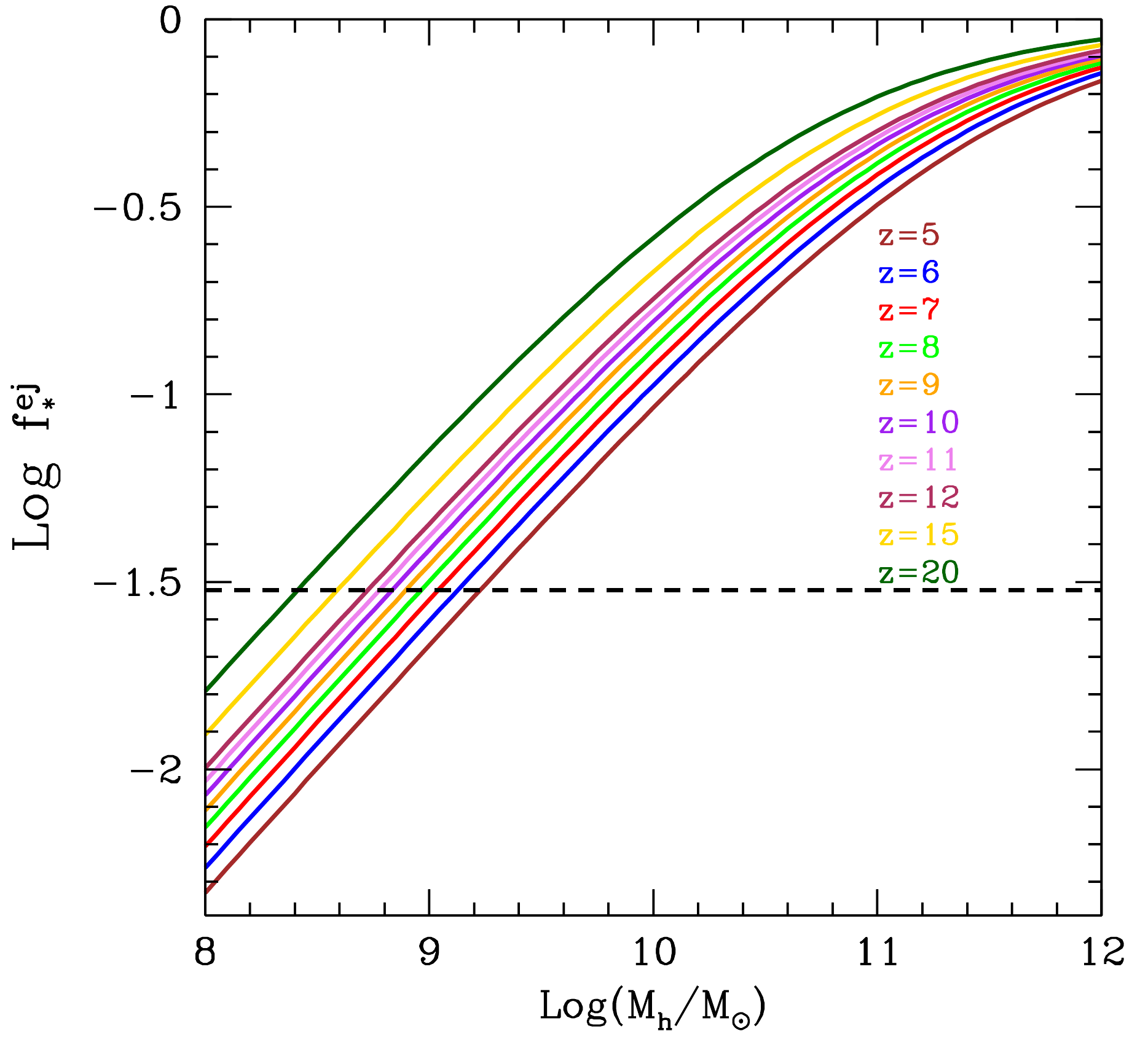}}
\caption{The ejection efficiency ($f_*^{ej}$) as a function of halo mass for $z \simeq 5-20$; this is the star-formation efficiency required to eject all the gas from the galaxy and quench further star formation. The horizontal line shows the upper limit of $f_* = 0.03$ required to match to high-$z$ observations (Sec. \ref{lf}): we assume galaxies with $f_*^{ej}>0.03$ saturate at an effective efficiency of $f_*^{eff} = f_* = 0.03$. We have assumed each SN imparts an explosion energy of $E_{51}=10^{51}{\rm erg}$, of which a fraction $f_w=0.1$ drives winds and the SN rate ($\nu$) is calculated assuming a Salpeter IMF between $0.1-100\Msun$. These values of $E_{51}$ and $\nu$ lead to $V_s = 611$ km s$^{-1}$ as detailed in the text. Further, we note that as a result of the deepening potential wells, halos of a given mass can sustain higher $f_*^{ej}$ values with increasing redshift.}
\label{ch5_feff} 
\end{figure}

We then define the {\it ejection efficiency}, $f_*^{ej}$, as the fraction of gas that must be 
converted into stars to ``blow-away" the remaining gas from the galaxy (i.e.  $E_{ej} \le E_{SN} $). This 
can be calculated as 
\begin{equation}
f_*^{ej}(z) = \frac{V_{vir}^2(z)}{V_{vir}^2(z) + f_w V_s^2}.
\label{fej}
\end{equation}
This represents the maximum fraction of gas that can be converted into stars {\it at a given instant of time}. The relation between $f_*^{ej}$ and the halo mass, as a function of redshift, is shown in Fig. \ref{ch5_feff}. The {\it effective star formation efficiency} can then be expressed as 
\begin{equation}
f_*^{eff} =min[f_*,f_*^{ej}],
\end{equation}
where $f_*$ is the saturation star formation efficiency. This represents the maximum fraction of gas that can be converted into stars in a galaxy, at a given time, without expelling all the remaining gas. Since $V_{vir}$ scales with the halo mass, ``efficient star-formers" (hosted by large dark matter halos) can continuously convert a fraction ($f_*$) of their gas into stars, while feedback-limited systems can form stars with a maximum efficiency dictated by $f_*^{ej}$ that decreases with decreasing halo mass. Matching the bright and faint ends of the 
evolving UV luminosity function (UV LF) at $z \simeq 5-10$ requires $f_* = 0.03$ and $f_w = 0.1$ as will be shown in the later Sec. \ref{lf}. 

\begin{figure}[h]
\center{\includegraphics[scale=0.65]{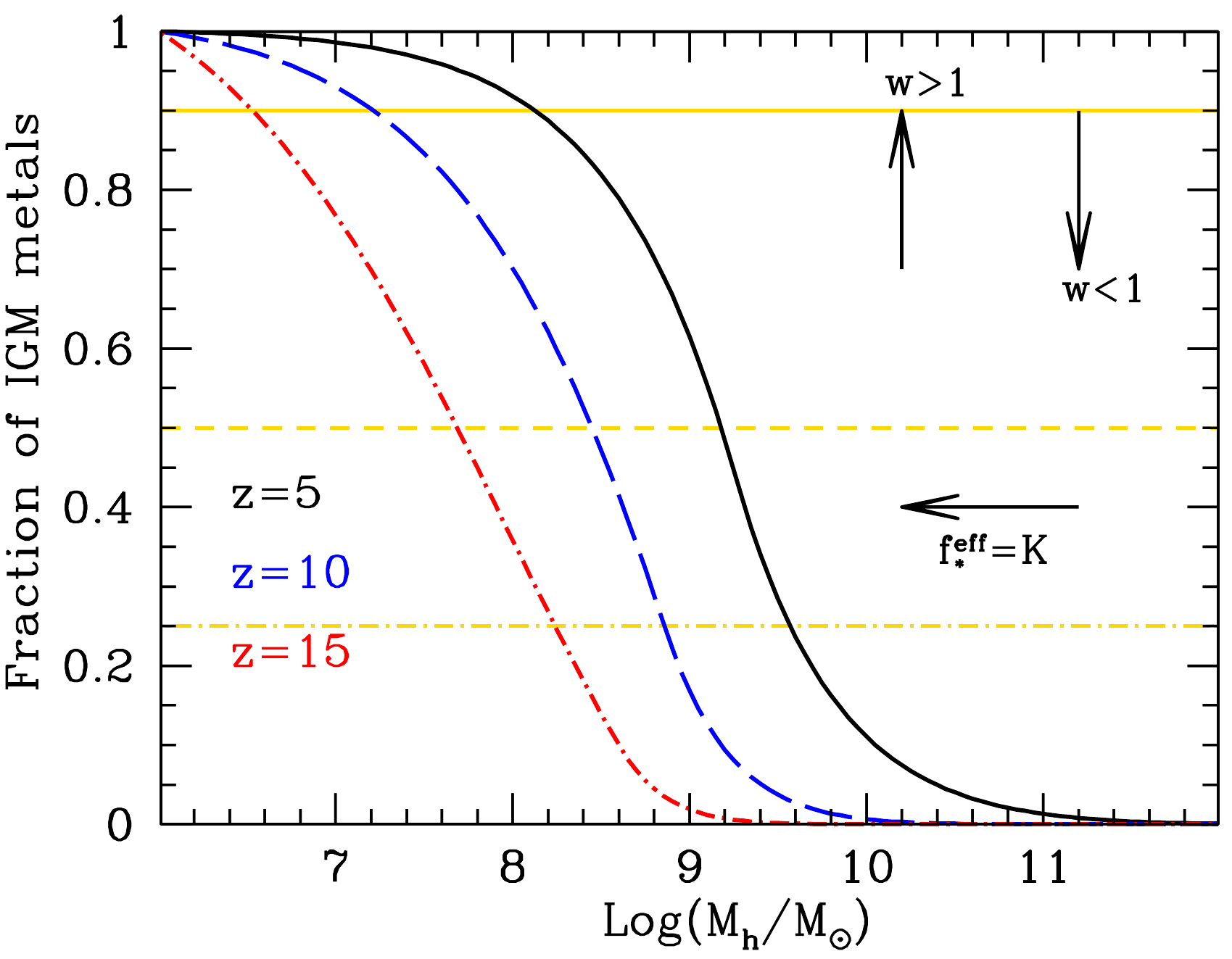}}
\caption{The cumulative fractional contribution to the IGM metal enrichment from galaxies below the halo mass value on the x-axis using the effective star formation efficiency as explained in the text in Sec. \ref{met_igm}. As marked, the solid, dashed and dot-dashed lines show the results at $z=5$, 10 and 15, respectively. The horizontal solid, dashed and dot-dashed lines show cumulative fractional contributions of 90\%, 50\% and 25\%, respectively. As shown, while galaxies below $M_h \sim 10^{7.8}\Msun$ contribute half of the total IGM metal budget at $z=15$, this value shifts to higher values of  $M_h \sim 10^{9.5}\Msun$ by $z=5$. Changing the model assumptions, such as assuming outflows to be preferentially metal-rich (w$>1$) or metal-poor (w$<1$) can increase and decrease the fractional contributions. Further, assuming a constant star formation efficiency for all halos, as opposed to a mass-dependent value, will result in a larger fractional contribution from lower masses as marked. }
\label{ch5_igmmet} 
\end{figure}

In addition to gas, these SN explosions also transport metals into the IGM thereby enriching it continually through time. We can carry out simple, order of magnitude, estimate to study the key IGM metal polluters as follows \citep[e.g.][]{dayal2014a}: assuming each halo to contain initially a cosmological gas mass fraction (i.e. $M_{g,i} = (\Omega_b/\Omega_c) M_h$) and using the effective star formation efficiency calculated above, we can obtain the stellar mass as $M_* = f_*^{eff} M_{g,i}$ and an outflowing gas mass $M_{out} = (M_{g,i}-M_*) [f_*^{eff}/f_*^{ej}]$ . This produces a metal mass equal to $M_z = y_Z \nu M_*$ resulting in a metallicity $Z = M_z [M_{g,i}-M_*]^{-1}$. Assuming perfect metal mixing in the ISM, this yields an outflowing metal mass $M_{outz}=M_{out}Z$. Multiplying this by the number density of the halo as obtained from the HMF, we obtain the cumulative (fractional) IGM metal contribution as a function of halo mass and redshift; note that, in the case of perfect metal-mixing, outflows do not impact the metallicity of the ISM ($Z = M_z/M_g$) since the metal and gas mass losses compensate for each other. The results of this calculation are shown in Fig. \ref{ch5_igmmet}. As seen, if galaxies residing in  halos with masses as low as $10^6\msun$ form stars, 90\% (50\%) of IGM metals are provided by $M_h \lsim 10^{8.1} \, (10^{9.5})\Msun$ halos at $z=5$ with these numbers decreasing to $M_h \lsim 10^{6.5} \, (10^{7.8})\Msun$ by $z=15$. If star formation is suppressed below the atomic cooling threshold, larger halos will naturally provide a larger fraction of IGM metals. This clearly shows the IGM metal pollution to be primarily driven by low-mass halos that lose their gas and metals in outflows. On the other hand, as a result of their larger potential wells, larger-mass halos keep most of their gas and metals bound within them at all $z$. As also shown, assuming a constant star formation efficiency, as opposed to a halo mass dependent one, would shift the metal contribution towards lower-mass halos. Relaxing assumptions of perfect mixing of metals and gas would result in a larger (smaller) metal enrichment if the outflowing gas were to be preferentially metal-rich ($w>1$) or metal-poor ($w<1$) where $w$ is the ratio between the metallicity of the outflowing gas and the ISM metallicity.

The metal-enrichment of the IGM, driven by a central source (SNII in our case), can be calculated in more detail using the physical scenario detailed in \citep{tegmark1993, madau2001, medvedev2013, ferrara2016}: the SNII explosion exerts pressure on the surrounding hot isothermal plasma with pressure $p_b$ and temperature $T_b$ resulting in the formation of a hot bubble, surrounded by a cool, dense spherical shell of mass $M_m$, radius $R_m$, volume $V_m$ and thickness $\delta R_m$, expanding into an ambient medium of baryonic density $\rho_b$ (the IGM in this case). The shell is driven by the thermal pressure of the internal gas which has to accelerate new material onto the expanding shell in addition to overcoming gravitational collapse. The equations of mass, energy and momentum conservation for this shell in an expanding Universe can be written as:
\begin{eqnarray}
\frac{\dot M_m}{M_m} & = & \frac{1}{R_m^3 \rho_b} \frac{d}{dt}(R_m^3 \rho_b) \\
\frac{d}{dt} \dot R_m & = & \frac{8\pi G p_b}{\Omega_b H^2 R_m} - \frac{3}{R_m} (\dot R_m - HR_m)^2 - (\Omega - 0.5 \Omega_m)\frac{H^2R_m}{2} \\
\dot E & = & L - p_b dV_m/dt = L-4\pi p_b R_m^2 \dot R_m.
\end{eqnarray} 
where $\Omega$ is the total density parameter and $L$ is the luminosity that incorporates all sources of heating and cooling of the plasma.

The main features of this expanding bubble can be summarised as follows: for bubble ages $t\ll t_H$, both the impact of gravity and the Hubble flow are negligible, resulting in $R \propto t^{3/5}$. As $t \rightarrow t_H$, the evolution becomes much more complicated given that SN explosion energy input has ceased which, combined with the impact of gravity, results in a slower, decelerating expansion with $R \propto t^{1/4}$. Finally, at $t>> t_H$, the shell gets frozen into the Hubble flow with $R \propto t^{2/3}$ for $\Omega=1$. However, at timescales much longer than the source lifetime, the metal-bubble volume, $V_m$, can be approximated by the Sedov-von Neumann-Taylor solution - the asymptotically adiabatic non cosmological limit to the equations above - resulting in
\begin{equation}
V_m = \frac{4\pi}{3} R_m^3 = \bigg( \frac{E_{51}\nu M_* t^2}{\bar \rho} \bigg)^{3/5},
\label{met_radius}
\end{equation}
where $E_{51}\nu M_*$ represents the total explosion energy input by SN during the formation of a stellar mass $M_*$, $t$ is the time for which the bubble expands and $\bar \rho$ is the mean IGM density. Using numerical simulations, \citep{pallottini2014} find essentially all simulated bubbles expand for $t \sim 250$ Myrs. Again, as above, using a typical $10^{9.5} \Msun$ halo at high-$z$ ($z \simeq 7$) with $M_* = 1.4 \times 10^6$, $\nu = [134 \, {\rm \Msun}]^{-1}$, $E_{51} =10^{51} \, {\rm ergs}$ and $\bar \rho (z=7)$, we find $R_m = 4.18 \, {\rm kpc}$ which is about 4 times larger than the virial radius of this halo ($R_{vir} = 1.04 \, {\rm kpc}$). 

We can take one step further and calculate the metal filling fraction of the Universe, $Q_m$. As noted in Sec. \ref{dm_assembly}, structures grow hierarchically in the $\Lambda$CDM framework resulting in a large number of low mass halos below the knee of the HMF which can be described by a power-law such that $dN/d \log M_h \propto M_h^{(n_{eff}-3)/6}$ with $n_{eff} = -2.2$ \citep[e.g.][]{ferrara2016}; we ignore the exponentially decreasing number density of high-mass halos that keep most of their gas and metals bound within their potential. As explained above, hot, metal-enriched gas driven by SN winds can escape its host halo, shock the IGM and form a blast-wave sweeping up a region of space which scales as $R_m \propto E^{3/5}$ in the Sedov-Taylor phase. The filling factor of these bubbles can then be calculated as \citep{madau2001, ferrara2016}
\begin{equation}
Q_m \propto \frac{ E^{3/5} dN/d log M_h}{E dN/d \log M_h} \propto (dN/d \log M_h)^{2/5} \propto M_h^{5.2/15}.
\label{filfrac_met}
\end{equation}

\subsection{The link between metal enrichment and reionization }
\label{link_reio_met}
The analytic calculations above, that find star-forming halos with the lowest masses to be the key metal-polluters, have now also been confirmed by semi-analytic models \citep[e.g.][]{bremer2018} - these show most of the metals to come from high-$z$ galaxies fainter than an absolute UV magnitude of $\muv =-15$, corresponding to the faintest galaxies seen so far using the Hubble frontier Field data \citep{atek2015, livermore2016}. We caution the reader that this result has explicitly excluded galaxy clustering and re-accretion of metal-rich ejecta onto galaxies (the galactic fountain scenario) given numerical simulations are required for such analysis. As will be seen in Sec. \ref{sources_reio}, these low-mass halos are also considered to be the key reionization sources, establishing a critical link between the joint processes of reionization and the metal enrichment of the IGM in the first billion years. 

The proper volume, $V_{II}$, of an ionized region can be calculated as \citep{shapiro1987} and see also Sec. \ref{temp_den_ion}: 
\begin{equation}
\frac{dV_{II}}{dt} = \frac{d{N_s}} {dt} \frac{M_* f_{esc}}{n_H} - \frac{V_{II}}{t_{rec}} + 3HV_{II},
\label{filfrac_h2}
\end{equation}
where $d{N_s}/dt$ is the production rate of ionizing photons and has an initial value (at $t_0=2$ Myrs) of $\dot N_s(0)= 10^{46.6}\, {\rm s^{-1}\, \Msun^{-1}}$ for a $0.1-100\Msun$ Salpeter IMF and $Z = 0.05 \Zsun$ using {\it Starburst99} \citep{leitherer1999, leitherer2010} and undergoes a time evolution as $\dot N_s(t) = \dot N_s(0)-3.92 \log (t/t_0) + 0.7$ \citep{dayal2017a}. Further, $f_{esc}$ is the escape fraction of ionizing photons from the ISM as explained in Sec. \ref{fesc}. Given that the source lifetime is generally less than the recombination timescale and the Hubble time, we can write
\begin{equation}
V_{II} \simeq \frac{dN_s} {dt} \frac{M_* f_{esc}}{n_H} = f_{esc} \frac{N_\gamma M_*}{\bar \rho},
\label{filfrac_h2_approx}
\end{equation}
where $N_\gamma = 5 \times 10^3$ is the number of \HI ionizing photons produced per baryon bound in stars \citep{ferrara2016}.  

We can now compare the IGM volume ionized by star formation in a galaxy to that metal-enriched by it as \citep{ferrara2016}
\begin{equation}
\frac{V_{II}}{V_z} = \bigg( \frac{E_{51}\nu M_* t^2}{\bar \rho} \bigg)^{3/5} \frac{\bar \rho}{f_{esc} N_{\gamma} M_*} \propto \bigg( \frac{\bar \rho}{M_*}\bigg)^{2/5} \propto n_*^{2/5},
\end{equation} 
where $n_*$ is the mean comoving number density of galaxies with stellar mass $M_*$. This implies that the metal enrichment tracks reionization more efficiently if the key sources are more numerous, i.e. have low masses. 

Finally, we look at the expected mean metallicity of the IGM once reionization is complete following the calculations in \citep{ferrara2016}: the mass of metals produced by a stellar population of mass $M_*$ can be expressed as
\begin{equation}
M_z = \nu y_z M_* = \nu y_z \mu m_p N_{b*} = \nu y_z \mu m_p \frac{N_s}{N_{\gamma}},
\end{equation}
where $\mu =1.22$ is the mean molecular weight and $N_{b*}$ is the number of baryons in stars and we have used $N_s = N_{\gamma} N_{b*}$. Then using the fact that $\kappa (z_{re}) n_H/(1-Y)$ photons are required by the end of reionization, we obtain
\begin{equation}
M_z = \nu y_z \mu m_p \frac{\kappa (z_{re}) n_H}{(1-Y) N_{\gamma}}.
\label{mz}
\end{equation} 
Here, $\kappa(z_{re})$ accounts for the fact that ionized hydrogen can recombine in over-dense regions, requiring multiple ionizing photons to maintain reionization which can be approximated as
\begin{equation}
\kappa(z_{re}) = \frac{t_H(z_{re})}{t_{rec}} \approx \frac{H_0 \Omega_m (1+z_{re})^{3/2}}{[\chi_{HI} n_H  \alpha_B(T) C]^{-1}} = 2.9 \bigg(\frac{1+z_{re}}{7} \bigg)^{3/2},
\label{kappa}
\end{equation} 
where $\alpha_B(T)$ is the hydrogen case B recombination co-efficient and we have used a clumping factor $C = \langle n_\HII^2\rangle/\langle n_\HII \rangle^2 \simeq 3$ where the brackets denote an average over a cosmologically representative volume; these terms are detailed in Sec. \ref{temp_den_ion} that follows. Combining Eqns. \ref{mz}, \ref{kappa} and the fact that $\langle Z \rangle = M_z/M_H = M_z/(m_p n_H)$, we obtain \citep{ferrara2016}
\begin{equation}
\langle Z \rangle = 5.5 \times 10^{-4} \bigg(\frac{1+z_{re}}{7} \bigg)^{3/2} \Zsun,
\end{equation}
implying that the IGM, on average, is polluted well above $Z_{crit}$ by the end of reionization. However, this does not preclude pockets of pristine gas in which a small fraction of PopIII stars can still form. Interestingly, an earlier reionization results in a higher average IGM metallicity value - this is because the higher the redshift, the denser is the Universe, requiring a larger number of ionizing photons to maintain reionization that results in a correspondingly higher value of the metal enrichment.

\newpage

\section{Ionizing radiation from early galaxies: the Epoch of Reionization}
\label{ch7}

As noted in the last section, star formation (both PopIII and II) results in Lyman continuum (LyC; h$\nu >1$ Ryd) photons that can ionize hydrogen, starting the Epoch of (hydrogen) Reionization; although quasars, powered by accretion of gas onto a black hole, also produce ionizing photons, we mostly focus our attention on star-forming galaxies in this review. Given the complex density fields of gas and dust in the ISM, however, only a small fraction of LyC photons actually emerge into the IGM. We start by discussing the escape fraction of hydrogen ionizing photons in Sec. \ref{fesc} before moving on to the growth of ionized regions (Sec. \ref{temp_den_ion}) and the impact of reionization on galaxy formation (Sec. \ref{uvb_fb}) and end by examining the key reionization sources (Sec. \ref{sources_reio}). 

\subsection{The escape of ionizing photons}
\label{fesc}
The escape fraction of  LyC photons, $f_{esc}$, denotes the ratio between the ionizing photons ($\lambda < 912$\AA\, in the rest-frame) that emerge out of the galactic environment, and ionize the IGM, and the total amount produced within a galaxy. The escape fraction is a crucial quantity for reionization: determining the history and topology of reionization, it directly impacts galaxy formation by establishing the UVB that can photo-evaporate gas from low-mass halos in ionized regions. On the other hand, the ionizing photons absorbed in the ISM ($\propto [1-f_{esc}$]) lead to the production of nebular emission lines. Significantly affecting the broadband fluxes \citep{zackrisson2008}, these can cause an appreciable variation in the inferred properties (e.g. the stellar mass and ages) of high-$z$ galaxies \citep{schaerer2009, debarros2014}. However, a key bottleneck in reionization studies is that the value of $f_{esc}$, and even its dependence on galaxy mass and redshift, remain poorly understood at best as now discussed. 

Pinning down $f_{esc}$ observationally has presented a massive challenge. This is because it is nearly impossible to directly detect LyC photons at $z \gsim 4$ given the increasing abundance of LLS (column density $\simeq 10^{17} {\rm cm^{-2}}$) that lead to a 5-fold decrease in the LyC transmission from $z \simeq 0$ to $z>3.5$ \citep{inoue2008, prochaska2010}. While LyC photons can be detected at slightly lower redshifts, inferring $f_{esc}$ still requires knowing their intrinsic production rate which is hard to undertake since Balmer lines (specially H$\alpha$; used to infer this value) shift into the near-IR at $z >3$. Therefore, what is generally measured is the flux ratio at $1500$\AA\, versus the LyC, $({\rm L_{1500}/L_{LyC})_{obs}}$, which can be linked to the intrinsic flux ratio of the underlying stellar population, ${\rm (L_{1500}/L_{LyC})_{int}}$, as \citep[e.g.][]{siana2007}
\begin{equation}
({\rm L_{1500}/L_{LyC})_{obs}} = ({\rm L_{1500}/L_{LyC})_{int}} \times 10^{-0.4({\rm A_{1500}-A_{LyC}})}  e^{\tau_{\rm HI,IGM}} e^{\tau_{\rm HI,ISM}}.
\end {equation}
Here, ${\rm A_{1500}-A_{LyC}}$ is the differential dust extinction (between 1500 and 912 \AA\, in the rest-frame) in magnitudes and $e^{\rm \tau_{HI,IGM}}$ and $e^{\rm \tau_{HI,ISM}}$ are the IGM and ISM optical depths to LyC photons, respectively. A \textit{relative} escape fraction ($f_{esc}^{rel}$) can then be defined as 
\begin{equation}
f_{esc}^{rel} = {\rm \frac{(L_{1500}/L_{LyC})_{int}}{(L_{1500}/L_{LyC})_{obs}}} e^{\rm \tau_{HI,IGM}},
\end{equation}
which can be turned into an \textit{absolute} $f_{esc}$ using the conversion
\begin{equation}
f_{esc} = 10^{-0.4 {\rm A(1500 \AA)}} f_{esc}^{rel},
\end{equation}
where the attenuation at 1500\AA, ${\rm A(1500\AA})$ can be related to the colour excess as ${\rm A(1500\AA})= 10.33 {\rm E(B-V)}$  \citep[e.g.][]{siana2007} using the Calzetti \citep{calzetti1997} dust reddening law inferred from local starbursts. 

We show a compilation of the available observational estimates of $f_{esc}$ at $z \simeq 0-6$ in Fig. \ref{fig_obs_fesc}, which are now detailed; this comparison must, however, be treated with caution given the lack of uniformity between the different samples shown. Combining estimates of $f_{esc}$ both from direct observations of LyC photons and those inferred indirectly from the observed ionizing background intensity, \citet{inoue2006} find the median $f_{esc}$ value to rise by a factor of about 10 from $f_{esc} \lsim 1\%$ at $z \lsim 1$ to $f_{esc}\simeq 10\%$ at $z \gsim 4$. An enormous amount of observational effort has been invested to supplement these numbers: using a ground-based custom-built narrow band filter to study the $z \simeq 3.09$ cluster SSA22, \citet{nestor2011} find $\langle f_{esc} \rangle = 12\%^{+11}_{-6}$ for LBGs, with LAEs showing a much higher value of $\langle f_{esc} \rangle \gsim 17\%$. Including spectroscopic data, these numbers were later refined to $\langle f_{esc}^{LBG} \rangle \sim 5-7\%$ and $\langle f_{esc}^{LAE} \rangle \sim 10-30\%$ \citep{nestor2013}. These results were interpreted as implying high (low) $f_{esc}$ values along lines-of-sight where SN feedback had (not) cleared the ISM of gas and dust. These are in good agreement with the value of $\langle f_{esc} \rangle = 5.9\%^{+14.5}_{-4.2}$ inferred for $z \sim 2.2$ galaxies \citep{matthee2017} using ground-based data and $\langle f_{esc} \rangle \lsim 3.1\%$ at $z \sim 2.38$ \citep{vasei2016} and $f_{esc} \lsim 2.7-3.4\%$ using HST data (assuming ${\rm E(B-V)}=0.1$) at $z \sim 3.1$ \citep{siana2015}. These values are in accord with the low values inferred for bright galaxies ($L \gsim 0.5 L^*_{z=3}$\footnote{$L^*_{z=3}$ is the characteristic luminosity at $z=3$ corresponding to ${\rm M_{UV}}=-21.0$ \citep{steidel1999} using the AB magnitude system \citep{oke-gunn1983}.}) with $f_{esc} <11.1\% (0.77\%)$ at $z \sim 3.3$ \citep{boutsia2011} (\citep{grazian2016} where we have assumed ${\rm E(B-V)}=0.1$) and $f_{esc} \lsim 5-20\%$ at $z \sim 3.4-4.5$ \citep{vanzella2010} depending on how the data are selected (e.g. by magnitude or redshift). Taken together, these results imply an escape fraction that effectively decreases with increasing mass or luminosity while broadly showing an increase with redshift.
  
Complicating this picture, however, observations have revealed outliers at all redshifts ($z\sim 2-5$) that show $f_{esc}$ values larger than 20\% and even as high as $100\%$. For example, \citet{naidu2017} find $f_{esc} = 60\%^{+40}_{-38}, 72\%^{+28}_{-48}, 62\%^{+38}_{-51}$ for 3 galaxies at $z \sim 2$; they detect 3 other active galactic nuclei (AGN) hosts that show values between $13-100\%$. At $z \sim 3$, \citet{mostardi2015} have inferred $f_{esc} \sim 14-19\%$ for an extremely young galaxy with age less than 50 Myr and \citet{shapley2016} have inferred $f_{esc} >51\%$ even for a typical $L_*$ galaxy. At $z \sim 3.2$, \citet{vanzella2016} have found a unique, compact Green Pea galaxy (Ion2) that shows an [OIII]4959,5007+H$\beta$ rest-frame equivalent width of 1600\AA\, and $f_{esc} \gsim 50\%$ depending on the exact IGM attenuation used; in a future work \citep{vanzella2017} they have supplemented this with the serendipitously discovered ``Ion3'' at $z \sim 4$ that also shows $f_{esc} >60\%$. Finally, using gravitational lensing, \citet{leelo2016} have inferred $f_{esc} \sim 19\pm 6\%$ at $z \sim 4-5$. 

These results involve a few caveats: first, the escape fractions critically depend on the assumed star formation history and age of the underlying stellar population (that determines the intrinsic $1500$\AA\,-LyC ratio), the inclusion of photons from binary stars, the ${\rm E(B-V)}$ value along the line of sight, and the precise model used to account for the IGM attenuation of LyC photons. Further, a number of studies \citep{iwata2009, vanzella2012b, siana2015} have found the LyC emission to be offset from the UV by as much as 1-1.5'' - results presented above that assume co-spatial fluxes to constrain $f_{esc}$ must therefore be used with caution. Moreover, other studies \citep{vanzella2012b, siana2015, grazian2016} find $\gsim 50$\% of bright LBGs to be contaminated by low-$z$ interlopers, requiring high resolution imaging and deep spectroscopy to distinguish LyC emitters from foreground galaxies. Finally, some works \citep[e.g.][]{cooke2014} caution that the standard LBG colour selection technique, requiring no flux blue-ward of LyC for a successful detection, is intrinsically biased against selecting LyC emitters. Indeed, this work finds $f_{esc} \simeq 33 \pm 7\%$ for LyC emitting galaxies (LCGs) that drops by half, to $f_{esc} \simeq 16\pm 4\%$, when considering the entire LBG+LCG population although crucially, these authors do not account for any IGM attenuation. 

\begin{figure*}
\center{\includegraphics[scale=0.65]{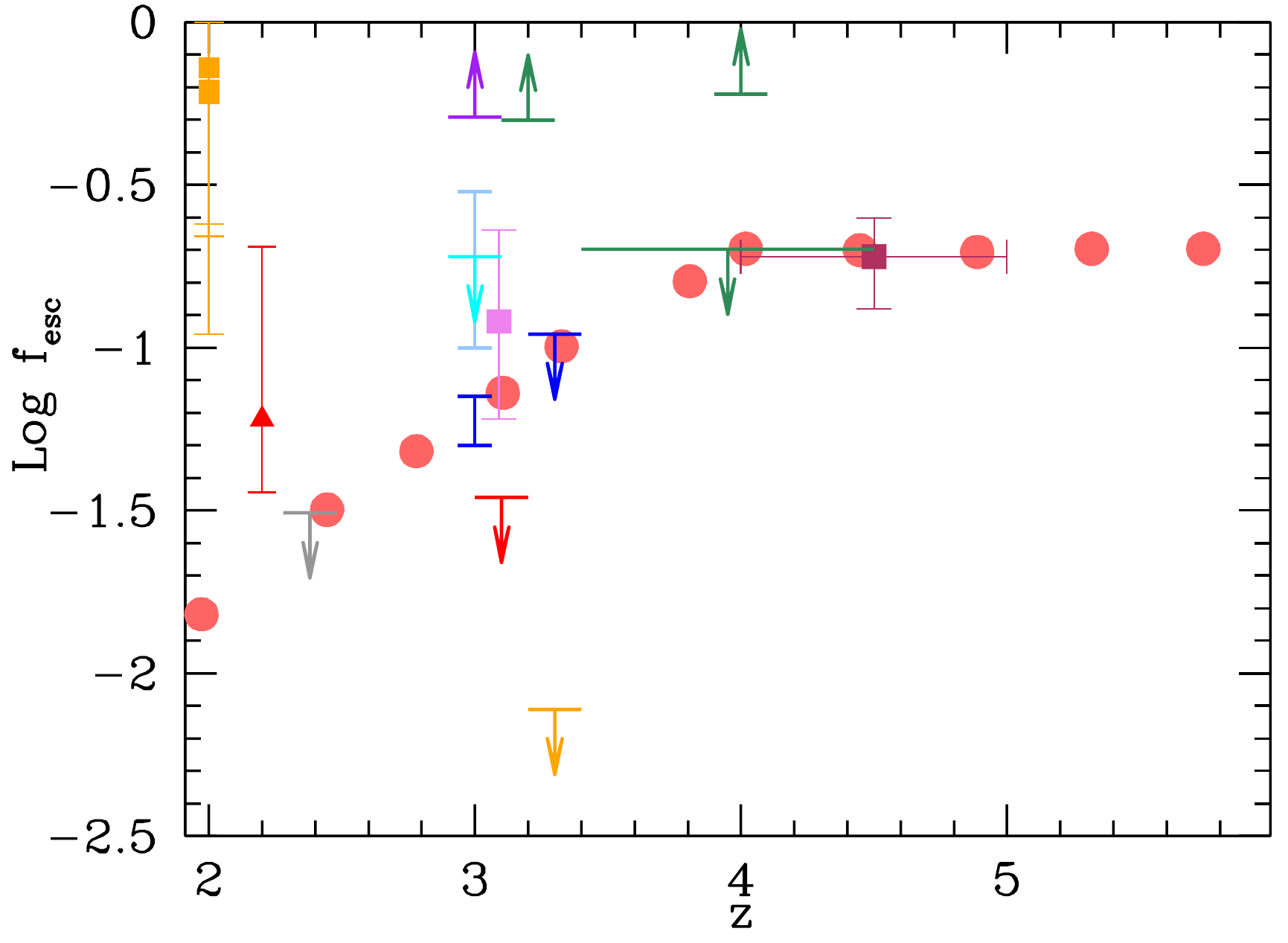}}
\caption{A summary of observationally inferred $f_{esc}$ values as a function of $z$. The plotted points show observational data collected by \citet[filled circles;][]{inoue2006}, \citet[violet filled squares;][]{nestor2011}, \citet[dark and light blue bars showing limits inferred for LBGs and LAEs, respectively;][]{nestor2013}, \citet[red filled triangle;][]{matthee2017}, \citet[orange filled squares;][]{naidu2017} and \citet[maroon filled square;][]{leelo2016}. The downward facing arrows showing upper limits are from the work by \citet[gray;][]{vasei2016}, \citet[red;][]{siana2015}, \citet[blue;][]{boutsia2011}, \citet[orange;][]{grazian2016}, \citet[green;][]{vanzella2010} and \citet[cyan;][]{mostardi2015}. Finally, the upward facing arrows showing lower limits are from the work by \citet[purple;][]{shapley2016} and Vanzella et al. 2016, 2017 [green; 327, 328]. Although showing a large scatter, these observations might be broadly taken to indicate a positive correlation of $f_{esc}$ with redshift.}
\label{fig_obs_fesc} 
\end{figure*}

On the theoretical front, constraining $f_{esc}$ requires coupling realistic renditions of ISM properties with a full radiative transfer code. The complexity of the problem has necessitated a number of theoretical approaches: we start with semi-empirical techniques that focus on inferring $f_{esc}$ values that yield the correct CMB electron scattering optical depth for a galaxy population matched to observations. For example, it has been shown that simultaneously reproducing the high-$z$ UV LF, the electron scattering optical depth $\tau \sim 0.08$ from the {\it WMAP 7-} and {\it WMAP 9-}year data \citep{komatsu2011, hinshaw2013} and Ly$\alpha$ forest statistics requires one of the following conditions be met \citep{kuhlen2012, mitra2013, robertson2013}: (i) the UV LF should either be extrapolated to magnitudes as faint as $\muv \sim -10$ or $-13$ compared to the current detection limits of $\muv \simeq -17$; or (ii) $f_{esc}$ should increase from about 4\% at $z \simeq 4$ to about 100\% at $z >10$; or (iii) a hybrid solution can be found wherein undetected galaxies contribute significantly and $f_{esc}$ evolves modestly. However, the lower value of $\tau = 0.05 - 0.06$ measured by {\it Planck} \citep{planck2015, planck2016} is fully consistent with lower, and more reasonable, $f_{esc}$ values ranging between 10\% \citep{mitra2015, sun2016} and 20\% \citep{robertson2015, dayal2017a}. Making the assumption of a minimum star formation rate required to blow outflows and clear channels for the escape of ionizing radiation \citep{sharma2016}, on the other hand, naturally find an (inverted) mass-dependence such that $f_{esc}$ increases with mass. However, \citet{ferrara2013} caution that $f_{esc}$ values can be large only if channels are opened within the first few Myrs after star formation, prior to SN explosion, since SN occur only after ionizing photon production has already started declining. Modelling low-mass $ T_{vir}\lsim 10^4$ K galaxies, these authors show that $f_{esc}=0$ before the ionizing front generated by star formation reaches the virial radius, and effectively increases to 1, at a star formation efficiency of $\sim 10^{-3}$, once the ionization front breaks out to $r \gsim R_{vir}$. 

Using SPH or AMR techniques, several authors have reported a wide range of $f_{esc}$ values for low mass halos such that $f_{esc} \simeq 0.15-0.6$ for $M_h \lsim 10^{6-7}\Msun$ \citep{wise2014, ma2015, xu2016} and decreases to $f_{esc} \simeq 0.05-0.4$ for $M_h \lsim 10^{8-9}\Msun$ \citep{wise2009, yajima2011, xu2016}. On the other hand, a number of works \citep{paardekooper2011, paardekooper2013, paardekooper2015} find much lower values of $f_{esc} \simeq 10^{-5} - 0.1$, with only 30\% of $>10^8 \Msun$ halos showing $f_{esc} \gsim 0.01$. The variance amongst these results clearly demonstrates the complexity of the dependence of $f_{esc}$ on the ISM density, clumping and turbulence structure as well as the availability of low-density lines-of-sight. The results for high-mass ($M_h \gsim 10^{11}\Msun$) halos are in better agreement, yielding $f_{esc}$ values ranging between $0.01-0.07$ at $z \simeq 3-9$ \citep{razoumov2007, razoumov2010, gnedin2008, yajima2011}. Interestingly, all of these works find $f_{esc}$ to decrease with an increase in the halo mass because of a more clumped or higher density ISM. The only exception is the work by \citet{gnedin2008} who find $f_{esc} \simeq 0.02$ for $M_h \gsim 10^{11} \Msun$ galaxies as a result of young stars lying outside the \HI disk; $f_{esc}$ decreases for low mass galaxies because of stars being deeper embedded inside thicker and/or more extended \HI disks. Finally, in terms of its redshift evolution, most works converge on $f_{esc} \simeq 0.01-0.02$ at $z \simeq 2-4$ increasing with $z$ to $\simeq 0.06-0.1$ at $z \gsim 3.6$ to as high as $0.8$ at $z \gsim 10.4$ \citep{razoumov2006, razoumov2007, yajima2011}. 

These $f_{esc}$ values are critically linked to a number of physical parameters: for example, $f_{esc}$ values are found to be positively linked to the star formation rate and are the highest during the most intense starbursts, varying by as much as 1-6 orders of magnitude as a result of SN creating channels for ionizing photons \citep{gnedin2008, wise2009, paardekooper2011, trebitsch2017}. The IMF is another critical quantity impacting $f_{esc}$. A Salpeter IMF results in lesser SN feedback and hence higher star formation rates implying a higher $f_{esc}$ \citep{wise2009, razoumov2010}. Indeed, \citet{wise2009} find $f_{esc}$ to increase from 0.25 for $M_h>10^8 \Msun$ halos and a top-heavy IMF to 0.4 using a Salpeter IMF. Although anisotropic line of sight and orientation (edge-on or face-on viewing angle) effects can vary $f_{esc}$ by about an order of magnitude \citep{razoumov2007, gnedin2008, yajima2011, paardekooper2011,paardekooper2013, xu2016}, all these calculations agree that the effects of dust are not critical at high redshifts - since \HI and gas densities are correlated, in regions of high \HI density where dust masses can be high, $f_{esc} \sim 0$ in any case, while regions of low \HI density (where $f_{esc}$ is high) do not contain much dust. Finally, the exact radius (0.5-2 $R_{vir}$) at which $f_{esc}$ is measured in not critical given most absorption takes place deep in the ISM, with clumps and filaments at large $R_{vir}$ having a marginal effect due to their small solid angle for absorption \citep{wise2009}. 

We end with a few caveats. The first is that most massive stars exist in binary systems, a fact that is only starting to be included in population synthesis models. Enhancing the population of massive stars (at $\gsim 3$ Myrs) after the burst, this binary fraction can lead to a larger $f_{esc}$, by as much as a factor of 3-10, by virtue of leakage channels produced after massive star formation \citep{Ma2016} and/or a slower decline in the ionizing luminosity \citep{rosdahl2018}. Further, a number of works \citep{conroy2012, kimm2014} have shown that run-away stars, that travel at speeds of more than 30 ${\rm km \, s^{-1}}$ and can comprise as much as 30\% of all massive stars in the MW, can travel 0.1-1 kpc from their dense birth clouds into low-density regions, increasing $f_{esc}$ by a factor of 2-8 depending on the halo mass, galaxy geometry and the mechanism producing the runaway. A number of works also caution that $f_{esc}$ values can be over-estimated by calculations that do not explicitly resolve high-density molecular clouds \citep{ma2015} or model the early phases of galaxy formation \citep{wise2014}. On the other hand, ignoring the high turbulence expected in high-$z$ galaxies \citep{wise2008, greif2008, sur2016} can under-estimate $f_{esc}$ by factors of 2-4 even in media where the optical depth is $\simeq 1$ \citep{safarzadeh2016}.

We end on the positive note that forthcoming observations might present novel ways of shedding light on $f_{esc}$: spectroscopy with the {\it JWST} might potentially be used to identify strong LyC leakers ($f_{esc} \gsim 50\%$) even up to $z \simeq 9$ combining estimates of the UV spectral slope ($\beta$; calculated between 1300-3000\AA\, in the rest frame) with the H$\beta$ emission line equivalent width, modulo caveats regarding the age, metallicity and dust attenuation \citep{zackrisson2013}. Further, the ${\rm [OIII]\lambda5007/[OII]\lambda3727}$ ratio seems positively correlated with $f_{esc}$ in density bound \HII\, regions as shown by studies of local LyC leakers \citep{nakajima2014} that can be extrapolated to high-$z$ \citep[see e.g.][and references therein]{faisst2016}. Finally, the H$\alpha$ and visible continuum surface brightness profiles, either around bright ($L \gsim 5 L_*$) galaxies or stacked faint {\it JWST} sources, could shed light on $f_{esc}$ \citep{masribas2017}.


\subsection{Growth and properties of ionized regions}
\label{temp_den_ion}

As LyC photons stream out of sources and escape into the IGM they start interacting with the gas composed mostly of hydrogen and helium atoms. The main effect is that an \HII\, region, bounded by an ionization front, will form around the source within which the gas is photo-heated to temperatures of the order of a few $10^4$ K \citep[e.g.][]{miralda1994}. For an ideal situation in which the IGM density is uniform the shape of the HII region will be spherical; in more realistic cases the shape will be largely determined by the distribution of density inhomogeneities producing a fluctuating opacity to the propagating photons. 

Let us assume that a source emits LyC photons at a rate, $\dot N_s$. Then the physical expansion velocity (in excess of the Hubble flow) is regulated by the difference between the ionization and recombination rates occurring within the \HII regions. Mathematically this yields \citep{shapiro1987,Meiksin1993}:

\begin{equation}
4\pi r_I^2 n_{\rm H} \biggl(\frac{dr_I}{dt} - H r_I\biggr) = \dot N_s
-\frac{4}{3} \pi r_I^3 n_{\rm H}^2 C(z) \alpha_B(T),
\label{rI}
\end{equation}
where $r_I$ is the proper radius of the ionization front, $n_{\rm H}$ is the mean hydrogen density of the IGM and $C(z)$ is the clumping factor. Finally, $\alpha_B(T)$ is the Case B radiative recombination rate to the
$n=2$ level of hydrogen, allowing for ionizations from the ground state by the diffuse radiation emitted following recombinations to the ground state; if diffuse radiation is treated separately, then the Case A recombination rate should be used. The previous equation makes three assumptions: {\it(i)} the thickness of the I-front is negligible with respect to $r_I$; {\it (ii)} the propagation velocity of the ionization front is supersonic with respect to the ambient gas and therefore the dynamical response of the gas to the sudden heating can be neglected (``R-type" front); and {\it (iii)} the velocity of the front is sub-luminal: this assumption might break down in the initial phases of the evolution (small $r_I$). While Eqn. \ref{rI} is very similar to the classical one for the Str\"omgren radius first derived by \citet{shapiro1987}, in an expanding universe the mean density decreases with time and ionized regions hardly have time to fill their Str\"omgren radius before they overlap with another bubble. 

As already mentioned, the presence of inhomogeneities in the gas distribution has essentially two different effects. First, it produces variations, which might be substantial, in the optical depth to ionizing radiation along different directions \citep{Maselli07, Bolton07, Eilers17} as a result of which some of the line of sights from the source may be totally blocked. The gas behind these obscuring over-densities cannot be reached by ionizing photons and therefore remains neutral. As a result the I-front surface loses its spherical symmetry and becomes jagged, expanding faster into low-density voids. Second, the recombination rate, which is $\propto n_H^2$, is strongly enhanced in over-dense regions resulting in a considerably decreased final ionized volume \citep{Iliev05a}. These IGM inhomogeneities are expected \citep{Miralda00, Chardin17} as a result of the typically lognormal distribution of the density field deduced from observations of post-reionization IGM (the so-called Ly$\alpha$ forest). However these are relatively mild and typically correspond to weakly non-linear fluctuations with over-densities $\simlt 30$. However, due to gravitational clustering, the environment of a given galaxy might contain a number of lower mass nonlinear structures, i.e. mini-halos. These structures, which might be $\approx 10^5$ times denser than the mean IGM \citep{Shapiro04}, act as powerful sinks of ionizing photons.  
Because the clumping factor may be dominated by rare dense structures, establishing its value from
numerical simulations is difficult and depends strongly on the numerical resolution adopted. A simple fitting formula has been provided in \citet{Iliev2007},
\begin{equation}
C(z) \simeq 26.2917 \exp[-0.1822z+0.003505z^2],
\label{Clump}
\end{equation}
which is applicable in the redshift range $6<z<30$. However, this clumping factor can be significantly reduced once the impact of the reionization UV  background is taken into account \citep{Iliev05b}, as shown in Sec. \ref{uvb_fb} that follows.

As individual bubbles around sources grow with time they are likely to overlap with other bubbles produced by nearby sources. This process, by which the entire cosmic volume becomes ionized, can be described in statistical terms. Analogous to the metal enrichment of the IGM, we introduce the porosity parameter, $Q_{\rm HII}$, i.e. the product of the number density of bubbles times the average volume of each of them\footnote{Sometimes the filling factor of the ionized regions,  $f_{\rm HII} = (1-e^{-Q_{\rm HII}})$, is also used; for $Q_{\rm HII}\ll 1$ they are almost equivalent.}. Hence we define the time (or, equivalently, redshift) when $Q_{\rm HII}=1$ as the reionization epoch.

The evolution of $Q_{\rm HII}(z)$ is simply given by \cite{Madau1999, Meiksin2009}:
\begin{equation}
- H(z)(1+z)\,\frac{dQ_{\rm HII}}{dz}=\frac{\dot n_S(z)}{n_{\rm H}(0)}
-\frac{C(z)Q_{\rm HII}}{t_{\rm rec}(z)},
\label{dQdz}
\end{equation}
where $n_{\rm H}(0)$ is the comoving number density of hydrogen at $z=0$ and $\dot n_S(z)$ is the production rate of ionizing photons per comoving volume. The recombination timescale, $t_{\rm rec}$, at redshift $z$ can be expressed as
\begin{equation}
t_{\rm rec}(z) = [\chi_{HI} n_{\rm H}(z) C(z) \alpha_B(T)]^{-1}.
\label{trec}
\end{equation}

Once $Q_{\rm HII}>1$, all the cosmic gas has been ionized. Towards the final phases of the reionization process the mean free path of the photons becomes comparable or larger than the Hubble radius $c/H(z)$. This implies that each patch of gas can see radiation coming from all the sources (the emissivity) and a roughly uniform UVB is established. In practice, though, the mean free path is limited to $\lambda_{mfp} = c/H(z) (dN_{LLS}/dz)^{-1}$, where $N_{LLS}$ is the number of optically thick over-dense LLS. At $z=4$, for example, \citet{Bolton2007} estimate that $dN_{LLS}/dz=3.3$. The intensity of the UVB is then simply written as $J(z) \simeq \dot n_S(z) \lambda_{mfp}$. Although the previous approach provides a broad-brush description of the reionization progress, and has been widely used given its simple analytical solution, the complexity of the problem can only be fully described by cosmological radiative transfer simulations. For more details, specially on the spatial distribution of ionized and neutral gas, we defer the reader to the extensive review by \citet{Meiksin2009}.

Finally, the above analytical treatment might not catch some deeper physical aspects related to the reionization topology. \citet{Furlanetto2016} pointed out that the naive view of discrete ionized bubbles growing independently of each other throughout reionization might be qualitatively
wrong. Reionization, in fact, is inherently a problem in percolation and phase transitions physics. 
According to this scenario, reionization exhibits two distinct percolation phase transitions. In the first, occurring when the neutral hydrogen fraction is $\chi_{HI} \approx 0.9$, the bulk of the ionized gas is quickly incorporated into a unique infinitely large ionized region. In the second (when $\chi_{HI} \approx 0.1$), a unique neutral region breaks into discrete neutral regions. The model predicts that the discrete ionized regions follow a power law size distribution, with nearly equal filling factors per logarithmic interval in bubble volume up to some characteristic size. This percolation theory could provide quite a precise description of reionization topology and evolution. It represents a promising research field over the next few years, particularly in view of the 21cm experiments that aim to detect high-$z$ \HI through its spin-flip transition emission.

\subsection{Impact of global ultraviolet background on galaxy formation}
\label{uvb_fb}
We now discuss the feedback-effects of the UVB created by galaxies during reionization on subsequent galaxy formation. This is critical since feedback could, in principle, inhibit the star-forming capabilities of the very galaxies driving reionization, delaying its progress and changing its morphology. 

During reionization, photoionization heating from the continually rising UVB can raise the gas temperature of about $2 \times 10^4$ K in ionized regions as noted above. This temperature is equal to the virial temperature (see Eqn. \ref{eqn_tvir}), thereby being equal to the gravitational potential, of a bound halo of  mass $\simeq 10^{8.25}\Msun$ at  $z \sim 15$. As expected, the equivalence between the external radiation field and the internal binding energy can have two effects: firstly, it can photo-evaporate gas from the small potential wells of the lowest mass galaxies (in the minihalo/low mass end of atomic cooling halo regime). Secondly, it can increase the gas pressure, associated with  the increase in IGM temperature, increasing the Jeans scale for gas collapse \citep[e.g.][]{gnedin2000, petkova2011, noh2014} and inhibiting gas from accreting onto galaxies. Both these effects can combine to suppress further star formation. However, since the baryonic content of a halo exposed to a UVB depends on multiple parameters, including the redshift, the thermal history and the intensity of the UVB, both the minimum halo mass that can host gas and the halo baryon fraction during reionization remain a matter of debate \citep{hoeft2006, okamoto2008, wise2012b, sobacchi2013a, hasegawa2013, sobacchi2013b}. The gas mass in a halo of total mass $M_t$ exposed to a UVB can be parameterised as \citep{gnedin2000}
\begin{equation}
\label{mg_reio}
M_g = \frac{f_b M_t}{[1+(2^{\alpha/3}-1)(M_{cr}/M_t)^\alpha]^{3/\alpha}},
\end{equation}
where $f_b = \Omega_b/\Omega_c$ is the cosmological baryon fraction, $\alpha$ is a free parameter and $M_{cr}$ is the ``characteristic" halo mass that can retain half its gas mass. The value of $M_{cr}$ remains a matter of contention: \citet{gnedin2000} find ($\alpha=1$) $M_c=M_F \simeq 10^{9.2}\Msun$ at $z=6$ \citep[see also][who find comparable mass suppression scales]{wise2008c, finlator2011, hasegawa2013} where $M_F$ is the filtering mass over which baryonic perturbations are smoothed in linear theory; this result is consistent with that found by \citet{naoz2009} and \citet{okamoto2008} at $z \gsim 6$. On the other hand, in the low-$z$ post-reionization regime, \citet{okamoto2008} find that, rather than the linear filtering scale, the characteristic mass {\it that can accrete gas} depends on the gas temperature at $R_{vir}$: if the gas temperature exceeds $T_{vir}(R_{vir})$, at a density that is a third of the halo virial over-density, no accretion takes place resulting in a much lower characteristic mass as compared to $M_F$ \citep[see also][]{wise2008c}. This result is supported by the work of \citet{hoeft2006} who also find the mass that can retain half its baryons as $M_{cr}<<M_F$ - this result is driven by the equilibrium temperature between photo-heating and radiative cooling at a characteristic over-density of $\delta \simeq 1000$; only halos capable of compressing gas to this density despite a UVB can accrete gas. These authors also note that halos are never completely baryon-free since cold-phase gas and baryons bound into stars will resist evaporation. The discrepancy in these results possibly arises because mass scales have either been evaluated at the mean density \citep{gnedin2000} or at the halo virial densities \citep{hoeft2006, okamoto2008}. 

\begin{figure*}
\center{\includegraphics[scale=0.52]{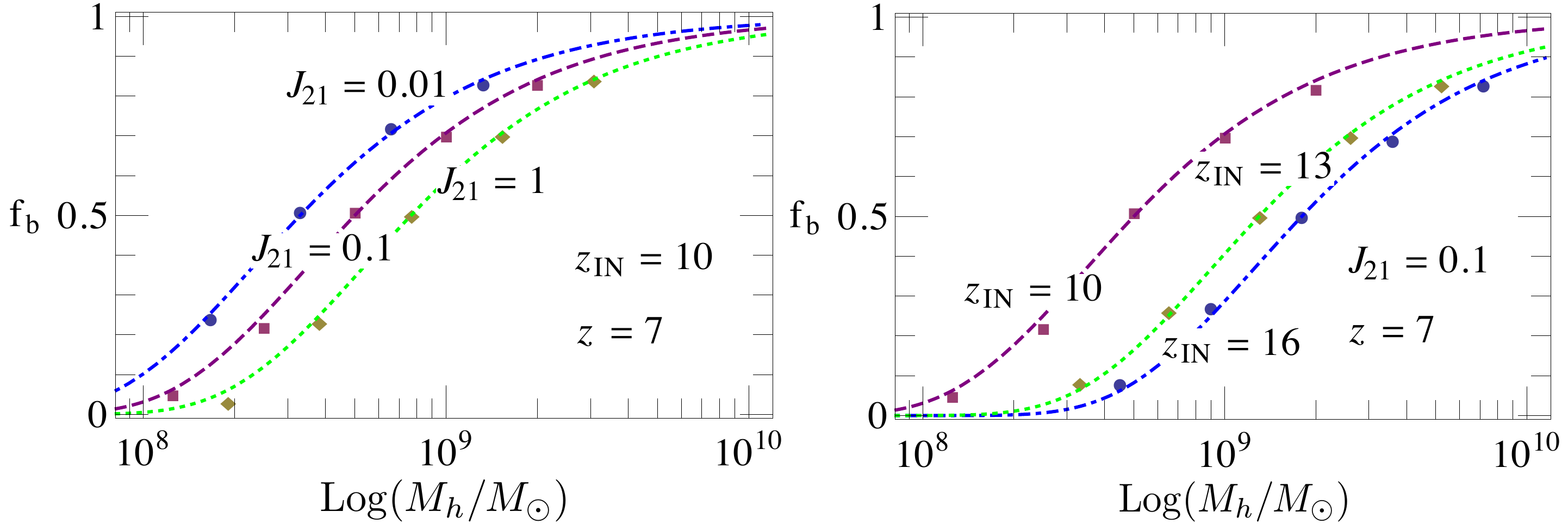}}
\vspace{-0.4cm}
\caption{The baryon fraction, $f_b$, as a function of the halo mass at $z=7$ for different reionization histories \citep{sobacchi2013a}. The {\it left panel} shows results for a UVB intensity ($J_{21}$) that varies over two orders of magnitude fixing the redshift ($z_{in}=10$) at which the halo is exposed to the UVB. As expected, $f_b$ decreases with an increasing $J_{21}$ value. The {\it right panel} shows results for a fixed $J_{21}=0.1$ but where $z_{in}$ varies from 10 to 16. As shown, the earlier a halo of a given mass is exposed to the UVB, the lower is its resulting $f_b$ value.}
\label{ch5_mcrit}
\end{figure*}

In reality however, a spherically collapsing gas cloud never encounters densities within an order of magnitude of the cosmic mean \citep{noh2014} - while a gas cloud encountering gas densities comparable to the virial density of a halo will be able to radiate away its energy and keep collapsing, the bottleneck occurs at densities an order of magnitude less (more) than the virial density (cosmic density) where gas can not cool efficiently. Using 1D collapse simulations, the above formalism has been extended to account for the delay in hydrodynamic response of baryons exposed to a UVB by calculating the baryon fraction as $f_b(z) = 2^{-M_{cr}(z)/M_h}$ \citep{sobacchi2013a} where
\begin{equation}
M_{cr}(z) = 2.8 \times 10^9 \Msun J_{21}^{0.17} \bigg(\frac{1+z}{10}\bigg)^{-2.1}\bigg[1-\bigg(\frac{1+z}{1+z_{in}} \bigg)^2 \bigg]^{2.5}.
\end{equation}
Here ${\rm J_{21}}/(10^{-21} {\rm erg\, s^{-1} Hz^{-1} cm^{-2} sr^{-1}})$ expresses the intensity of the UVB and $z_{in}$ is the redshift at which the halo of mass $M_h$ is exposed to the UVB. As shown in Fig. \ref{ch5_mcrit}, this formalism results in $f_b$ decreasing with an increasing intensity of the UVB for a fixed value of $z_{in}$ (left panel). Further, as expected, the earlier a halo is exposed to a UVB of a given intensity, the lower is its baryon fraction (right panel of the same figure).

\begin{figure*}
\center{\includegraphics[scale=0.5]{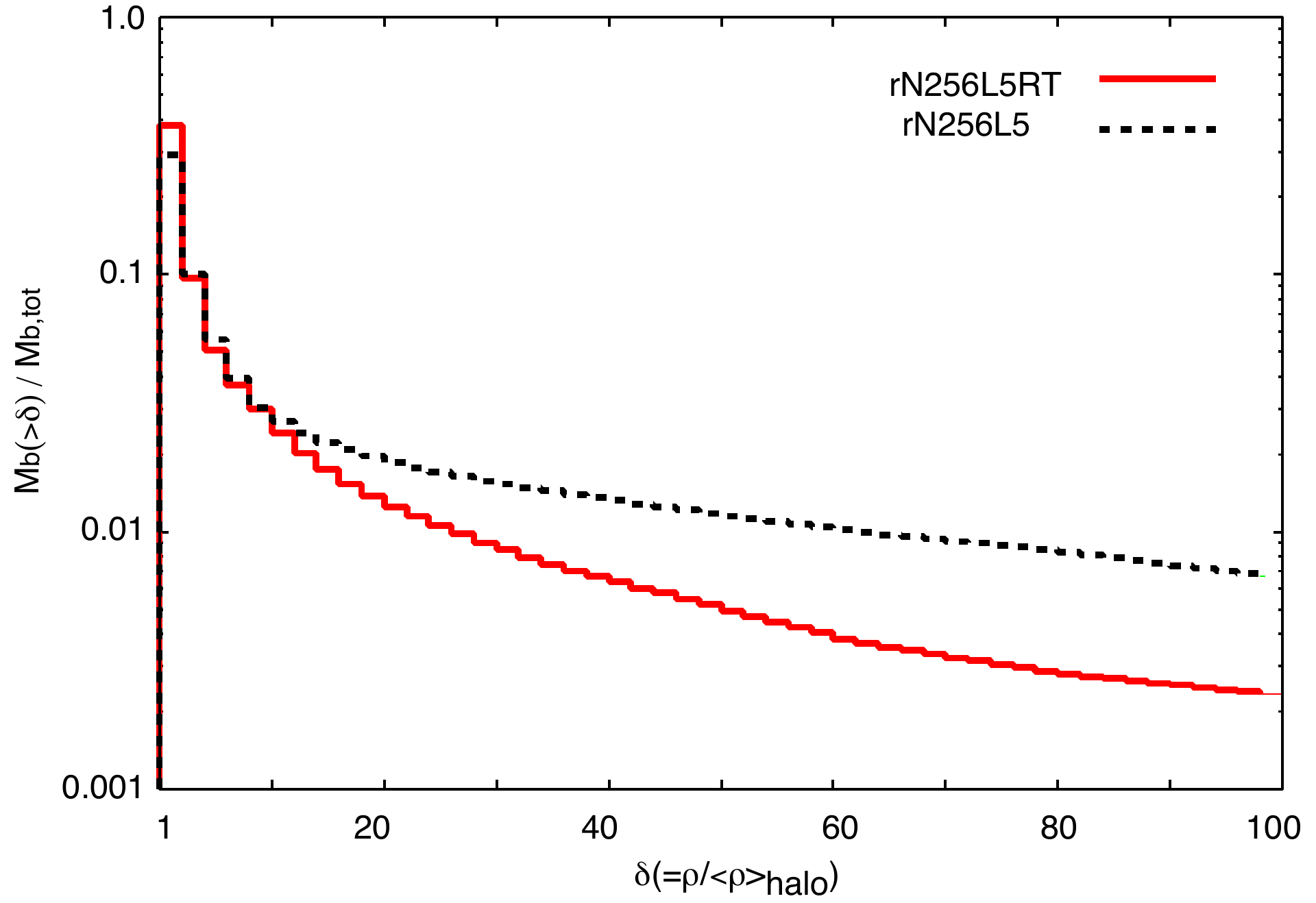}}
\caption{The impact of the UVB on ISM clumping \citep{hasegawa2013}. This plot shows the fraction of baryonic mass residing in regions with a baryon density exceeding the value on the x-axis for a typical $6.7 \times 10^9\Msun$ halo. As clearly seen, including UV feedback (red solid line) smoothes out the baryon mass in the most over-dense (inner-most) regions as compared to a case without UV feedback (dashed black line) by as much as a factor of 3, severely suppressing the star-formation rate.}
\label{fig_clump_ism} 
\end{figure*}

In addition to the IGM, the UVB can also impact the ISM density: radiation hydrodynamic simulations show that accounting for UV feedback inside the ISM can decrease the baryon mass residing in the most overdense regions. As shown in Fig. \ref{fig_clump_ism} this severely suppresses the star formation capabilities of low-mass halos \citep{hasegawa2013}. As UVB feedback affected halos merge, their lower baryon fraction values propagate through to the next generations of halos resulting in an observable impact on the SFR$-M_h$ \citep{finlator2011} and $M_*-M_h$ relations for higher mass halos \citep{dayal2017a}. 

In addition to such ``negative feedback", the UVB can also have a positive (accelerating) effect on reionization by decreasing the IGM clumping. Governing the recombination rate of the gas, the clumping factor is one of the key parameters determining the progress of reionization and the persistence of ionization in over-dense regions. Simulations have now confirmed that photo-ionization heating can act as an effective pressure term by increasing the Jeans mass and smoothing out density perturbations on small scales \citep{pawlik2009, finlator2012, shull2012}, especially if X-ray sources had already pre-heated the gas \citep[e.g.][]{haiman2011, knevitt2014}. This leads to a clumping factor that reduces with $z$ as \citep{pawlik2009, haardt2012}:
\begin{equation}
C(z)=  1+ 43 \, z^{-1.71}.
\end{equation}
As expected, a decrease in the recombination rate has a positive impact on the number of photons needed to complete and maintain reionization. 

In terms of the UVB impact on the reionization topology, \citet{mcquinn2007} have coupled N-body simulations with a RT code to show that, for a given neutral fraction, the UVB suppression of star formation in ionized regions results in \HII\, regions becoming larger (by as much a factor of 4). These \HII\, regions are also more spherical since reionization shifts to being driven by massive rare sources. Indeed, simulations show a more uniform distribution of \HII\, regions, peaked on smaller scales, and the large scale power reduced by tens of percents in the presence of a UVB \citep{sobacchi2013b}. However, both these authors point out that the reionization morphology is primarily dependent on the value of the average neutral hydrogen fraction, $\langle \chi_{HI}\rangle$, and only secondly on the nature and properties of the sources.

Finally, as noted in Sec. \ref{cosm_model}, CDM exhibits a number of problems in matching theoretical predictions with observations on small scales. One of the key manifestations of this is the observed lack of theoretically predicted satellites of the MW \citep[``the missing satellite problem'';] []{moore1999, klypin1999}. One possible explanation is the UVB feedback from reionization: using N-body simulations and abundance matching, \citet{busha2010} have shown that theoretical results match a number of observables (including the luminosity function, radial distribution and circular velocity distribution) for MW dwarfs if only halos with $T_{vir} \gsim 10^5$K can form stars assuming a reionization redshift of $z_{re}= 8^{+3}_{-2}$. Further, changing $z_{re}$ can change the dwarf satellite number density by as much as two orders of magnitude. Similar results have been obtained \citet{lunnan2012} using N-body simulations of the MW coupled with a semi-analytic galaxy formation model where suppressing star formation in halos with $V_{vir} \lsim 50 {\rm km\, s^{-1}}$ can change the satellite numbers by a factor of 3-4 as $z_{re}$ changes between $7.4-11.9$. Further, Keck and {\it HST} photometry of (6) MW dwarfs indicate 100\% of their stars having formed by $z \sim 3$ - the similarity of their ancient population suggests a synchronised truncation of star formation driven by an external influence such as reionization \citep{brown2014}. In terms of galaxy populations, the increasing deficit of baryonic mass with decreasing halo mass inferred, using {\it SDSS} and {\it ALFALFA} (Arecibo Legacy Fast ALFA survey) observations, might have a contribution from reionization feedback in addition to SN-driven outflows \citep{papastergis2012}.

 \begin{figure*}
{
\includegraphics[width=0.5\textwidth]{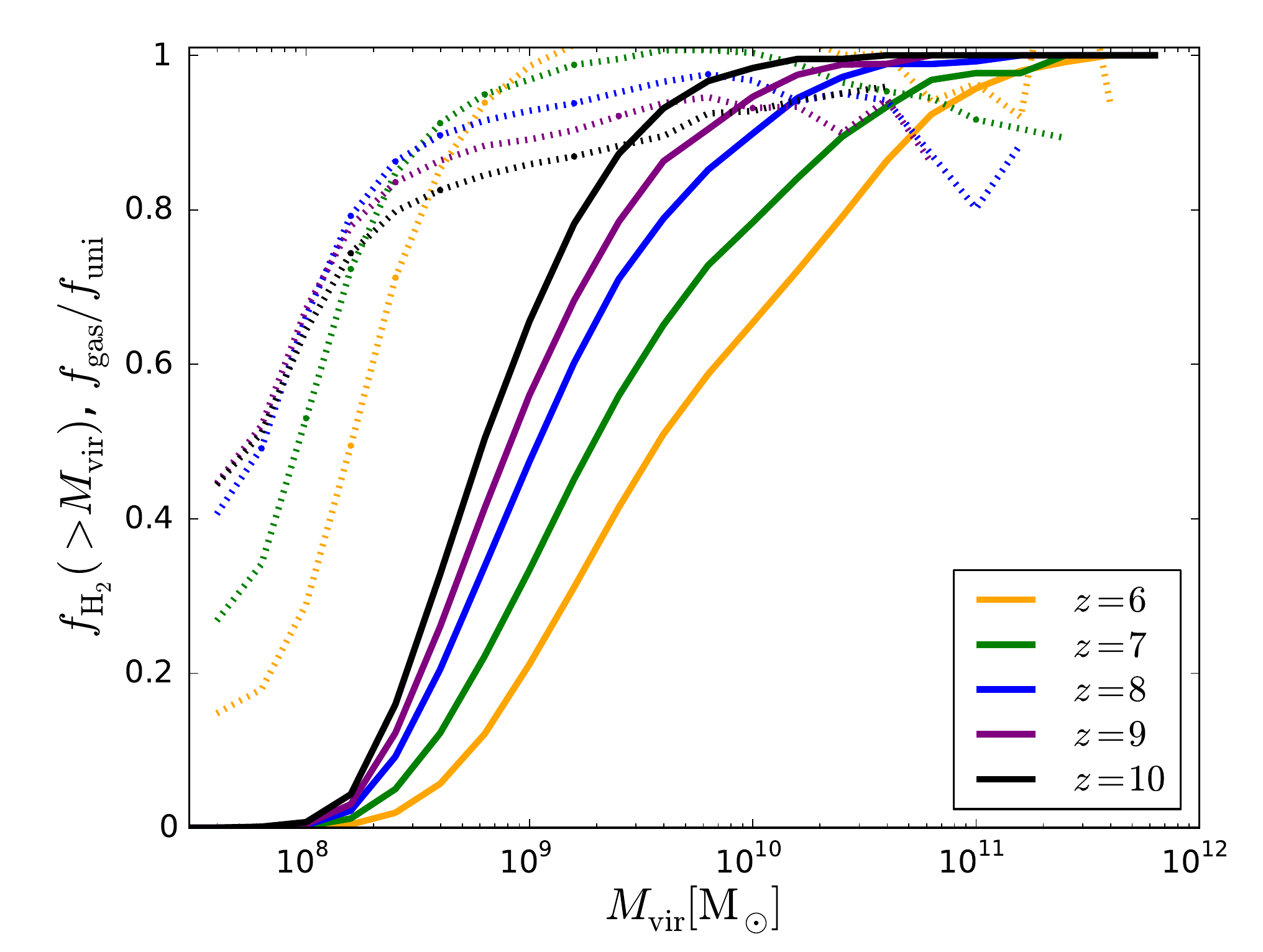}
\includegraphics[width=0.5\textwidth]{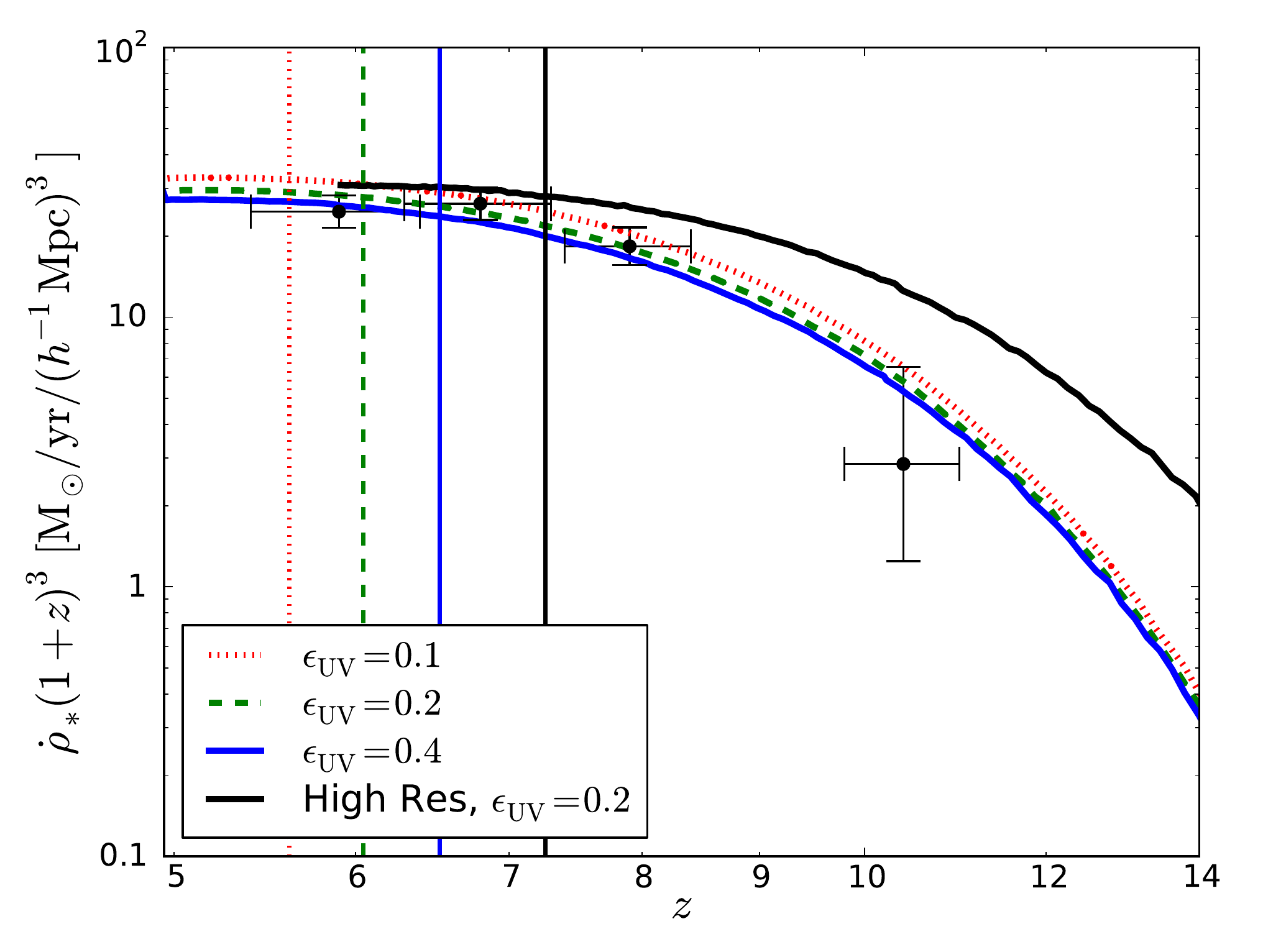}
}
\vspace{-1.5\baselineskip}
\caption{{\it Left panel:} The average gas fraction (dotted lines) and the molecular gas fraction (solid lines) as a function of $z$ using high-resolution ($20h^{-1}$ Mpc; $1024^3$) simulation runs from \citet{gnedin2014}. As shown, the low-mass halos most affected by UV feedback contain little molecular gas and therefore, do not form stars in any case. {\it Right panel:} the impact of UV feedback on the global star formation rate density for different simulation runs \citep{gnedin2014}. The black line corresponds to the high resolution simulation shown in the left panel, with the other lines showing results for the fiducial  ($40h^{-1}$ Mpc; $1024^3$) model for three different values of $\epsilon_{UV}$ - the escape fraction of ionizing radiation up to the resolution limit; black points show observational data from \citep{bouwens2015}. As shown, reionization does not impact the global star formation rate in these simulations.}
\label{ch5_gnedin_nouv}
\end{figure*}

\begin{figure}[h]
\center{\includegraphics[scale=1.15]{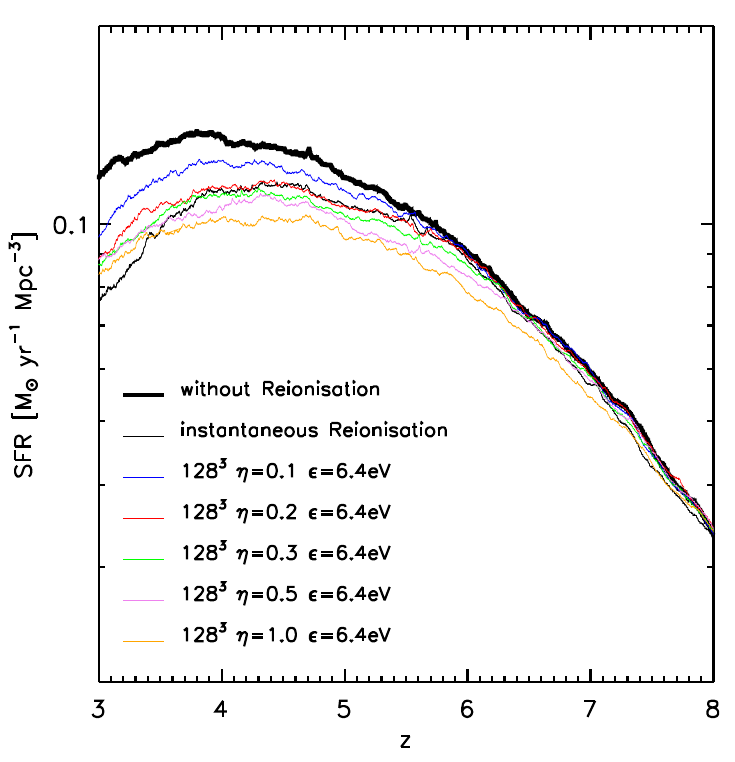}}
\caption{The impact of the UVB on the star formation rate density \citep[SFRD;][]{petkova2011}. As marked, the thin black line shows the SFRD, at its maximum value, in a model without any UV feedback. The think black line shows the SFRD in the case of instantaneous reionization at $z=6$ using the Haardt \& Madau background \citep{haardt1996}. The other lines show the values coupling GADGET with a moments method of radiative transfer for ISM escape fraction ($\eta$) values ranging between 10-100\%. As expected, the impact of UV feedback increases with increasing $\eta$. }
\label{fig_uvb_fb} 
\end{figure}

A number of calculations, summarized below, have been used to couple UV feedback and galaxy formation. Details concerning the box sizes and resolutions can be found in Table \ref{table31} in Sec. \ref{theo_tools}. Essentially, the problem is one of the physical scales that need to be modelled: while galaxies have physical scales of a few kpc, reionization and its impact needs to be modelled on scales of $\sim$100 Mpc (as required for a convergent topology \citep{iliev2014}), especially in the end stages when almost all galaxies are immersed in the UVB. A number of different techniques have therefore been adopted to circumvent the problem. These range from semi-analytic models \citep{sobacchi2013a, sobacchi2013b} to N-body simulations/semi-analytic models coupled with RT \citep{dixon2016, mutch2016} to 
high-resolution, small volume radiation-hydrodynamic simulations \citep{petkova2011, finlator2011, hasegawa2013, gnedin2014, pawlik2013, pawlik2015, aubert2015} to high-resolution, intermediate volume radiation-hydrodynamic simulations \citep{ocvirk2016}. These have led to often conflicting scenarios in which the UVB either has a negligible effect or fully suppresses star-formation in low mass galaxies as now summarised:

\textit{(i) Negligible effects}. A number of works \citep[e.g.][]{sobacchi2013b, raicevic2011} note that accounting for the delay in the hydrodynamic response of baryons to the impinging UVB is crucial to prevent (a spurious) overestimation of the impact of the UVB. Using a variety of approaches mentioned above, the star formation rate density (and by extension reionization) is found to be impervious to the UVB unless the key reionization sources are either molecular-cooling driven \citep{sobacchi2013b} rapidly losing their gas after SN explosions \citep{mutch2016, pawlik2015} or low-mass galaxies that contain little to no molecular gas in the first place \citep{gnedin2014, mutch2016} as also shown in Fig. \ref{ch5_gnedin_nouv}. The impact of the UVB would also be mitigated if the gas fuelling star-formation had already cooled into halos by the time it is exposed to the UVB \citep{raicevic2011}. Further, although cumulatively decreasing the gas content of early galaxies through time, the UVB can only suppress the number of ionizing photons-to-baryon ratio by about a factor $\lsim 2$ towards the end stages of reionization, when most galaxies are immersed in ionized regions. This corresponds to changing the end redshift of reionization by $\Delta z \lsim 0.5$ \citep{sobacchi2013b}. Further, \citet{gnedin2014} show that, given that the galaxies affected by UV feedback contain little to no molecular gas in the first place, UV feedback has no observable impact (slope changes by 0.1 for a unit shift in $z_{re}$) on the UV LF. \\

\textit{(ii) Strong effects}. On the contrary, several other works find UV feedback to have a large effect both on reionization and the properties of high-redshift galaxies. Combining SPH simulations with an optically thin variable Eddington tensor approximation, \citet{petkova2011} find that the increase in gas temperature, and the corresponding decrease in gas densities due to pressure effects, leads to gas cooling and collapsing more slowly. This results in a decrease in the SFR density at a given $z$ as shown in Fig. \ref{fig_uvb_fb}. This effect is found to be the most pronounced for the lowest mass halos ($T_{vir} \simeq T_{gas}$). Further, the SFR efficiency decreases with an increase in $f_{esc}$ ($\eta$ in Fig. \ref{fig_uvb_fb}) since a larger number of ionizing photons are available to heat the gas; although \citep{simpson2013} find similar results, they show that SN feedback is also responsible for dispersing the cold, dense gas in the halo core. This is in agreement with results from radiation hydrodynamic simulations that show the UVB can suppress star formation in halos with $T_{vir}<10^5$ K, corresponding to $M_h \lsim 10^{9.2}\Msun$ at $z=6$ \citep{finlator2011}. These authors caution, however, that the effects of SN feedback and the UVB couple positively at small scales - outflows decrease the star formation rate, resulting in a lower UVB, permitting low-mass halos to retain their gas. Further, they also find that, even, low-mass halos recently exposed to a UVB are resilient to its effects since most the gas has cooled to high densities and will form stars for many dynamical timescales. Although \citep{hasegawa2013} find similar trends of UV feedback suppressing the star formation rate density by a factor of 5-6 at $z=6$, their work results in much stronger feedback as compared to other works \citep{wise2009, finlator2011, petkova2011}. This arises from their inclusion of two ISM effects: (a) the internal UV field generated by star formation inside low mass ($M_h \lsim 10^9 \Msun$) halos causes the dispersal of high density regions, and (b) the merger of low-mass systems into larger ones destroys dense gas preventing ${\rm H_2}$ formation and cooling.

\subsection{The sources of reionization: galaxies, quasars or ..? }
\label{sources_reio}
We are now in a position to discuss the key reionization sources. Reionization should naturally begin with the formation of the first metal-free stars at $z \simeq 15$, with the key sources shifting to larger metal-enriched halos, powered by PopIII/II stars and even black holes, at later times. However, this naive picture is complicated by the fact that the sources of reionization depend on a number of poorly understood parameters including the minimum halo mass that can host enough cool gas to allow star formation, the star formation/black hole accretion rates and intrinsic stellar populations at high-redshifts, the escape fraction of LyC photons and the impact of the UVB on the gas mass (and hence star formation rates) of low-mass halos. It is precisely because of these reasons that the nature and properties of the reionization sources, ranging from star forming galaxies to QSOs to exotic scenarios (e.g. decaying dark matter) remain debated. 

Assuming galaxies to be the key reionization sources one can try and estimate the contribution of different halo masses to the ionizing photon budget at that redshift, the results of which are shown in Fig. \ref{fig_sources_ana}. 

\begin{figure}[h]
\center{\includegraphics[scale=0.65]{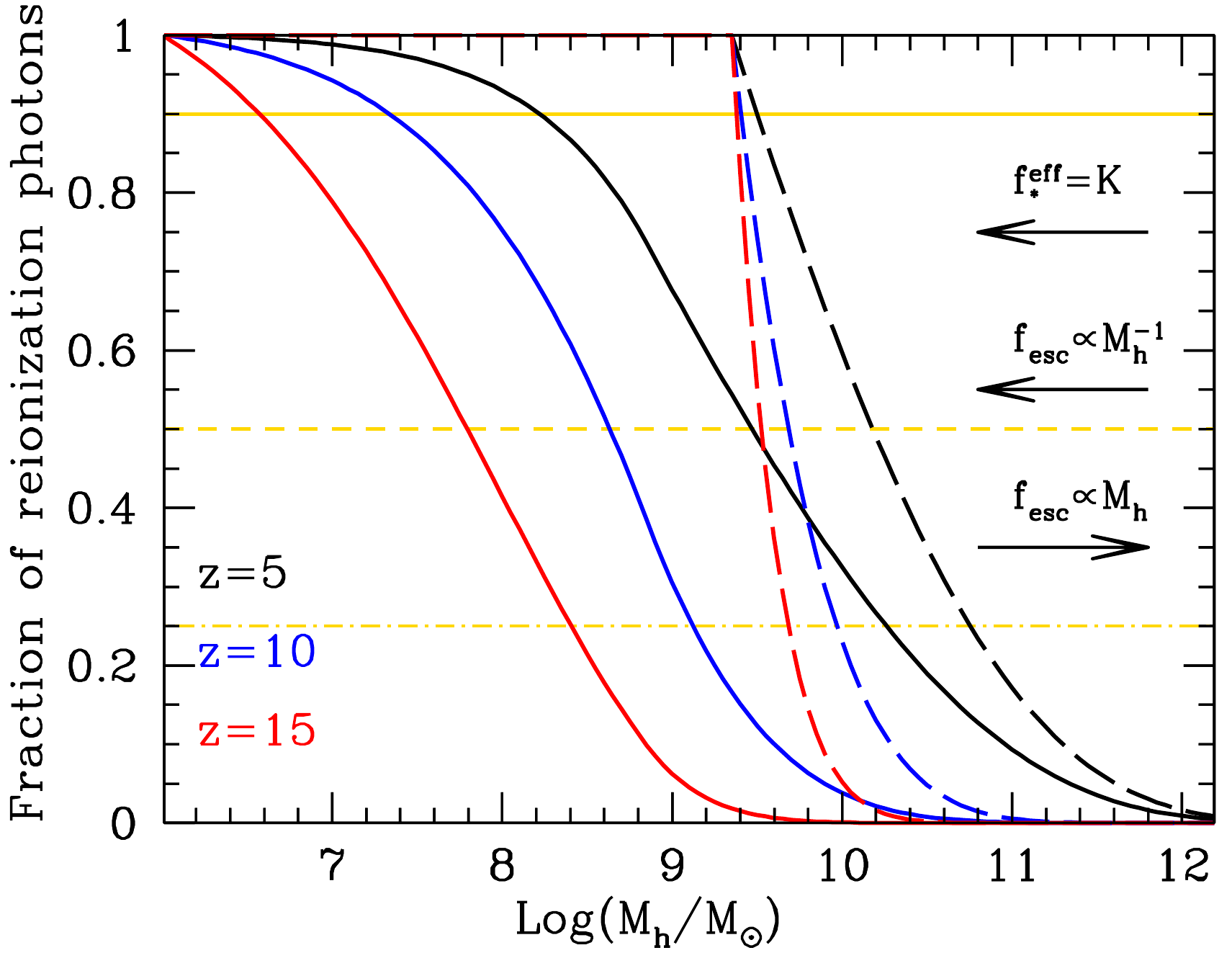}}
\caption{The cumulative fractional contribution to LyC photons from galaxies below the halo mass value on the x-axis. Assuming each halo to contain a cosmological baryon fraction, a constant $f_{esc}$ and a constant clumping factor $C=1$, we show results (for $z=5, 10$ and 15 as marked) using the maximal star formation efficiency-halo mass relation derived in Sec. \ref{met_igm}. The solid lines show results without any UVB feedback. Dashed lines show results for a {\it maximal} UV feedback scenario where all galaxies with $V_{vir}<50{\rm km\, s^{-1}}$ are assumed to be devoid of baryons due to photo-evaporation (i.e. $f_b=0$). The horizontal lines show the 25, 50 and 90\% fractions. As seen, while $M_h \lsim 10^{8.7}\Msun$ galaxies can provide half of the LyC photons using a halo mass dependent star formation efficiency, this value increases to $M_h \lsim 10^{9.8}\Msun$ if only galaxies above $V_{vir}>50 {\rm km\, s^{-1}}$ can support star formation. Changing the model assumptions, such as assuming $f_{esc}$ to increase (decrease) with the halo mass will result in a larger fractional contribution from higher (lower) $M_h$ values. Further, assuming a constant star formation efficiency for all halos, as opposed to a mass-dependent value, will result in a larger fractional contribution from lower masses as marked.}
\label{fig_sources_ana} 
\end{figure}

We start with the HMF and assign each galaxy a cosmological baryon fraction and a star formation efficiency that provides enough SN energy to unbind the rest of the gas, up to a maximum threshold which we take to be 0.03, as in Sec. \ref{met_igm}. Further, we assume constant values for both $f_{esc}$ and $C$. Allowing galaxies with halo masses as low as $10^6\msun$ to form stars, half of the LyC at $z = 15$ (5) come from halos with $M_h \lsim 10^{7.8}\, (10^{9.6})\Msun$; larger halos will naturally provide a larger fraction of LyC photons if star formation is suppressed below the atomic cooling threshold. Complicating the calculation, we assume a {\it maximal} UV feedback scenario that completely suppresses the baryon fraction (i.e. $f_b=0$) for all galaxies with $V_{vir}\lsim 50 {\rm km\, s^{-1}}$. This results an increase in the halo masses providing half of the LyC photons to $M_h \lsim 10^{9.6} (10^{10.4})\Msun$ at $z=15\, (5)$. As marked in the same plot, changing the model assumptions will naturally change these results: for example, assuming $f_{esc}$ to increase (decrease) with the halo mass will result in a larger fractional contribution from higher (lower) mass halos. Further, assuming a constant star formation efficiency for all halos will result in a larger fractional contribution from lower masses. These simple calculations demonstrate that, for almost all reasonable physical scenarios, most LyC photons are provided by low-mass halos at high-$z$. Higher mass halos, however, might become more important at later times due to an increase in the number density of the largest sources and UV suppression of star formation in low-mass halos \citep{choudhury2007, paardekooper2015, xu2016, chen2017}. 

\begin{figure}[h]
\center{\includegraphics[scale=0.55]{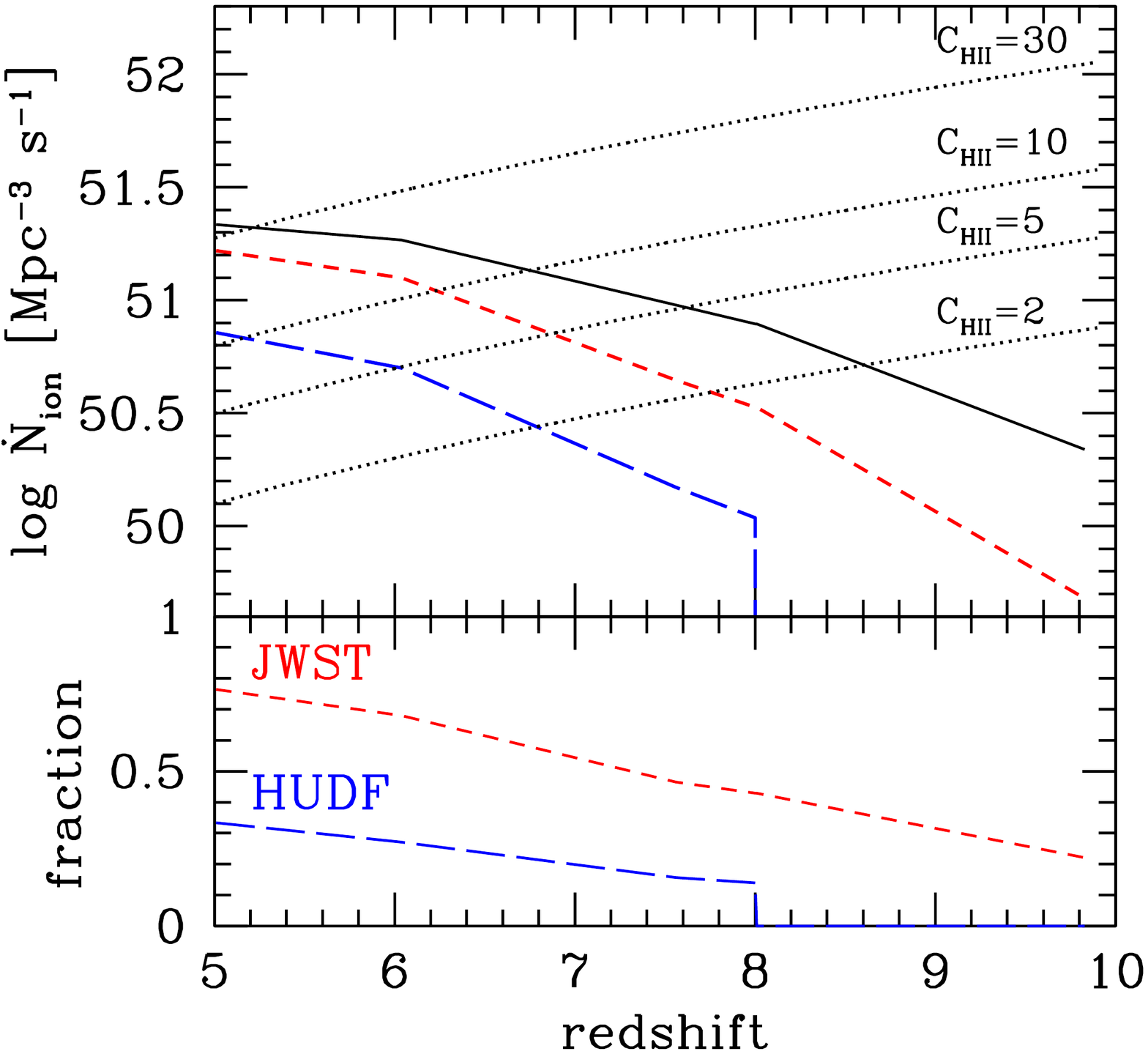}}
\caption{Results from hydrodynamic simulations showing the contributions of galaxies of different magnitudes to reionization \citep{salvaterra2011}. {\it Upper panel:} Redshift evolution of the total specific ionization rate (solid line). The short (long)-dashed line corresponds to galaxies detectable by {\it JWST (HST/WFC3)}; dotted lines show the specific IGM recombination rate for different values of the clumping factor. {\it Bottom panel}: Fraction of ionizing photons coming from galaxies identified by JWST and in the HUDF. These calculations have assumed a fixed value of $f_{esc}=0.2$. As shown, while the {\it HST} (using a magnitude limit of $\muv \sim -18$) can only detect (massive) galaxies providing about 40\% of reionization photons, the {\it JWST} is critically required to peer deep enough ($\muv \sim -15$) to be able to detect the sources providing 80\% of the total ionizing photon budget.}
\label{fig_salv_reios} 
\end{figure}

This broad picture agrees with the majority of theoretical and observational works. Observationally, different groups have combined arguments involving the UV luminosity density contributed by galaxies of different masses \citep{finkelstein2012}, the steep faint-end slope of the LBG luminosity function \citep[e.g.][]{bouwens2012}, $f_{esc}$ being higher for galaxies with bluer UV slopes \citep{duncan2015}, and the abundance and luminosity distribution of galaxies from the Hubble Ultra Deep Fields \citep{robertson2013, robertson2015} to conclude that galaxies with $\muv \simeq -10$ to  $-15$ alone can reionize the IGM. This is accord with theoretical results that claim galaxies alone could have reionized the Universe \citep{choudhury2007, petkova2011, finlator2011, yajima2011, pawlik2009, becker2013, gnedin2014, dayal2017, hassan2018}. Using roughly constant values of $f_{esc}$ across the entire galaxy population, semi-analytic models \citep{choudhury2007, raicevic2011, dayal2017}, semi-numerical models \citep{liu2016,qin2017}, SPH simulations \citep{razoumov2010, salvaterra2011}, combinations of SPH simulations and radiative transfer \citep{finlator2011, yajima2011, paardekooper2011, paardekooper2013,paardekooper2015} and AMR simulations \citep{wise2014} all converge on low-mass ($M_h \lsim 10^{9.5} \Msun$ and $\muv \gsim -17$) halos providing the bulk ($\gsim 50\%$) of LyC photons at $z \gsim 7$. We show one such result, from SPH simulations, in Fig. \ref{fig_salv_reios} \citep{salvaterra2011}. This figure clearly shows that the photon output required to maintain reionization increases with the IGM clumping factor. Furthermore, it shows that, with its magnitude limit of $\muv \sim-15$, the {\it JWST} will be crucial in shedding light on the sources that provide roughly 80\% of LyC photons.

However, the contribution from galaxies of different masses (or luminosities) sensitively depends on the details of the feedback prescriptions implemented and values used for $f_{esc}$ and $C$. For example, while reionization finishes as early as $z_{re} \sim 8$ for $f_{esc}=1$, it gets delayed to $z_{re} \sim 5.5$ using a lower $f_{esc}=0.2$ \citep{petkova2011}. Using AMR simulations, where $f_{esc}$ decreases from $\sim 50\%$ for $M_h \sim 10^{7-7.5}\Msun$ halos to $\sim 5\%$ for larger halos, \citep{xu2016, chen2017} find that while low-mass halos dominate the reionization budget at early epochs, larger $M_h \gsim 10^9\Msun$ halos can provide up to 50\% of the total ionization budget at later epochs \citep[$z \lsim 10$; see also][]{choudhury2007}. On the other hand, the few models that either assume $f_{esc}$ to scale with the star formation rate density (and hence with the halo mass) or quench star formation in any cell with $T\gsim 2 \times 10^4$K (resulting in a severe star formation suppression in $M_h \lsim 3 \times 10^9 \Msun$ halos) naturally find much larger halos, with $M_h \sim 10^{10-11}\Msun$ ($\muv \gsim -18$), dominating the total reionization budget. Interestingly, coupling N-body simulations with RT, \citet{iliev2012} have shown that \HI 21 cm line statistics (e.g. the power spectra) can be used to distinguish between reionization driven solely by high-mass ($\gsim 10^9 \Msun$) halos as opposed to one where both high- and low-mass ($10^{8-9}\Msun$) halos contribute. These calculations have been extended by \citet{dixon2016} who show that, while the overall shape and skewness of the 21 cm signal are quite robust to the exact nature of source suppression, box size and resolution, higher order statistics (e.g. the kurtosis of the 21 cm power spectrum) could be used to distinguish between reionization purely driven by low-mass versus high-mass halos.

\begin{figure}
\center{\includegraphics[scale=0.55]{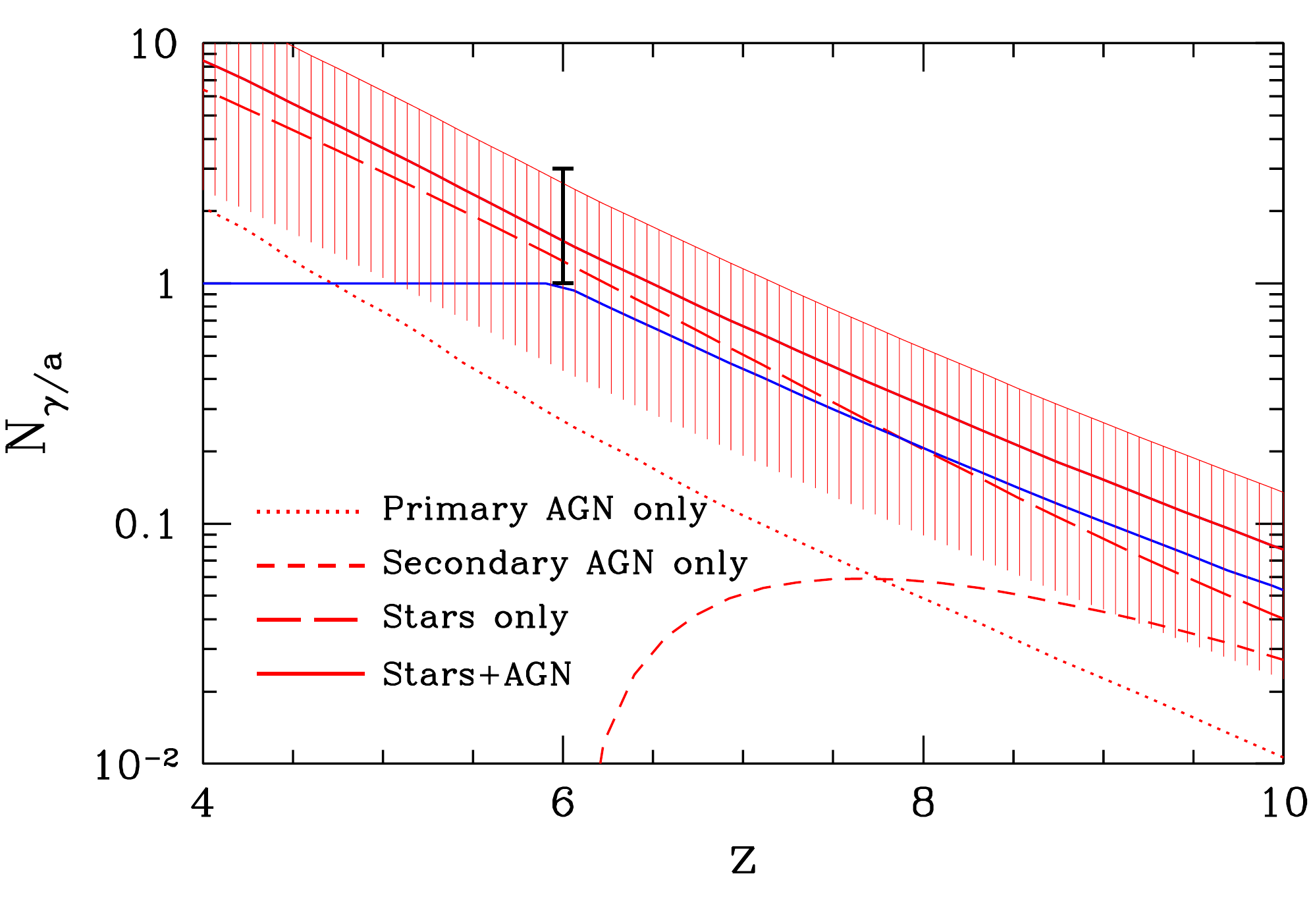}}
\caption{Total ionizations-to-atom ratio as a function of redshift assuming quasars with large seeds with high spins from \citet{volonteri2009}; the hatched region shows the estimated observational uncertainty associated with the model. Dotted, short-dashed, and long-dashed red lines show the contributions from primary and secondary ionization from quasars and from stars respectively. As shown, secondary ionizations from quasars can have a significant impact at $z \gsim 8$, with galaxies dominating the budget at lower redshifts. }
\label{fig_volonteri_qso} 
\end{figure}

Quasars might also significant contribute to reionization \citep{volonteri2009, madau2015, mitra2015, mitra2018}, specially at $z \gsim 8$ if ionizations by secondary electrons are accounted for, with galaxies taking over as the dominant reionization sources at $z \lsim 6$ as shown in Fig. \ref{fig_volonteri_qso}. The contribution of AGN to reionization has witnessed a resurgence after the recent claim of extremely high faint AGN number densities measured by \citet{giallongo2015} at $z \gsim 4$.
It has been shown that if such a claim is confirmed, and this high comoving emissivity persists up to $z \simeq 10$, then AGN alone could drive reionization with little/no contribution from galaxies \citep{madau2015}. However, in addition to the fact that later works have not been able to corroborate these high AGN number densities, an AGN-dominated reionization scenario is in tension with current constrains on the thermal history of the IGM \citep{daloisio2017}. Further, AGN-driven reionization results in a slower evolution of the HeII Ly$\alpha$ forest as compared to observations and is therefore ruled out at the $2-\sigma$ level; the data therefore favours a model where galaxies with $f_{esc} \sim 10\%$ dominate in the initial reionization stages with AGN being relevant only at $z \lsim 6$ \citep{mitra2018}. 

In terms of the imprint on reionization, \citet{hassan2018} have used a semi-numerical framework to show that AGN-dominated models produce larger ionized regions, reflected in the 21 cm power being twice at large at all scales. Further, using SPH simulations processed with RT \citet{kakiichi2017} show that, although UV obscured quasars can lead to a broader ionization front as compared to galaxies and unobscured quasars, they can imprint a distinctive morphology on the \HII\, regions only if their ionizing photon contribution is consistently larger than galaxies since the onset of reionization. Finally, \citep{bolgar2018} raise the point that the quasar impact depends on its duty cycle, leaving potentially observable imprints on the 21 cm signal in their direct environment.  

Finally, we list a few sources going beyond metal-rich star formation and quasars:

\textit{(a) PopIII stars}: Post-processing SPH simulations with RT, \citet{paardekooper2013} show that, although they could have started the process of reionization, given their short life-times and low numbers, PopIII stars contribute negligibly to reionization; similar results have been found by semi-analytic models \citep{mitra2015} and AMR simulations \citep{wise2008c}. These results are borne out by the findings of \citet{maio2010} who find that PopIII stars contribute $\lsim 10\%$ to the total star formation rate density at $z \lsim 15$ even for an ``extreme" PopIII Salpeter IMF between $0.1-100 \Msun$. 

\textit{(b) Runaway stars}: Contributing 50-90\% of the total ionizing radiation escaping galaxies and enhancing $f_{esc}$ values by as much as a factor of 7 in low-mass galaxies ($\muv  = -12$ to $-16$ at $z=10$),  runaway stars could have had a significant contribution to reionization \citep{conroy2012, kimm2014} although \citet{ma2015} find a much lower enhancement. 

\textit{(b) Decaying/annihilating DM}: While decaying/annihilating light dark matter particles (of mass 1-10 Mev) and sterile neutrinos (2-8 keV) can contribute to partial early reionization at $z \lsim 100$, their contribution to the electron scattering optical depth is minimal \citep[$\tau \lsim 0.01$;][]{mapelli2006} if the unresolved X-ray background emissivity constraints are to be respected \citep[e.g.][]{salvaterra2005, mcquinn2012}. The main contribution of dark matter annihilation/decays lies in an earlier and more homogeneous heating of the IGM. This can leave detectable imprints, specially on the height and positions of the 21 cm peaks at large scales \citep[$k \sim 0.1 {\rm Mpc^{-1}}$;][]{evoli2014}, which can be used to constrain the dark matter energy injection rates and lifetimes \citep{furlanetto2006b}.


\newpage

\section{Emerging galaxy properties }
\label{gal_prop}

Having described the underlying theory for early galaxy formation, we devote this section to a detailed comparison between theoretical results and observational data. This (iterative) process has been enormously successful, both, in interpreting the available measurements and improving theoretical models as now described.

\subsection{High-z galaxies: confronting observations and theory} 
\label{obs_gal}

The past decade has seen the ``golden age" in the hunt for high-$z$ galaxies, made possible by observatories including the {\it HST}, the 8.2m {\it Subaru}, the 8.2m {\it VLT}, the 85cm Spitzer and the 10m {\it Keck}. Providing coverage from the optical through to mid infra-red wavelengths, these facilities have allowed unprecedented constraints on the physical properties of early galaxy populations. We start by briefly summarising the two main approaches used to detect high-$z$ galaxies:
\begin{itemize}
\item{The Lyman break technique to identify Lyman Break Galaxies (LBGs): First proposed by \citet{meier1976}, this approach relies on successfully identifying the $912$\AA\, Lyman break caused by the absorption of LyC photons by neutral ISM gas. At $z\gsim5$, the Lyman Break effectively shifts to 1216\AA, with photons of lower wavelengths being absorbed by neutral hydrogen in both the ISM and the intervening IGM. This ``Gunn-Peterson" absorption trough \citep{gunn-peterson1965} in high-$z$ quasar spectra has been a crucial input in inferring the end redshift of reionization \citep[e.g.][]{becker2001,fan2006}. As shown in the following sections, LBG selections have yielded thousands of candidates with $z$ as high as $z \simeq 11$, thanks to two key instruments - the Advanced Camera for Surveys (ACS) and the Wide Field Camera3 (WFC3) onboard the {\it HST} \citep{ellis2012, oesch2013, oesch2016, bouwens2007, bouwens2010, bouwens2011a, bouwens2011b, bouwens2015, atek2015, livermore2016, bowler2014, bowler2015, mclure2010, mclure2011, mclure2013, bradley2012, bradley2014, castellano2010a, castellano2010b}, shedding light on the hitherto hidden early galaxies that had formed in the first billion years. }

\item{The Ly$\alpha$ line to identify Lyman Alpha Emitters (LAEs): As detailed in Sec. \ref{fesc}, most hydrogen ionizing photons are absorbed by \HI in the ISM of galaxies, with only a small fraction ($f_{esc}$) escaping into the IGM. Given the large densities associated with the ISM, this ionized gas recombines on timescales of a few Myr (Eqn. \ref{trec}) giving rise to the Ly$\alpha$ recombination line at $1216$\AA\, in the galaxy rest-frame. Using the case-B recombination approximation, where the recombined \HI is optically thick and there is a 2/3 probability of a recombination leading to a Ly$\alpha$ photon \citep{osterbrock1989}, the intrinsic Ly$\alpha$ luminosity can be calculated as $L_\alpha = [2/3] (1-f_{esc}) \dot N_s {\rm h} \nu_\alpha\, [{\rm erg\, s^{-1}}]$. As noted in Sec. \ref{link_reio_met}, $\dot N_s$ is the number of \HI ionizing photons produced by the stellar population and $\nu_\alpha$ is the Ly$\alpha$ frequency. Despite the elegant proposal of using the Ly$\alpha$ line as a beacon for high-$z$ galaxies as early as 1967 \citep{partridge1967a}, the first LAEs were 
only successfully identified three decades after the publication of this seminal paper \citep{dey1998, spinrad1998, hu1998}. Since these early detections, narrow band techniques have identified hundreds of LAEs upto $z \simeq 8$ \citep{ouchi2005, taniguchi2005, ouchi2008, ouchi2010, iye2006, kashikawa2006, ota2008, ota2010, taniguchi2009, kashikawa2011,matthee2014,matthee2015, ouchi2017}. However, a key drawback of the Ly$\alpha$ line is its extremely high optical depth against neutral hydrogen \citep{madau2000}: 
\begin{equation}
\tau_\alpha = \frac{\pi e^2 f  \lambda_\alpha}{m_e c H(z)} \chi_{HI}(z) n_{H}(z) \simeq 1.5\times 10^5  \chi_{HI}(z) h^{-1} \Omega_M^{-1/2} \bigg(\frac{\Omega_b h^2}{0.019}\bigg) \bigg( \frac{1+z}{8} \bigg)^{3/2},
\end{equation}
where $f$ is the oscillator strength, $ \lambda_\alpha$ is the Ly$\alpha$ wavelength and $e$ and $m_e$ represent the electron charge and mass, respectively. As seen from this equation, $\tau_\alpha \gg 1$ for a galaxy embedded in an IGM with $\chi_{HI}=0.1$ resulting in a severe attenuation of the Ly$\alpha$ luminosity. Such a galaxy would probably {\it not} be visible as a LAE. Statistically, this implies a decreasing fraction of star forming galaxies being visible in the Ly$\alpha$ with increasing $\chi_{HI}$. An indication of this effect is already available from the apparent Ly$\alpha$ LF (number density of galaxies as a function of the Ly$\alpha$ luminosity) which, showing effectively no evolution between $z \simeq 3.1-5.7$ \citep{ouchi2008}, shows a steep drop at higher $z$ with the knee luminosity ($L_*$) at $z =  6.6$ being about 50\% of the value at $z=5.7$ \citep{kashikawa2006, kashikawa2011}. This decline has also been confirmed by other groups \citep{ouchi2010, hu2010, santos2016, ota2017}. The increasing dearth of LAEs at high-$z$ has prompted searches for other emission lines which include CIII]$\lambda 1909$ \AA\, \citep{stark2015}, [OIII]+H$\beta$ \citep{smit2014, smit2015, roberts-borsani2016} and CIV]$\lambda 1548$ \AA\, \citep{stark2015b, schmidt2017}. Interestingly, such strong line emitters have been sign-posts for Ly$\alpha$ emission: for example \citet{roberts-borsani2016} have observed strong line emitters at $z \simeq 7-9$, the follow up spectroscopy of which shows all (four) sources to have a Ly$\alpha$ line. Suggesting up to 50\% of the galaxy population being in an extreme emission phase as a result of their young stellar populations and intense ionizing fields, these methods provide an excellent opportunity, both for direct detections using the Near-Infrared Spectrograph (NIRSpec) onboard the {\it JWST} as well as targets for Ly$\alpha$ emission}.
\end{itemize}

In terms of the relation between LAEs and LBGs, there is consensus that these are both drawn from the same underlying population \cite[e.g.][]{verhamme2008, dayal2012}; the fraction of time a galaxy is visible as an LAE or LBG depends on its underlying stellar mass \citep{hutter2015}. While the decrease in the fraction of LBGs showing Ly$\alpha$ emission (with increasing redshift) has been attributed to an increasingly neutral IGM \citep[e.g.][]{stark2011, pentericci2014}, this interpretation is complicated by a number of caveats. These include the EW criterion used to select LAEs \citep[e.g.][]{ouchi2008}, the degeneracy between the (redshift-dependent) effects of ISM dust and IGM attenuation on Ly$\alpha$ photons \citep{dayal2012}, a three dimensional degeneracy between $f_{esc}$, ISM and IGM attenuation as well as the ionization structure encountered along a given line-of-sight \citep{hutter2014}. 

We also briefly mention other techniques used to hunt for high-$z$ galaxies that include \citep[for a full discussion see e.g.][]{dunlop2013, stark2016}: 
{\it(a)  The Balmer break technique}: which relies on identifying the Balmer break at 3646\AA\, in the rest-frame of the galaxy. However, given the fact that high-$z$ stellar populations are generally younger than a Gyr and that the Balmer break only causes a factor of 2 drop in the flux, this technique has not yielded any galaxies not already identified by the LAE and LBG techniques \citep{wiklind2008}; 
{\it (b) Super Massive Black Holes}: the presence of SMBHs has helped identify QSOs at $z$ as high as $z \simeq 7.5$ \citep{willott2010, mortlock2011}. Observations at $z \sim 6-7$ have firmly established that the rarest, most massive ($10^{12-13}\Msun$) and brightest galaxies ($\muv \lsim -23.5$) host SMBHs with mass $\sim 10^{7-8}\Msun$ solar masses \citep{willott2015, kashikawa2015, jiang2016, ono2017}. 
While cross-correlating CANDELS and {\it Chandra} X-ray data, \citet{giallongo2015} have recently detected AGN activity in galaxies that are about 4 magnitudes fainter, with $\muv \sim -19$, this result has not been confirmed by later observations with the Hyper Suprime Camera on {\it Subaru} that, instead, roughly find a 100\% galaxy fraction down to $\muv \simeq -22$ at $z \simeq 4-6$ \citep[e.g.][]{ono2017}; {\it (c) Gamma Ray Bursts (GRBs)}: arising from the death throes of massive stars \citep{woosley2006}, their extreme luminosity has resulted in GRBs being observed up to $z \simeq 8.4$ with {\it Swift} \citep{salvaterra2009, tanvir2009}. The low surface brightness of their host galaxies, however, has limited the study of their physical properties. In what follows, we will solely focus on the LAE and LBG techniques that have, so far, been the most successful at collecting statistically significant samples of $z \gsim 5$ galaxies.


  \subsection{The high-$z$ UV Luminosity function}
  \label{lf}
Describing the number density of galaxies as a function of their observed UV luminosity\footnote{The intrinsic UV luminosity has an initial value (at $t_0=2$ Myrs) of ${\rm L_{UV}}= 10^{33.07}\, {\rm erg\, s^{-1}\, \AA^{-1} \,\Msun^{-1}}$ for a $0.1-100\Msun$ Salpeter IMF and $Z = 0.05 \Zsun$ using {\it Starburst99} \citep{leitherer1999, leitherer2010}. It undergoes a time evolution as ${\rm L_{UV}(t) = L_{UV}(0)}-1.33 \log (t/t_0)  + 0.462$ \citep{dayal2017a}.}, the UV LF and its $z$-evolution is one of the most robust observables available for high-$z$ galaxies. In the simplest form, the UV LF can be constructed by assigning each galaxy a cosmological ratio of baryons to DM ($\Omega_b/\Omega_c$) and multiplying this with a (constant) star-formation efficiency value, say $f_*$. This (no feedback) model naturally results in the UV LF tracing the same shape as the HMF as shown (by the dashed line) in Fig. \ref{fig_uvlf_theo}. This UV LF can therefore be parameterised using the Schechter function \citep{schechter1976} to obtain the number of galaxies as a function of luminosity ($dn/dL$) as: 
\begin{equation}
\frac{dn}{dL} = \phi (L) = \frac{\phi_*}{L_*} \bigg(\frac{L}{L_*}\bigg)^\alpha {\rm exp}^{-(L/L_*)}.
\label{eqn_uvlf}
\end{equation}
This UV LF is fully defined once the three free parameters - $\phi_*, L_*$ and $\alpha$, signifying the characteristic number density and luminosity, and the faint-end slope, respectively - are specified. As expected, this function is a power-law at the low-luminosity end (at $L<L_*$) that drops off exponentially above the characteristic (or ``knee") luminosity. Given the general notation of expressing most high-$z$ observables in terms of the UV magnitude, substituting $\phi(\muv) d\muv = \phi(L) d(-L)$ and $\muv-\rm {M_{UV*}} = -2.5 {\log} (L/L_*)$, where $\rm {M_{UV*}}$ is the characteristic (or knee) magnitude, the Schechter function can be written as 
\begin{equation}
\frac{dn}{d\muv} = \phi (\muv) = \frac{{\rm ln} 10}{2.5} \phi_* \bigg( 10^{0.4(\rm {M_{UV*}}-\muv)}\bigg)^{\alpha+1} {\rm exp}^{-10^{0.4(\rm {M_{UV*}}-\muv)}}.
\label{eqn_uvlf_mag}
\end{equation}

\begin{figure}[h]
\center{\includegraphics[scale=0.6]{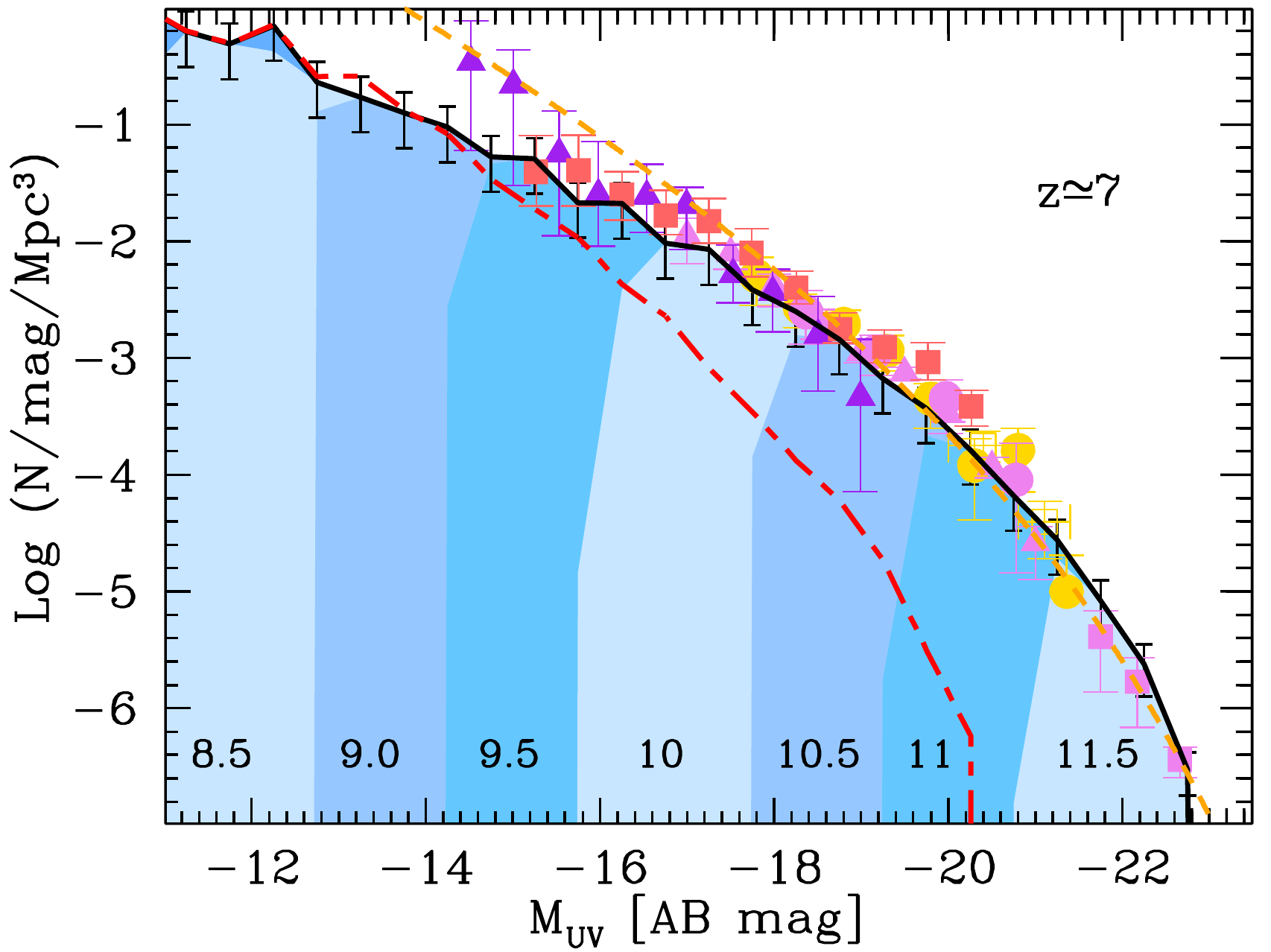}}
\caption{The LBG UV LF at $z \simeq 7$ \citep{dayal2014a}. The solid black line shows the results using the fiducial model, i.e. using the maximal star formation efficiency-halo mass relation derived in Sec. \ref{met_igm}. This model also includes the impact of mergers, smooth accretion from the IGM and SN-powered gas ejection on both the gas and stellar mass build-up of high-$z$ galaxies; the black error bars show poissonian errors arising from the luminosity dispersion in each bin. The dashed red line shows that the UV LF would have been severely under-estimated (with increasing brightness) had we not considered the gas brought in by mergers. The orange line shows the results obtained by multiplying the halo mass function (where each galaxy is assigned a cosmological gas-to-DM ratio) with a constant star-formation 
efficiency $f_* = 0.9\%$. Finally, the numbers within the shaded contours indicate the host halo mass. Points show observational results from: \citet[filled violet circles]{oesch2010}, \citet[filled yellow circles]{bouwens2011b}, \citet[empty yellow squares]{castellano2010a}, \citet[filled violet triangles]{mclure2013}, \citet[filled violet squares]{bowler2014}, \citet[filled red squares]{atek2015} and \citet[filled purple triangles]{livermore2016}.  }
\label{fig_uvlf_theo} 
\end{figure}

Starting with observations, a combination of space-based surveys (e.g. the Hubble Ultra Deep Field and the eXtreme Deep Field), carried out with the WFC3 onboard the HST \citep[e.g.][]{oesch2010,bouwens2011b, castellano2010a, mclure2010, mclure2013, bouwens2017b}, the Hubble CLASH lensing surveys \citep[e.g.][]{bradley2014, atek2015, livermore2016,mcleod2016} and ground-based observations \citep[e.g.][]{bradley2012, bowler2015, bouwens2016b} have allowed mapping out the high-$z$ LBG UV LF over 8 magnitudes from $\muv \sim -14.5$ to $\muv \sim -22.6$ at $z \simeq 7$ as shown in Fig. \ref{fig_uvlf_theo}. Observations indicate the bright-end at $z\simeq 7$ to be better fit by a double power-law that steepens to the Schechter function by $z\simeq 5$ which could possibly arise as a result of increasing dust extinction, the onset of AGN feedback or even lensing magnification biases \citep{bowler2015, ono2017}. Since the three UV LF free parameters are degenerate when inferred by fitting to observations, it is generally agreed that the faint-end steepens with $z$ \citep[][and references therein]{bouwens2017b} which has critical implications for reionization (see Sec. \ref{sources_reio}). Disentangling whether the $z$-evolution of the UV LF is driven by an evolution in $\phi_*$ \citep[density evolution;][]{mclure2010} or $L_*$ \citep[luminosity evolution;][]{bouwens2011b} has remained more challenging. However, theoretical works \citep{dayal2013} show that this evolution depends on the luminosity range probed: the steady brightening of the bright end of the LF is driven primarily by genuine physical luminosity evolution and arises due to a fairly steady increase in the UV luminosity (and hence star-formation rates) in the most massive LBGs; at the faint luminosity end, the evolution of the UV LF involves a mix of positive and negative luminosity evolution (as low-mass galaxies temporarily brighten then fade) coupled with both positive and negative density evolution (as new low-mass galaxies form, and other low-mass galaxies are consumed by merging). We note that, in addition to incompleteness corrections and cosmic variance, a number of observational biases, including magnification uncertainties \citep[e.g.][]{livermore2016, bouwens2017b} and the estimated size-luminosity relation \citep[][and see also Sec. \ref{sizes}]{grazian2012, vanzella2016, bouwens2017}, can have a significant impact on the observed number counts at the faint-end of the UV LF.

Reproducing the bright-end of the UV LF at $z \sim 5-10$ requires an increasing star formation efficiency (assuming a cosmological baryon fraction). This results in galaxies of the same UV luminosity residing in halos that are twice as massive at $z \simeq 5$ as at $z \simeq 8$ leading to a steepening in the predicted faint-end slope, from $\alpha_{HMF} = -2.11 \rightarrow -2.32$ from $z=5 \rightarrow 8$ \citep[see e.g.][]{dayal2013, dayal2014a, stark2016}. Further, as noted in Secs. \ref{Sup} and \ref{met_igm}, SN feedback is critical for reproducing the faint-end of the UV LF. Using the feedback model introduced in Sec. \ref{met_igm}, where $f_*^{eff} =min[f_*,f_*^{ej}]$, we find that the observed evolving UV LF from $z \sim 5-10$ is well fit by a model \citep[for details see][]{dayal2014a} where galaxies have a maximum star formation efficiency value of $f_* \sim 3\%$ (at any given time) and where roughly 10\% of the total SN energy coupling to the gas (shown by the solid line in Fig. \ref{fig_uvlf_theo}). We also revisit the question of whether galaxies on the UV LF gain their gas through mergers or gas accreted along with dark matter (smooth accretion) from the IGM by switching off gas being brought in by mergers in any part of the merger tree (short-long dashed line in Fig. \ref{fig_uvlf_theo}). As clearly shown, gas brought in by mergers is not very important at the faint end since the tiny progenitors of these galaxies are feedback limited in the fiducial model, resulting in largely dry mergers. However, the progenitors of increasingly luminous galaxies are critical to building up the large gas reservoirs that power star formation. Indeed, the LF at the bright end drops to 10\% of the fiducial model value ($\muv$ decreases by $\approx 2$ magnitudes) if the gas brought in by mergers is not taken into account. However, we caution the reader that smoothly accreted gas in a progenitor will be considered as ``accreted" gas for the successor halo - these numbers therefore represent an upper limit on the importance of accretion.

\begin{figure*}[h]
\center{\includegraphics[scale=0.59]{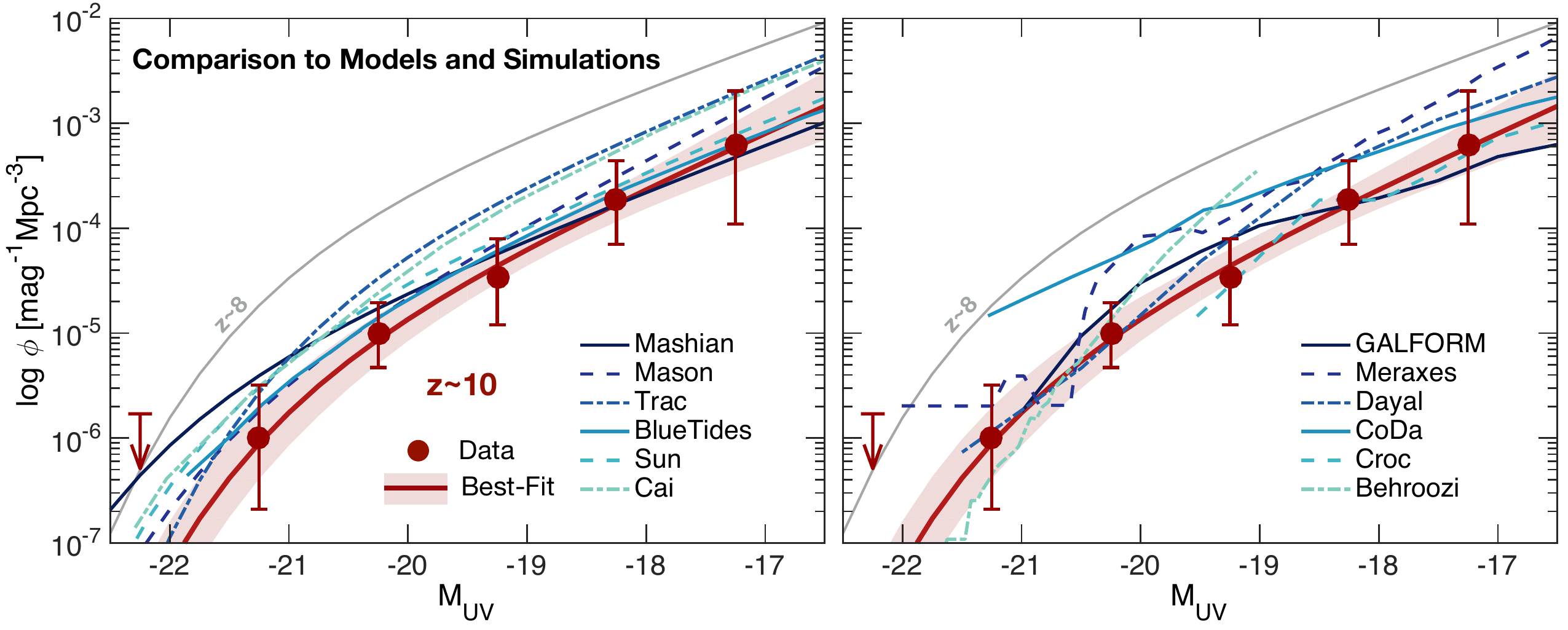}}
\caption{A comparison of the UV LF at $z \sim 10$ from observations (filled circles) compared with a range of theoretical models \citep[for details see][]{oesch2017}. Despite the diversity of physical approaches and processes implemented, all models predict the UV LFs that, do not differ by more than a factor of 3-4 over the luminosity range and, evolve significantly from
$z \sim 8$ (light gray line) to $z \sim 10$. }
\label{fig_z10_oesch} 
\end{figure*}

  \begin{figure*}[h]
\center{\includegraphics[scale=0.9]{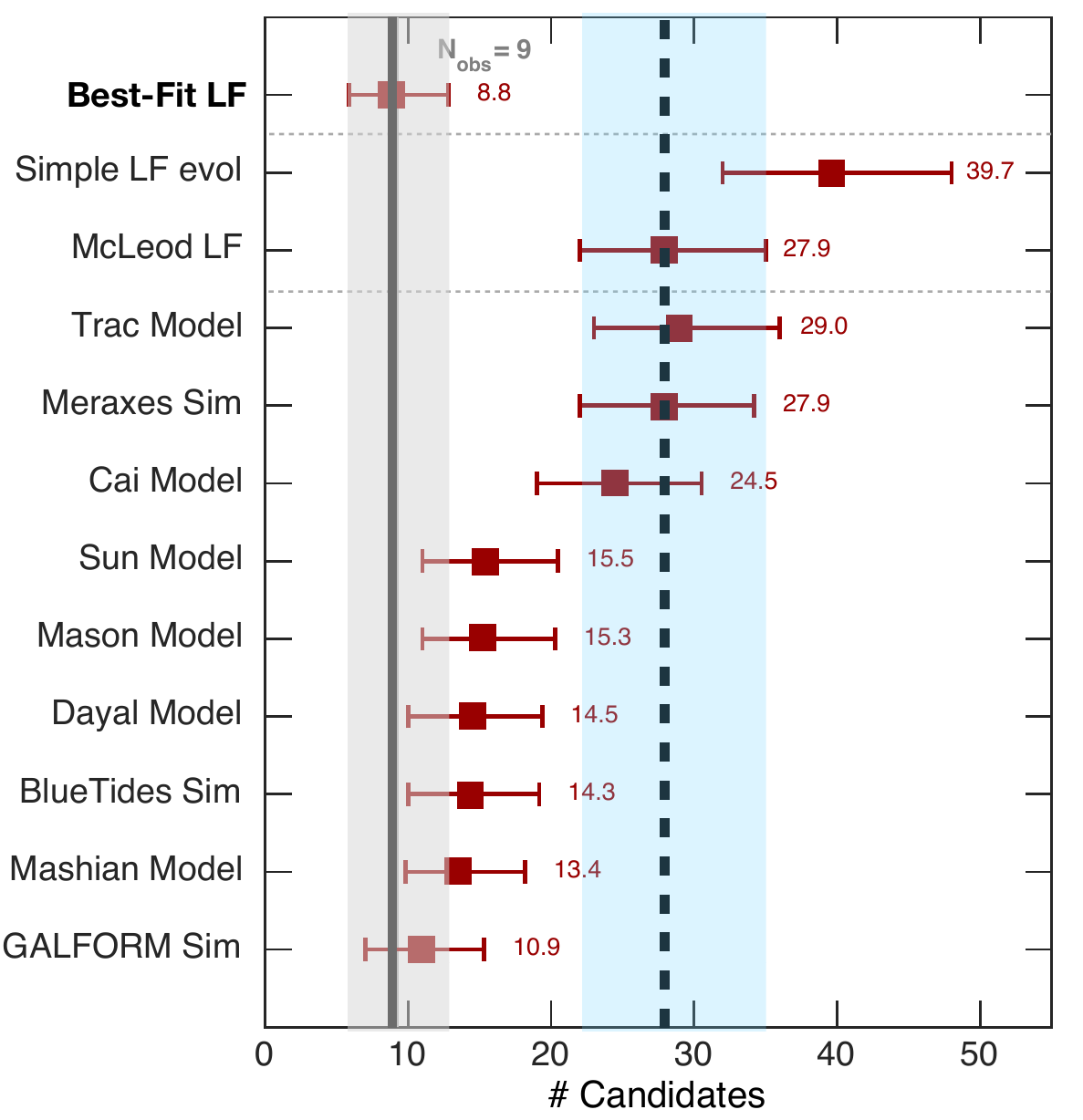}}
\caption{A comparison between the expected number of  galaxy candidates at $z \sim 10$ from theory and observations \citep{oesch2017}. Vertical lines show observational results from \citet[][solid line]{oesch2017} and \citet[][dashed line]{mcleod2016}, with the shaded regions showing the 1-$\sigma$ errors. Extrapolating the $z \sim 4-8$ LF evolutionary trends to $z \sim 10$ yields the  ``simple LF evol" model which is expected to be an upper limit on the number of $z \sim 10$ galaxies \citep{oesch2017}. All the other points show theoretical results. Despite using a wide range of modeling approaches and physical ingredients, essentially all models are in agreement with the range of $9-35$ galaxies expected from observations. }
\label{fig_z10_counts} 
\end{figure*}

 On the other hand, calculations of the stellar mass build-up show most of it is assembled in the largest progenitors of any galaxy \citep{dayal2013} : this results in a global picture of the progenitors of the most massive ($M_h \gsim 10^{11.5}\Msun$) high-$z$ galaxies being ``conveyor belts" of gas that is turned into stars in the deepest potential wells of their most massive progenitors. 
 
Given its import, the UV LF has been modelled in a number of ways, ranging from abundance matching to semi-analytics to hydrodynamic simulations, an illustrative example of which, at $z=10$, is shown in Fig. \ref{fig_z10_oesch}. Interestingly, despite the wide variety of approaches followed and parameters used, most of these models are in reasonable agreement with the data. This is probably driven by a number of physical effects that can have a degenerate effect on the UV LF including, but not limited to: the impact of the SN and UVB feedback in decreasing the gas content of low mass dark matter halos, the impact of dust in decreasing the observed luminosity from the most massive halos and the propagation of gas mass through the galaxy assembly history. As noted, a number of observational biases (i.e. the size correction, lensing magnification correction, luminosity limits probed) can impact the faint-end slope and need to be constrained in more detail. These complications are clearly reflected in the fact that both observations and theoretical results expect anywhere between 10-40 $z \simeq 10$ candidate LBGs based, mostly, on the value and luminosity extent of the faint-end slope, as shown in Fig. \ref{fig_z10_counts}. However, precisely as a result of these degeneracies, both the faint end slope and the cut-off mass/luminosity to which galaxies can keep forming stars remain a matter of debate \citep{dayal2014a, gnedin2016,liu2016, ocvirk2016}.

Moving on, the UV LF can also be constructed for LAEs \citep[see e.g. Fig. 14;][]{dunlop2013} that, as already noted, shows a steep decline at $z \gsim 6$ when compared to the LBG UV LF. This is also reflected in a steep decrease in the fraction of (faint; $\muv > -20.25$) LBGs detected through their Ly$\alpha$ emission as shown in Fig. \ref{fig_laefrac}; the observational trends, showing a large scatter, are more unclear for the brightest galaxies. While it might be tempting to attribute this solely to an increase in the average value of $\chi_{HI}$ with redshift, the interpretation is actually quite complex and could arise from a combination of one or more physical effects including: an evolution of the underlying halo mass function \citep{dijkstra2007, dayal2008}, evolution of galaxy physical properties (star formation rates and dust content) with redshift \citep{dijkstra2007, dayal2008, dayal2010, dayal2012}, the number density evolution of the Lyman Limit absorption systems \citep{bolton2013} and even the different CGM of massive and low mass halos \citep{weinberger2018}. Indeed, it has also been cautioned that simple color-cuts, in EW and UV magnitude, that do not mimic the actual observational strategies used to select LAEs and LBGs, might also bias theoretical results \citep{dijkstra2012} and that integrating down to extremely faint magnitudes might make most LBGs appear as LAEs \citep{steidel2011, dayal2012, leclercq2017}. The advent of the {\it JWST} will be crucial in shedding more light on the absence (or presence) of Ly$\alpha$ emission in LBGs. 

\begin{figure}[h]
\center{\includegraphics[scale=0.57]{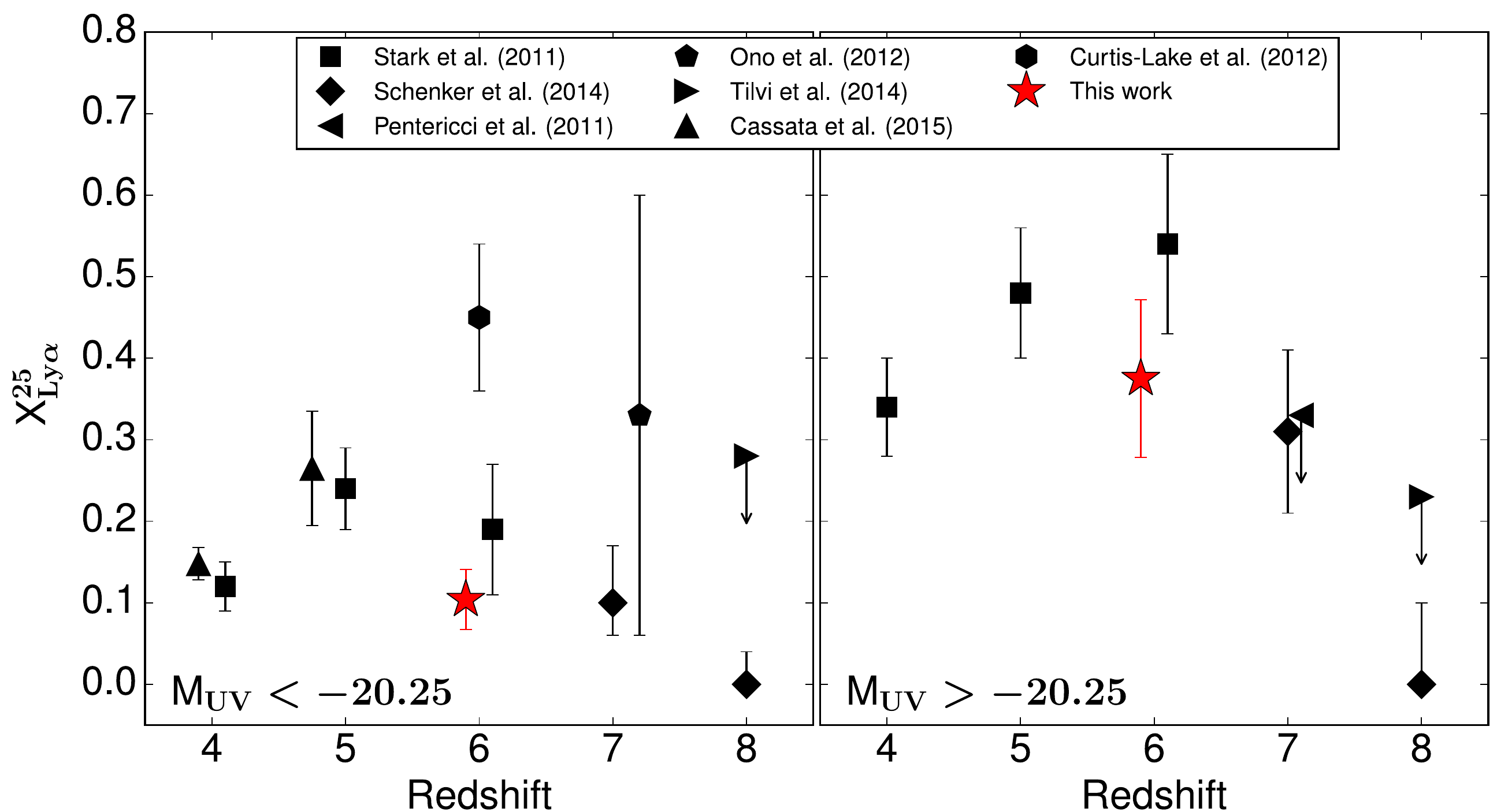}}
\caption{A compilation of the fraction of LBGs with Ly$\alpha$ lines with EW(Ly$\alpha$) $> 25{\rm \AA}$ ($X^{25}_{Ly\alpha}$) at $4 \lsim z \lsim 8$ \citep{debarros2017}. The {\it left} and {\it right} panels show results for the brightest ($\muv < -20.25$) and faintest ($\muv > -20.25$) galaxies. This includes results from \citet{stark2011}, \citet{schenker2014}, \citet{pentericci2011}, \citet{ono2012}, \citet{tilvi2014}, \citet{cassata2015} and \citet{curtis-lake2012}; the red stars show results from \citet{debarros2017}. As shown, all works converge on a steep drop in the fraction of faint LBGs showing Ly$\alpha$ emission (with an EW larger than $25{\rm \AA}$) at $z \gsim 6.5$ with a larger scatter seen for rare, bright sources.}
\label{fig_laefrac} 
\end{figure}

\subsection{Stellar mass density } 
\label{smd}
The stellar mass, $M_*$, is one of the most fundamental galaxy properties, encapsulating information about its entire star-formation history. However, achieving accurate stellar mass estimates of early galaxies from broad-band data has been difficult because it depends on the assumptions made regarding the IMF, the strength of nebular emission, the star formation history, stellar metallicity and dust; the latter three parameters are degenerate, adding to the complexity of the problem. Indeed, accounting for the presence of nebular emission lines can result in a downward revision of the stellar ages and masses by as much as a factor of 3 \citep{schaerer2009, labbe2013}. Although properly constraining $M_*$ ideally requires rest-frame near infra-red data, which will be provided by future instruments such as the Mid-Infrared instrument (MIRI) onboard the {\it JWST}, broad-band {\it HST} and {\it Spitzer} data are already being used to infer the stellar mass density (SMD; stellar mass per unit volume) of galaxies at redshifts as high as $z\simeq 10$, as shown in Fig. \ref{fig_smd}. 

In brief, the observed SMD values inferred from LBGs at $z\simeq 5-10$ have been collected using a multitude of mass and luminosity limits and making different physical assumptions regarding the presence of nebular emission lines: e.g. while a number of groups have computed the SMD down to a limiting magnitude of $\muv =-18$ \citep{gonzalez2011, labbe2013, stark2013, oesch2014}, some account for the impact of nebular emission \citep{labbe2013, stark2013} while the others do not. On the other hand, a number of works compute the SMD by integrating the Schechter (stellar mass) function between $M_* = 10^8-10^{13}\Msun$ \citep{duncan2014, grazian2015, song2016}. Interestingly, all of these results are in accord, within error bars, perhaps demonstrating the lack of leverage in the SMD inferred using these different means. As per these measurements, the SMD-$z$ relation has a slope that scales as $(1+z)^{-0.36}$ for $z \gsim 4$, although the error bars and paucity of data preclude any strong conclusions at $z \gsim 8$.

\begin{figure*}[h!]
\center{\includegraphics[scale=0.6]{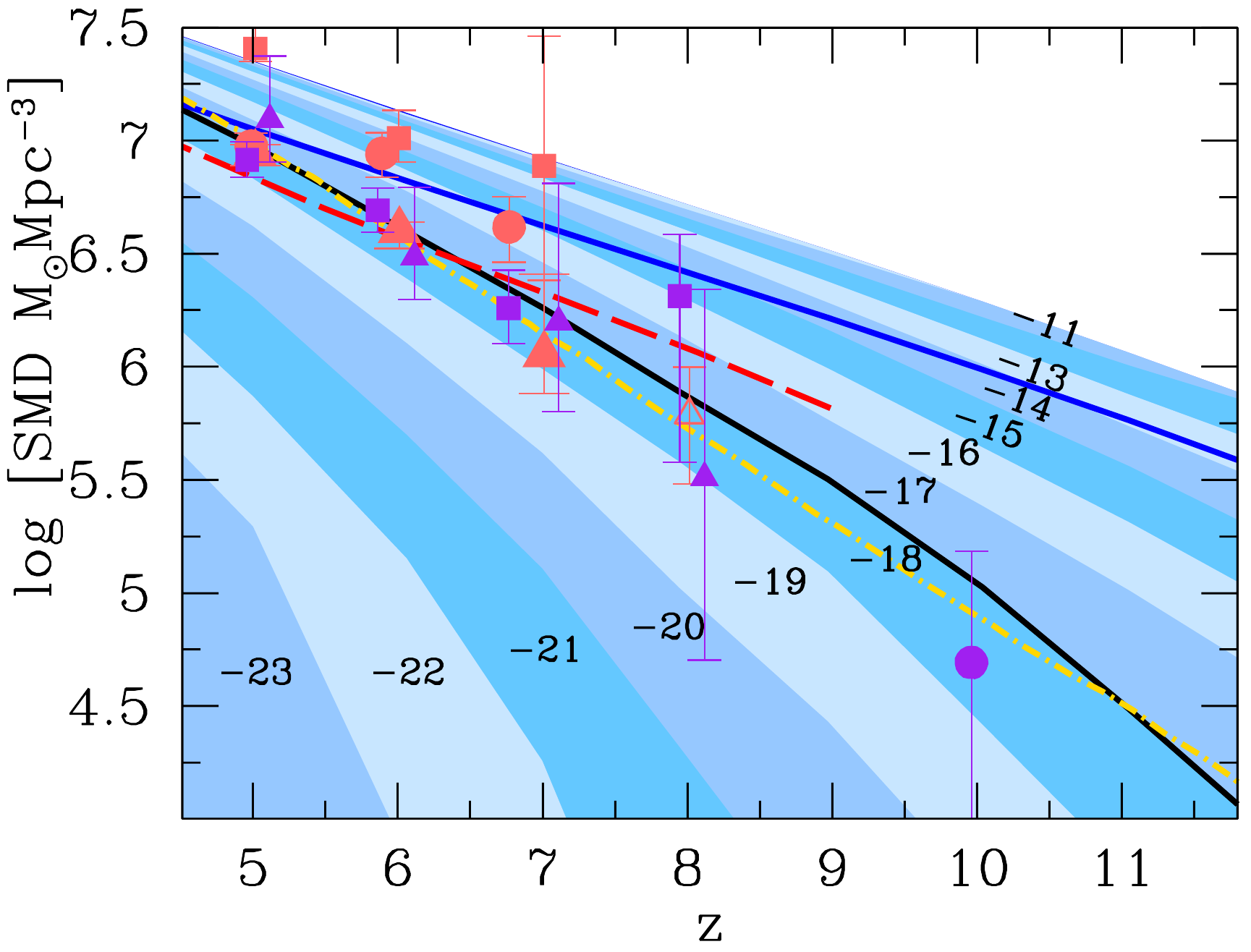}}
\caption{Redshift evolution of the cosmic stellar mass density. Points show the observed estimates, using a multitude of mass and magnitude cuts and both considering and ignoring the impact of nebular emission lines, from: \citet[red filled circles]{gonzalez2011}, \citet[red empty triangles]{labbe2013}, \citet[purple filled squares]{stark2013}, \citet[purple filled circles]{oesch2014}, \citet[red filled squares]{duncan2014}, \citet[red filled triangles]{grazian2015} and \citet[purple filled triangles]{song2016}; the error bars show the $1-\sigma$ error. The solid black line shows DELPHI results integrated down to a UV magnitude limit of $\muv = -18$ \citep{dayal2014a} with the shaded regions showing the contribution of the different magnitude bins marked to SMD through cosmic time; the solid blue line shows the magnitude limits at which galaxies provide 50\% to the total SMD at that redshift. The dot-dashed yellow line and the dashed red line show results from the DRAGONS project \citep{mutch2016} and the EAGLE simulations \citep{schaye2015}, respectively.}
\label{fig_smd} 
\end{figure*}

We compare these observations to the results from three theoretical models, one each demonstrating the semi-analytic \citep[DELPHI;][]{dayal2014a}, semi-numerical \citep[DRAGONS;][]{mutch2016} and hydrodynamic \citep[EAGLE;][]{schaye2015} approaches, as shown in Fig. \ref{fig_smd}. As shown, using a magnitude cut of $\muv\lsim -17.7$, within error bars, the results from all three models are in reasonable agreement with each other as well the data despite the different physical ingredients implemented, the baselines used to calibrate model parameters and even the numerical methods used. 

Conducting a census of the total stellar mass we find that, as a result of their enormous numbers, small, faint galaxies contain most of the stellar mass in the Universe at $5 \leq z \leq 12$ as shown in Fig. \ref{fig_smd}. Indeed, galaxies brighter than current observational limits ($ \muv \leq -18$) contain about $50\%$ of the total stellar mass at $z \simeq 5$. This value then steadily decreases with redshift such that observed galaxies contain a quarter of the total stellar mass at $z \simeq 6.5$ and only $10\%$ of the total stellar mass at $z \simeq 9$ (i.e. at redshifts $z \geq 6.5$, most of the stellar mass of the Universe is locked up in galaxies too small to have been detected so far). The next generation of space instruments such as the {\it JWST}, along with future developments in the use of gravitational lensing (e.g. in the Frontier Fields), which might extend stellar mass estimates down to $\muv \sim -14$, will play an important role in revealing about half (a quarter) of the stellar mass in the Universe up to $z \simeq 8$ ($z \simeq 9.5$). 

 
\subsection{Cosmic star formation rate density} 
\label{sfrd}
The cosmic star formation rate density (SFRD), inferred from the UV luminosity, is an important indicator of active star formation over the past few Myrs, given that UV luminosity decays as roughly $(t/2 \mathrm{Myr})^{-1.3}$. A number of observational works have focused on estimating the SFRD using sources with $\muv \lsim -17.7$ from the {\it HST} Ultra deep (HUDF) surveys \citep{bouwens2011b, oesch2010, oesch2013, ellis2013, mclure2013} which have now been pushed to even fainter magnitudes using the CLASH lensing surveys \citep{zheng2012, coe2013, bouwens2014b, mcleod2016}; these results are shown in Fig. \ref{fig_sfrd}. Essentially, observations integrate the UV LF to the chosen magnitude limit (e.g. a limit of $\muv = -17.7$ roughly corresponding to a SFR rate of $0.7 \Msun\, {\rm yr^{-1}}$). This integrated UV luminosity density is converted into a SFRD, generally, using the conversion factor from \citet{madau1998} \footnote{where a SFR of $1\Msun {\rm yr^{-1}} = 1.4 \times 10^{-28} L_{UV} \, [{\rm erg\, s^{-1} \, Hz^{-1}}]$ for a Salpeter IMF between $0.1-100\Msun$ using {\it STARBURST99}.}. 

\begin{figure*}[h]
\center{\includegraphics[scale=0.6]{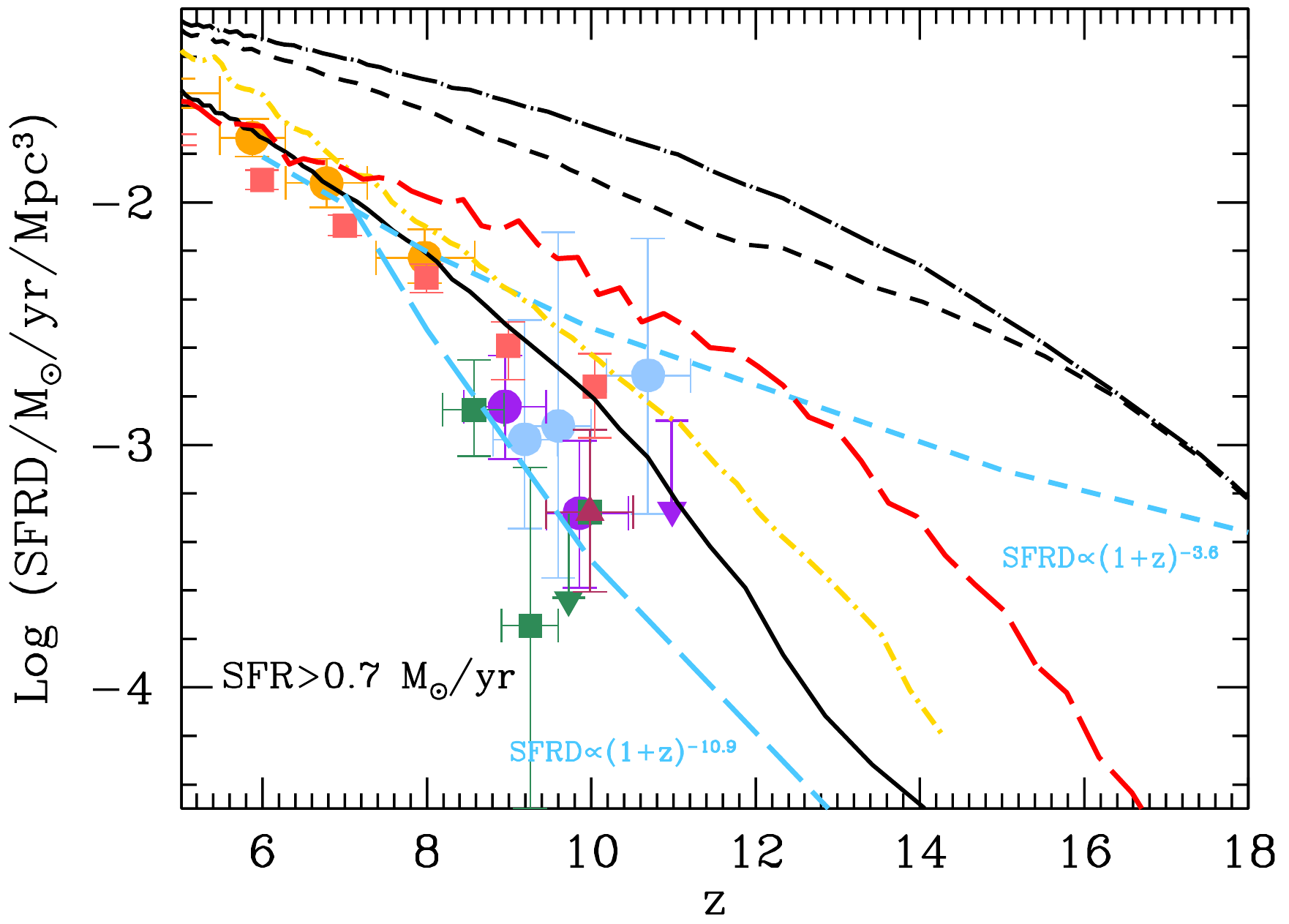}}
\caption{The evolution of the cosmic star formation rate density (SFRD) above a star-formation limit of $0.7 \Msun\, {\rm yr^{-1}}$ corresponding to $\muv \sim -17.7$. The points show data collected by \citet[orange filled circles]{bouwens2007, bouwens2012b}, the CLASH lensing surveys \citep[light blue filled circles;][]{zheng2012, coe2013, bouwens2014b}, \citet[purple filled circles]{ellis2013}, \citet[red filled squares]{mcleod2016}, \citet[green filled squares]{oesch2013} and \citet[maroon filled triangle]{oesch2014}. The downward facing arrows showing upper limits are from \citet[purple]{ellis2013} and \citet[green]{oesch2013}. The black lines shows DELPHI \citep{dayal2014a} results integrated down to a UV magnitude limit of $\muv = -17.7$ (solid line), $\muv = -15$ (dashed line) and for all galaxies (dot-dashed line). The dot-dashed gold line and the dashed red line show results from the DRAGONS project \citep{mutch2016} and the EAGLE simulations \citep{schaye2015}, respectively. We also show the SFRD-$z$ trend inferred from low-$z$ galaxies which evolves as $\propto (1+z)^{-3.6}$ (short-dashed blue line) with $z>8$ LBGs showing a much steeper fall-off $\propto (1+z)^{-10.9}$ (long-dashed blue line) \citep{oesch2014}. Although showing different slopes for the $z$-evolution of the SFRD, as of now, all three models (DELPHI, DRAGONS and EAGLE) are in accord with the observations within error bars.}
\label{fig_sfrd} 
\end{figure*}

It must be noted that inferring the UV luminosity density has been a herculean task at these high-$z$ 
due to a number of complexities including: dust corrections, the contamination from emission line galaxies, from local (L, M, T or Y) dwarfs and/or AGN, the impact of cosmic variance given the small fields and the uncertainties associated with the adopted lensing magnifications. Despite these complications, the data inferred from all fields is in excellent agreement as shown in the same figure. Given the paucity of statistics, however, the $z$-evolution of the SFRD has remained a matter of debate, with observational works finding a much steeper slope of SFRD$\propto (1+z)^{-10.9\pm2.5}$ at $z \gsim 8$ \citep[e.g.][]{oesch2014} as compared to the shallower SFRD$\propto (1+z)^{-3.6}$ inferred for lower redshift data \citep[e.g.][]{bouwens2007, bouwens2012}. Of course, it must be noted that the underlying metallicity, IMF, lack of binary stars and constancy of the SFR for about a 100 Myrs prior to observations are all implicit assumptions in this conversion. For example, this conversion would face an upward revision by a factor of about 2 if these galaxies had particularly young ages \citep[e.g.][]{bouwens2011b}. 

We then compare these observations to theoretical models (DELPHI, DRAGONS and EAGLE) finding that these results, whilst agreeing with observations at $z \lsim 8$, do not show the steep drop seen at higher-$z$. Interestingly, while the results from the semi-analytic \citep[DELPHI;][]{dayal2014a} and semi-numerical \citep[DRAGONS;][]{mutch2016} models are in excellent agreement with each other, possibly as a result of the similar physics implemented, the hydrodynamical simulation \citep[EAGLE;][]{schaye2015} results show a much shallower slope. Forthcoming data, with e.g. the {\it JWST} and {\it E-ELT}, will be instrumental in pinning down the observations and disentangling the impacts of the observational biases from the physical effects (e.g. gas accretion, mergers, the impact of SN and UV feedback) driving the star formation in these early systems. 

  \subsection{Dust content}
  \label{gas_dust}
Dust plays a critical role, both, in shaping the physics as well as visibility of early galaxies. For example, dust surfaces offer the main production sites for ${\rm H_2}$ formation \citep[e.g.][]{cazaux2004}, inducing the formation of molecular clouds and enhancing star formation activity \citep{hirashita2002, yamasawa2011}. Dust cooling also induces fragmentation of molecular clouds \citep[e.g.][]{omukai2005} thereby shaping the stellar IMF \citep{schneider2006}. Finally, dust grains absorb UV light, which is re-emitted in the Infra-red bands \citep[see e.g.][]{dayal2010dust}, affecting both the observed luminosity as well as the spectral slopes of early galaxies. Dust can be produced by three key sources in the high-$z$ universe: 

{\it (i) Core Collapse Supernovae (SNII):} exploding on timescales of less than 28 Myr \citep{matteucci1986}, stars with mass $M_s >8 \Msun$ end their lives as ``Core Collapse Supernovae". The classical nucleation theory \citep{feder1966} provides the backbone for the dust produced by such sources: solid materials can condense out of vapour in the supersaturated state once the partial pressure of the given species exceeds the vapour pressure in the condensed phase. This well-defined condensation barrier naturally results in clusters of a ``critical"  size which then grow into macroscopic dust grains by the accretion of material. A number of theoretical models have used this theory to predict that silicate and carbon grains form in the SN ejecta, a few hundred days after the explosion, with condensation efficiencies in the range $0.1-3$, obtaining $0.1-1\Msun$ of dust per supernova for progenitors in the range of $12-40\Msun$ \citep[e.g.][]{todini2001}. While there is now clear observational evidence for dust produced by supernovae \citep{wooden1993, elmhamdi2003, kozasa2009, gall2011}, the dust yields remain a matter of debate. This is because supernovae observations, although limited to only a few objects, show dust masses of the order of $10^{-5}-10^{-3}\Msun$ hinting at condensation efficiencies that are about two orders of magnitude lower than theoretical predictions. On the other hand, observations of young galactic SN remnants with {\it Spitzer} show freshly formed dust masses $\sim 0.015-0.05 \Msun$ \citep{rho2008, rho2009} with {\it Herschel} far-infrared observations yielding estimates as high as $0.4-0.7\Msun$ for SN1987A \citep{matsuura2011} with follow-up ALMA observations confirming this to be the newly formed dust component \citep{indebetouw2014}. One possible means of reconciling these apparently divergent results is to account for partial destruction of newly formed dust in the SN reverse shock, with dust destruction being more efficient for small grains \citep{bianchi2007} and with increasing ISM densities \citep{bianchi2007,yamasawa2011}. 

{\it (ii) Asymptotic Giant (AGB) Branch stars:} All stars with initial masses in the range of $0.8-8 \Msun$ on the main sequence evolve through the thermally pulsating AGN phase, losing mass at a very high rate, before becoming white dwarfs. Dust can form, by condensation of solid particles in the outflowing gas, in the cool, dense atmospheres during this stage, leading to optically thick circum-stellar dust shells. The dust mixtures formed depend on the relative abundances of carbon and oxygen (C/O) of the ejected matter. Refining the models for dust production from AGBs as a function of their initial stellar mass and metallicity, \citet{zhukovska2008} predict that while silicate dust is mostly produced by oxygen-rich stars (M type with ${\rm C/O} < 1$ and S type with ${\rm C/O} \simeq 1$) with an initial mass $M_s<1.5 \Msun$ and $M_s>4 \Msun$, carbon and SiC dust production is, instead, dominated by stars with an initial mass in the range of $1.5-4 \Msun$ during the carbon-rich phase (called C type with ${\rm C/O}>1$); iron dust is produced in all AGB stars. Interestingly, carbon dust of mass $10^{-5}-10^{-2}\Msun$ can be produced at all metallicities. However, comparable amounts of silicate dust can only be produced at $Z \sim \Zsun$ requiring the presence of sufficient Si, O, Mg and Fe, which have to be gathered from the pre-enriched ISM, given they can not be produced by the star itself \citep{valiante2009}. Given their evolutionary timescales of about a Gyr before the onset of dust production, AGB stars were thought to play a negligible role in the dust enrichment of early galaxies \citep{dwek2007}. However, it has now been shown that while both SN and AGBs can contribute to the dust enrichment in the early Universe after several hundred Myrs of star formation \citep{valiante2009}, high mass AGB stars ($M_s > 3\Msun$) can be the dominant dust producers only if each SN produces $\lsim 3 \times 10^{-3} \Msun$ of dust \citep{gall2011}. These arguments have been applied to LAEs given their predominantly young populations - using the {\it ALMA} (1.2mm) upper limit of $52.1 \mu Jy$ for Himiko, an exceptionally bright $z \simeq 6.6$ LAE, \citet{hirashita2014} (and see also \citet{hirashita2017}) infer an each SN to produce a dust mass $m_d <0.15-0.45 \Msun$ as shown in Fig. \ref{fig_snyield_hira}. 

\begin{figure*}[h]
\center{\includegraphics[scale=0.5]{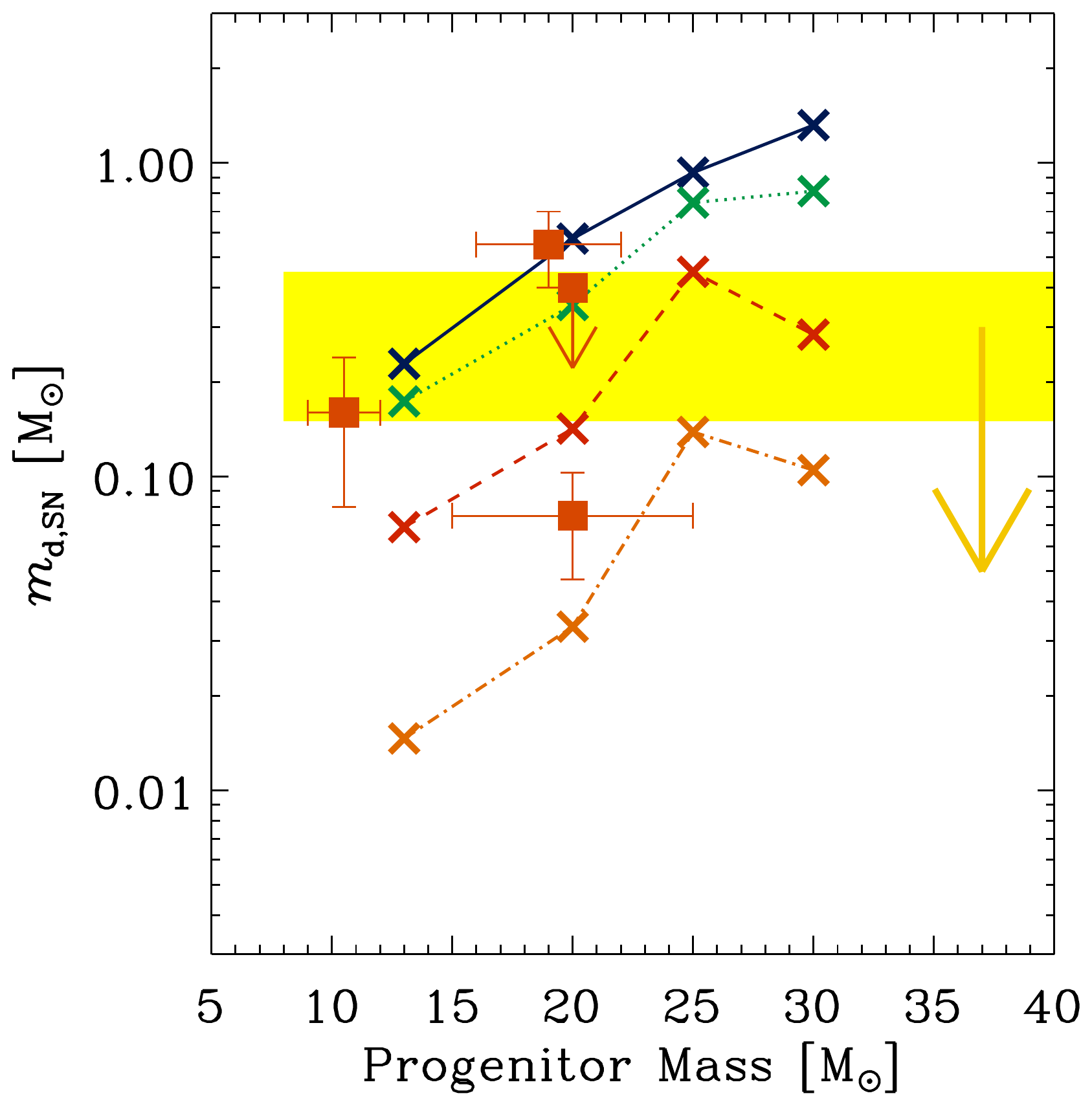}}
\caption{The dust mass formed in a SN as a function of the progenitor mass, in the range of $8-40\Msun$, at the zero age main sequence assuming all the dust mass in the ALMA-detected $z \simeq 6.6$ LAE -``Himiko" - is produced by SN \citep{hirashita2014}. As shown, this assumption results in an upper limit of $m_d=0.15-0.45\Msun$ of dust produced per SN, for different dust species, shown by the shaded area. Crosses connected by the solid, dotted, dashed and dot-dashed lines show theoretical predictions of \citep{nozawa2007} for hydrogen number densities corresponding to $n_H =0,0.1,10\, {\rm cm^{-3}}$ in the ambient medium. The filled squares with error bars show observed dust masses detected by {\it Herschel} for nearby SN remnants \citep{barlow2010, gomez2012, matsuura2011, indebetouw2014, otsuka2010}.}
\label{fig_snyield_hira} 
\end{figure*}

{\it (iiii) Grain growth in the ISM:} Finally, dust mass can also grow by the accretion of gas-phase metals onto pre-existing dust grains in cold, dense phase of the ISM \citep{draine2009, jones2011}. One of the key parameters determining such dust growth is the metallicity. To quantify this, \citet{asano2013} have introduced the concept of  a ``critical metallicity" at which the contribution of AGB stars and SN equal that of the dust mass growth in the ISM. 


In the simplest approximation, the time evolution of the dust mass, $dM_d/dt$, can be calculated as \citep[e.g.][]{dwek1998, valiante2009, inoue2011, asano2013, debennassuti2014} 
\begin{equation}
\frac{dM_d}{dt} = \dot Y_d + \dot G_{ac} + \dot I_d  - \mathcal{D}_{ISM} \psi(t)  - \dot D_{SN} - \dot O_d.
\end{equation}
Here, the rate of increase of dust mass depends on the rate at which dust produced by SN and AGB stars is injected into the ISM ($\dot Y_d$), the rate at which the dust mass grows in the ISM ($\dot G_{ac}$) and the rate of dust infall from the IGM ($\dot I_d$). These are countered by a number of effects that, instead, decrease the dust mass: $ \mathcal{D}_{ISM} \psi(t)$ expresses the rate of dust astration (assimilation) into further star formation where $\mathcal{D}_{ISM}$ is the ISM dust-to-gas ratio and $\psi(t)$ is the star formation rate. As noted, dust can also be destroyed by SN shocks in the diffuse ISM ($\dot D_{SN}$) and assuming a homogeneous mixture of ISM gas and dust, dust can be lost in outflows of gas into the IGM ($\dot O_d$). We now explain the key terms (1,2 and 6 on the RHS) in more detail. Firstly, the dust yield can be calculated as \citep{valiante2009}:
\begin{equation}
\dot Y_d = \int_{m_*(t)}^{m_{up}} m_d (M_s,Z) \phi(M_s) \psi(t-\tau_m) dM_s,
\end{equation}
where the lower integration limit refers to the mass of a star with lifetime $\tau_m=t$, $m_{up}$ is the upper mass limit of the stellar IMF and $m_d$ is the dust mass produced by a star of mass $M_s$ and metallicity $Z$. Further, the dust accretion rate can be expressed as \citep[e.g.][]{mancini2015}
\begin{equation}
\dot G_{ac} = X_c \frac{M_d(t)}{\tau_{acc}},
\end{equation}
where $X_c$ is the mass fraction of cold gas where dust can grow on a timescale $\tau_{acc}$ which critically depends on the density ($n_{mol}$), temperature ($T_{mol}$) and metallicity ($Z$) of the cold phase such that \citep{asano2013, debennassuti2014} 
\begin{equation}
\tau_{acc} \approx 20 {\rm Myr} \times \bigg(\frac{n_{mol}}{100 {\rm cm^{-3}}}\bigg)^{-1} \bigg(\frac{T_{mol}}{50 {\rm K}}\bigg)^{-1/2} \bigg(\frac{Z}{\Zsun}\bigg)^{-1}. 
\end{equation}
Finally, the rate of dust destruction can be written out as $\dot D_{SN} = M_d/\tau_{dest}$ where the destruction timescale $\tau_{dest}$ is \citep{mckee1989, lisenfeld1998}
\begin{equation}
\tau_{dest} (t) = \frac{M_g(t)}{\nu \dot M_*  \epsilon M_{sn} {\rm (100\, km \, s^{-1})} } ,
\end{equation}
where $M_g$ is the gas mass, $\dot M_*$ is the star formation rate and $\epsilon$ is the efficiency of dust destruction in a SN-shocked ISM which has typical values between $0.1-0.5$ for varying densities of the ISM and magnetic field strengths \citep{mckee1989, seab1983}. Further, $M_{sn} {\rm(100\,  km\, s^{-1}) }$ is the mass accelerated to 100 km ${\rm s^{-1}}$ by the SN blast wave and has a value of $6.8 \times 10^3 \, \Msun$ \citep{lisenfeld1998}.
However, the relative importance of the above mentioned processes of dust formation and destruction in early galaxies remain under debate. For example, \citet{ferrara2016b} have shown that ISM dust growth remains problematic, as in the diffuse ISM (where grains reside most of the time), accretion rates are very slow, dust temperatures are high and growth is impeded by the Coulomb barrier. They therefore conclude that the impact of dust destruction by shocks has probably been severely overestimated. 

\begin{table}
\begin{center}
\caption{Summary of the high-$z$ galaxies named in column 1 with the redshifts (column 2), UV magnitudes (column 3), stellar masses (column 4) and dust masses (column 5). All quantities are computed using the homogeneous stellar mass-magnitude relations and dust temperatures and emissivity values as in \citep{mancini2015}.}
\begin{tabular}{cccccc}
\hline
\hline
Name & $z$ & $\muv$ & ${\rm Log}(M_*/\Msun)$ & ${\rm Log}(M_d/\Msun)$ Reference \\
\hline
\hline
A1703-zD1 & 6.800 & -20.3 & $9.2\pm0.3$ & $< 7.36$ & \citep{schaerer2015} \\ 
z8-GND-5296 & 7.508 & -21.4 & $9.7\pm0.3$ & $< 8.28 $ & \citep{schaerer2015}  \\
HCM6A & 6.560 & -20.8 & $9.5\pm0.3$ & $< 7.71$ & \citep{kanekar2013} \\
IOK-1& 6.960 & -21.3 & $9.7\pm0.3$ & $< 7.43$ & \citep{ota2014} \\
Himiko & 6.595 & -21.7 & $9.9\pm0.3$ & $< 7.30$ & \citep{ouchi2013} \\
BDF-3299 & 7.109 & -20.44 & $9.3\pm0.3$ & $< 7.02$ & \citep{maiolino2015}\\
BDF-512 & 7.008 & -20.49 & $9.3\pm0.3$ & $< 7.36$ & \citep{maiolino2015} \\
BDF-46975 & 6.844 & -21.49 & $9.8\pm0.3$ & $< 7.38$ & \citep{maiolino2015} \\
\hline
\hline
A1689-zD1 & 7.500 & -19.7 & $9.0\pm0.3$ & $7.51 \pm 0.2$ & \citep{watson2015}\\
${\rm A2744_YD4}$ & 8.38 & - & $9.29^{+0.24}_{-0.11}$ & $6.74^{+0.65}_{-0.17}$ & \citep{laporte2017} \\
\hline
\end{tabular}
\label{table_dust_obs} 
\end{center}
\end{table}

In the absence of direct detection of far infra-red dust emission, the key quantity used to characterise the impact of dust has been the relation between the spectral slope ($\beta$; measured between 1500-3000\AA) in the galaxy rest frame and the observed UV magnitude - the ``Meurer relation" \citep{meurer1999}. Although calibrated using low-$z$ starburst galaxies and assuming $\beta =-2.23$, where $\beta$ is naturally a function of the age, metallicity and IMF of the underlying stellar population and the dust extinction curve (expressing the extinction as a function of wavelength), this relation has now been widely applied to $z \gsim 5$ galaxies to infer the $\beta-\muv$ - or the so-called ``color magnitude relation" \citep{finkelstein2012b, bouwens2012b, dunlop2013b, bouwens2014, rogers2014}. A number of theoretical works have now confirmed the existence of this relation \citep{dayal2012, wilkins2013, mancini2016}, highlighting the fact that the $\beta$ slope becomes increasing bluer (shallower) with decreasing luminosities and increasing redshifts and that the scatter \citep{rogers2014} arises because of a mix of intrinsically blue (dust-free) and dust enriched older objects, specially at $-18 \lsim \muv \lsim -19$ \citep{mancini2016}. Indeed, this decrease in the impact of dust on the UV luminosity reflects in the fact that while, theoretically, $z\simeq 8$ galaxies require almost no dust to match to the observed UV LF, matching to the $z \simeq 5$ UV LF requires dust extinction at all luminosities brighter than $\muv \simeq -18$ \citep{salvaterra2011, dayal2010, dayal2013, hutter2014, finkelstein2015, lignedin2016, wilkins2018}.

The presence of far infra-red dust emission, e.g. from {\it ALMA} for $z \simeq 5.25$ galaxies \citep{capak2015}, can also be used to get a hint on the dust attenuation through the infra-red excess-UV spectral slope (IRX-$\beta$) relation \citep{meurer1999}. Here, ${\rm IRX} = {\rm Log(F_{IR}/F_{1600})}$ is the ratio between the far infra-red dust emission and the UV flux at 1600\AA\, in the galaxy rest-frame. Theoretical works now show that, given their lower stellar masses and lower chemical maturity, high-$z$ galaxies are characterised by lower IRX values and bluer $\beta$ slopes compared to lower-$z$ counterparts for the same colours. \citet{capak2015} point out that the IRX values can also be used an an indicator of the dust formation mechanisms - for example, significant dust contribution from grain growth can induce a considerable scatter in the IRX at a given value of the $\beta$ slope. Given their blue slopes, most high-$z$ galaxies only had upper limits on the dust-mass until the discovery of dust in the $z = 7.5$ galaxy A1689-zD1 and ${\rm A2744_YD4}$ at $z=8.38$ as summarised in Table \ref{table_dust_obs}. Theoretical works have shown that even using the upper limit of the stellar dust yields (from both SN and AGBs) results in dust masses that are about two orders of magnitude lower than those observed for both A1689-zD1 and ${\rm A2744_YD4}$ and that the only way of reconciling the observed dust masses with theory is to include grain growth on timescales as short as 0.2 Myr which imply a density of about $10^4 {\rm cm^{-3}}$ in the cold phase of the ISM \citep{mancini2015} (and see also \citep{michalowski2015}). \citet{mancini2015} also make the point that although grain growth becomes the dominant process as early at $z \sim 10-12$, the average contribution from AGBs can be large as 40\% and can therefore not be ignored. However, \citet{behrens2018} point out that given that the far infra-red luminosity scales as $L_{IR} \propto M_{dust} T_{dust}^6$, these observations could be explained by reasonable dust masses if the temperature is ramped up by a factor 2-3. Further observations with {\it ALMA} will be imperative in shedding more light on the (currently debated) dust masses of early galaxies. 
  
\subsection{Spin and shapes}
\label{shape_spin}
We start by discussing the spin of high-$z$ galaxies before discussing their shapes. Galaxy formation theory \citep[e.g.][]{fall1980, mo1998} suggests that gas within a cooling radius can settle into a disk inside a dark matter halo. Assuming the gas and dark matter were initially well-mixed, such that the distribution of angular momentum of each mass element in the disk was the same as that in the halo, \citet{fall1980} deduced the angular momentum profile for an exponential disk and that the disk radius $R_d \propto \lambda$. Here, $\lambda$ is the pre-collapse angular momentum of the halo which is generally parameterised as \citep{peebles1969}
 \begin{equation}
 \lambda \equiv \frac{J |E|^{1/2}} {G M^{5/2}},
 \end{equation}
where $J, E$ and $M$ represent the total angular momentum, energy and mass of the system where the angular momentum is acquired through tidal interactions with neighbouring objects \citep{peebles1969}. As early as 1987, \citet{barnes1987} used CDM N-body simulations to show that $\lambda$ is insensitive to the initial primordial density perturbation spectrum and follows a log-normal distribution with a median value of $\lambda \sim 0.05$. They also confirmed that the angular momentum of pre-collapse halos grows linearly with time, as predicted by linear tidal-torque theory \citep{doro1970}. \citet{bullock2001} re-defined $\lambda$ so as to be easily calculated from N-body simulations as
\begin{equation}
\lambda' = \frac{J_{vir}}{\sqrt{2} M_{vir} V_{vir} R_{vir}},
\label{lambda_bullock}
\end{equation}
where $J_{vir}$ represents the angular momentum enclosed within the virial mass $M_{vir}$. Further, $\lambda ' = \lambda \times f(c_h)$ where $c_h$ is the concentration parameter such that $f(c_h) \simeq [2/3 + (c_h/21.5)^{0.7}]$ \citep{mo1998}. Robust against the exact choice of the outer radius, $\lambda '$ converges to $\lambda$ at the virial radius of a singular isothermal halo. N-body simulations too show $\lambda$ to follow a log-normal probability distribution function with a peak at $\lambda \sim 0.05$ \citep[e.g.][]{barnes1987}, $\lambda \sim 0.04 \pm 0.5$ \citep{bullock2001, vitvitska2002} and $\lambda \sim 0.04\pm0.55$ \citep{davis2009}, almost irrespective of mass or $z$ \citep{vitvitska2002, davis2009} as also shown in Fig. \ref{fig_spin_fnz}. This implies a relation between the disk and halo radius through $R_d = \lambda R_{vir}$ such that the size distribution should reflect the underlying dark matter halo size distribution because gas acquires most of its angular momentum with dark matter halos before collapse. Interestingly, the spin correlates with the environment, with galaxies with high spin parameters being more clustered in accord with the tidal-torque theory \citep[e.g.][]{barnes1987, davis2009}. 

\begin{figure*}[h]
\center{\includegraphics[scale=1.01]{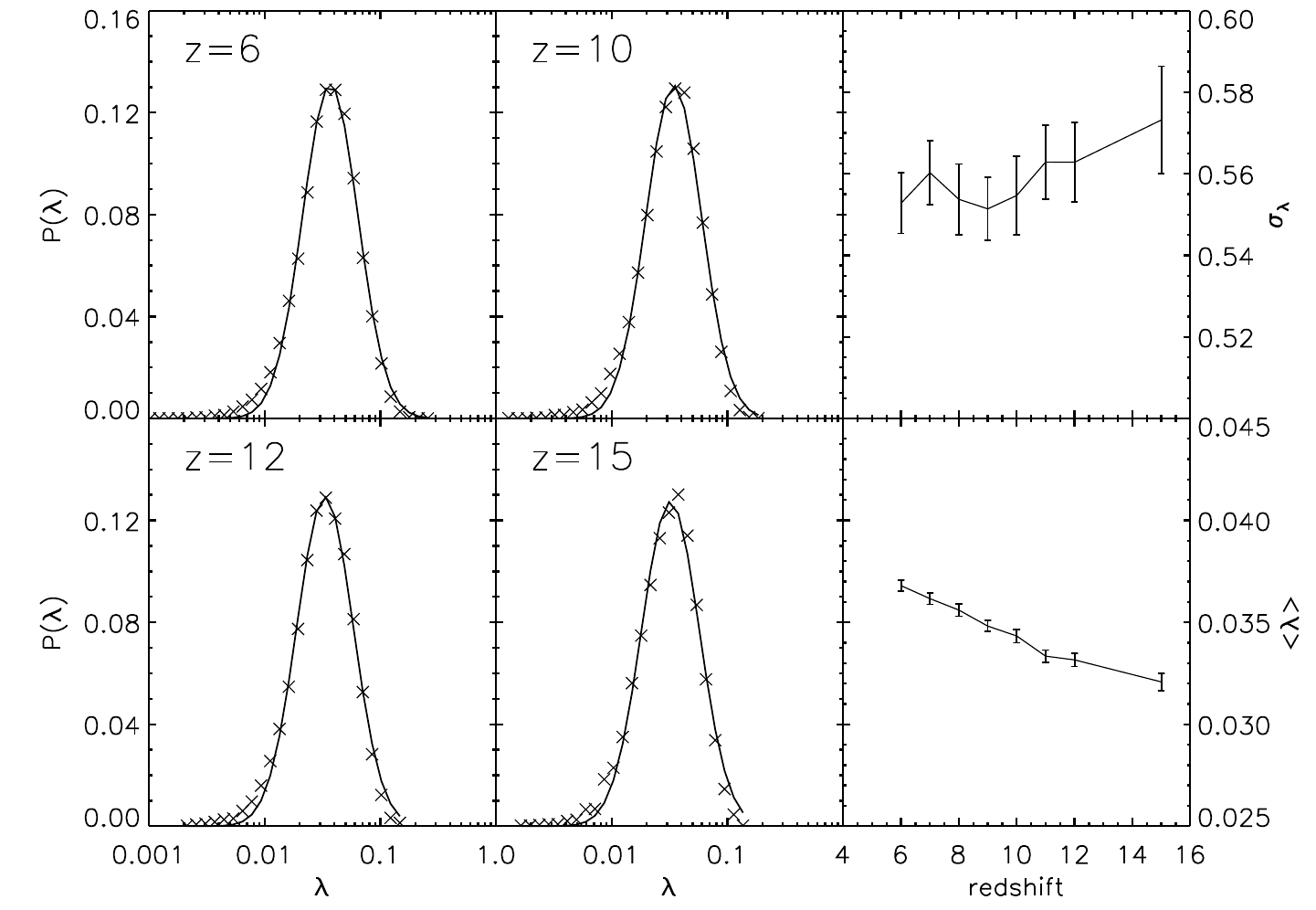}}
\caption{N-body simulations, for a box size of $2.46 h^{-1} {\rm Mpc}$ with a DM resolution mass of $7.3 \times 10^3\Msun$, run from $z \sim 100-6$ that show the spin parameter for all halos for $z =15-6$ as marked in the left two columns \citep{davis2009}. As shown by the solid line, the distribution is well-fit by a log-normal function. The right-most column shows the change in the mean and standard deviation of the log-normal fit for all the outputs studied.}
\label{fig_spin_fnz} 
\end{figure*}

Further, the cumulative mass distribution of the specific angular momentum has been found to have a {\it universal} shape that can be fit by a two-parameter ($\mu$ and $j_0$ below) function of the form \citep{bullock2001}
\begin{equation}
M(<j) = M_h \frac{\mu j}{j_0+j}; \,\,\,\, \mu>1,
\end{equation}
where $j$ represents the specific angular momentum of a halo of mass $M_h$. Naturally, the profile has a maximum specific angular momentum $j_{max} = j_0/(\mu-1)$, follows a power-law at $j \lsim j_0$ and flattens at $j \gsim j_0$. Here, $\mu$ acts as a shape parameter, leading to a pure power-law (pronounced bend) for $\mu>>1$ ($\mu \rightarrow 1$). Given that galaxies assemble their mass through a mix of relatively-quiescent accretion and mergers, this relation probably arises as a combination of the tidal-torque theory and the non-linear process of angular momentum transfer from satellite orbits during mergers \citep[see e.g.][]{bullock2001}; \citep{vitvitska2002} also show that $\lambda$ peaks sharply during major mergers and shows a steady decline during the gradual accretion of small satellites.  

\begin{figure*}
\center{\includegraphics[scale=0.35]{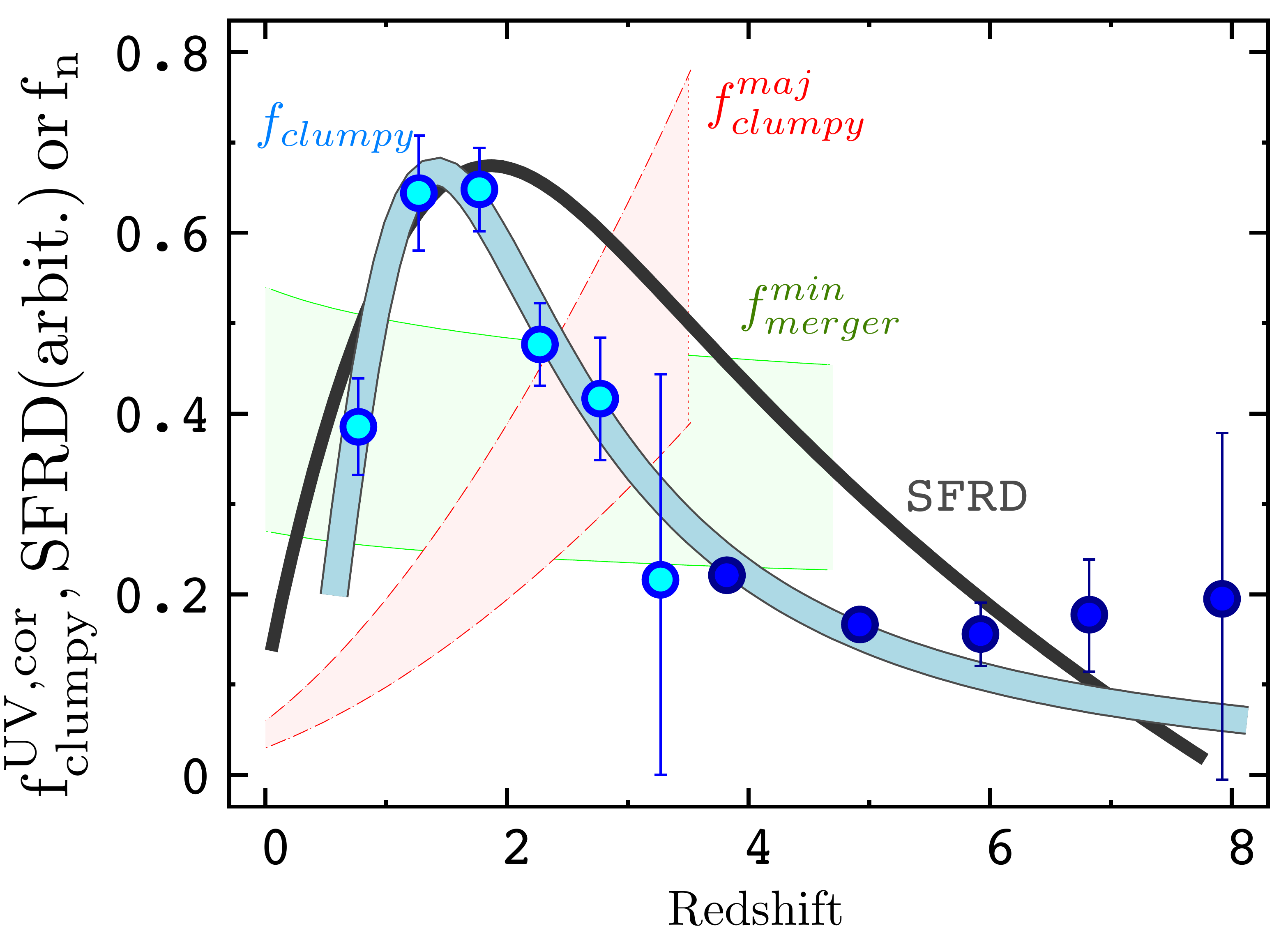}}
\caption{The redshift evolution of the ``clumpy" galaxy fraction at $z \simeq 0-8$ for star forming galaxies (filled cyan circles) and LBGs (filled blue circles) with $L_{UV}=0.3-1 L_{*z=3}$ \citep{shibuya2016} with Poissonian error bars. For comparison, the solid black line shows the cosmic SFRD from \citep{madau2014} which has been arbitrarily shifted and scaled along the y-axis. The shaded red and green regions show the galaxy major and minor merger fractions, respectively, assuming that the merger observability timescale ranges from 1 to 2 Gyr \citep{lotz2011}.}
\label{fig_clumpy} 
\end{figure*}

As for the shapes of high-$z$ galaxies, as early as 2004, \citet{ferguson2004} found a semi-major to semi-minor axis ratio of $b/a =0.65$ for 386 $z\sim 4$ galaxies selected via {\it HST}, with luminosities between $0.7-5L_*$, suggesting LBGs are not spheroids. This result was borne out by the larger sample (4700 LBGs between $z \simeq2.5-5$) collected by \citet{ravindranath2006} who found 40\% of LBGs to have exponential light-profiles, similar to disks, with only about 30\% showing the high concentrations seen for spheroids. A significant fraction (about 30\%) showed light profiles shallower than an exponential, exhibiting multiple cores or disturbed morphologies, indicative of close pairs or on-going mergers. As shown in Fig. \ref{fig_clumpy}, \citet{shibuya2016} show that while their observed LBGs are consistent with disk-like surface brightness profiles, the fraction of ``clumpy" galaxies in the UV, corrected for detection incompleteness, $f^{uv,cor}_{clumpy}$, whilst rising from $z \simeq 8$ to $1$ shows a subsequent decreases to $z \simeq 0$, tracking the cosmic star formation rate density, well described by the following function:
\begin{equation}
f^{uv,cor}_{clumpy} = a \times \frac{(1+z)^b}{1+[(1+z)/c]^d},
\end{equation}
where the best fit parameters are found to be $a=0.035, b=4.6, c=2.2$ and $d=6.7$. These results are in good agreement with those found by other groups \citep[see Fig. 4; ][]{shibuya2016} who find $f^{UV}_{clumpy} \simeq 20\%$ at $z \simeq 3-5$ \citep{conselice2009}, $f^{UV}_{clumpy} \simeq 40\%$ at $z \simeq 7$ \citep{jiang2013} and $f^{UV}_{clumpy}$ rising from $\simeq 10\%$ at $z \simeq 6-7$ to as much as $50\%$ by $z \simeq 8$ \citep{kawamata2015}. LAE observations too show the same clumpy structure, both at $z \sim 3.1$ \citep{bond2009} and, to a smaller extent, at $z \simeq 4-6$ \citep{taniguchi2009}. As shown in Fig. \ref{fig_clumpy}, the observed  redshift trend of $f^{uv,cor}_{clumpy}$ over $z \simeq 0-8$ requires minor mergers at $z \lsim 1$ and $z \gsim 2.5$ whilst requiring most mergers to be major at $z \simeq 1-2$. A more elegant explanation is therefore offered by the violent disk instability model \citep[e.g.][]{dekel2009b, dekel2013} where clumps form in unstable regions with a Toomre parameter value below the critical value of unity in thick, gas-rich disks. Depending on the inflows of cold gas, this model predicts the clumping fraction to evolve with the cosmic star formation rate density, as observed. Finally, in terms of galaxy properties, \citet{shibuya2016} find that $f^{UV}_{clumpy}$ increases with the UV luminosity, star formation rate density and the blue-ness of the spectral slope, although the error bars are quite significant. 

On the theoretical front, simulations show high-$z$ galaxies to be dominated by disks although the halo mass at which disks start dominating remains a matter of debate: while \citet{romanodiaz2011} find a 100\% disk fraction in $z \simeq 10$ galaxies for a total mass $\gsim 10^9\Msun$, \citet{feng2015} find disks to dominate (disk fraction of roughly $70\%$) in much larger $z =8-10$ galaxies with stellar masses $\gsim 10^{10}\Msun$. On the other hand, other works \citep[e.g.][]{pawlik2011} find disks in galaxies with masses as low as $8\times 10^7-1.2 \times 10^8\Msun$. Forthcoming observations with the {\it JWST} and {\it E-ELT} will be crucial in shedding more light on the shapes of early galaxies.   

\subsection{Sizes}
\label{sizes}
Indicating the dynamical state resulting from the processes (of feedback, merging, inflows, outflows) associated with galaxy assembly the size evolution of galaxies, measured through the half-light radius )$R_{hl}$) and the size-luminosity relation, are extremely useful parameters for constraining theoretical models. \citet{mo1998} have shown that, for an infinitesimally thin disk with an exponential surface density profile, the disk size $R_d$, which can be used as a proxy of $R_{hl}$, is related to $R_{vir}$ as
\begin{equation}
R_d = \frac{\lambda}{\sqrt 2} \frac{j_d}{\mu_d} R_{vir},
\end{equation}
\begin{figure*}[h!]
\center{\includegraphics[scale=0.65]{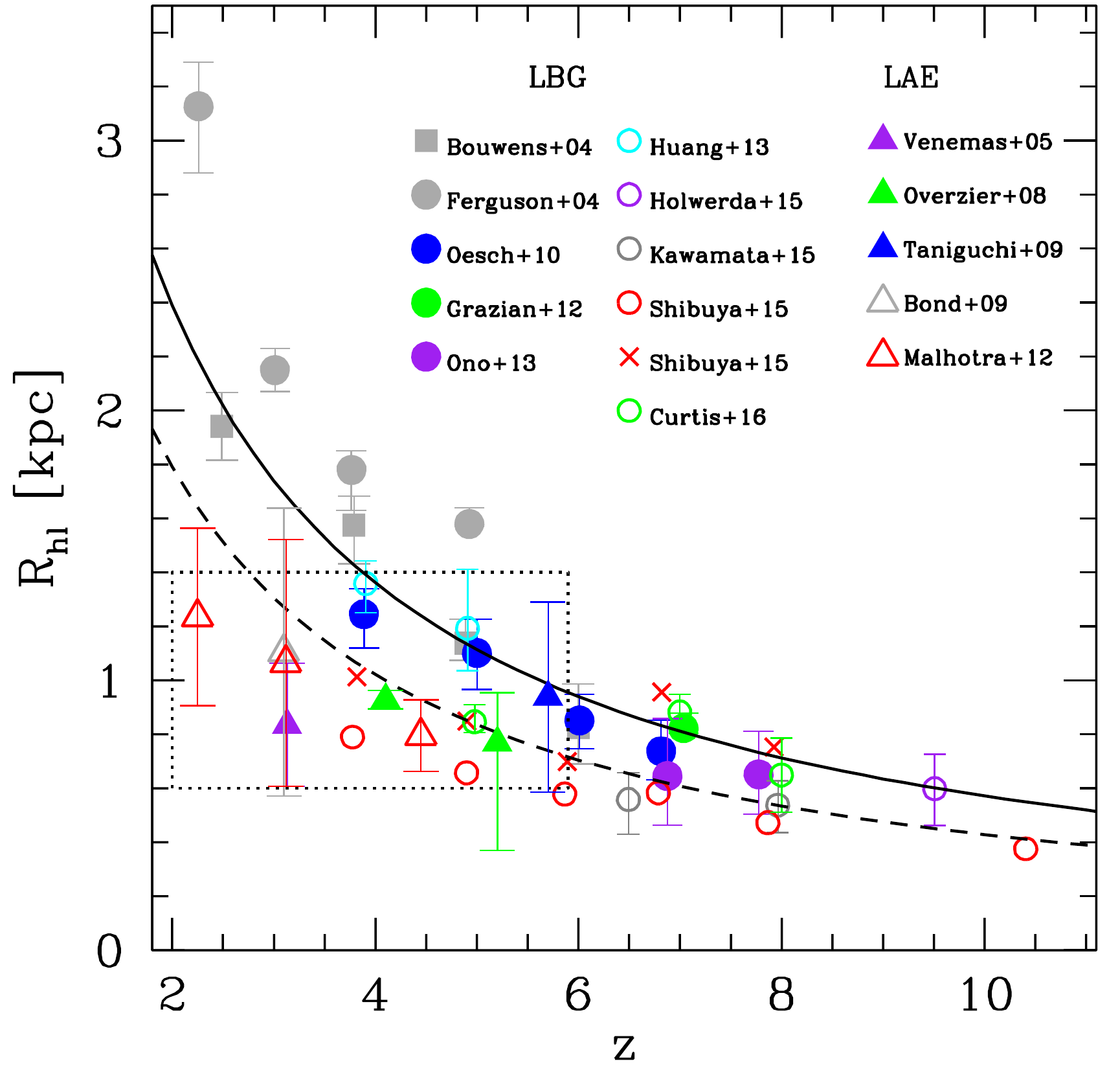}}
\caption{The observed half light radius for LAEs and LBGs as marked in the legend. The {\it LBG} data sets refer to galaxies with $0.3-1L_{*,z=3}$ from \citet{bouwens2004}, \citet{oesch2010}, \citet{grazian2012}, \citet{ono2013}, \citet{huang2013}, \citet{holwerda2015}, \citet[with circles and crosses showing median and average results respectively;][]{shibuya2015} and \citet{curtis-lake2016}, as marked. The \citet{ferguson2004} points are for bright LBGs with $L= 0.7-5L_{*,z=3}$ which might be responsible, at least in part, for their larger estimates of $R_{hl}$ compared to the other studies. The {\it LAE} data sets have been collected by \citet{venemans2005}, \citet{overzier2008}, \citet{taniguchi2009}, \citet{bond2009} and \citet{malhotra2012}, as marked. As shown by the solid and dashed black lines, the LBG size evolution is well-fit by a relation $R_{hl} \propto (1+z)^{-1.1}$ \citep{bouwens2004, shibuya2016} with different normalizations obtained by different groups. On the other hand, LAEs (demarcated by the dotted box) show a much shallower size evolution with $z$, with LAEs and LBGs effectively having the same sizes at $z \gsim 5$ \citep{malhotra2012}.}
\label{fig_rhl_fnz} 
\end{figure*}
where $\mu_d$ and $j_d$ are the fraction of mass and angular momentum in the disk relative to the halo. This relation can be used to derive the expected $z$-evolution of $R_d$ through the redshift evolution of $R_{vir}$ using the virial theorem (Sec. \ref{dm_assembly}) such that: 
\begin{equation}
R_d(z) \propto R_{vir}(z) \propto M_h^{1/3} H^{-2/3}(z),
\end{equation}
where we use $M_h = (4\pi/3) R_{vir}^3 \bar \rho$, assuming a collapse density of 200 times the critical density. Using $H(z) = H(0) (\Omega_m(1+z)^3+\Omega_\Lambda)^{1/2}$ and ignoring $\Omega_\Lambda$, which is reasonable at $z>2$, at a fixed halo mass we obtain
\begin{equation}
R_d \propto (1+z)^{-1}.
\end{equation}
The virial theorem can also be used to link $R_d$ to the galaxy velocity and to the luminosity through the Tully-Fisher \citep{tully-fisher1977} and Faber-Jackson \citep{faber-jackson1976} relations that find $L \propto V_{vir}$. For a given $L$, this yields
\begin{equation}
R_d \propto \frac{V_{vir}}{H(z)} \propto (1+z)^{-1.5}.
\end{equation} 
These two behaviours bracket the range of $R_d = fn(z)$ allowed using these simple arguments. 

\citet{liu2017} have used a 100 Mpc N-body simulation with a DM particle resolution mass of $2.6 \times 10^6 h^{-1}\Msun$ to calculate $R_d$ assuming $j_d/\mu_d=1$ and calculating $\lambda$ as in Eqn. \ref{lambda_bullock} above. Estimating $R_{hl} = 1.678 R_d$, these authors find $R_{hl} \propto (1+z)^{-m}$ with $m =2.0\pm 0.07$ for $z \sim 5-10$ LBGs with luminosities in the range $0.3-1L^*_{z=3}$. Further, they find $R_{hl} \propto L^\beta$ where $\beta$ steepens from $0.25 \pm 0.02$ at $z \sim 5$ to $0.36\pm0.03$ at $z \sim 10$ implying galaxies of a given luminosity become more compact with increasing $z$, as expected. These authors also point out that SN feedback is required in order to reproduce observed galaxy sizes at $z \sim 5$; the lack of SN feedback results in a younger stellar population that produces more luminosity, leading to a shallower $R_d - L$ relation. These theoretical results are slightly steeper than those inferred from observational campaigns, shown in Fig. \ref{fig_rhl_fnz}, which have been used to infer $m \sim -1.5$ \citep{ferguson2004, hathi2008}, $m = -1.12 \pm 0.17$ \citep{oesch2010}, $m = -1.2 \pm 0.1$ \citep{mosleh2012, ono2013, kawamata2015} and $m = -1$ \citep{bouwens2004, holwerda2015, shibuya2016, allen2016} in the range $(0.3-1)L^*_{z=3}$. However, \citet{curtis-lake2016} point out that using the modal value of the size-distribution results in a very shallow relation at $z \sim 4-8$ fit by $m = 0.2 \pm 0.26$ which is consistent with the null hypothesis of no size evolution. 

Observations have detected a number of other correlations including those linking the size and stellar mass where galaxies with increasing stellar masses have larger $R_{hl}$ values over $z \sim 0$ to 7 \citep{barden2005, graham2008, mosleh2012}. However, at high-$z$, the UV luminosity is a much more robust quantity so that we can express $R_{hl} \propto L_{UV}^\alpha$. Indeed, the theoretical value of $\alpha =0.25$ is in good agreement with the values inferred observationally: $\alpha \sim 0.3-0.5$ at $z \sim 7$ \citep{grazian2012}, $\alpha = 0.24$ at $z \sim 7$ \citep{holwerda2015}, $\alpha = 0.22-0.25$ at $z \sim 4,5$ \citep{huang2013} and $\alpha = 0.27$ at $z \sim 8$ \citep{shibuya2016}. Exploiting the power of the Hubble frontier Fields, {\it MUSE} and {\it VLT}, the $R_{hl}-L_{UV}$ tend has now been confirmed to hold down to magnitudes as faint as $\muv = -15$ to $-17$, yielding point source profiles with $R_{hl} \sim 160-240$pc at $z \sim 2-8$ \citep{bouwens2016} and $R_{hl} \sim 16-140$pc at $z \sim 6$ \citep{vanzella2016} with $\alpha \sim 0.5$. 

The size distribution of LBGs can be well fit by a log-normal distribution with a minor evolution in the peak value from $z \sim 2.3-7$ \citep{oesch2010, huang2013} with a tail of larger galaxies building up towards lower-$z$ driven either by hierarchical build up or an enhanced accretion of cold gas  \citep{ferguson2004, oesch2010}. The values of luminosity and the half light radius have also been used to infer the average star formation surface densities whose values are found to be $\sum_{SFR} \sim 1.9-2 \Msun\, {\rm yr^{-1} \, kpc^{-2}}$ at $z \sim 4-8$ for $(0.3-1)L^*_{z=3}$ LBGs, hinting at very similar star formation efficiencies for these LBGs \citep{oesch2010, ono2013}. However, these values are slightly lower than $\sum_{SFR} \sim 4.1\, (5.8)\, \Msun\, {\rm yr^{-1} \, kpc^{-2}}$ at $z \sim 6-7\, (8)$ inferred by \citet{kawamata2015}. 

On the other hand, galaxies detected as LAEs are found to be more compact than LBGs \citep{dow2007} and show a $z$-independent, constant small size $\sim 1.6$ kpc at $z\sim 2.5-6.5$ \citep{venemans2005, overzier2008, taniguchi2009, bond2009, malhotra2012}. At the same redshifts, LBGs are bigger by a factor of 1.2-2.5, with the sizes for these two populations converging at $z \gsim 5$ \citep{malhotra2012}. 

Finally, we note that the size-luminosity relation has enormous implications for the faint end of the UV LF. This is because the effects of completeness correction become increasingly severe if faint high-$z$ galaxies are extended sources. While the faint end of the UV LF could be as steep as $\alpha = -2$ if faint galaxies are extended ($\sim 0.2-0.3'$ at apparent J-band magnitudes of $=28-29$), it would be much shallower with $\alpha \sim -1.7$ if faint galaxies can be approximated as point objects \citep{grazian2012}. Indeed, the latest observations of faint ($\muv =-15$ to $-17$) LBGs at $z \sim 6-8$ suggest these to fit the latter scenario, resulting in smaller completeness corrections and shallower faint end slopes by a factor of 2-3 \citep{bouwens2017, vanzella2016}. The $R_{hl}-L_{UV}$ relation also has connotations for the $z$-evolution of the UV LF: although the surface brightness is constant for both populations, 
while the evolving sizes of LBGs (coupled with the constant surface brightness) result in an evolving UV LF, the constant small sizes of LAEs can explain their non-evolving UV LF at $z \sim 3-6$ \citep{malhotra2012}. Finally, \citet{zick2018} caution that the faint, compact sources seen in the Hubble frontier fields are most probably globular clusters hosted by faint galaxies rather than being faint galaxies themselves.  


\newpage

\newpage

\section{Open questions and future outlook}

Despite the enormous progress in our understanding of the inter-twined processes of galaxy formation and reionization in the first billion years, a number of issues remain unsolved as it should now appear clear from this review. While a number of these issues might remain unsolved or debated in the short term, where possible we provide an outlook on how others might be solved. \\

1. {\it What is the nature of dark matter?}\\
As shown in Sec. \ref{ch3}, while the cold dark matter paradigm has been exceptionally successful at explaining the large scale structure of the Universe, its small scale problems have left the field open to proposals of {\it alternative} dark matter models including warm dark matter, interactive cold dark matter, self-interacting dark matter and annihilating dark matter, to name a few. While detections of a 3.5 kev line from the Perseus cluster could possibly arise as a result of dark matter annihilations, given the lack of any experimental evidence of super-symmetery, the nature of dark matter remains one of the greatest puzzles in modern cosmology. \\

2. {\it IMF of the first stars} \\
As shown in Sec. \ref{first_sf}, the classic paradigm, of metal-free gas collapsing to form one massive star, underwent a change in 2010 when simulations started finding metal-free gas to be extremely prone to fragmentation. However, the final mass of the PopIII star, as well as the IMF remain open questions. Observations of metal-poor stars in the Milky Way and detections of PopIII stars in the early Universe, using for example the {\it JWST} would be invaluable on shedding more light on these issues. \\

3. {\it Inhomogeneous metal enrichment in the first billion years} \\ 
The metal enrichment of the ISM remains poorly understood in the early Universe. How and when metallicity relations - linking key galaxy observables such as the stellar mass, star formation rate, gas-phase metallicity and possibly even the cold gas mass - emerged remains an open question. In the near future, a combination of observations with {\it ALMA} (of the molecular component) and the {\it JWST} (of nebular emission lines to infer the gas-phase metallicity) could help extend lower-$z$ mass-metallicity relations to these early cosmic epochs. \\

4. {\it LyC escape fraction and its dependence on galaxy properties and redshift} \\
As shown in Sec. \ref{fesc}, the escape fraction of LyC photons, being a complex function of the ISM gas and dust content, supernova feedback, turbulence and line of sight, remains a major open question for reionization, with both theory and observations finding values ranging anywhere between $<1\%$ to $\gsim 50\%$. However, forthcoming observations, with the {\it JWST} might potentially yield strong LyC leakers up to $z \simeq 9$. In addition, indirect methods such as those combining estimates of the UV spectral slope ($\beta$) with the H$\beta$ emission line, using the [OIII]$\lambda5007/$[OII]$\lambda3727$ ratio as a proxy for $f_{esc}$,  or exploiting H$\alpha$ and visible continuum surface brightness profiles with the {\it JWST} could help shed more light on the mass- and redshift-dependence of $f_{esc}$. \\

5. {\it Topology and history of reionization} \\
As shown in Sec. \ref{ch7}, both the reionization history and topology (inside-out versus outside-in) depend on a number of inter-linked processes including the abundance- and redshift evolution of star forming galaxies, the supernova- and reionization-feedback dependent intrinsic star formation (and hence LyC photon production) rates, the escape fraction of LyC photons from the galaxies and the IGM clumping factor to name a few. As of now, the $z$-evolution of LAEs and the fraction of LBGs showing Ly$\alpha$ emission have been used to get hints on the, average and piecewise, reionization history between $z \sim 5.7-7$. However, over the next decade, 21cm experiments such as the {\it SKA} and {\it Hera} will be used, not just to obtain tomographical maps of the reionization process, but also shed light on the topology of reionization by cross-correlating 21cm and galaxy data \citep[e.g.][]{vrbanec2016, sobacchi2016, hutter2017, hutter2018, kubota2018}. \\ 

6. {\it Impact of the UVB background on galaxy formation}  \\
As shown in Sec. \ref{uvb_fb}, while the UV background, could, on the one hand, slow down the progress of reionization by photo-evaporating gas from low-mass halos, this effect could, to some extend, be countered by the UVB smoothing out the IGM clumping. However, the UV impact on both these processes critically depends on the emissivity from reionization sources, the strength of the UVB and the relative redshifts of the source and that at which the UVB impinges on a source, to name a few. As a result, the impact of the UVB on both star formation and the IGM clumping factor remain open questions. Although observations of the faintest galaxies with the {\it JWST} might shed light on the faint end of the UV LF, decoupling the impact of supernova versus UV feedback will, possibly, not be tractable. \\

7. {\it Role of AGN, intermediate mass black holes and micro quasars in reionization} \\
As shown in Sec. \ref{sources_reio}, while most works focus on galaxies to be the key reionization sources, the role low to intermediate mass black holes, hosted by faint AGN or micro quasars, remains an open question. Indeed, as a number of works have shown, the secondary ionizations from black hole activity and the suppression of the most numerous low-mass galaxies due to the rising UVB, could both allow significant, or at least non-negligible, reionization contributions from low mass black holes. \\

8. {\it How far could the faint-end of the UV LF extend ?} \\
Star formation in galaxies requires the dark matter halos to be able to, hold on to, and cool, a sufficient amount of, gas. However, given that both supernova and UV feedback can, to some extent, heat or blow-out the gas from low-mass halos, the lower limit of halos that can form stars, and hence contribute to reionization, remains an open question. Naturally this implies that both the faint-end slope and the cut-off magnitude (beyond which star formation is suppressed) remain open questions as of now. In the future, using {\it JWST} to observe lensed galaxies might possibly shed light on the faint-end turn-over point of the UV LF and its redshift evolution, if not the cut-off magnitude. \\

9. {\it Dust formation and obscuration} \\
As shown in Sec. \ref{gas_dust}, explaining the extremely high observed dust-to-stellar mass ratios ($\sim 0.2\%$) of some high-$z$ galaxies requires most of the dust be grown by accretion in the ISM, on timescales that are 10 times shorter than those inferred for the Milky Way, with only a small fraction provided by supernovae and AGB stars. Given that works show dust grain growth to be extremely inefficient in the ISM given its high ionization fields, the key dust sources, masses and dust impact on the observability of high-$z$ galaxies remain open questions. Pinning down the dust masses of high-$z$ galaxies, {\it ALMA} will be invaluable in shedding more light on this issue over the next years. \\ 

\newpage

\section{Acknowledgements}
PD acknowledges support from the ERC Starting Grant DELPHI H2020/717001, and the European Commission and University of Groningen CO-FUND Rosalind
Franklin program. AF acknowledges support from the ERC Advanced Grant
INTERSTELLAR H2020/740120. This research was also  partly supported by the
Munich Institute for Astro- and Particle Physics (MIAPP) of the DFG cluster of excellence ``Origin and Structure of the Universe". PD
acknowledges the generous hospitality of the Scuola Normale Superiore,
where part of this work was carried out.  The authors are grateful to the anonymous referee,
Rychard Bouwens, 
Volker Bromm,
Emma Curtis-Lake, 
Benedetta Ciardi, 
Avishai Dekel,
Nick Gnedin, 
Anne Hutter, 
Anupam Mazumdar, 
Pascal Oesch, 
Andrea Pallottini, 
Olmo Piana, 
Renske Smit, and
Eros Vanzella
for their invaluable scientific inputs. Finally PD dedicates this review to her grandmother - the first mathematician in the family.
\newpage

 \section*{References}
 \footnotesize{\bibliography{mybib}}

\begin{thebibliography}{648}
\expandafter\ifx\csname natexlab\endcsname\relax\def\natexlab#1{#1}\fi
\providecommand{\url}[1]{\texttt{#1}}
\providecommand{\href}[2]{#2}
\providecommand{\path}[1]{#1}
\providecommand{\DOIprefix}{doi:}
\providecommand{\ArXivprefix}{arXiv:}
\providecommand{\URLprefix}{URL: }
\providecommand{\Pubmedprefix}{pmid:}
\providecommand{\doi}[1]{\href{http://dx.doi.org/#1}{\path{#1}}}
\providecommand{\Pubmed}[1]{\href{pmid:#1}{\path{#1}}}
\providecommand{\bibinfo}[2]{#2}
\ifx\xfnm\relax \def\xfnm[#1]{\unskip,\space#1}\fi
\bibitem[{{Str{\"o}mberg}(1934)}]{stromberg1934}
\bibinfo{author}{G.~{Str{\"o}mberg}}, \bibinfo{journal}{\apj}
  \bibinfo{volume}{80} (\bibinfo{year}{1934}) \bibinfo{pages}{327}.
\bibitem[{{von Weizs{\"a}cker}(1951)}]{Weizscker1951}
\bibinfo{author}{C.~F. {von Weizs{\"a}cker}}, \bibinfo{journal}{\apj}
  \bibinfo{volume}{114} (\bibinfo{year}{1951}) \bibinfo{pages}{165}.
\bibitem[{{Hoyle}(1951)}]{Hoyle1951}
\bibinfo{author}{F.~{Hoyle}}, in: \bibinfo{booktitle}{Problems of Cosmical
  Aerodynamics},  p. \bibinfo{pages}{195}.
\bibitem[{{Hoyle}(1953)}]{Hoyle1953}
\bibinfo{author}{F.~{Hoyle}}, \bibinfo{journal}{\apj}  \bibinfo{volume}{118}
  (\bibinfo{year}{1953}) \bibinfo{pages}{513}.
\bibitem[{{Sciama}(1955)}]{Sciama1955}
\bibinfo{author}{D.~W. {Sciama}}, \bibinfo{journal}{\mnras}
  \bibinfo{volume}{115} (\bibinfo{year}{1955}) \bibinfo{pages}{3--14}.
\bibitem[{{Gamow}(1953)}]{Gamow1953}
\bibinfo{author}{G.~{Gamow}}, \bibinfo{journal}{\aj}  \bibinfo{volume}{58}
  (\bibinfo{year}{1953}) \bibinfo{pages}{39}.
\bibitem[{{Bondi}(1952)}]{Bondi1952}
\bibinfo{author}{H.~{Bondi}}, \bibinfo{title}{{Cosmology.}},
  \bibinfo{year}{1952}.
\bibitem[{{Vororitsov-Velyaminov}(1961)}]{Vororitsov1961}
\bibinfo{author}{B.~{Vororitsov-Velyaminov}}, \bibinfo{journal}{\aj}
  \bibinfo{volume}{66} (\bibinfo{year}{1961}) \bibinfo{pages}{551}.
\bibitem[{{Ostriker} and {Cowie}(1981)}]{Ostriker1981}
\bibinfo{author}{J.~P. {Ostriker}}, \bibinfo{author}{L.~L. {Cowie}},
  \bibinfo{journal}{\apjl}  \bibinfo{volume}{243} (\bibinfo{year}{1981})
  \bibinfo{pages}{L127--L131}.
\bibitem[{{Penzias} and {Wilson}(1965)}]{penzias-wilson1965}
\bibinfo{author}{A.~A. {Penzias}}, \bibinfo{author}{R.~W. {Wilson}},
  \bibinfo{journal}{Astrophysical Journal}  \bibinfo{volume}{142}
  (\bibinfo{year}{1965}) \bibinfo{pages}{419--421}.
\bibitem[{{Peebles}(1965)}]{Peebles1965}
\bibinfo{author}{P.~J.~E. {Peebles}}, \bibinfo{journal}{\apj}
  \bibinfo{volume}{142} (\bibinfo{year}{1965}) \bibinfo{pages}{1317}.
\bibitem[{{Harrison}(1967)}]{Harrison1967}
\bibinfo{author}{E.~R. {Harrison}}, \bibinfo{journal}{\aj}
  \bibinfo{volume}{72} (\bibinfo{year}{1967}) \bibinfo{pages}{303}.
\bibitem[{{Silk}(1968)}]{Silk1968}
\bibinfo{author}{J.~{Silk}}, \bibinfo{journal}{\nat}  \bibinfo{volume}{218}
  (\bibinfo{year}{1968}) \bibinfo{pages}{453--454}.
\bibitem[{{Harrison}(1970)}]{Harrison1970}
\bibinfo{author}{E.~R. {Harrison}}, \bibinfo{journal}{\prd}
  \bibinfo{volume}{1} (\bibinfo{year}{1970}) \bibinfo{pages}{2726--2730}.
\bibitem[{{Zeldovich}(1972)}]{Zeldovich1972}
\bibinfo{author}{Y.~B. {Zeldovich}}, \bibinfo{journal}{\mnras}
  \bibinfo{volume}{160} (\bibinfo{year}{1972}) \bibinfo{pages}{1P}.
\bibitem[{Zel'dovich(1970)}]{Zeldovich1970}
\bibinfo{author}{Y.~B. Zel'dovich}, \bibinfo{journal}{Astronomy and
  Astrophysics}  \bibinfo{volume}{5} (\bibinfo{year}{1970})
  \bibinfo{pages}{84}.
\bibitem[{{Gunn} and {Gott}(1972)}]{Gunn1972}
\bibinfo{author}{J.~E. {Gunn}}, \bibinfo{author}{J.~R. {Gott}, III},
  \bibinfo{journal}{\apj}  \bibinfo{volume}{176} (\bibinfo{year}{1972})
  \bibinfo{pages}{1}.
\bibitem[{{Press} and {Schechter}(1974)}]{press-sch1974}
\bibinfo{author}{W.~H. {Press}}, \bibinfo{author}{P.~{Schechter}},
  \bibinfo{journal}{Astrophysical Journal}  \bibinfo{volume}{187}
  (\bibinfo{year}{1974}) \bibinfo{pages}{425--438}.
\bibitem[{{Larson}(1974)}]{Larson1974}
\bibinfo{author}{R.~B. {Larson}}, \bibinfo{journal}{\mnras}
  \bibinfo{volume}{169} (\bibinfo{year}{1974}) \bibinfo{pages}{229--246}.
\bibitem[{{White} and {Rees}(1978)}]{White1978}
\bibinfo{author}{S.~D.~M. {White}}, \bibinfo{author}{M.~J. {Rees}},
  \bibinfo{journal}{\mnras}  \bibinfo{volume}{183} (\bibinfo{year}{1978})
  \bibinfo{pages}{341--358}.
\bibitem[{{Gott} and {Thuan}(1976)}]{Gott1976}
\bibinfo{author}{J.~R. {Gott}, III}, \bibinfo{author}{T.~X. {Thuan}},
  \bibinfo{journal}{\apj}  \bibinfo{volume}{204} (\bibinfo{year}{1976})
  \bibinfo{pages}{649--667}.
\bibitem[{{Silk}(1977)}]{Silk1977}
\bibinfo{author}{J.~{Silk}}, \bibinfo{journal}{\apj}  \bibinfo{volume}{211}
  (\bibinfo{year}{1977}) \bibinfo{pages}{638--648}.
\bibitem[{{Binney}(1977)}]{binney1977}
\bibinfo{author}{J.~{Binney}}, \bibinfo{journal}{\apj}  \bibinfo{volume}{215}
  (\bibinfo{year}{1977}) \bibinfo{pages}{483--491}.
\bibitem[{{Rees} and {Ostriker}(1977)}]{rees1977}
\bibinfo{author}{M.~J. {Rees}}, \bibinfo{author}{J.~P. {Ostriker}},
  \bibinfo{journal}{\mnras}  \bibinfo{volume}{179} (\bibinfo{year}{1977})
  \bibinfo{pages}{541--559}.
\bibitem[{{Rubin} et~al.(1978){Rubin}, {Thonnard}, and {Ford}}]{Rubin1978}
\bibinfo{author}{V.~C. {Rubin}}, \bibinfo{author}{N.~{Thonnard}},
  \bibinfo{author}{W.~K. {Ford}, Jr.}, \bibinfo{journal}{\apjl}
  \bibinfo{volume}{225} (\bibinfo{year}{1978}) \bibinfo{pages}{L107--L111}.
\bibitem[{{Guth}(1981)}]{Guth1981}
\bibinfo{author}{A.~H. {Guth}}, \bibinfo{journal}{\prd}  \bibinfo{volume}{23}
  (\bibinfo{year}{1981}) \bibinfo{pages}{347--356}.
\bibitem[{{Sato}(1981)}]{Sato1981}
\bibinfo{author}{K.~{Sato}}, \bibinfo{journal}{\mnras}  \bibinfo{volume}{195}
  (\bibinfo{year}{1981}) \bibinfo{pages}{467--479}.
\bibitem[{Mazumdar and Rocher(2011)}]{mazumdar2010}
\bibinfo{author}{A.~Mazumdar}, \bibinfo{author}{J.~Rocher},
  \bibinfo{journal}{Phys. Rept.}  \bibinfo{volume}{497} (\bibinfo{year}{2011})
  \bibinfo{pages}{85--215}.
\bibitem[{{Linde}(1982)}]{Linde1982}
\bibinfo{author}{A.~D. {Linde}}, \bibinfo{journal}{Physics Letters B}
  \bibinfo{volume}{116} (\bibinfo{year}{1982}) \bibinfo{pages}{335--339}.
\bibitem[{{Albrecht} and {Steinhardt}(1982)}]{Albrecht1982}
\bibinfo{author}{A.~{Albrecht}}, \bibinfo{author}{P.~J. {Steinhardt}},
  \bibinfo{journal}{Physical Review Letters}  \bibinfo{volume}{48}
  (\bibinfo{year}{1982}) \bibinfo{pages}{1220--1223}.
\bibitem[{{Linde}(1983)}]{Linde1983}
\bibinfo{author}{A.~D. {Linde}}, \bibinfo{journal}{Physics Letters B}
  \bibinfo{volume}{129} (\bibinfo{year}{1983}) \bibinfo{pages}{177--181}.
\bibitem[{{Bardeen} et~al.(1983){Bardeen}, {Steinhardt}, and
  {Turner}}]{Bardeen1983}
\bibinfo{author}{J.~M. {Bardeen}}, \bibinfo{author}{P.~J. {Steinhardt}},
  \bibinfo{author}{M.~S. {Turner}}, \bibinfo{journal}{\prd}
  \bibinfo{volume}{28} (\bibinfo{year}{1983}) \bibinfo{pages}{679--693}.
\bibitem[{{Peebles}(1982)}]{Peebles1982}
\bibinfo{author}{P.~J.~E. {Peebles}}, \bibinfo{journal}{\apjl}
  \bibinfo{volume}{263} (\bibinfo{year}{1982}) \bibinfo{pages}{L1--L5}.
\bibitem[{{Bardeen} et~al.(1986){Bardeen}, {Bond}, {Kaiser}, and
  {Szalay}}]{Bardeen1986}
\bibinfo{author}{J.~M. {Bardeen}}, \bibinfo{author}{J.~R. {Bond}},
  \bibinfo{author}{N.~{Kaiser}}, \bibinfo{author}{A.~S. {Szalay}},
  \bibinfo{journal}{\apj}  \bibinfo{volume}{304} (\bibinfo{year}{1986})
  \bibinfo{pages}{15--61}.
\bibitem[{{Sugiyama}(1995)}]{Sugiyama1995}
\bibinfo{author}{N.~{Sugiyama}}, \bibinfo{journal}{\apjs}
  \bibinfo{volume}{100} (\bibinfo{year}{1995}) \bibinfo{pages}{281}.
\bibitem[{{Navarro} and {Benz}(1991)}]{Navarro1991}
\bibinfo{author}{J.~F. {Navarro}}, \bibinfo{author}{W.~{Benz}},
  \bibinfo{journal}{\apj}  \bibinfo{volume}{380} (\bibinfo{year}{1991})
  \bibinfo{pages}{320--329}.
\bibitem[{{Katz} and {Gunn}(1991)}]{Katz1991}
\bibinfo{author}{N.~{Katz}}, \bibinfo{author}{J.~E. {Gunn}},
  \bibinfo{journal}{\apj}  \bibinfo{volume}{377} (\bibinfo{year}{1991})
  \bibinfo{pages}{365--381}.
\bibitem[{Steidel et~al.(1996)Steidel, Giavalisco, Pettini, Dickinson, and
  Adelberger}]{steidel1996}
\bibinfo{author}{C.~C. Steidel}, \bibinfo{author}{M.~Giavalisco},
  \bibinfo{author}{M.~Pettini}, \bibinfo{author}{M.~Dickinson},
  \bibinfo{author}{K.~L. Adelberger}, \bibinfo{journal}{Astrophysical Journal}
  \bibinfo{volume}{462} (\bibinfo{year}{1996}) \bibinfo{pages}{L17}.
\bibitem[{Ellis et~al.(2013)}]{ellis2012}
\bibinfo{author}{R.~S. Ellis}, et~al., \bibinfo{journal}{Astrophys. J.}
  \bibinfo{volume}{763} (\bibinfo{year}{2013}) \bibinfo{pages}{L7}.
\bibitem[{Atek et~al.(2015)}]{atek2015}
\bibinfo{author}{H.~Atek}, et~al., \bibinfo{journal}{Astrophys. J.}
  \bibinfo{volume}{814} (\bibinfo{year}{2015}) \bibinfo{pages}{69}.
\bibitem[{{Smit} et~al.(2014){Smit}, {Bouwens}, and {Labb{\'e}}}]{smit2014}
\bibinfo{author}{R.~{Smit}}, \bibinfo{author}{R.~J. {Bouwens}},
  \bibinfo{author}{I.~e.~a. {Labb{\'e}}}, \bibinfo{journal}{\apj}
  \bibinfo{volume}{784} (\bibinfo{year}{2014}) \bibinfo{pages}{58}.
\bibitem[{Zitrin et~al.(2015)Zitrin, Labbe, and Belli}]{zitrin2015}
\bibinfo{author}{A.~Zitrin}, \bibinfo{author}{I.~Labbe},
  \bibinfo{author}{S.~e.~a. Belli}, \bibinfo{journal}{Astrophysical Journal}
  \bibinfo{volume}{810} (\bibinfo{year}{2015}) \bibinfo{pages}{L12}.
\bibitem[{{Oke} and {Gunn}(1983)}]{oke-gunn1983}
\bibinfo{author}{J.~B. {Oke}}, \bibinfo{author}{J.~E. {Gunn}},
  \bibinfo{journal}{Astrophysical Journal}  \bibinfo{volume}{266}
  (\bibinfo{year}{1983}) \bibinfo{pages}{713--717}.
\bibitem[{{Maartens}(2011)}]{maartens2011}
\bibinfo{author}{R.~{Maartens}}, \bibinfo{journal}{Philosophical Transactions
  of the Royal Society of London Series A}  \bibinfo{volume}{369}
  (\bibinfo{year}{2011}) \bibinfo{pages}{5115--5137}.
\bibitem[{{Saadeh} et~al.(2016){Saadeh}, {Feeney}, {Pontzen}, {Peiris}, and
  {McEwen}}]{saadeh2016}
\bibinfo{author}{D.~{Saadeh}}, \bibinfo{author}{S.~M. {Feeney}},
  \bibinfo{author}{A.~{Pontzen}}, \bibinfo{author}{H.~V. {Peiris}},
  \bibinfo{author}{J.~D. {McEwen}}, \bibinfo{journal}{Physical Review Letters}
  \bibinfo{volume}{117} (\bibinfo{year}{2016}) \bibinfo{pages}{131302}.
\bibitem[{{Alpher} et~al.(1948){Alpher}, {Bethe}, and {Gamow}}]{alpher1948}
\bibinfo{author}{R.~A. {Alpher}}, \bibinfo{author}{H.~{Bethe}},
  \bibinfo{author}{G.~{Gamow}}, \bibinfo{journal}{Physical Review}
  \bibinfo{volume}{73} (\bibinfo{year}{1948}) \bibinfo{pages}{803--804}.
\bibitem[{{Planck Collaboration} et~al.(2016{\natexlab{a}}){Planck
  Collaboration}, {Ade}, {Aghanim}, {Arnaud}, and et~al.}]{planck2015}
\bibinfo{author}{{Planck Collaboration}}, \bibinfo{author}{P.~A.~R. {Ade}},
  \bibinfo{author}{N.~{Aghanim}}, \bibinfo{author}{M.~{Arnaud}},
  \bibinfo{author}{et~al.}, \bibinfo{journal}{\aap}  \bibinfo{volume}{594}
  (\bibinfo{year}{2016}{\natexlab{a}}) \bibinfo{pages}{A13}.
\bibitem[{{Planck Collaboration} et~al.(2016{\natexlab{b}}){Planck
  Collaboration}, {Aghanim}, {Ashdown}, and et. al.}]{planck2016}
\bibinfo{author}{{Planck Collaboration}}, \bibinfo{author}{N.~{Aghanim}},
  \bibinfo{author}{M.~{Ashdown}}, \bibinfo{author}{et. al.},
  \bibinfo{journal}{\aap}  \bibinfo{volume}{596}
  (\bibinfo{year}{2016}{\natexlab{b}}) \bibinfo{pages}{A107}.
\bibitem[{{Bennett} et~al.(2003){Bennett}, {Halpern}, {Hinshaw}, and et.
  al.}]{bennett2003}
\bibinfo{author}{C.~L. {Bennett}}, \bibinfo{author}{M.~{Halpern}},
  \bibinfo{author}{G.~{Hinshaw}}, \bibinfo{author}{et. al.},
  \bibinfo{journal}{\apjs}  \bibinfo{volume}{148} (\bibinfo{year}{2003})
  \bibinfo{pages}{1--27}.
\bibitem[{{Peebles}(1971)}]{peebles1971}
\bibinfo{author}{P.~J.~E. {Peebles}}, \bibinfo{title}{{Physical cosmology}},
  \bibinfo{year}{1971}.
\bibitem[{{Blumenthal} et~al.(1984){Blumenthal}, {Faber}, {Primack}, and
  {Rees}}]{blumenthal1984}
\bibinfo{author}{G.~R. {Blumenthal}}, \bibinfo{author}{S.~M. {Faber}},
  \bibinfo{author}{J.~R. {Primack}}, \bibinfo{author}{M.~J. {Rees}},
  \bibinfo{journal}{\nat}  \bibinfo{volume}{311} (\bibinfo{year}{1984})
  \bibinfo{pages}{517--525}.
\bibitem[{{Bond} and {Szalay}(1983)}]{bond-szalay1983}
\bibinfo{author}{J.~R. {Bond}}, \bibinfo{author}{A.~S. {Szalay}},
  \bibinfo{journal}{\apj}  \bibinfo{volume}{274} (\bibinfo{year}{1983})
  \bibinfo{pages}{443--468}.
\bibitem[{{Cole} et~al.(2005){Cole}, {Percival}, {Peacock}, and et.
  al.}]{cole2005}
\bibinfo{author}{S.~{Cole}}, \bibinfo{author}{W.~J. {Percival}},
  \bibinfo{author}{J.~A. {Peacock}}, \bibinfo{author}{et. al.},
  \bibinfo{journal}{\mnras}  \bibinfo{volume}{362} (\bibinfo{year}{2005})
  \bibinfo{pages}{505--534}.
\bibitem[{{Hinshaw} et~al.(2013){Hinshaw}, {Larson}, {Komatsu}, and et.
  al.}]{hinshaw2013}
\bibinfo{author}{G.~{Hinshaw}}, \bibinfo{author}{D.~{Larson}},
  \bibinfo{author}{E.~{Komatsu}}, \bibinfo{author}{et. al.},
  \bibinfo{journal}{\apjs}  \bibinfo{volume}{208} (\bibinfo{year}{2013})
  \bibinfo{pages}{19}.
\bibitem[{{Planck Collaboration} et~al.(2014){Planck Collaboration}, {Ade},
  {Aghanim}, {Armitage-Caplan}, and et~al.}]{planck2014}
\bibinfo{author}{{Planck Collaboration}}, \bibinfo{author}{P.~A.~R. {Ade}},
  \bibinfo{author}{N.~{Aghanim}}, \bibinfo{author}{C.~{Armitage-Caplan}},
  \bibinfo{author}{et~al.}, \bibinfo{journal}{\aap}  \bibinfo{volume}{571}
  (\bibinfo{year}{2014}) \bibinfo{pages}{A16}.
\bibitem[{{Slosar} et~al.(2013){Slosar}, {Ir{\v s}i{\v c}}, {Kirkby}, and et.
  al.}]{slosar2013}
\bibinfo{author}{A.~{Slosar}}, \bibinfo{author}{V.~{Ir{\v s}i{\v c}}},
  \bibinfo{author}{D.~{Kirkby}}, \bibinfo{author}{et. al.},
  \bibinfo{journal}{\jcap}  \bibinfo{volume}{4} (\bibinfo{year}{2013})
  \bibinfo{pages}{26}.
\bibitem[{{Weinberg} et~al.(2015){Weinberg}, {Bullock}, {Governato}, {Kuzio de
  Naray}, and {Peter}}]{weinberg2013}
\bibinfo{author}{D.~H. {Weinberg}}, \bibinfo{author}{J.~S. {Bullock}},
  \bibinfo{author}{F.~{Governato}}, \bibinfo{author}{R.~{Kuzio de Naray}},
  \bibinfo{author}{A.~H.~G. {Peter}}, \bibinfo{journal}{Proceedings of the
  National Academy of Science}  \bibinfo{volume}{112} (\bibinfo{year}{2015})
  \bibinfo{pages}{12249--12255}.
\bibitem[{{Moore} et~al.(1999){Moore}, {Quinn}, {Governato}, {Stadel}, and
  {Lake}}]{moore1999}
\bibinfo{author}{B.~{Moore}}, \bibinfo{author}{T.~{Quinn}},
  \bibinfo{author}{F.~{Governato}}, \bibinfo{author}{J.~{Stadel}},
  \bibinfo{author}{G.~{Lake}}, \bibinfo{journal}{\mnras}  \bibinfo{volume}{310}
  (\bibinfo{year}{1999}) \bibinfo{pages}{1147--1152}.
\bibitem[{{Klypin} et~al.(1999){Klypin}, {Kravtsov}, {Valenzuela}, and
  {Prada}}]{klypin1999}
\bibinfo{author}{A.~{Klypin}}, \bibinfo{author}{A.~V. {Kravtsov}},
  \bibinfo{author}{O.~{Valenzuela}}, \bibinfo{author}{F.~{Prada}},
  \bibinfo{journal}{\apj}  \bibinfo{volume}{522} (\bibinfo{year}{1999})
  \bibinfo{pages}{82--92}.
\bibitem[{{Boylan-Kolchin} et~al.(2011){Boylan-Kolchin}, {Bullock}, and
  {Kaplinghat}}]{boylan2011}
\bibinfo{author}{M.~{Boylan-Kolchin}}, \bibinfo{author}{J.~S. {Bullock}},
  \bibinfo{author}{M.~{Kaplinghat}}, \bibinfo{journal}{Monthly Notices of the
  Royal Astronomical Society}  \bibinfo{volume}{415} (\bibinfo{year}{2011})
  \bibinfo{pages}{L40--L44}.
\bibitem[{{Boylan-Kolchin} et~al.(2012){Boylan-Kolchin}, {Bullock}, and
  {Kaplinghat}}]{boylan2012}
\bibinfo{author}{M.~{Boylan-Kolchin}}, \bibinfo{author}{J.~S. {Bullock}},
  \bibinfo{author}{M.~{Kaplinghat}}, \bibinfo{journal}{\mnras}
  \bibinfo{volume}{422} (\bibinfo{year}{2012}) \bibinfo{pages}{1203--1218}.
\bibitem[{{Moore} et~al.(1999){Moore}, {Ghigna}, {Governato}, {Lake}, {Quinn},
  {Stadel}, and {Tozzi}}]{moore1999b}
\bibinfo{author}{B.~{Moore}}, \bibinfo{author}{S.~{Ghigna}},
  \bibinfo{author}{F.~{Governato}}, \bibinfo{author}{G.~{Lake}},
  \bibinfo{author}{T.~{Quinn}}, \bibinfo{author}{J.~{Stadel}},
  \bibinfo{author}{P.~{Tozzi}}, \bibinfo{journal}{\apjl}  \bibinfo{volume}{524}
  (\bibinfo{year}{1999}) \bibinfo{pages}{L19--L22}.
\bibitem[{{Navarro} et~al.(1997){Navarro}, {Frenk}, and {White}}]{navarro1997}
\bibinfo{author}{J.~F. {Navarro}}, \bibinfo{author}{C.~S. {Frenk}},
  \bibinfo{author}{S.~D.~M. {White}}, \bibinfo{journal}{\apj}
  \bibinfo{volume}{490} (\bibinfo{year}{1997}) \bibinfo{pages}{493--508}.
\bibitem[{{Subramanian} et~al.(2000){Subramanian}, {Cen}, and
  {Ostriker}}]{subramanian2000}
\bibinfo{author}{K.~{Subramanian}}, \bibinfo{author}{R.~{Cen}},
  \bibinfo{author}{J.~P. {Ostriker}}, \bibinfo{journal}{\apj}
  \bibinfo{volume}{538} (\bibinfo{year}{2000}) \bibinfo{pages}{528--542}.
\bibitem[{{Wyse}(2001)}]{wyse2001}
\bibinfo{author}{R.~F.~G. {Wyse}}, in: \bibinfo{editor}{J.~G. {Funes}},
  \bibinfo{editor}{E.~M. {Corsini}} (Eds.), \bibinfo{booktitle}{Galaxy Disks
  and Disk Galaxies}, volume \bibinfo{volume}{230} of
  \textit{\bibinfo{series}{Astronomical Society of the Pacific Conference
  Series}},  pp. \bibinfo{pages}{71--80}.
\bibitem[{{Teyssier} et~al.(2013){Teyssier}, {Pontzen}, {Dubois}, and
  {Read}}]{teyssier2013}
\bibinfo{author}{R.~{Teyssier}}, \bibinfo{author}{A.~{Pontzen}},
  \bibinfo{author}{Y.~{Dubois}}, \bibinfo{author}{J.~I. {Read}},
  \bibinfo{journal}{\mnras}  \bibinfo{volume}{429} (\bibinfo{year}{2013})
  \bibinfo{pages}{3068--3078}.
\bibitem[{{Bode} et~al.(2001){Bode}, {Ostriker}, and {Turok}}]{bode2001}
\bibinfo{author}{P.~{Bode}}, \bibinfo{author}{J.~P. {Ostriker}},
  \bibinfo{author}{N.~{Turok}}, \bibinfo{journal}{\apj}  \bibinfo{volume}{556}
  (\bibinfo{year}{2001}) \bibinfo{pages}{93--107}.
\bibitem[{Bulbul et~al.(2014)Bulbul, Markevitch, Foster, Smith, Loewenstein
  et~al.}]{bulbul2014}
\bibinfo{author}{E.~Bulbul}, \bibinfo{author}{M.~Markevitch},
  \bibinfo{author}{A.~Foster}, \bibinfo{author}{R.~K. Smith},
  \bibinfo{author}{M.~Loewenstein}, et~al., \bibinfo{journal}{Astrophys.J.}
  \bibinfo{volume}{789} (\bibinfo{year}{2014}) \bibinfo{pages}{13}.
\bibitem[{Boyarsky et~al.(2014)Boyarsky, Ruchayskiy, Iakubovskyi, and
  Franse}]{boyarsky2014}
\bibinfo{author}{A.~Boyarsky}, \bibinfo{author}{O.~Ruchayskiy},
  \bibinfo{author}{D.~Iakubovskyi}, \bibinfo{author}{J.~Franse},
  \bibinfo{journal}{Phys.Rev.Lett.}  \bibinfo{volume}{113}
  (\bibinfo{year}{2014}) \bibinfo{pages}{251301}.
\bibitem[{{Cappelluti} et~al.(2018){Cappelluti}, {Bulbul}, {Foster}, and et.
  al.}]{cappelluti2017}
\bibinfo{author}{N.~{Cappelluti}}, \bibinfo{author}{E.~{Bulbul}},
  \bibinfo{author}{A.~{Foster}}, \bibinfo{author}{et. al.},
  \bibinfo{journal}{\apj}  \bibinfo{volume}{854} (\bibinfo{year}{2018})
  \bibinfo{pages}{179}.
\bibitem[{{Macci{\`o}} et~al.(2012){Macci{\`o}}, {Stinson}, {Brook}, and et.
  al.}]{maccio2012}
\bibinfo{author}{A.~V. {Macci{\`o}}}, \bibinfo{author}{G.~{Stinson}},
  \bibinfo{author}{C.~B. {Brook}}, \bibinfo{author}{et. al.},
  \bibinfo{journal}{\apjl}  \bibinfo{volume}{744} (\bibinfo{year}{2012})
  \bibinfo{pages}{L9}.
\bibitem[{{Schneider} et~al.(2014){Schneider}, {Anderhalden}, {Macci{\`o}}, and
  {Diemand}}]{schneider_wdm2014}
\bibinfo{author}{A.~{Schneider}}, \bibinfo{author}{D.~{Anderhalden}},
  \bibinfo{author}{A.~V. {Macci{\`o}}}, \bibinfo{author}{J.~{Diemand}},
  \bibinfo{journal}{\mnras}  \bibinfo{volume}{441} (\bibinfo{year}{2014})
  \bibinfo{pages}{L6--L10}.
\bibitem[{{Hui} et~al.(2017){Hui}, {Ostriker}, {Tremaine}, and
  {Witten}}]{hui2017}
\bibinfo{author}{L.~{Hui}}, \bibinfo{author}{J.~P. {Ostriker}},
  \bibinfo{author}{S.~{Tremaine}}, \bibinfo{author}{E.~{Witten}},
  \bibinfo{journal}{\prd}  \bibinfo{volume}{95} (\bibinfo{year}{2017})
  \bibinfo{pages}{043541}.
\bibitem[{{Du} et~al.(2017){Du}, {Behrens}, and {Niemeyer}}]{du2017}
\bibinfo{author}{X.~{Du}}, \bibinfo{author}{C.~{Behrens}},
  \bibinfo{author}{J.~C. {Niemeyer}}, \bibinfo{journal}{\mnras}
  \bibinfo{volume}{465} (\bibinfo{year}{2017}) \bibinfo{pages}{941--951}.
\bibitem[{{Spergel} and {Steinhardt}(2000)}]{spergel2000}
\bibinfo{author}{D.~N. {Spergel}}, \bibinfo{author}{P.~J. {Steinhardt}},
  \bibinfo{journal}{Physical Review Letters}  \bibinfo{volume}{84}
  (\bibinfo{year}{2000}) \bibinfo{pages}{3760--3763}.
\bibitem[{{Rocha} et~al.(2013){Rocha}, {Peter}, {Bullock}, {Kaplinghat},
  {Garrison-Kimmel}, {O{\~n}orbe}, and {Moustakas}}]{rocha2013}
\bibinfo{author}{M.~{Rocha}}, \bibinfo{author}{A.~H.~G. {Peter}},
  \bibinfo{author}{J.~S. {Bullock}}, \bibinfo{author}{M.~{Kaplinghat}},
  \bibinfo{author}{S.~{Garrison-Kimmel}}, \bibinfo{author}{J.~{O{\~n}orbe}},
  \bibinfo{author}{L.~A. {Moustakas}}, \bibinfo{journal}{\mnras}
  \bibinfo{volume}{430} (\bibinfo{year}{2013}) \bibinfo{pages}{81--104}.
\bibitem[{{Vogelsberger} et~al.(2014){Vogelsberger}, {Zavala}, {Simpson}, and
  {Jenkins}}]{vogelsberger2014_wdm}
\bibinfo{author}{M.~{Vogelsberger}}, \bibinfo{author}{J.~{Zavala}},
  \bibinfo{author}{C.~{Simpson}}, \bibinfo{author}{A.~{Jenkins}},
  \bibinfo{journal}{\mnras}  \bibinfo{volume}{444} (\bibinfo{year}{2014})
  \bibinfo{pages}{3684--3698}.
\bibitem[{{Wang} et~al.(2014){Wang}, {Peter}, {Strigari}, {Zentner}, {Arant},
  {Garrison-Kimmel}, and {Rocha}}]{wang2014}
\bibinfo{author}{M.-Y. {Wang}}, \bibinfo{author}{A.~H.~G. {Peter}},
  \bibinfo{author}{L.~E. {Strigari}}, \bibinfo{author}{A.~R. {Zentner}},
  \bibinfo{author}{B.~{Arant}}, \bibinfo{author}{S.~{Garrison-Kimmel}},
  \bibinfo{author}{M.~{Rocha}}, \bibinfo{journal}{\mnras}
  \bibinfo{volume}{445} (\bibinfo{year}{2014}) \bibinfo{pages}{614--629}.
\bibitem[{{B{\oe}hm} et~al.(2001){B{\oe}hm}, {Fayet}, and
  {Schaeffer}}]{boehm2001}
\bibinfo{author}{C.~{B{\oe}hm}}, \bibinfo{author}{P.~{Fayet}},
  \bibinfo{author}{R.~{Schaeffer}}, \bibinfo{journal}{Physics Letters B}
  \bibinfo{volume}{518} (\bibinfo{year}{2001}) \bibinfo{pages}{8--14}.
\bibitem[{{Dvorkin} et~al.(2014){Dvorkin}, {Blum}, and
  {Kamionkowski}}]{dvorkin2014}
\bibinfo{author}{C.~{Dvorkin}}, \bibinfo{author}{K.~{Blum}},
  \bibinfo{author}{M.~{Kamionkowski}}, \bibinfo{journal}{\prd}
  \bibinfo{volume}{89} (\bibinfo{year}{2014}) \bibinfo{pages}{023519}.
\bibitem[{{Jeans}(1902)}]{jeans1902}
\bibinfo{author}{J.~H. {Jeans}}, \bibinfo{journal}{Philosophical Transactions
  of the Royal Society of London Series A}  \bibinfo{volume}{199}
  (\bibinfo{year}{1902}) \bibinfo{pages}{1--53}.
\bibitem[{{Lifshitz}(1946)}]{lifshitz1946}
\bibinfo{author}{E.~M. {Lifshitz}}, \bibinfo{journal}{Zhurnal Eksperimentalnoi
  i Teoreticheskoi Fiziki}  \bibinfo{volume}{16} (\bibinfo{year}{1946})
  \bibinfo{pages}{587--602}.
\bibitem[{{Peebles}(1993)}]{peebles1993}
\bibinfo{author}{P.~J.~E. {Peebles}}, \bibinfo{title}{{Principles of Physical
  Cosmology}},  \bibinfo{year}{1993}.
\bibitem[{{Mo} et~al.(2010){Mo}, {van den Bosch}, and {White}}]{mo2010}
\bibinfo{author}{H.~{Mo}}, \bibinfo{author}{F.~C. {van den Bosch}},
  \bibinfo{author}{S.~{White}}, \bibinfo{title}{{Galaxy Formation and
  Evolution}},  \bibinfo{year}{2010}.
\bibitem[{{Peebles}(1980)}]{peebles1980}
\bibinfo{author}{P.~J.~E. {Peebles}}, \bibinfo{title}{{The large-scale
  structure of the universe}},  \bibinfo{year}{1980}.
\bibitem[{Barkana and Loeb(2001)}]{barkana-loeb2001}
\bibinfo{author}{R.~Barkana}, \bibinfo{author}{A.~Loeb},
  \bibinfo{journal}{Phys. Rept.}  \bibinfo{volume}{349} (\bibinfo{year}{2001})
  \bibinfo{pages}{125--238}.
\bibitem[{Ciardi and Ferrara(2005)}]{ciardi-ferrara2005}
\bibinfo{author}{B.~Ciardi}, \bibinfo{author}{A.~Ferrara},
  \bibinfo{journal}{Space Sci. Rev.}  \bibinfo{volume}{116}
  (\bibinfo{year}{2005}) \bibinfo{pages}{625--705}.
\bibitem[{{Bryan} and {Norman}(1998)}]{bryan1998}
\bibinfo{author}{G.~L. {Bryan}}, \bibinfo{author}{M.~L. {Norman}},
  \bibinfo{journal}{Astrophys. J.}  \bibinfo{volume}{495}
  (\bibinfo{year}{1998}) \bibinfo{pages}{80--99}.
\bibitem[{{Taylor}(2011)}]{taylor2011}
\bibinfo{author}{J.~E. {Taylor}}, \bibinfo{journal}{Advances in Astronomy}
  \bibinfo{volume}{2011} (\bibinfo{year}{2011}) \bibinfo{pages}{604898}.
\bibitem[{{Reed} et~al.(2007){Reed}, {Bower}, {Frenk}, {Jenkins}, and
  {Theuns}}]{reed2007}
\bibinfo{author}{D.~S. {Reed}}, \bibinfo{author}{R.~{Bower}},
  \bibinfo{author}{C.~S. {Frenk}}, \bibinfo{author}{A.~{Jenkins}},
  \bibinfo{author}{T.~{Theuns}}, \bibinfo{journal}{Monthly Notices of the Royal
  Astronomical Society}  \bibinfo{volume}{374} (\bibinfo{year}{2007})
  \bibinfo{pages}{2--15}.
\bibitem[{{Klypin} et~al.(2011){Klypin}, {Trujillo-Gomez}, and
  {Primack}}]{klypin2011}
\bibinfo{author}{A.~A. {Klypin}}, \bibinfo{author}{S.~{Trujillo-Gomez}},
  \bibinfo{author}{J.~{Primack}}, \bibinfo{journal}{\apj}
  \bibinfo{volume}{740} (\bibinfo{year}{2011}) \bibinfo{pages}{102}.
\bibitem[{{Carroll} et~al.(1992){Carroll}, {Press}, and {Turner}}]{carroll1992}
\bibinfo{author}{S.~M. {Carroll}}, \bibinfo{author}{W.~H. {Press}},
  \bibinfo{author}{E.~L. {Turner}}, \bibinfo{journal}{\araa}
  \bibinfo{volume}{30} (\bibinfo{year}{1992}) \bibinfo{pages}{499--542}.
\bibitem[{{Sheth} and {Tormen}(1999)}]{sheth-tormen1999}
\bibinfo{author}{R.~K. {Sheth}}, \bibinfo{author}{G.~{Tormen}},
  \bibinfo{journal}{Monthly Notices of the Royal Astronomical Society}
  \bibinfo{volume}{308} (\bibinfo{year}{1999}) \bibinfo{pages}{119--126}.
\bibitem[{{Sheth} and {Tormen}(2002)}]{sheth-tormen2002}
\bibinfo{author}{R.~K. {Sheth}}, \bibinfo{author}{G.~{Tormen}},
  \bibinfo{journal}{Monthly Notices of the Royal Astronomical Society}
  \bibinfo{volume}{329} (\bibinfo{year}{2002}) \bibinfo{pages}{61--75}.
\bibitem[{{Watson} et~al.(2013){Watson}, {Iliev}, {D'Aloisio}, and et.
  al.}]{watson2013}
\bibinfo{author}{W.~A. {Watson}}, \bibinfo{author}{I.~T. {Iliev}},
  \bibinfo{author}{A.~{D'Aloisio}}, \bibinfo{author}{et. al.},
  \bibinfo{journal}{\mnras}  \bibinfo{volume}{433} (\bibinfo{year}{2013})
  \bibinfo{pages}{1230--1245}.
\bibitem[{{Davis} et~al.(1985){Davis}, {Efstathiou}, {Frenk}, and
  {White}}]{davis1985}
\bibinfo{author}{M.~{Davis}}, \bibinfo{author}{G.~{Efstathiou}},
  \bibinfo{author}{C.~S. {Frenk}}, \bibinfo{author}{S.~D.~M. {White}},
  \bibinfo{journal}{\apj}  \bibinfo{volume}{292} (\bibinfo{year}{1985})
  \bibinfo{pages}{371--394}.
\bibitem[{{Klypin} and {Holtzman}(1997)}]{klypin1997}
\bibinfo{author}{A.~{Klypin}}, \bibinfo{author}{J.~{Holtzman}},
  \bibinfo{journal}{ArXiv Astrophysics e-prints}   (\bibinfo{year}{1997}).
\bibitem[{{Trac} et~al.(2015){Trac}, {Cen}, and {Mansfield}}]{trac2015}
\bibinfo{author}{H.~{Trac}}, \bibinfo{author}{R.~{Cen}},
  \bibinfo{author}{P.~{Mansfield}}, \bibinfo{journal}{\apj}
  \bibinfo{volume}{813} (\bibinfo{year}{2015}) \bibinfo{pages}{54}.
\bibitem[{{Allen} et~al.(2011){Allen}, {Evrard}, and {Mantz}}]{allen2011}
\bibinfo{author}{S.~W. {Allen}}, \bibinfo{author}{A.~E. {Evrard}},
  \bibinfo{author}{A.~B. {Mantz}}, \bibinfo{journal}{\araa}
  \bibinfo{volume}{49} (\bibinfo{year}{2011}) \bibinfo{pages}{409--470}.
\bibitem[{{Harrison} and {Coles}(2012)}]{harrison2012}
\bibinfo{author}{I.~{Harrison}}, \bibinfo{author}{P.~{Coles}},
  \bibinfo{journal}{\mnras}  \bibinfo{volume}{421} (\bibinfo{year}{2012})
  \bibinfo{pages}{L19--L23}.
\bibitem[{{Bertschinger}(1985)}]{bertschinger1985}
\bibinfo{author}{E.~{Bertschinger}}, \bibinfo{journal}{\apjs}
  \bibinfo{volume}{58} (\bibinfo{year}{1985}) \bibinfo{pages}{39--65}.
\bibitem[{{Navarro} et~al.(1996){Navarro}, {Frenk}, and {White}}]{navarro1996}
\bibinfo{author}{J.~F. {Navarro}}, \bibinfo{author}{C.~S. {Frenk}},
  \bibinfo{author}{S.~D.~M. {White}}, \bibinfo{journal}{\apj}
  \bibinfo{volume}{462} (\bibinfo{year}{1996}) \bibinfo{pages}{563}.
\bibitem[{{Ludlow} and {Angulo}(2017)}]{ludlow2017}
\bibinfo{author}{A.~D. {Ludlow}}, \bibinfo{author}{R.~E. {Angulo}},
  \bibinfo{journal}{\mnras}  \bibinfo{volume}{465} (\bibinfo{year}{2017})
  \bibinfo{pages}{L84--L88}.
\bibitem[{{Springel} et~al.(2008){Springel}, {Wang}, {Vogelsberger}, and et.
  al.}]{springel2008}
\bibinfo{author}{V.~{Springel}}, \bibinfo{author}{J.~{Wang}},
  \bibinfo{author}{M.~{Vogelsberger}}, \bibinfo{author}{et. al.},
  \bibinfo{journal}{\mnras}  \bibinfo{volume}{391} (\bibinfo{year}{2008})
  \bibinfo{pages}{1685--1711}.
\bibitem[{{Stadel} et~al.(2009){Stadel}, {Potter}, {Moore}, and et.
  al.}]{stadel2009}
\bibinfo{author}{J.~{Stadel}}, \bibinfo{author}{D.~{Potter}},
  \bibinfo{author}{B.~{Moore}}, \bibinfo{author}{et. al.},
  \bibinfo{journal}{\mnras}  \bibinfo{volume}{398} (\bibinfo{year}{2009})
  \bibinfo{pages}{L21--L25}.
\bibitem[{{Dutton} and {Macci{\`o}}(2014)}]{dutton2014}
\bibinfo{author}{A.~A. {Dutton}}, \bibinfo{author}{A.~V. {Macci{\`o}}},
  \bibinfo{journal}{\mnras}  \bibinfo{volume}{441} (\bibinfo{year}{2014})
  \bibinfo{pages}{3359--3374}.
\bibitem[{{Einasto}(1965)}]{einasto1965}
\bibinfo{author}{J.~{Einasto}}, \bibinfo{journal}{Trudy Astrofizicheskogo
  Instituta Alma-Ata}  \bibinfo{volume}{5} (\bibinfo{year}{1965})
  \bibinfo{pages}{87--100}.
\bibitem[{{Burkert}(1995)}]{burkert1995}
\bibinfo{author}{A.~{Burkert}}, \bibinfo{journal}{\apjl}  \bibinfo{volume}{447}
  (\bibinfo{year}{1995}) \bibinfo{pages}{L25}.
\bibitem[{{Salucci} and {Burkert}(2000)}]{salucci2000}
\bibinfo{author}{P.~{Salucci}}, \bibinfo{author}{A.~{Burkert}},
  \bibinfo{journal}{\apjl}  \bibinfo{volume}{537} (\bibinfo{year}{2000})
  \bibinfo{pages}{L9--L12}.
\bibitem[{{Kroupa} et~al.(1993){Kroupa}, {Tout}, and {Gilmore}}]{kroupa1993}
\bibinfo{author}{P.~{Kroupa}}, \bibinfo{author}{C.~A. {Tout}},
  \bibinfo{author}{G.~{Gilmore}}, \bibinfo{journal}{\mnras}
  \bibinfo{volume}{262} (\bibinfo{year}{1993}) \bibinfo{pages}{545--587}.
\bibitem[{{Chabrier}(2003)}]{chabrier2003}
\bibinfo{author}{G.~{Chabrier}}, \bibinfo{journal}{\apjl}
  \bibinfo{volume}{586} (\bibinfo{year}{2003}) \bibinfo{pages}{L133--L136}.
\bibitem[{{Gao} et~al.(2008){Gao}, {Navarro}, {Cole}, {Frenk}, {White},
  {Springel}, {Jenkins}, and {Neto}}]{gao2008}
\bibinfo{author}{L.~{Gao}}, \bibinfo{author}{J.~F. {Navarro}},
  \bibinfo{author}{S.~{Cole}}, \bibinfo{author}{C.~S. {Frenk}},
  \bibinfo{author}{S.~D.~M. {White}}, \bibinfo{author}{V.~{Springel}},
  \bibinfo{author}{A.~{Jenkins}}, \bibinfo{author}{A.~F. {Neto}},
  \bibinfo{journal}{\mnras}  \bibinfo{volume}{387} (\bibinfo{year}{2008})
  \bibinfo{pages}{536--544}.
\bibitem[{{Zhao} et~al.(2009){Zhao}, {Jing}, {Mo}, and {B{\"o}rner}}]{zhao2009}
\bibinfo{author}{D.~H. {Zhao}}, \bibinfo{author}{Y.~P. {Jing}},
  \bibinfo{author}{H.~J. {Mo}}, \bibinfo{author}{G.~{B{\"o}rner}},
  \bibinfo{journal}{\apj}  \bibinfo{volume}{707} (\bibinfo{year}{2009})
  \bibinfo{pages}{354--369}.
\bibitem[{{Ludlow} et~al.(2014){Ludlow}, {Navarro}, {Angulo}, {Boylan-Kolchin},
  {Springel}, {Frenk}, and {White}}]{ludlow2014}
\bibinfo{author}{A.~D. {Ludlow}}, \bibinfo{author}{J.~F. {Navarro}},
  \bibinfo{author}{R.~E. {Angulo}}, \bibinfo{author}{M.~{Boylan-Kolchin}},
  \bibinfo{author}{V.~{Springel}}, \bibinfo{author}{C.~{Frenk}},
  \bibinfo{author}{S.~D.~M. {White}}, \bibinfo{journal}{\mnras}
  \bibinfo{volume}{441} (\bibinfo{year}{2014}) \bibinfo{pages}{378--388}.
\bibitem[{{Chan} et~al.(2015){Chan}, {Kere{\v s}}, {O{\~n}orbe}, {Hopkins},
  {Muratov}, {Faucher-Gigu{\`e}re}, and {Quataert}}]{chan2015}
\bibinfo{author}{T.~K. {Chan}}, \bibinfo{author}{D.~{Kere{\v s}}},
  \bibinfo{author}{J.~{O{\~n}orbe}}, \bibinfo{author}{P.~F. {Hopkins}},
  \bibinfo{author}{A.~L. {Muratov}}, \bibinfo{author}{C.-A.
  {Faucher-Gigu{\`e}re}}, \bibinfo{author}{E.~{Quataert}},
  \bibinfo{journal}{\mnras}  \bibinfo{volume}{454} (\bibinfo{year}{2015})
  \bibinfo{pages}{2981--3001}.
\bibitem[{{Governato} et~al.(2010){Governato}, {Brook}, {Mayer}, and et.
  al.}]{governato2010}
\bibinfo{author}{F.~{Governato}}, \bibinfo{author}{C.~{Brook}},
  \bibinfo{author}{L.~{Mayer}}, \bibinfo{author}{et. al.},
  \bibinfo{journal}{\nat}  \bibinfo{volume}{463} (\bibinfo{year}{2010})
  \bibinfo{pages}{203--206}.
\bibitem[{{Governato} et~al.(2012){Governato}, {Zolotov}, {Pontzen}, and et.
  al.}]{governato2012}
\bibinfo{author}{F.~{Governato}}, \bibinfo{author}{A.~{Zolotov}},
  \bibinfo{author}{A.~{Pontzen}}, \bibinfo{author}{et. al.},
  \bibinfo{journal}{\mnras}  \bibinfo{volume}{422} (\bibinfo{year}{2012})
  \bibinfo{pages}{1231--1240}.
\bibitem[{{Di Cintio} et~al.(2014){Di Cintio}, {Brook}, {Dutton}, {Macci{\`o}},
  {Stinson}, and {Knebe}}]{dicintio2014}
\bibinfo{author}{A.~{Di Cintio}}, \bibinfo{author}{C.~B. {Brook}},
  \bibinfo{author}{A.~A. {Dutton}}, \bibinfo{author}{A.~V. {Macci{\`o}}},
  \bibinfo{author}{G.~S. {Stinson}}, \bibinfo{author}{A.~{Knebe}},
  \bibinfo{journal}{\mnras}  \bibinfo{volume}{441} (\bibinfo{year}{2014})
  \bibinfo{pages}{2986--2995}.
\bibitem[{{Pontzen} and {Governato}(2014)}]{pontzen2014}
\bibinfo{author}{A.~{Pontzen}}, \bibinfo{author}{F.~{Governato}},
  \bibinfo{journal}{\nat}  \bibinfo{volume}{506} (\bibinfo{year}{2014})
  \bibinfo{pages}{171--178}.
\bibitem[{{Ferrara}(2008)}]{Ferrara2008}
\bibinfo{author}{A.~{Ferrara}}, \bibinfo{title}{{Cosmological Feedbacks from
  the First Stars}}, \bibinfo{year}{2008},  pp. \bibinfo{pages}{161--258}.
\bibitem[{{Kroupa}(2001)}]{kroupa2001}
\bibinfo{author}{P.~{Kroupa}}, \bibinfo{journal}{\mnras}  \bibinfo{volume}{322}
  (\bibinfo{year}{2001}) \bibinfo{pages}{231--246}.
\bibitem[{{McKee} and {Ostriker}(2007)}]{McKee2007}
\bibinfo{author}{C.~F. {McKee}}, \bibinfo{author}{E.~C. {Ostriker}},
  \bibinfo{journal}{\araa}  \bibinfo{volume}{45} (\bibinfo{year}{2007})
  \bibinfo{pages}{565--687}.
\bibitem[{{Krumholz}(2015)}]{Krumholz2015}
\bibinfo{author}{M.~R. {Krumholz}}, \bibinfo{journal}{ArXiv e-prints}
  (\bibinfo{year}{2015}).
\bibitem[{{Spitzer}(1978)}]{Spitzer1978}
\bibinfo{author}{L.~{Spitzer}}, \bibinfo{title}{{Physical processes in the
  interstellar medium}},  \bibinfo{year}{1978}.
\bibitem[{{Ostriker} and {McKee}(1988)}]{Ostriker1988}
\bibinfo{author}{J.~P. {Ostriker}}, \bibinfo{author}{C.~F. {McKee}},
  \bibinfo{journal}{Reviews of Modern Physics}  \bibinfo{volume}{60}
  (\bibinfo{year}{1988}) \bibinfo{pages}{1--68}.
\bibitem[{{Veilleux} et~al.(2005){Veilleux}, {Cecil}, and
  {Bland-Hawthorn}}]{Veilleux2005}
\bibinfo{author}{S.~{Veilleux}}, \bibinfo{author}{G.~{Cecil}},
  \bibinfo{author}{J.~{Bland-Hawthorn}}, \bibinfo{journal}{\araa}
  \bibinfo{volume}{43} (\bibinfo{year}{2005}) \bibinfo{pages}{769--826}.
\bibitem[{{Conselice}(2014)}]{Conselice2014}
\bibinfo{author}{C.~J. {Conselice}}, \bibinfo{journal}{\araa}
  \bibinfo{volume}{52} (\bibinfo{year}{2014}) \bibinfo{pages}{291--337}.
\bibitem[{{Somerville} and {Dav{\'e}}(2015)}]{somerville2015}
\bibinfo{author}{R.~S. {Somerville}}, \bibinfo{author}{R.~{Dav{\'e}}},
  \bibinfo{journal}{\araa}  \bibinfo{volume}{53} (\bibinfo{year}{2015})
  \bibinfo{pages}{51--113}.
\bibitem[{{Behroozi} and {Silk}(2015)}]{behroozi2015}
\bibinfo{author}{P.~S. {Behroozi}}, \bibinfo{author}{J.~{Silk}},
  \bibinfo{journal}{\apj}  \bibinfo{volume}{799} (\bibinfo{year}{2015})
  \bibinfo{pages}{32}.
\bibitem[{{Mashian} et~al.(2016){Mashian}, {Oesch}, and {Loeb}}]{mashian2016}
\bibinfo{author}{N.~{Mashian}}, \bibinfo{author}{P.~A. {Oesch}},
  \bibinfo{author}{A.~{Loeb}}, \bibinfo{journal}{\mnras}  \bibinfo{volume}{455}
  (\bibinfo{year}{2016}) \bibinfo{pages}{2101--2109}.
\bibitem[{{Knebe} et~al.(2011){Knebe}, {Knollmann}, {Muldrew}, and et.
  al.}]{knebe2011}
\bibinfo{author}{A.~{Knebe}}, \bibinfo{author}{S.~R. {Knollmann}},
  \bibinfo{author}{S.~I. {Muldrew}}, \bibinfo{author}{et. al.},
  \bibinfo{journal}{\mnras}  \bibinfo{volume}{415} (\bibinfo{year}{2011})
  \bibinfo{pages}{2293--2318}.
\bibitem[{{Springel}(2010)}]{springel2010}
\bibinfo{author}{V.~{Springel}}, \bibinfo{journal}{\mnras}
  \bibinfo{volume}{401} (\bibinfo{year}{2010}) \bibinfo{pages}{791--851}.
\bibitem[{{Schaye} et~al.(2015){Schaye}, {Crain}, {Bower}, and et.
  al.}]{schaye2015}
\bibinfo{author}{J.~{Schaye}}, \bibinfo{author}{R.~A. {Crain}},
  \bibinfo{author}{R.~G. {Bower}}, \bibinfo{author}{et. al.},
  \bibinfo{journal}{\mnras}  \bibinfo{volume}{446} (\bibinfo{year}{2015})
  \bibinfo{pages}{521--554}.
\bibitem[{{Iliev} et~al.(2014){Iliev}, {Mellema}, {Ahn}, {Shapiro}, {Mao}, and
  {Pen}}]{iliev2014}
\bibinfo{author}{I.~T. {Iliev}}, \bibinfo{author}{G.~{Mellema}},
  \bibinfo{author}{K.~{Ahn}}, \bibinfo{author}{P.~R. {Shapiro}},
  \bibinfo{author}{Y.~{Mao}}, \bibinfo{author}{U.-L. {Pen}},
  \bibinfo{journal}{\mnras}  \bibinfo{volume}{439} (\bibinfo{year}{2014})
  \bibinfo{pages}{725--743}.
\bibitem[{{Trac} and {Gnedin}(2011)}]{trac2011}
\bibinfo{author}{H.~Y. {Trac}}, \bibinfo{author}{N.~Y. {Gnedin}},
  \bibinfo{journal}{Advanced Science Letters}  \bibinfo{volume}{4}
  (\bibinfo{year}{2011}) \bibinfo{pages}{228--243}.
\bibitem[{{Hopkins} et~al.(2014){Hopkins}, {Kere{\v s}}, {O{\~n}orbe},
  {Faucher-Gigu{\`e}re}, {Quataert}, {Murray}, and {Bullock}}]{hopkins2014}
\bibinfo{author}{P.~F. {Hopkins}}, \bibinfo{author}{D.~{Kere{\v s}}},
  \bibinfo{author}{J.~{O{\~n}orbe}}, \bibinfo{author}{C.-A.
  {Faucher-Gigu{\`e}re}}, \bibinfo{author}{E.~{Quataert}},
  \bibinfo{author}{N.~{Murray}}, \bibinfo{author}{J.~S. {Bullock}},
  \bibinfo{journal}{\mnras}  \bibinfo{volume}{445} (\bibinfo{year}{2014})
  \bibinfo{pages}{581--603}.
\bibitem[{{Wise} et~al.(2012){Wise}, {Abel}, {Turk}, {Norman}, and
  {Smith}}]{wise2012b}
\bibinfo{author}{J.~H. {Wise}}, \bibinfo{author}{T.~{Abel}},
  \bibinfo{author}{M.~J. {Turk}}, \bibinfo{author}{M.~L. {Norman}},
  \bibinfo{author}{B.~D. {Smith}}, \bibinfo{journal}{\mnras}
  \bibinfo{volume}{427} (\bibinfo{year}{2012}) \bibinfo{pages}{311--326}.
\bibitem[{{Aubert} et~al.(2015){Aubert}, {Deparis}, and {Ocvirk}}]{aubert2015}
\bibinfo{author}{D.~{Aubert}}, \bibinfo{author}{N.~{Deparis}},
  \bibinfo{author}{P.~{Ocvirk}}, \bibinfo{journal}{\mnras}
  \bibinfo{volume}{454} (\bibinfo{year}{2015}) \bibinfo{pages}{1012--1037}.
\bibitem[{{Hasegawa} and {Semelin}(2013)}]{hasegawa2013}
\bibinfo{author}{K.~{Hasegawa}}, \bibinfo{author}{B.~{Semelin}},
  \bibinfo{journal}{\mnras}  \bibinfo{volume}{428} (\bibinfo{year}{2013})
  \bibinfo{pages}{154--166}.
\bibitem[{{Finlator} et~al.(2011){Finlator}, {Dav{\'e}}, and
  {{\"O}zel}}]{finlator2011}
\bibinfo{author}{K.~{Finlator}}, \bibinfo{author}{R.~{Dav{\'e}}},
  \bibinfo{author}{F.~{{\"O}zel}}, \bibinfo{journal}{\apj}
  \bibinfo{volume}{743} (\bibinfo{year}{2011}) \bibinfo{pages}{169}.
\bibitem[{{Pallottini} et~al.(2017){Pallottini}, {Ferrara}, {Gallerani},
  {Vallini}, {Maiolino}, and {Salvadori}}]{pallottini2017}
\bibinfo{author}{A.~{Pallottini}}, \bibinfo{author}{A.~{Ferrara}},
  \bibinfo{author}{S.~{Gallerani}}, \bibinfo{author}{L.~{Vallini}},
  \bibinfo{author}{R.~{Maiolino}}, \bibinfo{author}{S.~{Salvadori}},
  \bibinfo{journal}{\mnras}  \bibinfo{volume}{465} (\bibinfo{year}{2017})
  \bibinfo{pages}{2540--2558}.
\bibitem[{{Petkova} and {Springel}(2011)}]{petkova2011}
\bibinfo{author}{M.~{Petkova}}, \bibinfo{author}{V.~{Springel}},
  \bibinfo{journal}{\mnras}  \bibinfo{volume}{412} (\bibinfo{year}{2011})
  \bibinfo{pages}{935--946}.
\bibitem[{{Norman} et~al.(2015){Norman}, {Reynolds}, {So}, {Harkness}, and
  {Wise}}]{norman2015}
\bibinfo{author}{M.~L. {Norman}}, \bibinfo{author}{D.~R. {Reynolds}},
  \bibinfo{author}{G.~C. {So}}, \bibinfo{author}{R.~P. {Harkness}},
  \bibinfo{author}{J.~H. {Wise}}, \bibinfo{journal}{\apjs}
  \bibinfo{volume}{216} (\bibinfo{year}{2015}) \bibinfo{pages}{16}.
\bibitem[{{O'Shea} et~al.(2015){O'Shea}, {Wise}, {Xu}, and
  {Norman}}]{oshea2015}
\bibinfo{author}{B.~W. {O'Shea}}, \bibinfo{author}{J.~H. {Wise}},
  \bibinfo{author}{H.~{Xu}}, \bibinfo{author}{M.~L. {Norman}},
  \bibinfo{journal}{\apjl}  \bibinfo{volume}{807} (\bibinfo{year}{2015})
  \bibinfo{pages}{L12}.
\bibitem[{{Xu} et~al.(2016){Xu}, {Wise}, {Norman}, {Ahn}, and
  {O'Shea}}]{xu2016}
\bibinfo{author}{H.~{Xu}}, \bibinfo{author}{J.~H. {Wise}},
  \bibinfo{author}{M.~L. {Norman}}, \bibinfo{author}{K.~{Ahn}},
  \bibinfo{author}{B.~W. {O'Shea}}, \bibinfo{journal}{\apj}
  \bibinfo{volume}{833} (\bibinfo{year}{2016}) \bibinfo{pages}{84}.
\bibitem[{{Gnedin} and {Kaurov}(2014)}]{gnedin2014}
\bibinfo{author}{N.~Y. {Gnedin}}, \bibinfo{author}{A.~A. {Kaurov}},
  \bibinfo{journal}{\apj}  \bibinfo{volume}{793} (\bibinfo{year}{2014})
  \bibinfo{pages}{30}.
\bibitem[{{Mutch} et~al.(2016){Mutch}, {Geil}, {Poole}, {Angel}, {Duffy},
  {Mesinger}, and {Wyithe}}]{mutch2016}
\bibinfo{author}{S.~J. {Mutch}}, \bibinfo{author}{P.~M. {Geil}},
  \bibinfo{author}{G.~B. {Poole}}, \bibinfo{author}{P.~W. {Angel}},
  \bibinfo{author}{A.~R. {Duffy}}, \bibinfo{author}{A.~{Mesinger}},
  \bibinfo{author}{J.~S.~B. {Wyithe}}, \bibinfo{journal}{\mnras}
  \bibinfo{volume}{462} (\bibinfo{year}{2016}) \bibinfo{pages}{250--276}.
\bibitem[{{Dayal} et~al.(2014){Dayal}, {Ferrara}, {Dunlop}, and
  {Pacucci}}]{dayal2014a}
\bibinfo{author}{P.~{Dayal}}, \bibinfo{author}{A.~{Ferrara}},
  \bibinfo{author}{J.~S. {Dunlop}}, \bibinfo{author}{F.~{Pacucci}},
  \bibinfo{journal}{\mnras}  \bibinfo{volume}{445} (\bibinfo{year}{2014})
  \bibinfo{pages}{2545--2557}.
\bibitem[{{Ocvirk} et~al.(2016){Ocvirk}, {Gillet}, {Shapiro}, and et.
  al.}]{ocvirk2016}
\bibinfo{author}{P.~{Ocvirk}}, \bibinfo{author}{N.~{Gillet}},
  \bibinfo{author}{P.~R. {Shapiro}}, \bibinfo{author}{et. al.},
  \bibinfo{journal}{\mnras}  \bibinfo{volume}{463} (\bibinfo{year}{2016})
  \bibinfo{pages}{1462--1485}.
\bibitem[{{Pawlik} et~al.(2017){Pawlik}, {Rahmati}, {Schaye}, {Jeon}, and
  {Dalla Vecchia}}]{pawlik2017}
\bibinfo{author}{A.~H. {Pawlik}}, \bibinfo{author}{A.~{Rahmati}},
  \bibinfo{author}{J.~{Schaye}}, \bibinfo{author}{M.~{Jeon}},
  \bibinfo{author}{C.~{Dalla Vecchia}}, \bibinfo{journal}{\mnras}
  \bibinfo{volume}{466} (\bibinfo{year}{2017}) \bibinfo{pages}{960--973}.
\bibitem[{{Yajima} et~al.(2011){Yajima}, {Choi}, and {Nagamine}}]{yajima2011}
\bibinfo{author}{H.~{Yajima}}, \bibinfo{author}{J.-H. {Choi}},
  \bibinfo{author}{K.~{Nagamine}}, \bibinfo{journal}{\mnras}
  \bibinfo{volume}{412} (\bibinfo{year}{2011}) \bibinfo{pages}{411--422}.
\bibitem[{{Vogelsberger} et~al.(2014){Vogelsberger}, {Genel}, {Springel}, and
  et. al.}]{vogelsberger2014}
\bibinfo{author}{M.~{Vogelsberger}}, \bibinfo{author}{S.~{Genel}},
  \bibinfo{author}{V.~{Springel}}, \bibinfo{author}{et. al.},
  \bibinfo{journal}{\mnras}  \bibinfo{volume}{444} (\bibinfo{year}{2014})
  \bibinfo{pages}{1518--1547}.
\bibitem[{{Feng} et~al.(2016){Feng}, {Di-Matteo}, {Croft}, {Bird}, {Battaglia},
  and {Wilkins}}]{feng2016}
\bibinfo{author}{Y.~{Feng}}, \bibinfo{author}{T.~{Di-Matteo}},
  \bibinfo{author}{R.~A. {Croft}}, \bibinfo{author}{S.~{Bird}},
  \bibinfo{author}{N.~{Battaglia}}, \bibinfo{author}{S.~{Wilkins}},
  \bibinfo{journal}{\mnras}  \bibinfo{volume}{455} (\bibinfo{year}{2016})
  \bibinfo{pages}{2778--2791}.
\bibitem[{{Lacey} et~al.(2016){Lacey}, {Baugh}, {Frenk}, and et.
  al.}]{lacey2016}
\bibinfo{author}{C.~G. {Lacey}}, \bibinfo{author}{C.~M. {Baugh}},
  \bibinfo{author}{C.~S. {Frenk}}, \bibinfo{author}{et. al.},
  \bibinfo{journal}{\mnras}  \bibinfo{volume}{462} (\bibinfo{year}{2016})
  \bibinfo{pages}{3854--3911}.
\bibitem[{{Springel} and {Hernquist}(2003)}]{springel2003}
\bibinfo{author}{V.~{Springel}}, \bibinfo{author}{L.~{Hernquist}},
  \bibinfo{journal}{\mnras}  \bibinfo{volume}{339} (\bibinfo{year}{2003})
  \bibinfo{pages}{289--311}.
\bibitem[{{Haardt} and {Madau}(2001)}]{haardt2001}
\bibinfo{author}{F.~{Haardt}}, \bibinfo{author}{P.~{Madau}},  in:
  \bibinfo{editor}{D.~M. {Neumann}}, \bibinfo{editor}{J.~T.~V. {Tran}} (Eds.),
  \bibinfo{booktitle}{Clusters of Galaxies and the High Redshift Universe
  Observed in X-rays}.
\bibitem[{{Faucher-Gigu{\`e}re} et~al.(2009){Faucher-Gigu{\`e}re}, {Lidz},
  {Zaldarriaga}, and {Hernquist}}]{faucher2009}
\bibinfo{author}{C.-A. {Faucher-Gigu{\`e}re}}, \bibinfo{author}{A.~{Lidz}},
  \bibinfo{author}{M.~{Zaldarriaga}}, \bibinfo{author}{L.~{Hernquist}},
  \bibinfo{journal}{\apj}  \bibinfo{volume}{703} (\bibinfo{year}{2009})
  \bibinfo{pages}{1416--1443}.
\bibitem[{{Bond}(1981)}]{bond1981}
\bibinfo{author}{H.~E. {Bond}}, \bibinfo{journal}{\apj}  \bibinfo{volume}{248}
  (\bibinfo{year}{1981}) \bibinfo{pages}{606--611}.
\bibitem[{{Yoshida} et~al.(2007){Yoshida}, {Omukai}, and
  {Hernquist}}]{yoshida2007}
\bibinfo{author}{N.~{Yoshida}}, \bibinfo{author}{K.~{Omukai}},
  \bibinfo{author}{L.~{Hernquist}}, \bibinfo{journal}{\apjl}
  \bibinfo{volume}{667} (\bibinfo{year}{2007}) \bibinfo{pages}{L117--L120}.
\bibitem[{{Bromm} and {Larson}(2004)}]{bromm-larson2004}
\bibinfo{author}{V.~{Bromm}}, \bibinfo{author}{R.~B. {Larson}},
  \bibinfo{journal}{\araa}  \bibinfo{volume}{42} (\bibinfo{year}{2004})
  \bibinfo{pages}{79--118}.
\bibitem[{{Glover}(2005)}]{glover2005}
\bibinfo{author}{S.~{Glover}}, \bibinfo{journal}{\ssr}  \bibinfo{volume}{117}
  (\bibinfo{year}{2005}) \bibinfo{pages}{445--508}.
\bibitem[{{Bromm} et~al.(2009){Bromm}, {Yoshida}, {Hernquist}, and
  {McKee}}]{bromm2009}
\bibinfo{author}{V.~{Bromm}}, \bibinfo{author}{N.~{Yoshida}},
  \bibinfo{author}{L.~{Hernquist}}, \bibinfo{author}{C.~F. {McKee}},
  \bibinfo{journal}{\nat}  \bibinfo{volume}{459} (\bibinfo{year}{2009})
  \bibinfo{pages}{49--54}.
\bibitem[{{Bromm} and {Yoshida}(2011)}]{bromm-yoshida2011}
\bibinfo{author}{V.~{Bromm}}, \bibinfo{author}{N.~{Yoshida}},
  \bibinfo{journal}{\araa}  \bibinfo{volume}{49} (\bibinfo{year}{2011})
  \bibinfo{pages}{373--407}.
\bibitem[{{Wise}(2012)}]{wise2012}
\bibinfo{author}{J.~H. {Wise}}, \bibinfo{journal}{ArXiv e-prints}
  (\bibinfo{year}{2012}).
\bibitem[{{Fall} and {Efstathiou}(1980)}]{fall1980}
\bibinfo{author}{S.~M. {Fall}}, \bibinfo{author}{G.~{Efstathiou}},
  \bibinfo{journal}{\mnras}  \bibinfo{volume}{193} (\bibinfo{year}{1980})
  \bibinfo{pages}{189--206}.
\bibitem[{{Fardal} et~al.(2001){Fardal}, {Katz}, {Gardner}, {Hernquist},
  {Weinberg}, and {Dave}}]{Fardal2001}
\bibinfo{author}{M.~A. {Fardal}}, \bibinfo{author}{N.~{Katz}},
  \bibinfo{author}{J.~P. {Gardner}}, \bibinfo{author}{L.~{Hernquist}},
  \bibinfo{author}{D.~H. {Weinberg}}, \bibinfo{author}{R.~{Dave}},
  \bibinfo{journal}{\apj}  \bibinfo{volume}{562} (\bibinfo{year}{2001})
  \bibinfo{pages}{605--617}.
\bibitem[{{Birnboim} and {Dekel}(2003)}]{birnboim2003}
\bibinfo{author}{Y.~{Birnboim}}, \bibinfo{author}{A.~{Dekel}},
  \bibinfo{journal}{\mnras}  \bibinfo{volume}{345} (\bibinfo{year}{2003})
  \bibinfo{pages}{349--364}.
\bibitem[{{Kere{\v s}} et~al.(2005){Kere{\v s}}, {Katz}, {Weinberg}, and
  {Dav{\'e}}}]{keres2005}
\bibinfo{author}{D.~{Kere{\v s}}}, \bibinfo{author}{N.~{Katz}},
  \bibinfo{author}{D.~H. {Weinberg}}, \bibinfo{author}{R.~{Dav{\'e}}},
  \bibinfo{journal}{\mnras}  \bibinfo{volume}{363} (\bibinfo{year}{2005})
  \bibinfo{pages}{2--28}.
\bibitem[{{Ocvirk} et~al.(2008){Ocvirk}, {Pichon}, and {Teyssier}}]{ocvirk2008}
\bibinfo{author}{P.~{Ocvirk}}, \bibinfo{author}{C.~{Pichon}},
  \bibinfo{author}{R.~{Teyssier}}, \bibinfo{journal}{\mnras}
  \bibinfo{volume}{390} (\bibinfo{year}{2008}) \bibinfo{pages}{1326--1338}.
\bibitem[{{Dekel} and {Birnboim}(2006)}]{dekel2006}
\bibinfo{author}{A.~{Dekel}}, \bibinfo{author}{Y.~{Birnboim}},
  \bibinfo{journal}{\mnras}  \bibinfo{volume}{368} (\bibinfo{year}{2006})
  \bibinfo{pages}{2--20}.
\bibitem[{{Dekel} et~al.(2009){Dekel}, {Birnboim}, {Engel}, {Freundlich},
  {Goerdt}, {Mumcuoglu}, {Neistein}, {Pichon}, {Teyssier}, and
  {Zinger}}]{dekel2009}
\bibinfo{author}{A.~{Dekel}}, \bibinfo{author}{Y.~{Birnboim}},
  \bibinfo{author}{G.~{Engel}}, \bibinfo{author}{J.~{Freundlich}},
  \bibinfo{author}{T.~{Goerdt}}, \bibinfo{author}{M.~{Mumcuoglu}},
  \bibinfo{author}{E.~{Neistein}}, \bibinfo{author}{C.~{Pichon}},
  \bibinfo{author}{R.~{Teyssier}}, \bibinfo{author}{E.~{Zinger}},
  \bibinfo{journal}{Nature}  \bibinfo{volume}{457} (\bibinfo{year}{2009})
  \bibinfo{pages}{451--454}.
\bibitem[{{Kere{\v s}} et~al.(2009){Kere{\v s}}, {Katz}, {Fardal}, {Dav{\'e}},
  and {Weinberg}}]{keres2009}
\bibinfo{author}{D.~{Kere{\v s}}}, \bibinfo{author}{N.~{Katz}},
  \bibinfo{author}{M.~{Fardal}}, \bibinfo{author}{R.~{Dav{\'e}}},
  \bibinfo{author}{D.~H. {Weinberg}}, \bibinfo{journal}{\mnras}
  \bibinfo{volume}{395} (\bibinfo{year}{2009}) \bibinfo{pages}{160--179}.
\bibitem[{{van de Voort} et~al.(2011){van de Voort}, {Schaye}, {Booth}, {Haas},
  and {Dalla Vecchia}}]{vandevoort2011}
\bibinfo{author}{F.~{van de Voort}}, \bibinfo{author}{J.~{Schaye}},
  \bibinfo{author}{C.~M. {Booth}}, \bibinfo{author}{M.~R. {Haas}},
  \bibinfo{author}{C.~{Dalla Vecchia}}, \bibinfo{journal}{\mnras}
  \bibinfo{volume}{414} (\bibinfo{year}{2011}) \bibinfo{pages}{2458--2478}.
\bibitem[{{van de Voort} et~al.(2012){van de Voort}, {Schaye}, {Altay}, and
  {Theuns}}]{vandevoort2012}
\bibinfo{author}{F.~{van de Voort}}, \bibinfo{author}{J.~{Schaye}},
  \bibinfo{author}{G.~{Altay}}, \bibinfo{author}{T.~{Theuns}},
  \bibinfo{journal}{\mnras}  \bibinfo{volume}{421} (\bibinfo{year}{2012})
  \bibinfo{pages}{2809--2819}.
\bibitem[{{Dekel} et~al.(2013){Dekel}, {Zolotov}, {Tweed}, {Cacciato},
  {Ceverino}, and {Primack}}]{dekel2013}
\bibinfo{author}{A.~{Dekel}}, \bibinfo{author}{A.~{Zolotov}},
  \bibinfo{author}{D.~{Tweed}}, \bibinfo{author}{M.~{Cacciato}},
  \bibinfo{author}{D.~{Ceverino}}, \bibinfo{author}{J.~R. {Primack}},
  \bibinfo{journal}{\mnras}  \bibinfo{volume}{435} (\bibinfo{year}{2013})
  \bibinfo{pages}{999--1019}.
\bibitem[{{Benson} and {Bower}(2011)}]{benson2011}
\bibinfo{author}{A.~J. {Benson}}, \bibinfo{author}{R.~{Bower}},
  \bibinfo{journal}{\mnras}  \bibinfo{volume}{410} (\bibinfo{year}{2011})
  \bibinfo{pages}{2653--2661}.
\bibitem[{{Brooks} et~al.(2009){Brooks}, {Governato}, {Quinn}, {Brook}, and
  {Wadsley}}]{brooks2009}
\bibinfo{author}{A.~M. {Brooks}}, \bibinfo{author}{F.~{Governato}},
  \bibinfo{author}{T.~{Quinn}}, \bibinfo{author}{C.~B. {Brook}},
  \bibinfo{author}{J.~{Wadsley}}, \bibinfo{journal}{\apj}
  \bibinfo{volume}{694} (\bibinfo{year}{2009}) \bibinfo{pages}{396--410}.
\bibitem[{{Kauffmann} et~al.(2003){Kauffmann}, {Heckman}, {White}, and et.
  al.}]{kauffmann2003b}
\bibinfo{author}{G.~{Kauffmann}}, \bibinfo{author}{T.~M. {Heckman}},
  \bibinfo{author}{S.~D.~M. {White}}, \bibinfo{author}{et. al.},
  \bibinfo{journal}{\mnras}  \bibinfo{volume}{341} (\bibinfo{year}{2003})
  \bibinfo{pages}{54--69}.
\bibitem[{{Faucher-Gigu{\`e}re} and {Kere{\v s}}(2011)}]{faucher2011}
\bibinfo{author}{C.-A. {Faucher-Gigu{\`e}re}}, \bibinfo{author}{D.~{Kere{\v
  s}}}, \bibinfo{journal}{\mnras}  \bibinfo{volume}{412} (\bibinfo{year}{2011})
  \bibinfo{pages}{L118--L122}.
\bibitem[{{Goerdt} et~al.(2012){Goerdt}, {Dekel}, {Sternberg}, {Gnat}, and
  {Ceverino}}]{goerdt2012}
\bibinfo{author}{T.~{Goerdt}}, \bibinfo{author}{A.~{Dekel}},
  \bibinfo{author}{A.~{Sternberg}}, \bibinfo{author}{O.~{Gnat}},
  \bibinfo{author}{D.~{Ceverino}}, \bibinfo{journal}{\mnras}
  \bibinfo{volume}{424} (\bibinfo{year}{2012}) \bibinfo{pages}{2292--2315}.
\bibitem[{{Rubin} et~al.(2012){Rubin}, {Prochaska}, {Koo}, and
  {Phillips}}]{rubin2012}
\bibinfo{author}{K.~H.~R. {Rubin}}, \bibinfo{author}{J.~X. {Prochaska}},
  \bibinfo{author}{D.~C. {Koo}}, \bibinfo{author}{A.~C. {Phillips}},
  \bibinfo{journal}{\apjl}  \bibinfo{volume}{747} (\bibinfo{year}{2012})
  \bibinfo{pages}{L26}.
\bibitem[{{Rakic} et~al.(2012){Rakic}, {Schaye}, {Steidel}, and
  {Rudie}}]{rakic2012}
\bibinfo{author}{O.~{Rakic}}, \bibinfo{author}{J.~{Schaye}},
  \bibinfo{author}{C.~C. {Steidel}}, \bibinfo{author}{G.~C. {Rudie}},
  \bibinfo{journal}{\apj}  \bibinfo{volume}{751} (\bibinfo{year}{2012})
  \bibinfo{pages}{94}.
\bibitem[{{Rauch} et~al.(2011){Rauch}, {Becker}, {Haehnelt}, {Gauthier},
  {Ravindranath}, and {Sargent}}]{rauch2011}
\bibinfo{author}{M.~{Rauch}}, \bibinfo{author}{G.~D. {Becker}},
  \bibinfo{author}{M.~G. {Haehnelt}}, \bibinfo{author}{J.-R. {Gauthier}},
  \bibinfo{author}{S.~{Ravindranath}}, \bibinfo{author}{W.~L.~W. {Sargent}},
  \bibinfo{journal}{\mnras}  \bibinfo{volume}{418} (\bibinfo{year}{2011})
  \bibinfo{pages}{1115--1126}.
\bibitem[{{Rauch} et~al.(2013){Rauch}, {Becker}, {Haehnelt}, {Gauthier}, and
  {Sargent}}]{rauch2013}
\bibinfo{author}{M.~{Rauch}}, \bibinfo{author}{G.~D. {Becker}},
  \bibinfo{author}{M.~G. {Haehnelt}}, \bibinfo{author}{J.-R. {Gauthier}},
  \bibinfo{author}{W.~L.~W. {Sargent}}, \bibinfo{journal}{\mnras}
  \bibinfo{volume}{429} (\bibinfo{year}{2013}) \bibinfo{pages}{429--443}.
\bibitem[{{Martin} et~al.(2015){Martin}, {Matuszewski}, {Morrissey}, {Neill},
  {Moore}, {Cantalupo}, {Prochaska}, and {Chang}}]{martin2015}
\bibinfo{author}{D.~C. {Martin}}, \bibinfo{author}{M.~{Matuszewski}},
  \bibinfo{author}{P.~{Morrissey}}, \bibinfo{author}{J.~D. {Neill}},
  \bibinfo{author}{A.~{Moore}}, \bibinfo{author}{S.~{Cantalupo}},
  \bibinfo{author}{J.~X. {Prochaska}}, \bibinfo{author}{D.~{Chang}},
  \bibinfo{journal}{\nat}  \bibinfo{volume}{524} (\bibinfo{year}{2015})
  \bibinfo{pages}{192--195}.
\bibitem[{{Martin} et~al.(2016){Martin}, {Matuszewski}, {Morrissey}, {Neill},
  {Moore}, {Steidel}, and {Trainor}}]{martin2016}
\bibinfo{author}{D.~C. {Martin}}, \bibinfo{author}{M.~{Matuszewski}},
  \bibinfo{author}{P.~{Morrissey}}, \bibinfo{author}{J.~D. {Neill}},
  \bibinfo{author}{A.~{Moore}}, \bibinfo{author}{C.~C. {Steidel}},
  \bibinfo{author}{R.~{Trainor}}, \bibinfo{journal}{\apjl}
  \bibinfo{volume}{824} (\bibinfo{year}{2016}) \bibinfo{pages}{L5}.
\bibitem[{{Steidel} et~al.(2000){Steidel}, {Adelberger}, {Shapley}, {Pettini},
  {Dickinson}, and {Giavalisco}}]{steidel2000}
\bibinfo{author}{C.~C. {Steidel}}, \bibinfo{author}{K.~L. {Adelberger}},
  \bibinfo{author}{A.~E. {Shapley}}, \bibinfo{author}{M.~{Pettini}},
  \bibinfo{author}{M.~{Dickinson}}, \bibinfo{author}{M.~{Giavalisco}},
  \bibinfo{journal}{\apj}  \bibinfo{volume}{532} (\bibinfo{year}{2000})
  \bibinfo{pages}{170--182}.
\bibitem[{{Dijkstra} et~al.(2006){Dijkstra}, {Haiman}, and
  {Spaans}}]{dijkstra2006}
\bibinfo{author}{M.~{Dijkstra}}, \bibinfo{author}{Z.~{Haiman}},
  \bibinfo{author}{M.~{Spaans}}, \bibinfo{journal}{\apj}  \bibinfo{volume}{649}
  (\bibinfo{year}{2006}) \bibinfo{pages}{37--47}.
\bibitem[{{Dijkstra} and {Loeb}(2009)}]{dijkstra2009}
\bibinfo{author}{M.~{Dijkstra}}, \bibinfo{author}{A.~{Loeb}},
  \bibinfo{journal}{\mnras}  \bibinfo{volume}{400} (\bibinfo{year}{2009})
  \bibinfo{pages}{1109--1120}.
\bibitem[{{Goerdt} et~al.(2010){Goerdt}, {Dekel}, {Sternberg}, {Ceverino},
  {Teyssier}, and {Primack}}]{goerdt2010}
\bibinfo{author}{T.~{Goerdt}}, \bibinfo{author}{A.~{Dekel}},
  \bibinfo{author}{A.~{Sternberg}}, \bibinfo{author}{D.~{Ceverino}},
  \bibinfo{author}{R.~{Teyssier}}, \bibinfo{author}{J.~R. {Primack}},
  \bibinfo{journal}{\mnras}  \bibinfo{volume}{407} (\bibinfo{year}{2010})
  \bibinfo{pages}{613--631}.
\bibitem[{{Rosdahl} and {Blaizot}(2012)}]{rosdahl2012}
\bibinfo{author}{J.~{Rosdahl}}, \bibinfo{author}{J.~{Blaizot}},
  \bibinfo{journal}{\mnras}  \bibinfo{volume}{423} (\bibinfo{year}{2012})
  \bibinfo{pages}{344--366}.
\bibitem[{{Nilsson} et~al.(2006){Nilsson}, {Fynbo}, {M{\o}ller},
  {Sommer-Larsen}, and {Ledoux}}]{nilsson2006}
\bibinfo{author}{K.~K. {Nilsson}}, \bibinfo{author}{J.~P.~U. {Fynbo}},
  \bibinfo{author}{P.~{M{\o}ller}}, \bibinfo{author}{J.~{Sommer-Larsen}},
  \bibinfo{author}{C.~{Ledoux}}, \bibinfo{journal}{\aap}  \bibinfo{volume}{452}
  (\bibinfo{year}{2006}) \bibinfo{pages}{L23--L26}.
\bibitem[{{Smith} and {Jarvis}(2007)}]{smith2007}
\bibinfo{author}{D.~J.~B. {Smith}}, \bibinfo{author}{M.~J. {Jarvis}},
  \bibinfo{journal}{\mnras}  \bibinfo{volume}{378} (\bibinfo{year}{2007})
  \bibinfo{pages}{L49--L53}.
\bibitem[{{Smith} et~al.(2008){Smith}, {Jarvis}, {Lacy}, and
  {Mart{\'{\i}}nez-Sansigre}}]{smith2008}
\bibinfo{author}{D.~J.~B. {Smith}}, \bibinfo{author}{M.~J. {Jarvis}},
  \bibinfo{author}{M.~{Lacy}}, \bibinfo{author}{A.~{Mart{\'{\i}}nez-Sansigre}},
  \bibinfo{journal}{\mnras}  \bibinfo{volume}{389} (\bibinfo{year}{2008})
  \bibinfo{pages}{799--805}.
\bibitem[{{Erb} et~al.(2011){Erb}, {Bogosavljevi{\'c}}, and
  {Steidel}}]{erb2011}
\bibinfo{author}{D.~K. {Erb}}, \bibinfo{author}{M.~{Bogosavljevi{\'c}}},
  \bibinfo{author}{C.~C. {Steidel}}, \bibinfo{journal}{\apjl}
  \bibinfo{volume}{740} (\bibinfo{year}{2011}) \bibinfo{pages}{L31}.
\bibitem[{{Cantalupo}(2017)}]{cantalupo2017}
\bibinfo{author}{S.~{Cantalupo}}, in: \bibinfo{editor}{A.~{Fox}},
  \bibinfo{editor}{R.~{Dav{\'e}}} (Eds.), \bibinfo{booktitle}{Gas Accretion
  onto Galaxies}, volume \bibinfo{volume}{430} of
  \textit{\bibinfo{series}{Astrophysics and Space Science Library}},  p.
  \bibinfo{pages}{195}.
\bibitem[{{Fumagalli} et~al.(2011){Fumagalli}, {Prochaska}, {Kasen}, {Dekel},
  {Ceverino}, and {Primack}}]{fumagalli2011}
\bibinfo{author}{M.~{Fumagalli}}, \bibinfo{author}{J.~X. {Prochaska}},
  \bibinfo{author}{D.~{Kasen}}, \bibinfo{author}{A.~{Dekel}},
  \bibinfo{author}{D.~{Ceverino}}, \bibinfo{author}{J.~R. {Primack}},
  \bibinfo{journal}{\mnras}  \bibinfo{volume}{418} (\bibinfo{year}{2011})
  \bibinfo{pages}{1796--1821}.
\bibitem[{{Fumagalli} et~al.(2016){Fumagalli}, {O'Meara}, and
  {Prochaska}}]{fumagalli2016b}
\bibinfo{author}{M.~{Fumagalli}}, \bibinfo{author}{J.~M. {O'Meara}},
  \bibinfo{author}{J.~X. {Prochaska}}, \bibinfo{journal}{\mnras}
  \bibinfo{volume}{455} (\bibinfo{year}{2016}) \bibinfo{pages}{4100--4121}.
\bibitem[{{Lehner} et~al.(2016){Lehner}, {O'Meara}, {Howk}, {Prochaska}, and
  {Fumagalli}}]{lehner2016}
\bibinfo{author}{N.~{Lehner}}, \bibinfo{author}{J.~M. {O'Meara}},
  \bibinfo{author}{J.~C. {Howk}}, \bibinfo{author}{J.~X. {Prochaska}},
  \bibinfo{author}{M.~{Fumagalli}}, \bibinfo{journal}{ArXiv e-prints}
  (\bibinfo{year}{2016}).
\bibitem[{{Ribaudo} et~al.(2011){Ribaudo}, {Lehner}, {Howk}, {Werk}, {Tripp},
  {Prochaska}, {Meiring}, and {Tumlinson}}]{ribaudo2011}
\bibinfo{author}{J.~{Ribaudo}}, \bibinfo{author}{N.~{Lehner}},
  \bibinfo{author}{J.~C. {Howk}}, \bibinfo{author}{J.~K. {Werk}},
  \bibinfo{author}{T.~M. {Tripp}}, \bibinfo{author}{J.~X. {Prochaska}},
  \bibinfo{author}{J.~D. {Meiring}}, \bibinfo{author}{J.~{Tumlinson}},
  \bibinfo{journal}{\apj}  \bibinfo{volume}{743} (\bibinfo{year}{2011})
  \bibinfo{pages}{207}.
\bibitem[{{Crighton} et~al.(2013){Crighton}, {Hennawi}, and
  {Prochaska}}]{crighton2013}
\bibinfo{author}{N.~H.~M. {Crighton}}, \bibinfo{author}{J.~F. {Hennawi}},
  \bibinfo{author}{J.~X. {Prochaska}}, \bibinfo{journal}{\apjl}
  \bibinfo{volume}{776} (\bibinfo{year}{2013}) \bibinfo{pages}{L18}.
\bibitem[{{Fumagalli} et~al.(2016){Fumagalli}, {Cantalupo}, {Dekel}, {Morris},
  {O'Meara}, {Prochaska}, and {Theuns}}]{fumagalli2016a}
\bibinfo{author}{M.~{Fumagalli}}, \bibinfo{author}{S.~{Cantalupo}},
  \bibinfo{author}{A.~{Dekel}}, \bibinfo{author}{S.~L. {Morris}},
  \bibinfo{author}{J.~M. {O'Meara}}, \bibinfo{author}{J.~X. {Prochaska}},
  \bibinfo{author}{T.~{Theuns}}, \bibinfo{journal}{\mnras}
  \bibinfo{volume}{462} (\bibinfo{year}{2016}) \bibinfo{pages}{1978--1988}.
\bibitem[{{Giavalisco} et~al.(2011){Giavalisco}, {Vanzella}, {Salimbeni}, and
  et. al.}]{giavalisco2011}
\bibinfo{author}{M.~{Giavalisco}}, \bibinfo{author}{E.~{Vanzella}},
  \bibinfo{author}{S.~{Salimbeni}}, \bibinfo{author}{et. al.},
  \bibinfo{journal}{\apj}  \bibinfo{volume}{743} (\bibinfo{year}{2011})
  \bibinfo{pages}{95}.
\bibitem[{{Martin} et~al.(2012){Martin}, {Shapley}, {Coil}, {Kornei}, {Bundy},
  {Weiner}, {Noeske}, and {Schiminovich}}]{martin2012}
\bibinfo{author}{C.~L. {Martin}}, \bibinfo{author}{A.~E. {Shapley}},
  \bibinfo{author}{A.~L. {Coil}}, \bibinfo{author}{K.~A. {Kornei}},
  \bibinfo{author}{K.~{Bundy}}, \bibinfo{author}{B.~J. {Weiner}},
  \bibinfo{author}{K.~G. {Noeske}}, \bibinfo{author}{D.~{Schiminovich}},
  \bibinfo{journal}{\apj}  \bibinfo{volume}{760} (\bibinfo{year}{2012})
  \bibinfo{pages}{127}.
\bibitem[{{Tegmark} et~al.(1997){Tegmark}, {Silk}, {Rees}, {Blanchard}, {Abel},
  and {Palla}}]{tegmark1997}
\bibinfo{author}{M.~{Tegmark}}, \bibinfo{author}{J.~{Silk}},
  \bibinfo{author}{M.~J. {Rees}}, \bibinfo{author}{A.~{Blanchard}},
  \bibinfo{author}{T.~{Abel}}, \bibinfo{author}{F.~{Palla}},
  \bibinfo{journal}{\apj}  \bibinfo{volume}{474} (\bibinfo{year}{1997})
  \bibinfo{pages}{1}.
\bibitem[{{Saslaw} and {Zipoy}(1967)}]{saslaw1967}
\bibinfo{author}{W.~C. {Saslaw}}, \bibinfo{author}{D.~{Zipoy}},
  \bibinfo{journal}{\nat}  \bibinfo{volume}{216} (\bibinfo{year}{1967})
  \bibinfo{pages}{976--978}.
\bibitem[{{McDowell}(1961)}]{mcdowell1961}
\bibinfo{author}{M.~R.~C. {McDowell}}, \bibinfo{journal}{The Observatory}
  \bibinfo{volume}{81} (\bibinfo{year}{1961}) \bibinfo{pages}{240--243}.
\bibitem[{{Seager} et~al.(2000){Seager}, {Sasselov}, and {Scott}}]{seager2000}
\bibinfo{author}{S.~{Seager}}, \bibinfo{author}{D.~D. {Sasselov}},
  \bibinfo{author}{D.~{Scott}}, \bibinfo{journal}{\apjs}  \bibinfo{volume}{128}
  (\bibinfo{year}{2000}) \bibinfo{pages}{407--430}.
\bibitem[{{Mac Low} and {Shull}(1986)}]{maclow1986}
\bibinfo{author}{M.-M. {Mac Low}}, \bibinfo{author}{J.~M. {Shull}},
  \bibinfo{journal}{\apj}  \bibinfo{volume}{302} (\bibinfo{year}{1986})
  \bibinfo{pages}{585--589}.
\bibitem[{{Abel} et~al.(2000){Abel}, {Bryan}, and {Norman}}]{abel2000}
\bibinfo{author}{T.~{Abel}}, \bibinfo{author}{G.~L. {Bryan}},
  \bibinfo{author}{M.~L. {Norman}}, \bibinfo{journal}{\apj}
  \bibinfo{volume}{540} (\bibinfo{year}{2000}) \bibinfo{pages}{39--44}.
\bibitem[{{Abel} et~al.(2002){Abel}, {Bryan}, and {Norman}}]{abel2002}
\bibinfo{author}{T.~{Abel}}, \bibinfo{author}{G.~L. {Bryan}},
  \bibinfo{author}{M.~L. {Norman}}, \bibinfo{journal}{Science}
  \bibinfo{volume}{295} (\bibinfo{year}{2002}) \bibinfo{pages}{93--98}.
\bibitem[{{Bromm} et~al.(1999){Bromm}, {Coppi}, and {Larson}}]{bromm1999}
\bibinfo{author}{V.~{Bromm}}, \bibinfo{author}{P.~S. {Coppi}},
  \bibinfo{author}{R.~B. {Larson}}, \bibinfo{journal}{\apjl}
  \bibinfo{volume}{527} (\bibinfo{year}{1999}) \bibinfo{pages}{L5--L8}.
\bibitem[{{Bromm} et~al.(2002){Bromm}, {Coppi}, and {Larson}}]{bromm2002}
\bibinfo{author}{V.~{Bromm}}, \bibinfo{author}{P.~S. {Coppi}},
  \bibinfo{author}{R.~B. {Larson}}, \bibinfo{journal}{\apj}
  \bibinfo{volume}{564} (\bibinfo{year}{2002}) \bibinfo{pages}{23--51}.
\bibitem[{{Ebert}(1955)}]{ebert1955}
\bibinfo{author}{R.~{Ebert}}, \bibinfo{journal}{\zap}  \bibinfo{volume}{37}
  (\bibinfo{year}{1955}) \bibinfo{pages}{217}.
\bibitem[{{Bonnor}(1956)}]{bonnor1956}
\bibinfo{author}{W.~B. {Bonnor}}, \bibinfo{journal}{\mnras}
  \bibinfo{volume}{116} (\bibinfo{year}{1956}) \bibinfo{pages}{351}.
\bibitem[{{Palla} et~al.(1983){Palla}, {Salpeter}, and {Stahler}}]{palla1983}
\bibinfo{author}{F.~{Palla}}, \bibinfo{author}{E.~E. {Salpeter}},
  \bibinfo{author}{S.~W. {Stahler}}, \bibinfo{journal}{\apj}
  \bibinfo{volume}{271} (\bibinfo{year}{1983}) \bibinfo{pages}{632--641}.
\bibitem[{{Yoshida} et~al.(2006){Yoshida}, {Omukai}, {Hernquist}, and
  {Abel}}]{yoshida2006}
\bibinfo{author}{N.~{Yoshida}}, \bibinfo{author}{K.~{Omukai}},
  \bibinfo{author}{L.~{Hernquist}}, \bibinfo{author}{T.~{Abel}},
  \bibinfo{journal}{\apj}  \bibinfo{volume}{652} (\bibinfo{year}{2006})
  \bibinfo{pages}{6--25}.
\bibitem[{{Yoshida} et~al.(2008){Yoshida}, {Omukai}, and
  {Hernquist}}]{yoshida2008}
\bibinfo{author}{N.~{Yoshida}}, \bibinfo{author}{K.~{Omukai}},
  \bibinfo{author}{L.~{Hernquist}}, \bibinfo{journal}{Science}
  \bibinfo{volume}{321} (\bibinfo{year}{2008}) \bibinfo{pages}{669}.
\bibitem[{{McKee} and {Tan}(2008)}]{mckee2008}
\bibinfo{author}{C.~F. {McKee}}, \bibinfo{author}{J.~C. {Tan}},
  \bibinfo{journal}{\apj}  \bibinfo{volume}{681} (\bibinfo{year}{2008})
  \bibinfo{pages}{771--797}.
\bibitem[{{Clark} et~al.(2011){Clark}, {Glover}, {Smith}, {Greif}, {Klessen},
  and {Bromm}}]{clark2011b}
\bibinfo{author}{P.~C. {Clark}}, \bibinfo{author}{S.~C.~O. {Glover}},
  \bibinfo{author}{R.~J. {Smith}}, \bibinfo{author}{T.~H. {Greif}},
  \bibinfo{author}{R.~S. {Klessen}}, \bibinfo{author}{V.~{Bromm}},
  \bibinfo{journal}{Science}  \bibinfo{volume}{331} (\bibinfo{year}{2011})
  \bibinfo{pages}{1040}.
\bibitem[{{Bromm} and {Loeb}(2004)}]{bromm2004b}
\bibinfo{author}{V.~{Bromm}}, \bibinfo{author}{A.~{Loeb}},
  \bibinfo{journal}{\na}  \bibinfo{volume}{9} (\bibinfo{year}{2004})
  \bibinfo{pages}{353--364}.
\bibitem[{{Bate} et~al.(1995){Bate}, {Bonnell}, and {Price}}]{bate1995}
\bibinfo{author}{M.~R. {Bate}}, \bibinfo{author}{I.~A. {Bonnell}},
  \bibinfo{author}{N.~M. {Price}}, \bibinfo{journal}{\mnras}
  \bibinfo{volume}{277} (\bibinfo{year}{1995}) \bibinfo{pages}{362--376}.
\bibitem[{{Stacy} et~al.(2010){Stacy}, {Greif}, and {Bromm}}]{stacy2010}
\bibinfo{author}{A.~{Stacy}}, \bibinfo{author}{T.~H. {Greif}},
  \bibinfo{author}{V.~{Bromm}}, \bibinfo{journal}{\mnras}
  \bibinfo{volume}{403} (\bibinfo{year}{2010}) \bibinfo{pages}{45--60}.
\bibitem[{{Clark} et~al.(2011){Clark}, {Glover}, {Klessen}, and
  {Bromm}}]{clark2011}
\bibinfo{author}{P.~C. {Clark}}, \bibinfo{author}{S.~C.~O. {Glover}},
  \bibinfo{author}{R.~S. {Klessen}}, \bibinfo{author}{V.~{Bromm}},
  \bibinfo{journal}{\apj}  \bibinfo{volume}{727} (\bibinfo{year}{2011})
  \bibinfo{pages}{110}.
\bibitem[{{Turk} et~al.(2009){Turk}, {Abel}, and {O'Shea}}]{turk2009}
\bibinfo{author}{M.~J. {Turk}}, \bibinfo{author}{T.~{Abel}},
  \bibinfo{author}{B.~{O'Shea}}, \bibinfo{journal}{Science}
  \bibinfo{volume}{325} (\bibinfo{year}{2009}) \bibinfo{pages}{601}.
\bibitem[{{Machida} et~al.(2009){Machida}, {Omukai}, {Matsumoto}, and
  {Inutsuka}}]{machida2009}
\bibinfo{author}{M.~N. {Machida}}, \bibinfo{author}{K.~{Omukai}},
  \bibinfo{author}{T.~{Matsumoto}}, \bibinfo{author}{S.-I. {Inutsuka}},
  \bibinfo{journal}{\mnras}  \bibinfo{volume}{399} (\bibinfo{year}{2009})
  \bibinfo{pages}{1255--1263}.
\bibitem[{{Greif} et~al.(2011){Greif}, {Springel}, {White}, {Glover}, {Clark},
  {Smith}, {Klessen}, and {Bromm}}]{greif2011}
\bibinfo{author}{T.~H. {Greif}}, \bibinfo{author}{V.~{Springel}},
  \bibinfo{author}{S.~D.~M. {White}}, \bibinfo{author}{S.~C.~O. {Glover}},
  \bibinfo{author}{P.~C. {Clark}}, \bibinfo{author}{R.~J. {Smith}},
  \bibinfo{author}{R.~S. {Klessen}}, \bibinfo{author}{V.~{Bromm}},
  \bibinfo{journal}{\apj}  \bibinfo{volume}{737} (\bibinfo{year}{2011})
  \bibinfo{pages}{75}.
\bibitem[{{Prieto} et~al.(2011){Prieto}, {Padoan}, {Jimenez}, and
  {Infante}}]{prieto2011}
\bibinfo{author}{J.~{Prieto}}, \bibinfo{author}{P.~{Padoan}},
  \bibinfo{author}{R.~{Jimenez}}, \bibinfo{author}{L.~{Infante}},
  \bibinfo{journal}{\apjl}  \bibinfo{volume}{731} (\bibinfo{year}{2011})
  \bibinfo{pages}{L38}.
\bibitem[{{Xu} et~al.(2013){Xu}, {Wise}, and {Norman}}]{xu2013}
\bibinfo{author}{H.~{Xu}}, \bibinfo{author}{J.~H. {Wise}},
  \bibinfo{author}{M.~L. {Norman}}, \bibinfo{journal}{\apj}
  \bibinfo{volume}{773} (\bibinfo{year}{2013}) \bibinfo{pages}{83}.
\bibitem[{{Turk} et~al.(2012){Turk}, {Oishi}, {Abel}, and {Bryan}}]{turk2012}
\bibinfo{author}{M.~J. {Turk}}, \bibinfo{author}{J.~S. {Oishi}},
  \bibinfo{author}{T.~{Abel}}, \bibinfo{author}{G.~L. {Bryan}},
  \bibinfo{journal}{\apj}  \bibinfo{volume}{745} (\bibinfo{year}{2012})
  \bibinfo{pages}{154}.
\bibitem[{{Hirano} and {Bromm}(2017)}]{hirano2017}
\bibinfo{author}{S.~{Hirano}}, \bibinfo{author}{V.~{Bromm}},
  \bibinfo{journal}{\mnras}  \bibinfo{volume}{470} (\bibinfo{year}{2017})
  \bibinfo{pages}{898--914}.
\bibitem[{{Omukai} and {Nishi}(1998)}]{omukai1998}
\bibinfo{author}{K.~{Omukai}}, \bibinfo{author}{R.~{Nishi}},
  \bibinfo{journal}{\apj}  \bibinfo{volume}{508} (\bibinfo{year}{1998})
  \bibinfo{pages}{141--150}.
\bibitem[{{O'Shea} and {Norman}(2007)}]{oshea2007}
\bibinfo{author}{B.~W. {O'Shea}}, \bibinfo{author}{M.~L. {Norman}},
  \bibinfo{journal}{\apj}  \bibinfo{volume}{654} (\bibinfo{year}{2007})
  \bibinfo{pages}{66--92}.
\bibitem[{{Wise} et~al.(2008){Wise}, {Turk}, and {Abel}}]{wise2008}
\bibinfo{author}{J.~H. {Wise}}, \bibinfo{author}{M.~J. {Turk}},
  \bibinfo{author}{T.~{Abel}}, \bibinfo{journal}{\apj}  \bibinfo{volume}{682}
  (\bibinfo{year}{2008}) \bibinfo{pages}{745--757}.
\bibitem[{{Omukai} and {Palla}(2003)}]{omukai2003}
\bibinfo{author}{K.~{Omukai}}, \bibinfo{author}{F.~{Palla}},
  \bibinfo{journal}{\apj}  \bibinfo{volume}{589} (\bibinfo{year}{2003})
  \bibinfo{pages}{677--687}.
\bibitem[{{Hirano} et~al.(2014){Hirano}, {Hosokawa}, {Yoshida}, {Umeda},
  {Omukai}, {Chiaki}, and {Yorke}}]{hirano2014}
\bibinfo{author}{S.~{Hirano}}, \bibinfo{author}{T.~{Hosokawa}},
  \bibinfo{author}{N.~{Yoshida}}, \bibinfo{author}{H.~{Umeda}},
  \bibinfo{author}{K.~{Omukai}}, \bibinfo{author}{G.~{Chiaki}},
  \bibinfo{author}{H.~W. {Yorke}}, \bibinfo{journal}{\apj}
  \bibinfo{volume}{781} (\bibinfo{year}{2014}) \bibinfo{pages}{60}.
\bibitem[{{Susa} et~al.(2014){Susa}, {Hasegawa}, and {Tominaga}}]{susa2014}
\bibinfo{author}{H.~{Susa}}, \bibinfo{author}{K.~{Hasegawa}},
  \bibinfo{author}{N.~{Tominaga}}, \bibinfo{journal}{\apj}
  \bibinfo{volume}{792} (\bibinfo{year}{2014}) \bibinfo{pages}{32}.
\bibitem[{{Hosokawa} et~al.(2011){Hosokawa}, {Omukai}, {Yoshida}, and
  {Yorke}}]{hosokawa2011}
\bibinfo{author}{T.~{Hosokawa}}, \bibinfo{author}{K.~{Omukai}},
  \bibinfo{author}{N.~{Yoshida}}, \bibinfo{author}{H.~W. {Yorke}},
  \bibinfo{journal}{Science}  \bibinfo{volume}{334} (\bibinfo{year}{2011})
  \bibinfo{pages}{1250}.
\bibitem[{{Stacy} et~al.(2012){Stacy}, {Greif}, and {Bromm}}]{stacy2012}
\bibinfo{author}{A.~{Stacy}}, \bibinfo{author}{T.~H. {Greif}},
  \bibinfo{author}{V.~{Bromm}}, \bibinfo{journal}{\mnras}
  \bibinfo{volume}{422} (\bibinfo{year}{2012}) \bibinfo{pages}{290--309}.
\bibitem[{{Stacy} et~al.(2016){Stacy}, {Bromm}, and {Lee}}]{stacy2016}
\bibinfo{author}{A.~{Stacy}}, \bibinfo{author}{V.~{Bromm}},
  \bibinfo{author}{A.~T. {Lee}}, \bibinfo{journal}{\mnras}
  \bibinfo{volume}{462} (\bibinfo{year}{2016}) \bibinfo{pages}{1307--1328}.
\bibitem[{{Wise} and {Abel}(2008)}]{wise2008b}
\bibinfo{author}{J.~H. {Wise}}, \bibinfo{author}{T.~{Abel}},
  \bibinfo{journal}{\apj}  \bibinfo{volume}{685} (\bibinfo{year}{2008})
  \bibinfo{pages}{40--56}.
\bibitem[{{Greif} et~al.(2010){Greif}, {Glover}, {Bromm}, and
  {Klessen}}]{greif2010}
\bibinfo{author}{T.~H. {Greif}}, \bibinfo{author}{S.~C.~O. {Glover}},
  \bibinfo{author}{V.~{Bromm}}, \bibinfo{author}{R.~S. {Klessen}},
  \bibinfo{journal}{\apj}  \bibinfo{volume}{716} (\bibinfo{year}{2010})
  \bibinfo{pages}{510--520}.
\bibitem[{{Chatzopoulos} and {Wheeler}(2012)}]{chatz2012}
\bibinfo{author}{E.~{Chatzopoulos}}, \bibinfo{author}{J.~C. {Wheeler}},
  \bibinfo{journal}{\apj}  \bibinfo{volume}{760} (\bibinfo{year}{2012})
  \bibinfo{pages}{154}.
\bibitem[{{Fryer} et~al.(2001){Fryer}, {Woosley}, and {Heger}}]{fryer2001}
\bibinfo{author}{C.~L. {Fryer}}, \bibinfo{author}{S.~E. {Woosley}},
  \bibinfo{author}{A.~{Heger}}, \bibinfo{journal}{\apj}  \bibinfo{volume}{550}
  (\bibinfo{year}{2001}) \bibinfo{pages}{372--382}.
\bibitem[{{Heger} and {Woosley}(2002)}]{heger2002}
\bibinfo{author}{A.~{Heger}}, \bibinfo{author}{S.~E. {Woosley}},
  \bibinfo{journal}{\apj}  \bibinfo{volume}{567} (\bibinfo{year}{2002})
  \bibinfo{pages}{532--543}.
\bibitem[{{Bromm} and {Loeb}(2003)}]{bromm2003}
\bibinfo{author}{V.~{Bromm}}, \bibinfo{author}{A.~{Loeb}},
  \bibinfo{journal}{\nat}  \bibinfo{volume}{425} (\bibinfo{year}{2003})
  \bibinfo{pages}{812--814}.
\bibitem[{{Chiaki} et~al.(2018){Chiaki}, {Susa}, and {Hirano}}]{chiaki2018}
\bibinfo{author}{G.~{Chiaki}}, \bibinfo{author}{H.~{Susa}},
  \bibinfo{author}{S.~{Hirano}}, \bibinfo{journal}{\mnras}
  \bibinfo{volume}{475} (\bibinfo{year}{2018}) \bibinfo{pages}{4378--4395}.
\bibitem[{{Greif} et~al.(2007){Greif}, {Johnson}, {Bromm}, and
  {Klessen}}]{greif2007}
\bibinfo{author}{T.~H. {Greif}}, \bibinfo{author}{J.~L. {Johnson}},
  \bibinfo{author}{V.~{Bromm}}, \bibinfo{author}{R.~S. {Klessen}},
  \bibinfo{journal}{\apj}  \bibinfo{volume}{670} (\bibinfo{year}{2007})
  \bibinfo{pages}{1--14}.
\bibitem[{{Cen} and {Riquelme}(2008)}]{cen2008}
\bibinfo{author}{R.~{Cen}}, \bibinfo{author}{M.~A. {Riquelme}},
  \bibinfo{journal}{\apj}  \bibinfo{volume}{674} (\bibinfo{year}{2008})
  \bibinfo{pages}{644--652}.
\bibitem[{{Bromm} et~al.(2001){Bromm}, {Ferrara}, {Coppi}, and
  {Larson}}]{bromm2001}
\bibinfo{author}{V.~{Bromm}}, \bibinfo{author}{A.~{Ferrara}},
  \bibinfo{author}{P.~S. {Coppi}}, \bibinfo{author}{R.~B. {Larson}},
  \bibinfo{journal}{\mnras}  \bibinfo{volume}{328} (\bibinfo{year}{2001})
  \bibinfo{pages}{969--976}.
\bibitem[{{Schneider} et~al.(2002){Schneider}, {Ferrara}, {Natarajan}, and
  {Omukai}}]{schneider2002}
\bibinfo{author}{R.~{Schneider}}, \bibinfo{author}{A.~{Ferrara}},
  \bibinfo{author}{P.~{Natarajan}}, \bibinfo{author}{K.~{Omukai}},
  \bibinfo{journal}{\apj}  \bibinfo{volume}{571} (\bibinfo{year}{2002})
  \bibinfo{pages}{30--39}.
\bibitem[{{Schneider} et~al.(2006){Schneider}, {Omukai}, {Inoue}, and
  {Ferrara}}]{schneider2006}
\bibinfo{author}{R.~{Schneider}}, \bibinfo{author}{K.~{Omukai}},
  \bibinfo{author}{A.~K. {Inoue}}, \bibinfo{author}{A.~{Ferrara}},
  \bibinfo{journal}{\mnras}  \bibinfo{volume}{369} (\bibinfo{year}{2006})
  \bibinfo{pages}{1437--1444}.
\bibitem[{{Santoro} and {Shull}(2006)}]{santoro2006}
\bibinfo{author}{F.~{Santoro}}, \bibinfo{author}{J.~M. {Shull}},
  \bibinfo{journal}{\apj}  \bibinfo{volume}{643} (\bibinfo{year}{2006})
  \bibinfo{pages}{26--37}.
\bibitem[{{Smith} and {Sigurdsson}(2007)}]{smith_zcrit2007}
\bibinfo{author}{B.~D. {Smith}}, \bibinfo{author}{S.~{Sigurdsson}},
  \bibinfo{journal}{\apjl}  \bibinfo{volume}{661} (\bibinfo{year}{2007})
  \bibinfo{pages}{L5--L8}.
\bibitem[{{Schneider} et~al.(2012){Schneider}, {Omukai}, {Bianchi}, and
  {Valiante}}]{schneider2012}
\bibinfo{author}{R.~{Schneider}}, \bibinfo{author}{K.~{Omukai}},
  \bibinfo{author}{S.~{Bianchi}}, \bibinfo{author}{R.~{Valiante}},
  \bibinfo{journal}{\mnras}  \bibinfo{volume}{419} (\bibinfo{year}{2012})
  \bibinfo{pages}{1566--1575}.
\bibitem[{{Chiaki} et~al.(2014){Chiaki}, {Schneider}, {Nozawa}, {Omukai},
  {Limongi}, {Yoshida}, and {Chieffi}}]{chiaki2014}
\bibinfo{author}{G.~{Chiaki}}, \bibinfo{author}{R.~{Schneider}},
  \bibinfo{author}{T.~{Nozawa}}, \bibinfo{author}{K.~{Omukai}},
  \bibinfo{author}{M.~{Limongi}}, \bibinfo{author}{N.~{Yoshida}},
  \bibinfo{author}{A.~{Chieffi}}, \bibinfo{journal}{\mnras}
  \bibinfo{volume}{439} (\bibinfo{year}{2014}) \bibinfo{pages}{3121--3127}.
\bibitem[{{Karlsson} et~al.(2008){Karlsson}, {Johnson}, and
  {Bromm}}]{karlsson2008}
\bibinfo{author}{T.~{Karlsson}}, \bibinfo{author}{J.~L. {Johnson}},
  \bibinfo{author}{V.~{Bromm}}, \bibinfo{journal}{\apj}  \bibinfo{volume}{679}
  (\bibinfo{year}{2008}) \bibinfo{pages}{6--16}.
\bibitem[{{Maio} et~al.(2011){Maio}, {Khochfar}, {Johnson}, and
  {Ciardi}}]{maio2011}
\bibinfo{author}{U.~{Maio}}, \bibinfo{author}{S.~{Khochfar}},
  \bibinfo{author}{J.~L. {Johnson}}, \bibinfo{author}{B.~{Ciardi}},
  \bibinfo{journal}{\mnras}  \bibinfo{volume}{414} (\bibinfo{year}{2011})
  \bibinfo{pages}{1145--1157}.
\bibitem[{{Jeon} et~al.(2015){Jeon}, {Bromm}, {Pawlik}, and
  {Milosavljevi{\'c}}}]{jeon2015}
\bibinfo{author}{M.~{Jeon}}, \bibinfo{author}{V.~{Bromm}},
  \bibinfo{author}{A.~H. {Pawlik}}, \bibinfo{author}{M.~{Milosavljevi{\'c}}},
  \bibinfo{journal}{\mnras}  \bibinfo{volume}{452} (\bibinfo{year}{2015})
  \bibinfo{pages}{1152--1170}.
\bibitem[{{Mackey} et~al.(2003){Mackey}, {Bromm}, and {Hernquist}}]{mackey2003}
\bibinfo{author}{J.~{Mackey}}, \bibinfo{author}{V.~{Bromm}},
  \bibinfo{author}{L.~{Hernquist}}, \bibinfo{journal}{\apj}
  \bibinfo{volume}{586} (\bibinfo{year}{2003}) \bibinfo{pages}{1--11}.
\bibitem[{{Woosley} and {Weaver}(1986)}]{woosley1986}
\bibinfo{author}{S.~E. {Woosley}}, \bibinfo{author}{T.~A. {Weaver}},
  \bibinfo{journal}{\araa}  \bibinfo{volume}{24} (\bibinfo{year}{1986})
  \bibinfo{pages}{205--253}.
\bibitem[{{Muratov} et~al.(2013){Muratov}, {Gnedin}, {Gnedin}, and
  {Zemp}}]{muratov2013}
\bibinfo{author}{A.~L. {Muratov}}, \bibinfo{author}{O.~Y. {Gnedin}},
  \bibinfo{author}{N.~Y. {Gnedin}}, \bibinfo{author}{M.~{Zemp}},
  \bibinfo{journal}{\apj}  \bibinfo{volume}{773} (\bibinfo{year}{2013})
  \bibinfo{pages}{19}.
\bibitem[{{Schaerer}(2003)}]{schaerer2003}
\bibinfo{author}{D.~{Schaerer}}, \bibinfo{journal}{\aap}  \bibinfo{volume}{397}
  (\bibinfo{year}{2003}) \bibinfo{pages}{527--538}.
\bibitem[{{Kitayama} et~al.(2004){Kitayama}, {Yoshida}, {Susa}, and
  {Umemura}}]{kitayama2004}
\bibinfo{author}{T.~{Kitayama}}, \bibinfo{author}{N.~{Yoshida}},
  \bibinfo{author}{H.~{Susa}}, \bibinfo{author}{M.~{Umemura}},
  \bibinfo{journal}{\apj}  \bibinfo{volume}{613} (\bibinfo{year}{2004})
  \bibinfo{pages}{631--645}.
\bibitem[{{Whalen} et~al.(2004){Whalen}, {Abel}, and {Norman}}]{whalen2004}
\bibinfo{author}{D.~{Whalen}}, \bibinfo{author}{T.~{Abel}},
  \bibinfo{author}{M.~L. {Norman}}, \bibinfo{journal}{\apj}
  \bibinfo{volume}{610} (\bibinfo{year}{2004}) \bibinfo{pages}{14--22}.
\bibitem[{{Alvarez} et~al.(2006){Alvarez}, {Bromm}, and
  {Shapiro}}]{alvarez2006}
\bibinfo{author}{M.~A. {Alvarez}}, \bibinfo{author}{V.~{Bromm}},
  \bibinfo{author}{P.~R. {Shapiro}}, \bibinfo{journal}{\apj}
  \bibinfo{volume}{639} (\bibinfo{year}{2006}) \bibinfo{pages}{621--632}.
\bibitem[{{Johnson} et~al.(2007){Johnson}, {Greif}, and {Bromm}}]{johnson2007}
\bibinfo{author}{J.~L. {Johnson}}, \bibinfo{author}{T.~H. {Greif}},
  \bibinfo{author}{V.~{Bromm}}, \bibinfo{journal}{\apj}  \bibinfo{volume}{665}
  (\bibinfo{year}{2007}) \bibinfo{pages}{85--95}.
\bibitem[{{Ciardi} et~al.(2000){Ciardi}, {Ferrara}, and {Abel}}]{ciardi2000}
\bibinfo{author}{B.~{Ciardi}}, \bibinfo{author}{A.~{Ferrara}},
  \bibinfo{author}{T.~{Abel}}, \bibinfo{journal}{\apj}  \bibinfo{volume}{533}
  (\bibinfo{year}{2000}) \bibinfo{pages}{594--600}.
\bibitem[{{Draine} and {Bertoldi}(1996)}]{draine1996}
\bibinfo{author}{B.~T. {Draine}}, \bibinfo{author}{F.~{Bertoldi}},
  \bibinfo{journal}{\apj}  \bibinfo{volume}{468} (\bibinfo{year}{1996})
  \bibinfo{pages}{269}.
\bibitem[{{Regan} and {Haehnelt}(2009)}]{regan2009}
\bibinfo{author}{J.~A. {Regan}}, \bibinfo{author}{M.~G. {Haehnelt}},
  \bibinfo{journal}{\mnras}  \bibinfo{volume}{396} (\bibinfo{year}{2009})
  \bibinfo{pages}{343--353}.
\bibitem[{{Volonteri}(2012)}]{volonteri2012}
\bibinfo{author}{M.~{Volonteri}}, \bibinfo{journal}{Science}
  \bibinfo{volume}{337} (\bibinfo{year}{2012}) \bibinfo{pages}{544}.
\bibitem[{{Fan} et~al.(2006){Fan}, {Carilli}, and {Keating}}]{fan2006}
\bibinfo{author}{X.~{Fan}}, \bibinfo{author}{C.~L. {Carilli}},
  \bibinfo{author}{B.~{Keating}}, \bibinfo{journal}{\araa}
  \bibinfo{volume}{44} (\bibinfo{year}{2006}) \bibinfo{pages}{415--462}.
\bibitem[{{Oh} et~al.(2001){Oh}, {Nollett}, {Madau}, and {Wasserburg}}]{oh2001}
\bibinfo{author}{S.~P. {Oh}}, \bibinfo{author}{K.~M. {Nollett}},
  \bibinfo{author}{P.~{Madau}}, \bibinfo{author}{G.~J. {Wasserburg}},
  \bibinfo{journal}{\apjl}  \bibinfo{volume}{562} (\bibinfo{year}{2001})
  \bibinfo{pages}{L1--L4}.
\bibitem[{{Machacek} et~al.(2003){Machacek}, {Bryan}, and
  {Abel}}]{machacek2003}
\bibinfo{author}{M.~E. {Machacek}}, \bibinfo{author}{G.~L. {Bryan}},
  \bibinfo{author}{T.~{Abel}}, \bibinfo{journal}{\mnras}  \bibinfo{volume}{338}
  (\bibinfo{year}{2003}) \bibinfo{pages}{273--286}.
\bibitem[{{Oh} and {Haiman}(2003)}]{oh2003}
\bibinfo{author}{S.~P. {Oh}}, \bibinfo{author}{Z.~{Haiman}},
  \bibinfo{journal}{\mnras}  \bibinfo{volume}{346} (\bibinfo{year}{2003})
  \bibinfo{pages}{456--472}.
\bibitem[{{Frebel} and {Norris}(2015)}]{frebel2015}
\bibinfo{author}{A.~{Frebel}}, \bibinfo{author}{J.~E. {Norris}},
  \bibinfo{journal}{\araa}  \bibinfo{volume}{53} (\bibinfo{year}{2015})
  \bibinfo{pages}{631--688}.
\bibitem[{{Partridge} and {Peebles}(1967)}]{partridge1967a}
\bibinfo{author}{R.~B. {Partridge}}, \bibinfo{author}{P.~J.~E. {Peebles}},
  \bibinfo{journal}{Astrophysical Journal}  \bibinfo{volume}{147}
  (\bibinfo{year}{1967}) \bibinfo{pages}{868}.
\bibitem[{{Kashikawa} et~al.(2012){Kashikawa}, {Nagao}, {Toshikawa}, and et.
  al.}]{kashikawa2012}
\bibinfo{author}{N.~{Kashikawa}}, \bibinfo{author}{T.~{Nagao}},
  \bibinfo{author}{J.~{Toshikawa}}, \bibinfo{author}{et. al.},
  \bibinfo{journal}{\apj}  \bibinfo{volume}{761} (\bibinfo{year}{2012})
  \bibinfo{pages}{85}.
\bibitem[{{Johnson} et~al.(2009){Johnson}, {Greif}, {Bromm}, {Klessen}, and
  {Ippolito}}]{johnson2009}
\bibinfo{author}{J.~L. {Johnson}}, \bibinfo{author}{T.~H. {Greif}},
  \bibinfo{author}{V.~{Bromm}}, \bibinfo{author}{R.~S. {Klessen}},
  \bibinfo{author}{J.~{Ippolito}}, \bibinfo{journal}{\mnras}
  \bibinfo{volume}{399} (\bibinfo{year}{2009}) \bibinfo{pages}{37--47}.
\bibitem[{{Zackrisson} et~al.(2011){Zackrisson}, {Rydberg}, {Schaerer},
  {{\"O}stlin}, and {Tuli}}]{zackrisson2011a}
\bibinfo{author}{E.~{Zackrisson}}, \bibinfo{author}{C.-E. {Rydberg}},
  \bibinfo{author}{D.~{Schaerer}}, \bibinfo{author}{G.~{{\"O}stlin}},
  \bibinfo{author}{M.~{Tuli}}, \bibinfo{journal}{\apj}  \bibinfo{volume}{740}
  (\bibinfo{year}{2011}) \bibinfo{pages}{13}.
\bibitem[{{Pawlik} et~al.(2011){Pawlik}, {Milosavljevi{\'c}}, and
  {Bromm}}]{pawlik2011}
\bibinfo{author}{A.~H. {Pawlik}}, \bibinfo{author}{M.~{Milosavljevi{\'c}}},
  \bibinfo{author}{V.~{Bromm}}, \bibinfo{journal}{\apj}  \bibinfo{volume}{731}
  (\bibinfo{year}{2011}) \bibinfo{pages}{54}.
\bibitem[{{Zackrisson} et~al.(2011){Zackrisson}, {Inoue}, {Rydberg}, and
  {Duval}}]{zackrisson2011b}
\bibinfo{author}{E.~{Zackrisson}}, \bibinfo{author}{A.~K. {Inoue}},
  \bibinfo{author}{C.-E. {Rydberg}}, \bibinfo{author}{F.~{Duval}},
  \bibinfo{journal}{\mnras}  \bibinfo{volume}{418} (\bibinfo{year}{2011})
  \bibinfo{pages}{L104--L108}.
\bibitem[{{Windhorst} et~al.(2018){Windhorst}, {Timmes}, {Wyithe}, and et.
  al.}]{windhorst2018}
\bibinfo{author}{R.~A. {Windhorst}}, \bibinfo{author}{F.~X. {Timmes}},
  \bibinfo{author}{J.~S.~B. {Wyithe}}, \bibinfo{author}{et. al.},
  \bibinfo{journal}{\apjs}  \bibinfo{volume}{234} (\bibinfo{year}{2018})
  \bibinfo{pages}{41}.
\bibitem[{{Mesinger} et~al.(2006){Mesinger}, {Johnson}, and
  {Haiman}}]{mesinger2006}
\bibinfo{author}{A.~{Mesinger}}, \bibinfo{author}{B.~D. {Johnson}},
  \bibinfo{author}{Z.~{Haiman}}, \bibinfo{journal}{\apj}  \bibinfo{volume}{637}
  (\bibinfo{year}{2006}) \bibinfo{pages}{80--90}.
\bibitem[{{Visbal} et~al.(2015){Visbal}, {Haiman}, and {Bryan}}]{visbal2015}
\bibinfo{author}{E.~{Visbal}}, \bibinfo{author}{Z.~{Haiman}},
  \bibinfo{author}{G.~L. {Bryan}}, \bibinfo{journal}{\mnras}
  \bibinfo{volume}{450} (\bibinfo{year}{2015}) \bibinfo{pages}{2506--2513}.
\bibitem[{{Kolb} and {Turner}(1990)}]{kolb1990}
\bibinfo{author}{E.~W. {Kolb}}, \bibinfo{author}{M.~S. {Turner}},
  \bibinfo{title}{{The early universe.}},  \bibinfo{year}{1990}.
\bibitem[{{Iocco} et~al.(2007){Iocco}, {Mangano}, {Miele}, {Pisanti}, and
  {Serpico}}]{iocco2007}
\bibinfo{author}{F.~{Iocco}}, \bibinfo{author}{G.~{Mangano}},
  \bibinfo{author}{G.~{Miele}}, \bibinfo{author}{O.~{Pisanti}},
  \bibinfo{author}{P.~D. {Serpico}}, \bibinfo{journal}{\prd}
  \bibinfo{volume}{75} (\bibinfo{year}{2007}) \bibinfo{pages}{087304}.
\bibitem[{{Naoz} et~al.(2006){Naoz}, {Noter}, and {Barkana}}]{naoz2006}
\bibinfo{author}{S.~{Naoz}}, \bibinfo{author}{S.~{Noter}},
  \bibinfo{author}{R.~{Barkana}}, \bibinfo{journal}{\mnras}
  \bibinfo{volume}{373} (\bibinfo{year}{2006}) \bibinfo{pages}{L98--L102}.
\bibitem[{{Asplund} et~al.(2005){Asplund}, {Grevesse}, and
  {Sauval}}]{asplund2005}
\bibinfo{author}{M.~{Asplund}}, \bibinfo{author}{N.~{Grevesse}},
  \bibinfo{author}{A.~J. {Sauval}}, in: \bibinfo{editor}{T.~G. {Barnes}, III},
  \bibinfo{editor}{F.~N. {Bash}} (Eds.), \bibinfo{booktitle}{Cosmic Abundances
  as Records of Stellar Evolution and Nucleosynthesis}, volume
  \bibinfo{volume}{336} of \textit{\bibinfo{series}{Astronomical Society of the
  Pacific Conference Series}},  p.~\bibinfo{pages}{25}.
\bibitem[{{Ferrara}(2016)}]{ferrara2016}
\bibinfo{author}{A.~{Ferrara}}, \bibinfo{journal}{Understanding the Epoch of
  Cosmic Reionization: Challenges and Progress}  \bibinfo{volume}{423}
  (\bibinfo{year}{2016}) \bibinfo{pages}{163}.
\bibitem[{{Maio} et~al.(2010){Maio}, {Ciardi}, {Dolag}, {Tornatore}, and
  {Khochfar}}]{maio2010}
\bibinfo{author}{U.~{Maio}}, \bibinfo{author}{B.~{Ciardi}},
  \bibinfo{author}{K.~{Dolag}}, \bibinfo{author}{L.~{Tornatore}},
  \bibinfo{author}{S.~{Khochfar}}, \bibinfo{journal}{\mnras}
  \bibinfo{volume}{407} (\bibinfo{year}{2010}) \bibinfo{pages}{1003--1015}.
\bibitem[{{Tornatore} et~al.(2007){Tornatore}, {Ferrara}, and
  {Schneider}}]{tornatore2007}
\bibinfo{author}{L.~{Tornatore}}, \bibinfo{author}{A.~{Ferrara}},
  \bibinfo{author}{R.~{Schneider}}, \bibinfo{journal}{\mnras}
  \bibinfo{volume}{382} (\bibinfo{year}{2007}) \bibinfo{pages}{945--950}.
\bibitem[{{Oppenheimer} et~al.(2009){Oppenheimer}, {Dav{\'e}}, and
  {Finlator}}]{oppenheimer2009}
\bibinfo{author}{B.~D. {Oppenheimer}}, \bibinfo{author}{R.~{Dav{\'e}}},
  \bibinfo{author}{K.~{Finlator}}, \bibinfo{journal}{\mnras}
  \bibinfo{volume}{396} (\bibinfo{year}{2009}) \bibinfo{pages}{729--758}.
\bibitem[{{Pallottini} et~al.(2014){Pallottini}, {Ferrara}, {Gallerani},
  {Salvadori}, and {D'Odorico}}]{pallottini2014}
\bibinfo{author}{A.~{Pallottini}}, \bibinfo{author}{A.~{Ferrara}},
  \bibinfo{author}{S.~{Gallerani}}, \bibinfo{author}{S.~{Salvadori}},
  \bibinfo{author}{V.~{D'Odorico}}, \bibinfo{journal}{\mnras}
  \bibinfo{volume}{440} (\bibinfo{year}{2014}) \bibinfo{pages}{2498--2518}.
\bibitem[{{Salpeter}(1955)}]{salpeter1955}
\bibinfo{author}{E.~E. {Salpeter}}, \bibinfo{journal}{\apj}
  \bibinfo{volume}{121} (\bibinfo{year}{1955}) \bibinfo{pages}{161}.
\bibitem[{{Tsujimoto} et~al.(1995){Tsujimoto}, {Nomoto}, {Yoshii}, {Hashimoto},
  {Yanagida}, and {Thielemann}}]{tsujimoto1995}
\bibinfo{author}{T.~{Tsujimoto}}, \bibinfo{author}{K.~{Nomoto}},
  \bibinfo{author}{Y.~{Yoshii}}, \bibinfo{author}{M.~{Hashimoto}},
  \bibinfo{author}{S.~{Yanagida}}, \bibinfo{author}{F.-K. {Thielemann}},
  \bibinfo{journal}{\mnras}  \bibinfo{volume}{277} (\bibinfo{year}{1995})
  \bibinfo{pages}{945--958}.
\bibitem[{{Woosley} and {Weaver}(1995)}]{woosley1995}
\bibinfo{author}{S.~E. {Woosley}}, \bibinfo{author}{T.~A. {Weaver}},
  \bibinfo{journal}{\apjs}  \bibinfo{volume}{101} (\bibinfo{year}{1995})
  \bibinfo{pages}{181}.
\bibitem[{{Nomoto} et~al.(2006){Nomoto}, {Tominaga}, {Umeda}, {Kobayashi}, and
  {Maeda}}]{nomoto2006}
\bibinfo{author}{K.~{Nomoto}}, \bibinfo{author}{N.~{Tominaga}},
  \bibinfo{author}{H.~{Umeda}}, \bibinfo{author}{C.~{Kobayashi}},
  \bibinfo{author}{K.~{Maeda}}, \bibinfo{journal}{Nuclear Physics A}
  \bibinfo{volume}{777} (\bibinfo{year}{2006}) \bibinfo{pages}{424--458}.
\bibitem[{{Gibson} et~al.(1997){Gibson}, {Loewenstein}, and
  {Mushotzky}}]{gibson1997}
\bibinfo{author}{B.~K. {Gibson}}, \bibinfo{author}{M.~{Loewenstein}},
  \bibinfo{author}{R.~F. {Mushotzky}}, \bibinfo{journal}{\mnras}
  \bibinfo{volume}{290} (\bibinfo{year}{1997}) \bibinfo{pages}{623--628}.
\bibitem[{{Gull}(1973)}]{Gull73}
\bibinfo{author}{S.~F. {Gull}}, \bibinfo{journal}{\mnras}
  \bibinfo{volume}{161} (\bibinfo{year}{1973}) \bibinfo{pages}{47--69}.
\bibitem[{{Madau} et~al.(2001){Madau}, {Ferrara}, and {Rees}}]{madau2001}
\bibinfo{author}{P.~{Madau}}, \bibinfo{author}{A.~{Ferrara}},
  \bibinfo{author}{M.~J. {Rees}}, \bibinfo{journal}{\apj}
  \bibinfo{volume}{555} (\bibinfo{year}{2001}) \bibinfo{pages}{92--105}.
\bibitem[{{Slavin} et~al.(1993){Slavin}, {Shull}, and {Begelman}}]{Slavin93}
\bibinfo{author}{J.~D. {Slavin}}, \bibinfo{author}{J.~M. {Shull}},
  \bibinfo{author}{M.~C. {Begelman}}, \bibinfo{journal}{\apj}
  \bibinfo{volume}{407} (\bibinfo{year}{1993}) \bibinfo{pages}{83--99}.
\bibitem[{{Tenorio-Tagle}(1996)}]{Tenorio96}
\bibinfo{author}{G.~{Tenorio-Tagle}}, \bibinfo{journal}{\aj}
  \bibinfo{volume}{111} (\bibinfo{year}{1996}) \bibinfo{pages}{1641}.
\bibitem[{{Pezzulli} and {Fraternali}(2016)}]{Pezzulli16}
\bibinfo{author}{G.~{Pezzulli}}, \bibinfo{author}{F.~{Fraternali}},
  \bibinfo{journal}{\mnras}  \bibinfo{volume}{455} (\bibinfo{year}{2016})
  \bibinfo{pages}{2308--2322}.
\bibitem[{{Roy} and {Kunth}(1995)}]{Roy95}
\bibinfo{author}{J.-R. {Roy}}, \bibinfo{author}{D.~{Kunth}},
  \bibinfo{journal}{\aap}  \bibinfo{volume}{294} (\bibinfo{year}{1995})
  \bibinfo{pages}{432--442}.
\bibitem[{{de Avillez} and {Mac Low}(2002)}]{deAvillez02}
\bibinfo{author}{M.~A. {de Avillez}}, \bibinfo{author}{M.-M. {Mac Low}},
  \bibinfo{journal}{\apj}  \bibinfo{volume}{581} (\bibinfo{year}{2002})
  \bibinfo{pages}{1047--1060}.
\bibitem[{{Klessen} and {Lin}(2003)}]{Klessen03}
\bibinfo{author}{R.~S. {Klessen}}, \bibinfo{author}{D.~N. {Lin}},
  \bibinfo{journal}{\pre}  \bibinfo{volume}{67} (\bibinfo{year}{2003})
  \bibinfo{pages}{046311}.
\bibitem[{{Pan} and {Scannapieco}(2010)}]{Pan10}
\bibinfo{author}{L.~{Pan}}, \bibinfo{author}{E.~{Scannapieco}},
  \bibinfo{journal}{\apj}  \bibinfo{volume}{721} (\bibinfo{year}{2010})
  \bibinfo{pages}{1765--1782}.
\bibitem[{{Feng} and {Krumholz}(2014)}]{Feng14}
\bibinfo{author}{Y.~{Feng}}, \bibinfo{author}{M.~R. {Krumholz}},
  \bibinfo{journal}{\nat}  \bibinfo{volume}{513} (\bibinfo{year}{2014})
  \bibinfo{pages}{523--525}.
\bibitem[{{Armillotta} et~al.(2018){Armillotta}, {Krumholz}, and
  {Fujimoto}}]{Armillotta18}
\bibinfo{author}{L.~{Armillotta}}, \bibinfo{author}{M.~R. {Krumholz}},
  \bibinfo{author}{Y.~{Fujimoto}}, \bibinfo{journal}{ArXiv e-prints}
  (\bibinfo{year}{2018}).
\bibitem[{{Cen} and {Riquelme}(2008)}]{Cen08}
\bibinfo{author}{R.~{Cen}}, \bibinfo{author}{M.~A. {Riquelme}},
  \bibinfo{journal}{\apj}  \bibinfo{volume}{674} (\bibinfo{year}{2008})
  \bibinfo{pages}{644--652}.
\bibitem[{{Jeon} et~al.(2015){Jeon}, {Bromm}, {Pawlik}, and
  {Milosavljevi{\'c}}}]{Jeon15}
\bibinfo{author}{M.~{Jeon}}, \bibinfo{author}{V.~{Bromm}},
  \bibinfo{author}{A.~H. {Pawlik}}, \bibinfo{author}{M.~{Milosavljevi{\'c}}},
  \bibinfo{journal}{\mnras}  \bibinfo{volume}{452} (\bibinfo{year}{2015})
  \bibinfo{pages}{1152--1170}.
\bibitem[{{Smith} et~al.(2015){Smith}, {Wise}, {O'Shea}, {Norman}, and
  {Khochfar}}]{Smith15}
\bibinfo{author}{B.~D. {Smith}}, \bibinfo{author}{J.~H. {Wise}},
  \bibinfo{author}{B.~W. {O'Shea}}, \bibinfo{author}{M.~L. {Norman}},
  \bibinfo{author}{S.~{Khochfar}}, \bibinfo{journal}{\mnras}
  \bibinfo{volume}{452} (\bibinfo{year}{2015}) \bibinfo{pages}{2822--2836}.
\bibitem[{{Sarmento} et~al.(2018){Sarmento}, {Scannapieco}, and
  {Cohen}}]{Sarmento18}
\bibinfo{author}{R.~{Sarmento}}, \bibinfo{author}{E.~{Scannapieco}},
  \bibinfo{author}{S.~{Cohen}}, \bibinfo{journal}{\apj}  \bibinfo{volume}{854}
  (\bibinfo{year}{2018}) \bibinfo{pages}{75}.
\bibitem[{{Benson} et~al.(2003){Benson}, {Bower}, {Frenk}, {Lacey}, {Baugh},
  and {Cole}}]{benson2003}
\bibinfo{author}{A.~J. {Benson}}, \bibinfo{author}{R.~G. {Bower}},
  \bibinfo{author}{C.~S. {Frenk}}, \bibinfo{author}{C.~G. {Lacey}},
  \bibinfo{author}{C.~M. {Baugh}}, \bibinfo{author}{S.~{Cole}},
  \bibinfo{journal}{\apj}  \bibinfo{volume}{599} (\bibinfo{year}{2003})
  \bibinfo{pages}{38--49}.
\bibitem[{{Springel} and {Hernquist}(2003)}]{springel-hernquist2003}
\bibinfo{author}{V.~{Springel}}, \bibinfo{author}{L.~{Hernquist}},
  \bibinfo{journal}{\mnras}  \bibinfo{volume}{339} (\bibinfo{year}{2003})
  \bibinfo{pages}{312--334}.
\bibitem[{{Mac Low} and {Ferrara}(1999)}]{maclow1999}
\bibinfo{author}{M.-M. {Mac Low}}, \bibinfo{author}{A.~{Ferrara}},
  \bibinfo{journal}{\apj}  \bibinfo{volume}{513} (\bibinfo{year}{1999})
  \bibinfo{pages}{142--155}.
\bibitem[{{Tegmark} et~al.(1993){Tegmark}, {Silk}, and {Evrard}}]{tegmark1993}
\bibinfo{author}{M.~{Tegmark}}, \bibinfo{author}{J.~{Silk}},
  \bibinfo{author}{A.~{Evrard}}, \bibinfo{journal}{\apj}  \bibinfo{volume}{417}
  (\bibinfo{year}{1993}) \bibinfo{pages}{54}.
\bibitem[{{Medvedev} and {Loeb}(2013)}]{medvedev2013}
\bibinfo{author}{M.~V. {Medvedev}}, \bibinfo{author}{A.~{Loeb}},
  \bibinfo{journal}{\apj}  \bibinfo{volume}{768} (\bibinfo{year}{2013})
  \bibinfo{pages}{113}.
\bibitem[{{Bremer} et~al.(2018){Bremer}, {Dayal}, and
  {Ryan-Weber}}]{bremer2018}
\bibinfo{author}{J.~{Bremer}}, \bibinfo{author}{P.~{Dayal}},
  \bibinfo{author}{E.~V. {Ryan-Weber}}, \bibinfo{journal}{\mnras}
  (\bibinfo{year}{2018}).
\bibitem[{{Livermore} et~al.(2017){Livermore}, {Finkelstein}, and
  {Lotz}}]{livermore2016}
\bibinfo{author}{R.~C. {Livermore}}, \bibinfo{author}{S.~L. {Finkelstein}},
  \bibinfo{author}{J.~M. {Lotz}}, \bibinfo{journal}{\apj}
  \bibinfo{volume}{835} (\bibinfo{year}{2017}) \bibinfo{pages}{113}.
\bibitem[{{Shapiro} and {Giroux}(1987)}]{shapiro1987}
\bibinfo{author}{P.~R. {Shapiro}}, \bibinfo{author}{M.~L. {Giroux}},
  \bibinfo{journal}{\apjl}  \bibinfo{volume}{321} (\bibinfo{year}{1987})
  \bibinfo{pages}{L107--L112}.
\bibitem[{{Leitherer} et~al.(1999){Leitherer}, {Schaerer}, {Goldader}, and et.
  al.}]{leitherer1999}
\bibinfo{author}{C.~{Leitherer}}, \bibinfo{author}{D.~{Schaerer}},
  \bibinfo{author}{J.~D. {Goldader}}, \bibinfo{author}{et. al.},
  \bibinfo{journal}{\apjs}  \bibinfo{volume}{123} (\bibinfo{year}{1999})
  \bibinfo{pages}{3--40}.
\bibitem[{{Leitherer} et~al.(2010){Leitherer}, {Ortiz Ot{\'a}lvaro},
  {Bresolin}, {Kudritzki}, {Lo Faro}, {Pauldrach}, {Pettini}, and
  {Rix}}]{leitherer2010}
\bibinfo{author}{C.~{Leitherer}}, \bibinfo{author}{P.~A. {Ortiz Ot{\'a}lvaro}},
  \bibinfo{author}{F.~{Bresolin}}, \bibinfo{author}{R.-P. {Kudritzki}},
  \bibinfo{author}{B.~{Lo Faro}}, \bibinfo{author}{A.~W.~A. {Pauldrach}},
  \bibinfo{author}{M.~{Pettini}}, \bibinfo{author}{S.~A. {Rix}},
  \bibinfo{journal}{\apjs}  \bibinfo{volume}{189} (\bibinfo{year}{2010})
  \bibinfo{pages}{309--335}.
\bibitem[{{Dayal} et~al.(2017){Dayal}, {Choudhury}, {Bromm}, and
  {Pacucci}}]{dayal2017a}
\bibinfo{author}{P.~{Dayal}}, \bibinfo{author}{T.~R. {Choudhury}},
  \bibinfo{author}{V.~{Bromm}}, \bibinfo{author}{F.~{Pacucci}},
  \bibinfo{journal}{\apj}  \bibinfo{volume}{836} (\bibinfo{year}{2017})
  \bibinfo{pages}{16}.
\bibitem[{{Zackrisson} et~al.(2008){Zackrisson}, {Bergvall}, and
  {Leitet}}]{zackrisson2008}
\bibinfo{author}{E.~{Zackrisson}}, \bibinfo{author}{N.~{Bergvall}},
  \bibinfo{author}{E.~{Leitet}}, \bibinfo{journal}{\apjl}
  \bibinfo{volume}{676} (\bibinfo{year}{2008}) \bibinfo{pages}{L9}.
\bibitem[{{Schaerer} and {de Barros}(2009)}]{schaerer2009}
\bibinfo{author}{D.~{Schaerer}}, \bibinfo{author}{S.~{de Barros}},
  \bibinfo{journal}{\aap}  \bibinfo{volume}{502} (\bibinfo{year}{2009})
  \bibinfo{pages}{423--426}.
\bibitem[{{de Barros} et~al.(2014){de Barros}, {Schaerer}, and
  {Stark}}]{debarros2014}
\bibinfo{author}{S.~{de Barros}}, \bibinfo{author}{D.~{Schaerer}},
  \bibinfo{author}{D.~P. {Stark}}, \bibinfo{journal}{\aap}
  \bibinfo{volume}{563} (\bibinfo{year}{2014}) \bibinfo{pages}{A81}.
\bibitem[{{Inoue} and {Iwata}(2008)}]{inoue2008}
\bibinfo{author}{A.~K. {Inoue}}, \bibinfo{author}{I.~{Iwata}},
  \bibinfo{journal}{\mnras}  \bibinfo{volume}{387} (\bibinfo{year}{2008})
  \bibinfo{pages}{1681--1692}.
\bibitem[{{Prochaska} et~al.(2010){Prochaska}, {O'Meara}, and
  {Worseck}}]{prochaska2010}
\bibinfo{author}{J.~X. {Prochaska}}, \bibinfo{author}{J.~M. {O'Meara}},
  \bibinfo{author}{G.~{Worseck}}, \bibinfo{journal}{\apj}
  \bibinfo{volume}{718} (\bibinfo{year}{2010}) \bibinfo{pages}{392--416}.
\bibitem[{{Siana} et~al.(2007){Siana}, {Teplitz}, {Colbert}, and et.
  al.}]{siana2007}
\bibinfo{author}{B.~{Siana}}, \bibinfo{author}{H.~I. {Teplitz}},
  \bibinfo{author}{J.~{Colbert}}, \bibinfo{author}{et. al.},
  \bibinfo{journal}{\apj}  \bibinfo{volume}{668} (\bibinfo{year}{2007})
  \bibinfo{pages}{62--73}.
\bibitem[{{Calzetti}(1997)}]{calzetti1997}
\bibinfo{author}{D.~{Calzetti}}, \bibinfo{journal}{\aj}  \bibinfo{volume}{113}
  (\bibinfo{year}{1997}) \bibinfo{pages}{162--184}.
\bibitem[{{Inoue} et~al.(2006){Inoue}, {Iwata}, and {Deharveng}}]{inoue2006}
\bibinfo{author}{A.~K. {Inoue}}, \bibinfo{author}{I.~{Iwata}},
  \bibinfo{author}{J.-M. {Deharveng}}, \bibinfo{journal}{\mnras}
  \bibinfo{volume}{371} (\bibinfo{year}{2006}) \bibinfo{pages}{L1--L5}.
\bibitem[{{Nestor} et~al.(2011){Nestor}, {Shapley}, {Steidel}, and
  {Siana}}]{nestor2011}
\bibinfo{author}{D.~B. {Nestor}}, \bibinfo{author}{A.~E. {Shapley}},
  \bibinfo{author}{C.~C. {Steidel}}, \bibinfo{author}{B.~{Siana}},
  \bibinfo{journal}{\apj}  \bibinfo{volume}{736} (\bibinfo{year}{2011})
  \bibinfo{pages}{18}.
\bibitem[{{Nestor} et~al.(2013){Nestor}, {Shapley}, {Kornei}, {Steidel}, and
  {Siana}}]{nestor2013}
\bibinfo{author}{D.~B. {Nestor}}, \bibinfo{author}{A.~E. {Shapley}},
  \bibinfo{author}{K.~A. {Kornei}}, \bibinfo{author}{C.~C. {Steidel}},
  \bibinfo{author}{B.~{Siana}}, \bibinfo{journal}{\apj}  \bibinfo{volume}{765}
  (\bibinfo{year}{2013}) \bibinfo{pages}{47}.
\bibitem[{{Matthee} et~al.(2017){Matthee}, {Sobral}, {Best}, {Khostovan},
  {Oteo}, {Bouwens}, and {R{\"o}ttgering}}]{matthee2017}
\bibinfo{author}{J.~{Matthee}}, \bibinfo{author}{D.~{Sobral}},
  \bibinfo{author}{P.~{Best}}, \bibinfo{author}{A.~A. {Khostovan}},
  \bibinfo{author}{I.~{Oteo}}, \bibinfo{author}{R.~{Bouwens}},
  \bibinfo{author}{H.~{R{\"o}ttgering}}, \bibinfo{journal}{\mnras}
  \bibinfo{volume}{465} (\bibinfo{year}{2017}) \bibinfo{pages}{3637--3655}.
\bibitem[{{Vasei} et~al.(2016){Vasei}, {Siana}, {Shapley}, {Quider}, {Alavi},
  {Rafelski}, {Steidel}, {Pettini}, and {Lewis}}]{vasei2016}
\bibinfo{author}{K.~{Vasei}}, \bibinfo{author}{B.~{Siana}},
  \bibinfo{author}{A.~E. {Shapley}}, \bibinfo{author}{A.~M. {Quider}},
  \bibinfo{author}{A.~{Alavi}}, \bibinfo{author}{M.~{Rafelski}},
  \bibinfo{author}{C.~C. {Steidel}}, \bibinfo{author}{M.~{Pettini}},
  \bibinfo{author}{G.~F. {Lewis}}, \bibinfo{journal}{\apj}
  \bibinfo{volume}{831} (\bibinfo{year}{2016}) \bibinfo{pages}{38}.
\bibitem[{{Siana} et~al.(2015){Siana}, {Shapley}, {Kulas}, and et.
  al.}]{siana2015}
\bibinfo{author}{B.~{Siana}}, \bibinfo{author}{A.~E. {Shapley}},
  \bibinfo{author}{K.~R. {Kulas}}, \bibinfo{author}{et. al.},
  \bibinfo{journal}{\apj}  \bibinfo{volume}{804} (\bibinfo{year}{2015})
  \bibinfo{pages}{17}.
\bibitem[{{Steidel} et~al.(1999){Steidel}, {Adelberger}, {Giavalisco},
  {Dickinson}, and {Pettini}}]{steidel1999}
\bibinfo{author}{C.~C. {Steidel}}, \bibinfo{author}{K.~L. {Adelberger}},
  \bibinfo{author}{M.~{Giavalisco}}, \bibinfo{author}{M.~{Dickinson}},
  \bibinfo{author}{M.~{Pettini}}, \bibinfo{journal}{\apj}
  \bibinfo{volume}{519} (\bibinfo{year}{1999}) \bibinfo{pages}{1--17}.
\bibitem[{{Boutsia} et~al.(2011){Boutsia}, {Grazian}, {Giallongo}, and et.
  al.}]{boutsia2011}
\bibinfo{author}{K.~{Boutsia}}, \bibinfo{author}{A.~{Grazian}},
  \bibinfo{author}{E.~{Giallongo}}, \bibinfo{author}{et. al.},
  \bibinfo{journal}{\apj}  \bibinfo{volume}{736} (\bibinfo{year}{2011})
  \bibinfo{pages}{41}.
\bibitem[{{Grazian} et~al.(2016){Grazian}, {Giallongo}, {Gerbasi}, and et.
  al.}]{grazian2016}
\bibinfo{author}{A.~{Grazian}}, \bibinfo{author}{E.~{Giallongo}},
  \bibinfo{author}{R.~{Gerbasi}}, \bibinfo{author}{et. al.},
  \bibinfo{journal}{\aap}  \bibinfo{volume}{585} (\bibinfo{year}{2016})
  \bibinfo{pages}{A48}.
\bibitem[{{Vanzella} et~al.(2010){Vanzella}, {Giavalisco}, and
  {Inoue}}]{vanzella2010}
\bibinfo{author}{E.~{Vanzella}}, \bibinfo{author}{M.~{Giavalisco}},
  \bibinfo{author}{A.~K. e.~a. {Inoue}}, \bibinfo{journal}{\apj}
  \bibinfo{volume}{725} (\bibinfo{year}{2010}) \bibinfo{pages}{1011--1031}.
\bibitem[{{Naidu} et~al.(2017){Naidu}, {Oesch}, {Reddy}, and et.
  al.}]{naidu2017}
\bibinfo{author}{R.~P. {Naidu}}, \bibinfo{author}{P.~A. {Oesch}},
  \bibinfo{author}{N.~{Reddy}}, \bibinfo{author}{et. al.},
  \bibinfo{journal}{\apj}  \bibinfo{volume}{847} (\bibinfo{year}{2017})
  \bibinfo{pages}{12}.
\bibitem[{{Mostardi} et~al.(2015){Mostardi}, {Shapley}, {Steidel}, {Trainor},
  {Reddy}, and {Siana}}]{mostardi2015}
\bibinfo{author}{R.~E. {Mostardi}}, \bibinfo{author}{A.~E. {Shapley}},
  \bibinfo{author}{C.~C. {Steidel}}, \bibinfo{author}{R.~F. {Trainor}},
  \bibinfo{author}{N.~A. {Reddy}}, \bibinfo{author}{B.~{Siana}},
  \bibinfo{journal}{\apj}  \bibinfo{volume}{810} (\bibinfo{year}{2015})
  \bibinfo{pages}{107}.
\bibitem[{{Shapley} et~al.(2016){Shapley}, {Steidel}, {Strom},
  {Bogosavljevi{\'c}}, {Reddy}, {Siana}, {Mostardi}, and {Rudie}}]{shapley2016}
\bibinfo{author}{A.~E. {Shapley}}, \bibinfo{author}{C.~C. {Steidel}},
  \bibinfo{author}{A.~L. {Strom}}, \bibinfo{author}{M.~{Bogosavljevi{\'c}}},
  \bibinfo{author}{N.~A. {Reddy}}, \bibinfo{author}{B.~{Siana}},
  \bibinfo{author}{R.~E. {Mostardi}}, \bibinfo{author}{G.~C. {Rudie}},
  \bibinfo{journal}{\apjl}  \bibinfo{volume}{826} (\bibinfo{year}{2016})
  \bibinfo{pages}{L24}.
\bibitem[{{Vanzella} et~al.(2016){Vanzella}, {de Barros}, {Vasei}, and et.
  al.}]{vanzella2016}
\bibinfo{author}{E.~{Vanzella}}, \bibinfo{author}{S.~{de Barros}},
  \bibinfo{author}{K.~{Vasei}}, \bibinfo{author}{et. al.},
  \bibinfo{journal}{\apj}  \bibinfo{volume}{825} (\bibinfo{year}{2016})
  \bibinfo{pages}{41}.
\bibitem[{{Vanzella} et~al.(2018){Vanzella}, {Nonino}, {Cupani}, and et.
  al.}]{vanzella2017}
\bibinfo{author}{E.~{Vanzella}}, \bibinfo{author}{M.~{Nonino}},
  \bibinfo{author}{G.~{Cupani}}, \bibinfo{author}{et. al.},
  \bibinfo{journal}{\mnras}  \bibinfo{volume}{476} (\bibinfo{year}{2018})
  \bibinfo{pages}{L15--L19}.
\bibitem[{{Leethochawalit} et~al.(2016){Leethochawalit}, {Jones}, {Ellis},
  {Stark}, and {Zitrin}}]{leelo2016}
\bibinfo{author}{N.~{Leethochawalit}}, \bibinfo{author}{T.~A. {Jones}},
  \bibinfo{author}{R.~S. {Ellis}}, \bibinfo{author}{D.~P. {Stark}},
  \bibinfo{author}{A.~{Zitrin}}, \bibinfo{journal}{\apj}  \bibinfo{volume}{831}
  (\bibinfo{year}{2016}) \bibinfo{pages}{152}.
\bibitem[{{Iwata} et~al.(2009){Iwata}, {Inoue}, {Matsuda}, and et.
  al.}]{iwata2009}
\bibinfo{author}{I.~{Iwata}}, \bibinfo{author}{A.~K. {Inoue}},
  \bibinfo{author}{Y.~{Matsuda}}, \bibinfo{author}{et. al.},
  \bibinfo{journal}{\apj}  \bibinfo{volume}{692} (\bibinfo{year}{2009})
  \bibinfo{pages}{1287--1293}.
\bibitem[{{Vanzella} et~al.(2012){Vanzella}, {Guo}, {Giavalisco}, and et.
  al.}]{vanzella2012b}
\bibinfo{author}{E.~{Vanzella}}, \bibinfo{author}{Y.~{Guo}},
  \bibinfo{author}{M.~{Giavalisco}}, \bibinfo{author}{et. al.},
  \bibinfo{journal}{\apj}  \bibinfo{volume}{751} (\bibinfo{year}{2012})
  \bibinfo{pages}{70}.
\bibitem[{{Cooke} et~al.(2014){Cooke}, {Ryan-Weber}, {Garel}, and
  {D{\'{\i}}az}}]{cooke2014}
\bibinfo{author}{J.~{Cooke}}, \bibinfo{author}{E.~V. {Ryan-Weber}},
  \bibinfo{author}{T.~{Garel}}, \bibinfo{author}{C.~G. {D{\'{\i}}az}},
  \bibinfo{journal}{\mnras}  \bibinfo{volume}{441} (\bibinfo{year}{2014})
  \bibinfo{pages}{837--851}.
\bibitem[{{Komatsu} et~al.(2011){Komatsu}, {Smith}, {Dunkley}, and et.
  al.}]{komatsu2011}
\bibinfo{author}{E.~{Komatsu}}, \bibinfo{author}{K.~M. {Smith}},
  \bibinfo{author}{J.~{Dunkley}}, \bibinfo{author}{et. al.},
  \bibinfo{journal}{\apjs}  \bibinfo{volume}{192} (\bibinfo{year}{2011})
  \bibinfo{pages}{18}.
\bibitem[{{Kuhlen} and {Faucher-Gigu{\`e}re}(2012)}]{kuhlen2012}
\bibinfo{author}{M.~{Kuhlen}}, \bibinfo{author}{C.-A. {Faucher-Gigu{\`e}re}},
  \bibinfo{journal}{\mnras}  \bibinfo{volume}{423} (\bibinfo{year}{2012})
  \bibinfo{pages}{862--876}.
\bibitem[{{Mitra} et~al.(2013){Mitra}, {Ferrara}, and {Choudhury}}]{mitra2013}
\bibinfo{author}{S.~{Mitra}}, \bibinfo{author}{A.~{Ferrara}},
  \bibinfo{author}{T.~R. {Choudhury}}, \bibinfo{journal}{\mnras}
  \bibinfo{volume}{428} (\bibinfo{year}{2013}) \bibinfo{pages}{L1--L5}.
\bibitem[{{Robertson} et~al.(2013){Robertson}, {Furlanetto}, {Schneider}, and
  et. al.}]{robertson2013}
\bibinfo{author}{B.~E. {Robertson}}, \bibinfo{author}{S.~R. {Furlanetto}},
  \bibinfo{author}{E.~{Schneider}}, \bibinfo{author}{et. al.},
  \bibinfo{journal}{\apj}  \bibinfo{volume}{768} (\bibinfo{year}{2013})
  \bibinfo{pages}{71}.
\bibitem[{{Mitra} et~al.(2015){Mitra}, {Choudhury}, and {Ferrara}}]{mitra2015}
\bibinfo{author}{S.~{Mitra}}, \bibinfo{author}{T.~R. {Choudhury}},
  \bibinfo{author}{A.~{Ferrara}}, \bibinfo{journal}{\mnras}
  \bibinfo{volume}{454} (\bibinfo{year}{2015}) \bibinfo{pages}{L76--L80}.
\bibitem[{{Sun} and {Furlanetto}(2016)}]{sun2016}
\bibinfo{author}{G.~{Sun}}, \bibinfo{author}{S.~R. {Furlanetto}},
  \bibinfo{journal}{\mnras}  \bibinfo{volume}{460} (\bibinfo{year}{2016})
  \bibinfo{pages}{417--433}.
\bibitem[{{Robertson} et~al.(2015){Robertson}, {Ellis}, {Furlanetto}, and
  {Dunlop}}]{robertson2015}
\bibinfo{author}{B.~E. {Robertson}}, \bibinfo{author}{R.~S. {Ellis}},
  \bibinfo{author}{S.~R. {Furlanetto}}, \bibinfo{author}{J.~S. {Dunlop}},
  \bibinfo{journal}{\apjl}  \bibinfo{volume}{802} (\bibinfo{year}{2015})
  \bibinfo{pages}{L19}.
\bibitem[{{Sharma} et~al.(2016){Sharma}, {Theuns}, {Frenk}, {Bower}, {Crain},
  {Schaller}, and {Schaye}}]{sharma2016}
\bibinfo{author}{M.~{Sharma}}, \bibinfo{author}{T.~{Theuns}},
  \bibinfo{author}{C.~{Frenk}}, \bibinfo{author}{R.~{Bower}},
  \bibinfo{author}{R.~{Crain}}, \bibinfo{author}{M.~{Schaller}},
  \bibinfo{author}{J.~{Schaye}}, \bibinfo{journal}{\mnras}
  \bibinfo{volume}{458} (\bibinfo{year}{2016}) \bibinfo{pages}{L94--L98}.
\bibitem[{{Ferrara} and {Loeb}(2013)}]{ferrara2013}
\bibinfo{author}{A.~{Ferrara}}, \bibinfo{author}{A.~{Loeb}},
  \bibinfo{journal}{\mnras}  \bibinfo{volume}{431} (\bibinfo{year}{2013})
  \bibinfo{pages}{2826--2833}.
\bibitem[{{Wise} et~al.(2014){Wise}, {Demchenko}, {Halicek}, {Norman}, {Turk},
  {Abel}, and {Smith}}]{wise2014}
\bibinfo{author}{J.~H. {Wise}}, \bibinfo{author}{V.~G. {Demchenko}},
  \bibinfo{author}{M.~T. {Halicek}}, \bibinfo{author}{M.~L. {Norman}},
  \bibinfo{author}{M.~J. {Turk}}, \bibinfo{author}{T.~{Abel}},
  \bibinfo{author}{B.~D. {Smith}}, \bibinfo{journal}{\mnras}
  \bibinfo{volume}{442} (\bibinfo{year}{2014}) \bibinfo{pages}{2560--2579}.
\bibitem[{{Ma} et~al.(2015){Ma}, {Kasen}, {Hopkins}, {Faucher-Gigu{\`e}re},
  {Quataert}, {Kere{\v s}}, and {Murray}}]{ma2015}
\bibinfo{author}{X.~{Ma}}, \bibinfo{author}{D.~{Kasen}}, \bibinfo{author}{P.~F.
  {Hopkins}}, \bibinfo{author}{C.-A. {Faucher-Gigu{\`e}re}},
  \bibinfo{author}{E.~{Quataert}}, \bibinfo{author}{D.~{Kere{\v s}}},
  \bibinfo{author}{N.~{Murray}}, \bibinfo{journal}{\mnras}
  \bibinfo{volume}{453} (\bibinfo{year}{2015}) \bibinfo{pages}{960--975}.
\bibitem[{{Wise} and {Cen}(2009)}]{wise2009}
\bibinfo{author}{J.~H. {Wise}}, \bibinfo{author}{R.~{Cen}},
  \bibinfo{journal}{\apj}  \bibinfo{volume}{693} (\bibinfo{year}{2009})
  \bibinfo{pages}{984--999}.
\bibitem[{{Paardekooper} et~al.(2011){Paardekooper}, {Pelupessy}, {Altay}, and
  {Kruip}}]{paardekooper2011}
\bibinfo{author}{J.-P. {Paardekooper}}, \bibinfo{author}{F.~I. {Pelupessy}},
  \bibinfo{author}{G.~{Altay}}, \bibinfo{author}{C.~J.~H. {Kruip}},
  \bibinfo{journal}{\aap}  \bibinfo{volume}{530} (\bibinfo{year}{2011})
  \bibinfo{pages}{A87}.
\bibitem[{{Paardekooper} et~al.(2013){Paardekooper}, {Khochfar}, and {Dalla
  Vecchia}}]{paardekooper2013}
\bibinfo{author}{J.-P. {Paardekooper}}, \bibinfo{author}{S.~{Khochfar}},
  \bibinfo{author}{C.~{Dalla Vecchia}}, \bibinfo{journal}{\mnras}
  \bibinfo{volume}{429} (\bibinfo{year}{2013}) \bibinfo{pages}{L94--L98}.
\bibitem[{{Paardekooper} et~al.(2015){Paardekooper}, {Khochfar}, and {Dalla
  Vecchia}}]{paardekooper2015}
\bibinfo{author}{J.-P. {Paardekooper}}, \bibinfo{author}{S.~{Khochfar}},
  \bibinfo{author}{C.~{Dalla Vecchia}}, \bibinfo{journal}{\mnras}
  \bibinfo{volume}{451} (\bibinfo{year}{2015}) \bibinfo{pages}{2544--2563}.
\bibitem[{{Razoumov} and {Sommer-Larsen}(2007)}]{razoumov2007}
\bibinfo{author}{A.~O. {Razoumov}}, \bibinfo{author}{J.~{Sommer-Larsen}},
  \bibinfo{journal}{\apj}  \bibinfo{volume}{668} (\bibinfo{year}{2007})
  \bibinfo{pages}{674--681}.
\bibitem[{{Razoumov} and {Sommer-Larsen}(2010)}]{razoumov2010}
\bibinfo{author}{A.~O. {Razoumov}}, \bibinfo{author}{J.~{Sommer-Larsen}},
  \bibinfo{journal}{\apj}  \bibinfo{volume}{710} (\bibinfo{year}{2010})
  \bibinfo{pages}{1239--1246}.
\bibitem[{{Gnedin} et~al.(2008){Gnedin}, {Kravtsov}, and {Chen}}]{gnedin2008}
\bibinfo{author}{N.~Y. {Gnedin}}, \bibinfo{author}{A.~V. {Kravtsov}},
  \bibinfo{author}{H.-W. {Chen}}, \bibinfo{journal}{\apj}
  \bibinfo{volume}{672} (\bibinfo{year}{2008}) \bibinfo{pages}{765--775}.
\bibitem[{{Razoumov} and {Sommer-Larsen}(2006)}]{razoumov2006}
\bibinfo{author}{A.~O. {Razoumov}}, \bibinfo{author}{J.~{Sommer-Larsen}},
  \bibinfo{journal}{\apjl}  \bibinfo{volume}{651} (\bibinfo{year}{2006})
  \bibinfo{pages}{L89--L92}.
\bibitem[{{Trebitsch} et~al.(2017){Trebitsch}, {Blaizot}, {Rosdahl},
  {Devriendt}, and {Slyz}}]{trebitsch2017}
\bibinfo{author}{M.~{Trebitsch}}, \bibinfo{author}{J.~{Blaizot}},
  \bibinfo{author}{J.~{Rosdahl}}, \bibinfo{author}{J.~{Devriendt}},
  \bibinfo{author}{A.~{Slyz}}, \bibinfo{journal}{\mnras}  \bibinfo{volume}{470}
  (\bibinfo{year}{2017}) \bibinfo{pages}{224--239}.
\bibitem[{{Ma} et~al.(2016){Ma}, {Hopkins}, {Kasen}, {Quataert},
  {Faucher-Gigu{\`e}re}, {Kere{\v s}}, {Murray}, and {Strom}}]{Ma2016}
\bibinfo{author}{X.~{Ma}}, \bibinfo{author}{P.~F. {Hopkins}},
  \bibinfo{author}{D.~{Kasen}}, \bibinfo{author}{E.~{Quataert}},
  \bibinfo{author}{C.-A. {Faucher-Gigu{\`e}re}}, \bibinfo{author}{D.~{Kere{\v
  s}}}, \bibinfo{author}{N.~{Murray}}, \bibinfo{author}{A.~{Strom}},
  \bibinfo{journal}{\mnras}  \bibinfo{volume}{459} (\bibinfo{year}{2016})
  \bibinfo{pages}{3614--3619}.
\bibitem[{{Rosdahl} et~al.(2018){Rosdahl}, {Katz}, {Blaizot}, and et.
  al.}]{rosdahl2018}
\bibinfo{author}{J.~{Rosdahl}}, \bibinfo{author}{H.~{Katz}},
  \bibinfo{author}{J.~{Blaizot}}, \bibinfo{author}{et. al.},
  \bibinfo{journal}{\mnras}  \bibinfo{volume}{479} (\bibinfo{year}{2018})
  \bibinfo{pages}{994--1016}.
\bibitem[{{Conroy} and {Kratter}(2012)}]{conroy2012}
\bibinfo{author}{C.~{Conroy}}, \bibinfo{author}{K.~M. {Kratter}},
  \bibinfo{journal}{\apj}  \bibinfo{volume}{755} (\bibinfo{year}{2012})
  \bibinfo{pages}{123}.
\bibitem[{{Kimm} and {Cen}(2014)}]{kimm2014}
\bibinfo{author}{T.~{Kimm}}, \bibinfo{author}{R.~{Cen}},
  \bibinfo{journal}{\apj}  \bibinfo{volume}{788} (\bibinfo{year}{2014})
  \bibinfo{pages}{121}.
\bibitem[{{Greif} et~al.(2008){Greif}, {Johnson}, {Klessen}, and
  {Bromm}}]{greif2008}
\bibinfo{author}{T.~H. {Greif}}, \bibinfo{author}{J.~L. {Johnson}},
  \bibinfo{author}{R.~S. {Klessen}}, \bibinfo{author}{V.~{Bromm}},
  \bibinfo{journal}{\mnras}  \bibinfo{volume}{387} (\bibinfo{year}{2008})
  \bibinfo{pages}{1021--1036}.
\bibitem[{{Sur} et~al.(2016){Sur}, {Scannapieco}, and {Ostriker}}]{sur2016}
\bibinfo{author}{S.~{Sur}}, \bibinfo{author}{E.~{Scannapieco}},
  \bibinfo{author}{E.~C. {Ostriker}}, \bibinfo{journal}{\apj}
  \bibinfo{volume}{818} (\bibinfo{year}{2016}) \bibinfo{pages}{28}.
\bibitem[{{Safarzadeh} and {Scannapieco}(2016)}]{safarzadeh2016}
\bibinfo{author}{M.~{Safarzadeh}}, \bibinfo{author}{E.~{Scannapieco}},
  \bibinfo{journal}{\apjl}  \bibinfo{volume}{832} (\bibinfo{year}{2016})
  \bibinfo{pages}{L9}.
\bibitem[{{Zackrisson} et~al.(2013){Zackrisson}, {Inoue}, and
  {Jensen}}]{zackrisson2013}
\bibinfo{author}{E.~{Zackrisson}}, \bibinfo{author}{A.~K. {Inoue}},
  \bibinfo{author}{H.~{Jensen}}, \bibinfo{journal}{\apj}  \bibinfo{volume}{777}
  (\bibinfo{year}{2013}) \bibinfo{pages}{39}.
\bibitem[{{Nakajima} and {Ouchi}(2014)}]{nakajima2014}
\bibinfo{author}{K.~{Nakajima}}, \bibinfo{author}{M.~{Ouchi}},
  \bibinfo{journal}{\mnras}  \bibinfo{volume}{442} (\bibinfo{year}{2014})
  \bibinfo{pages}{900--916}.
\bibitem[{{Faisst}(2016)}]{faisst2016}
\bibinfo{author}{A.~L. {Faisst}}, \bibinfo{journal}{\apj}
  \bibinfo{volume}{829} (\bibinfo{year}{2016}) \bibinfo{pages}{99}.
\bibitem[{{Mas-Ribas} et~al.(2017){Mas-Ribas}, {Hennawi}, {Dijkstra}, {Davies},
  {Stern}, and {Rix}}]{masribas2017}
\bibinfo{author}{L.~{Mas-Ribas}}, \bibinfo{author}{J.~F. {Hennawi}},
  \bibinfo{author}{M.~{Dijkstra}}, \bibinfo{author}{F.~B. {Davies}},
  \bibinfo{author}{J.~{Stern}}, \bibinfo{author}{H.-W. {Rix}},
  \bibinfo{journal}{\apj}  \bibinfo{volume}{846} (\bibinfo{year}{2017})
  \bibinfo{pages}{11}.
\bibitem[{{Miralda-Escud{\'e}} and {Rees}(1994)}]{miralda1994}
\bibinfo{author}{J.~{Miralda-Escud{\'e}}}, \bibinfo{author}{M.~J. {Rees}},
  \bibinfo{journal}{\mnras}  \bibinfo{volume}{266} (\bibinfo{year}{1994})
  \bibinfo{pages}{343--352}.
\bibitem[{{Meiksin} and {Madau}(1993)}]{Meiksin1993}
\bibinfo{author}{A.~{Meiksin}}, \bibinfo{author}{P.~{Madau}},
  \bibinfo{journal}{\apj}  \bibinfo{volume}{412} (\bibinfo{year}{1993})
  \bibinfo{pages}{34--55}.
\bibitem[{{Maselli} et~al.(2007){Maselli}, {Gallerani}, {Ferrara}, and
  {Choudhury}}]{Maselli07}
\bibinfo{author}{A.~{Maselli}}, \bibinfo{author}{S.~{Gallerani}},
  \bibinfo{author}{A.~{Ferrara}}, \bibinfo{author}{T.~R. {Choudhury}},
  \bibinfo{journal}{\mnras}  \bibinfo{volume}{376} (\bibinfo{year}{2007})
  \bibinfo{pages}{L34--L38}.
\bibitem[{{Bolton} and {Haehnelt}(2007)}]{Bolton07}
\bibinfo{author}{J.~S. {Bolton}}, \bibinfo{author}{M.~G. {Haehnelt}},
  \bibinfo{journal}{\mnras}  \bibinfo{volume}{381} (\bibinfo{year}{2007})
  \bibinfo{pages}{L35--L39}.
\bibitem[{{Eilers} et~al.(2017){Eilers}, {Davies}, {Hennawi}, {Prochaska},
  {Luki{\'c}}, and {Mazzucchelli}}]{Eilers17}
\bibinfo{author}{A.-C. {Eilers}}, \bibinfo{author}{F.~B. {Davies}},
  \bibinfo{author}{J.~F. {Hennawi}}, \bibinfo{author}{J.~X. {Prochaska}},
  \bibinfo{author}{Z.~{Luki{\'c}}}, \bibinfo{author}{C.~{Mazzucchelli}},
  \bibinfo{journal}{\apj}  \bibinfo{volume}{840} (\bibinfo{year}{2017})
  \bibinfo{pages}{24}.
\bibitem[{{Iliev} et~al.(2005){Iliev}, {Scannapieco}, and {Shapiro}}]{Iliev05a}
\bibinfo{author}{I.~T. {Iliev}}, \bibinfo{author}{E.~{Scannapieco}},
  \bibinfo{author}{P.~R. {Shapiro}}, \bibinfo{journal}{\apj}
  \bibinfo{volume}{624} (\bibinfo{year}{2005}) \bibinfo{pages}{491--504}.
\bibitem[{{Miralda-Escud{\'e}} et~al.(2000){Miralda-Escud{\'e}}, {Haehnelt},
  and {Rees}}]{Miralda00}
\bibinfo{author}{J.~{Miralda-Escud{\'e}}}, \bibinfo{author}{M.~{Haehnelt}},
  \bibinfo{author}{M.~J. {Rees}}, \bibinfo{journal}{\apj}
  \bibinfo{volume}{530} (\bibinfo{year}{2000}) \bibinfo{pages}{1--16}.
\bibitem[{{Chardin} et~al.(2017){Chardin}, {Puchwein}, and
  {Haehnelt}}]{Chardin17}
\bibinfo{author}{J.~{Chardin}}, \bibinfo{author}{E.~{Puchwein}},
  \bibinfo{author}{M.~G. {Haehnelt}}, \bibinfo{journal}{\mnras}
  \bibinfo{volume}{465} (\bibinfo{year}{2017}) \bibinfo{pages}{3429--3445}.
\bibitem[{{Shapiro} et~al.(2004){Shapiro}, {Iliev}, and {Raga}}]{Shapiro04}
\bibinfo{author}{P.~R. {Shapiro}}, \bibinfo{author}{I.~T. {Iliev}},
  \bibinfo{author}{A.~C. {Raga}}, \bibinfo{journal}{\mnras}
  \bibinfo{volume}{348} (\bibinfo{year}{2004}) \bibinfo{pages}{753--782}.
\bibitem[{{Iliev} et~al.(2007){Iliev}, {Mellema}, {Shapiro}, and
  {Pen}}]{Iliev2007}
\bibinfo{author}{I.~T. {Iliev}}, \bibinfo{author}{G.~{Mellema}},
  \bibinfo{author}{P.~R. {Shapiro}}, \bibinfo{author}{U.-L. {Pen}},
  \bibinfo{journal}{\mnras}  \bibinfo{volume}{376} (\bibinfo{year}{2007})
  \bibinfo{pages}{534--548}.
\bibitem[{{Iliev} et~al.(2005){Iliev}, {Shapiro}, and {Raga}}]{Iliev05b}
\bibinfo{author}{I.~T. {Iliev}}, \bibinfo{author}{P.~R. {Shapiro}},
  \bibinfo{author}{A.~C. {Raga}}, \bibinfo{journal}{\mnras}
  \bibinfo{volume}{361} (\bibinfo{year}{2005}) \bibinfo{pages}{405--414}.
\bibitem[{{Madau} et~al.(1999){Madau}, {Haardt}, and {Rees}}]{Madau1999}
\bibinfo{author}{P.~{Madau}}, \bibinfo{author}{F.~{Haardt}},
  \bibinfo{author}{M.~J. {Rees}}, \bibinfo{journal}{\apj}
  \bibinfo{volume}{514} (\bibinfo{year}{1999}) \bibinfo{pages}{648--659}.
\bibitem[{{Meiksin}(2009)}]{Meiksin2009}
\bibinfo{author}{A.~A. {Meiksin}}, \bibinfo{journal}{Reviews of Modern Physics}
   \bibinfo{volume}{81} (\bibinfo{year}{2009}) \bibinfo{pages}{1405--1469}.
\bibitem[{{Bolton} and {Haehnelt}(2007)}]{Bolton2007}
\bibinfo{author}{J.~S. {Bolton}}, \bibinfo{author}{M.~G. {Haehnelt}},
  \bibinfo{journal}{\mnras}  \bibinfo{volume}{382} (\bibinfo{year}{2007})
  \bibinfo{pages}{325--341}.
\bibitem[{{Furlanetto} and {Oh}(2016)}]{Furlanetto2016}
\bibinfo{author}{S.~R. {Furlanetto}}, \bibinfo{author}{S.~P. {Oh}},
  \bibinfo{journal}{\mnras}  \bibinfo{volume}{457} (\bibinfo{year}{2016})
  \bibinfo{pages}{1813--1827}.
\bibitem[{{Gnedin}(2000)}]{gnedin2000}
\bibinfo{author}{N.~Y. {Gnedin}}, \bibinfo{journal}{\apj}
  \bibinfo{volume}{542} (\bibinfo{year}{2000}) \bibinfo{pages}{535--541}.
\bibitem[{{Noh} and {McQuinn}(2014)}]{noh2014}
\bibinfo{author}{Y.~{Noh}}, \bibinfo{author}{M.~{McQuinn}},
  \bibinfo{journal}{\mnras}  \bibinfo{volume}{444} (\bibinfo{year}{2014})
  \bibinfo{pages}{503--514}.
\bibitem[{{Hoeft} et~al.(2006){Hoeft}, {Yepes}, {Gottl{\"o}ber}, and
  {Springel}}]{hoeft2006}
\bibinfo{author}{M.~{Hoeft}}, \bibinfo{author}{G.~{Yepes}},
  \bibinfo{author}{S.~{Gottl{\"o}ber}}, \bibinfo{author}{V.~{Springel}},
  \bibinfo{journal}{\mnras}  \bibinfo{volume}{371} (\bibinfo{year}{2006})
  \bibinfo{pages}{401--414}.
\bibitem[{{Okamoto} et~al.(2008){Okamoto}, {Gao}, and {Theuns}}]{okamoto2008}
\bibinfo{author}{T.~{Okamoto}}, \bibinfo{author}{L.~{Gao}},
  \bibinfo{author}{T.~{Theuns}}, \bibinfo{journal}{\mnras}
  \bibinfo{volume}{390} (\bibinfo{year}{2008}) \bibinfo{pages}{920--928}.
\bibitem[{{Sobacchi} and {Mesinger}(2013{\natexlab{a}})}]{sobacchi2013a}
\bibinfo{author}{E.~{Sobacchi}}, \bibinfo{author}{A.~{Mesinger}},
  \bibinfo{journal}{\mnras}  \bibinfo{volume}{432}
  (\bibinfo{year}{2013}{\natexlab{a}}) \bibinfo{pages}{L51--L55}.
\bibitem[{{Sobacchi} and {Mesinger}(2013{\natexlab{b}})}]{sobacchi2013b}
\bibinfo{author}{E.~{Sobacchi}}, \bibinfo{author}{A.~{Mesinger}},
  \bibinfo{journal}{\mnras}  \bibinfo{volume}{432}
  (\bibinfo{year}{2013}{\natexlab{b}}) \bibinfo{pages}{3340--3348}.
\bibitem[{{Wise} and {Abel}(2008)}]{wise2008c}
\bibinfo{author}{J.~H. {Wise}}, \bibinfo{author}{T.~{Abel}},
  \bibinfo{journal}{\apj}  \bibinfo{volume}{684} (\bibinfo{year}{2008})
  \bibinfo{pages}{1--17}.
\bibitem[{{Naoz} et~al.(2009){Naoz}, {Barkana}, and {Mesinger}}]{naoz2009}
\bibinfo{author}{S.~{Naoz}}, \bibinfo{author}{R.~{Barkana}},
  \bibinfo{author}{A.~{Mesinger}}, \bibinfo{journal}{\mnras}
  \bibinfo{volume}{399} (\bibinfo{year}{2009}) \bibinfo{pages}{369--376}.
\bibitem[{{Pawlik} et~al.(2009){Pawlik}, {Schaye}, and {van
  Scherpenzeel}}]{pawlik2009}
\bibinfo{author}{A.~H. {Pawlik}}, \bibinfo{author}{J.~{Schaye}},
  \bibinfo{author}{E.~{van Scherpenzeel}}, \bibinfo{journal}{\mnras}
  \bibinfo{volume}{394} (\bibinfo{year}{2009}) \bibinfo{pages}{1812--1824}.
\bibitem[{{Finlator} et~al.(2012){Finlator}, {Oh}, {{\"O}zel}, and
  {Dav{\'e}}}]{finlator2012}
\bibinfo{author}{K.~{Finlator}}, \bibinfo{author}{S.~P. {Oh}},
  \bibinfo{author}{F.~{{\"O}zel}}, \bibinfo{author}{R.~{Dav{\'e}}},
  \bibinfo{journal}{\mnras}  \bibinfo{volume}{427} (\bibinfo{year}{2012})
  \bibinfo{pages}{2464--2479}.
\bibitem[{{Shull} et~al.(2012){Shull}, {Harness}, {Trenti}, and
  {Smith}}]{shull2012}
\bibinfo{author}{J.~M. {Shull}}, \bibinfo{author}{A.~{Harness}},
  \bibinfo{author}{M.~{Trenti}}, \bibinfo{author}{B.~D. {Smith}},
  \bibinfo{journal}{\apj}  \bibinfo{volume}{747} (\bibinfo{year}{2012})
  \bibinfo{pages}{100}.
\bibitem[{{Haiman}(2011)}]{haiman2011}
\bibinfo{author}{Z.~{Haiman}}, \bibinfo{journal}{\nat}  \bibinfo{volume}{472}
  (\bibinfo{year}{2011}) \bibinfo{pages}{47--48}.
\bibitem[{{Knevitt} et~al.(2014){Knevitt}, {Wynn}, {Power}, and
  {Bolton}}]{knevitt2014}
\bibinfo{author}{G.~{Knevitt}}, \bibinfo{author}{G.~A. {Wynn}},
  \bibinfo{author}{C.~{Power}}, \bibinfo{author}{J.~S. {Bolton}},
  \bibinfo{journal}{\mnras}  \bibinfo{volume}{445} (\bibinfo{year}{2014})
  \bibinfo{pages}{2034--2048}.
\bibitem[{{Haardt} and {Madau}(2012)}]{haardt2012}
\bibinfo{author}{F.~{Haardt}}, \bibinfo{author}{P.~{Madau}},
  \bibinfo{journal}{\apj}  \bibinfo{volume}{746} (\bibinfo{year}{2012})
  \bibinfo{pages}{125}.
\bibitem[{{McQuinn} et~al.(2007){McQuinn}, {Lidz}, {Zahn}, {Dutta},
  {Hernquist}, and {Zaldarriaga}}]{mcquinn2007}
\bibinfo{author}{M.~{McQuinn}}, \bibinfo{author}{A.~{Lidz}},
  \bibinfo{author}{O.~{Zahn}}, \bibinfo{author}{S.~{Dutta}},
  \bibinfo{author}{L.~{Hernquist}}, \bibinfo{author}{M.~{Zaldarriaga}},
  \bibinfo{journal}{\mnras}  \bibinfo{volume}{377} (\bibinfo{year}{2007})
  \bibinfo{pages}{1043--1063}.
\bibitem[{{Busha} et~al.(2010){Busha}, {Alvarez}, {Wechsler}, {Abel}, and
  {Strigari}}]{busha2010}
\bibinfo{author}{M.~T. {Busha}}, \bibinfo{author}{M.~A. {Alvarez}},
  \bibinfo{author}{R.~H. {Wechsler}}, \bibinfo{author}{T.~{Abel}},
  \bibinfo{author}{L.~E. {Strigari}}, \bibinfo{journal}{\apj}
  \bibinfo{volume}{710} (\bibinfo{year}{2010}) \bibinfo{pages}{408--420}.
\bibitem[{{Lunnan} et~al.(2012){Lunnan}, {Vogelsberger}, {Frebel}, {Hernquist},
  {Lidz}, and {Boylan-Kolchin}}]{lunnan2012}
\bibinfo{author}{R.~{Lunnan}}, \bibinfo{author}{M.~{Vogelsberger}},
  \bibinfo{author}{A.~{Frebel}}, \bibinfo{author}{L.~{Hernquist}},
  \bibinfo{author}{A.~{Lidz}}, \bibinfo{author}{M.~{Boylan-Kolchin}},
  \bibinfo{journal}{\apj}  \bibinfo{volume}{746} (\bibinfo{year}{2012})
  \bibinfo{pages}{109}.
\bibitem[{{Brown} et~al.(2014){Brown}, {Tumlinson}, {Geha}, and et.
  al.}]{brown2014}
\bibinfo{author}{T.~M. {Brown}}, \bibinfo{author}{J.~{Tumlinson}},
  \bibinfo{author}{M.~{Geha}}, \bibinfo{author}{et. al.},
  \bibinfo{journal}{\apj}  \bibinfo{volume}{796} (\bibinfo{year}{2014})
  \bibinfo{pages}{91}.
\bibitem[{{Papastergis} et~al.(2012){Papastergis}, {Cattaneo}, {Huang},
  {Giovanelli}, and {Haynes}}]{papastergis2012}
\bibinfo{author}{E.~{Papastergis}}, \bibinfo{author}{A.~{Cattaneo}},
  \bibinfo{author}{S.~{Huang}}, \bibinfo{author}{R.~{Giovanelli}},
  \bibinfo{author}{M.~P. {Haynes}}, \bibinfo{journal}{\apj}
  \bibinfo{volume}{759} (\bibinfo{year}{2012}) \bibinfo{pages}{138}.
\bibitem[{{Bouwens} et~al.(2015){Bouwens}, {Illingworth}, {Oesch}, and et.
  al.}]{bouwens2015}
\bibinfo{author}{R.~J. {Bouwens}}, \bibinfo{author}{G.~D. {Illingworth}},
  \bibinfo{author}{P.~A. {Oesch}}, \bibinfo{author}{et. al.},
  \bibinfo{journal}{Astrophysical Journal}  \bibinfo{volume}{803}
  (\bibinfo{year}{2015}) \bibinfo{pages}{34}.
\bibitem[{{Haardt} and {Madau}(1996)}]{haardt1996}
\bibinfo{author}{F.~{Haardt}}, \bibinfo{author}{P.~{Madau}},
  \bibinfo{journal}{\apj}  \bibinfo{volume}{461} (\bibinfo{year}{1996})
  \bibinfo{pages}{20}.
\bibitem[{{Dixon} et~al.(2016){Dixon}, {Iliev}, {Mellema}, {Ahn}, and
  {Shapiro}}]{dixon2016}
\bibinfo{author}{K.~L. {Dixon}}, \bibinfo{author}{I.~T. {Iliev}},
  \bibinfo{author}{G.~{Mellema}}, \bibinfo{author}{K.~{Ahn}},
  \bibinfo{author}{P.~R. {Shapiro}}, \bibinfo{journal}{\mnras}
  \bibinfo{volume}{456} (\bibinfo{year}{2016}) \bibinfo{pages}{3011--3029}.
\bibitem[{{Pawlik} et~al.(2013){Pawlik}, {Milosavljevi{\'c}}, and
  {Bromm}}]{pawlik2013}
\bibinfo{author}{A.~H. {Pawlik}}, \bibinfo{author}{M.~{Milosavljevi{\'c}}},
  \bibinfo{author}{V.~{Bromm}}, \bibinfo{journal}{\apj}  \bibinfo{volume}{767}
  (\bibinfo{year}{2013}) \bibinfo{pages}{59}.
\bibitem[{{Pawlik} et~al.(2015){Pawlik}, {Schaye}, and {Dalla
  Vecchia}}]{pawlik2015}
\bibinfo{author}{A.~H. {Pawlik}}, \bibinfo{author}{J.~{Schaye}},
  \bibinfo{author}{C.~{Dalla Vecchia}}, \bibinfo{journal}{\mnras}
  \bibinfo{volume}{451} (\bibinfo{year}{2015}) \bibinfo{pages}{1586--1605}.
\bibitem[{{Rai{\v c}evi{\'c}} et~al.(2011){Rai{\v c}evi{\'c}}, {Theuns}, and
  {Lacey}}]{raicevic2011}
\bibinfo{author}{M.~{Rai{\v c}evi{\'c}}}, \bibinfo{author}{T.~{Theuns}},
  \bibinfo{author}{C.~{Lacey}}, \bibinfo{journal}{\mnras}
  \bibinfo{volume}{410} (\bibinfo{year}{2011}) \bibinfo{pages}{775--787}.
\bibitem[{{Simpson} et~al.(2013){Simpson}, {Bryan}, {Johnston}, {Smith}, {Mac
  Low}, {Sharma}, and {Tumlinson}}]{simpson2013}
\bibinfo{author}{C.~M. {Simpson}}, \bibinfo{author}{G.~L. {Bryan}},
  \bibinfo{author}{K.~V. {Johnston}}, \bibinfo{author}{B.~D. {Smith}},
  \bibinfo{author}{M.-M. {Mac Low}}, \bibinfo{author}{S.~{Sharma}},
  \bibinfo{author}{J.~{Tumlinson}}, \bibinfo{journal}{\mnras}
  \bibinfo{volume}{432} (\bibinfo{year}{2013}) \bibinfo{pages}{1989--2011}.
\bibitem[{{Choudhury} and {Ferrara}(2007)}]{choudhury2007}
\bibinfo{author}{T.~R. {Choudhury}}, \bibinfo{author}{A.~{Ferrara}},
  \bibinfo{journal}{\mnras}  \bibinfo{volume}{380} (\bibinfo{year}{2007})
  \bibinfo{pages}{L6--L10}.
\bibitem[{{Chen} et~al.(2017){Chen}, {Norman}, {Xu}, and {Wise}}]{chen2017}
\bibinfo{author}{P.~{Chen}}, \bibinfo{author}{M.~L. {Norman}},
  \bibinfo{author}{H.~{Xu}}, \bibinfo{author}{J.~H. {Wise}},
  \bibinfo{journal}{ArXiv e-prints}   (\bibinfo{year}{2017}).
\bibitem[{{Salvaterra} et~al.(2011){Salvaterra}, {Ferrara}, and
  {Dayal}}]{salvaterra2011}
\bibinfo{author}{R.~{Salvaterra}}, \bibinfo{author}{A.~{Ferrara}},
  \bibinfo{author}{P.~{Dayal}}, \bibinfo{journal}{\mnras}
  \bibinfo{volume}{414} (\bibinfo{year}{2011}) \bibinfo{pages}{847--859}.
\bibitem[{{Finkelstein} et~al.(2012){Finkelstein}, {Papovich}, {Ryan}, and et.
  al.}]{finkelstein2012}
\bibinfo{author}{S.~L. {Finkelstein}}, \bibinfo{author}{C.~{Papovich}},
  \bibinfo{author}{R.~E. {Ryan}}, \bibinfo{author}{et. al.},
  \bibinfo{journal}{\apj}  \bibinfo{volume}{758} (\bibinfo{year}{2012})
  \bibinfo{pages}{93}.
\bibitem[{{Bouwens} et~al.(2012){Bouwens}, {Illingworth}, {Oesch}, and et.
  al.}]{bouwens2012}
\bibinfo{author}{R.~J. {Bouwens}}, \bibinfo{author}{G.~D. {Illingworth}},
  \bibinfo{author}{P.~A. {Oesch}}, \bibinfo{author}{et. al.},
  \bibinfo{journal}{\apjl}  \bibinfo{volume}{752} (\bibinfo{year}{2012})
  \bibinfo{pages}{L5}.
\bibitem[{{Duncan} and {Conselice}(2015)}]{duncan2015}
\bibinfo{author}{K.~{Duncan}}, \bibinfo{author}{C.~J. {Conselice}},
  \bibinfo{journal}{\mnras}  \bibinfo{volume}{451} (\bibinfo{year}{2015})
  \bibinfo{pages}{2030--2049}.
\bibitem[{{Becker} and {Bolton}(2013)}]{becker2013}
\bibinfo{author}{G.~D. {Becker}}, \bibinfo{author}{J.~S. {Bolton}},
  \bibinfo{journal}{\mnras}  \bibinfo{volume}{436} (\bibinfo{year}{2013})
  \bibinfo{pages}{1023--1039}.
\bibitem[{{Dayal} et~al.(2017){Dayal}, {Choudhury}, {Pacucci}, and
  {Bromm}}]{dayal2017}
\bibinfo{author}{P.~{Dayal}}, \bibinfo{author}{T.~R. {Choudhury}},
  \bibinfo{author}{F.~{Pacucci}}, \bibinfo{author}{V.~{Bromm}},
  \bibinfo{journal}{\mnras}  \bibinfo{volume}{472} (\bibinfo{year}{2017})
  \bibinfo{pages}{4414--4421}.
\bibitem[{{Hassan} et~al.(2018){Hassan}, {Dav{\'e}}, {Mitra}, {Finlator},
  {Ciardi}, and {Santos}}]{hassan2018}
\bibinfo{author}{S.~{Hassan}}, \bibinfo{author}{R.~{Dav{\'e}}},
  \bibinfo{author}{S.~{Mitra}}, \bibinfo{author}{K.~{Finlator}},
  \bibinfo{author}{B.~{Ciardi}}, \bibinfo{author}{M.~G. {Santos}},
  \bibinfo{journal}{\mnras}  \bibinfo{volume}{473} (\bibinfo{year}{2018})
  \bibinfo{pages}{227--240}.
\bibitem[{{Liu} et~al.(2016){Liu}, {Mutch}, {Angel}, {Duffy}, {Geil}, {Poole},
  {Mesinger}, and {Wyithe}}]{liu2016}
\bibinfo{author}{C.~{Liu}}, \bibinfo{author}{S.~J. {Mutch}},
  \bibinfo{author}{P.~W. {Angel}}, \bibinfo{author}{A.~R. {Duffy}},
  \bibinfo{author}{P.~M. {Geil}}, \bibinfo{author}{G.~B. {Poole}},
  \bibinfo{author}{A.~{Mesinger}}, \bibinfo{author}{J.~S.~B. {Wyithe}},
  \bibinfo{journal}{\mnras}  \bibinfo{volume}{462} (\bibinfo{year}{2016})
  \bibinfo{pages}{235--249}.
\bibitem[{{Qin} et~al.(2017){Qin}, {Mutch}, {Poole}, {Liu}, {Angel}, {Duffy},
  {Geil}, {Mesinger}, and {Wyithe}}]{qin2017}
\bibinfo{author}{Y.~{Qin}}, \bibinfo{author}{S.~J. {Mutch}},
  \bibinfo{author}{G.~B. {Poole}}, \bibinfo{author}{C.~{Liu}},
  \bibinfo{author}{P.~W. {Angel}}, \bibinfo{author}{A.~R. {Duffy}},
  \bibinfo{author}{P.~M. {Geil}}, \bibinfo{author}{A.~{Mesinger}},
  \bibinfo{author}{J.~S.~B. {Wyithe}}, \bibinfo{journal}{\mnras}
  \bibinfo{volume}{472} (\bibinfo{year}{2017}) \bibinfo{pages}{2009--2027}.
\bibitem[{{Iliev} et~al.(2012){Iliev}, {Mellema}, {Shapiro}, {Pen}, {Mao},
  {Koda}, and {Ahn}}]{iliev2012}
\bibinfo{author}{I.~T. {Iliev}}, \bibinfo{author}{G.~{Mellema}},
  \bibinfo{author}{P.~R. {Shapiro}}, \bibinfo{author}{U.-L. {Pen}},
  \bibinfo{author}{Y.~{Mao}}, \bibinfo{author}{J.~{Koda}},
  \bibinfo{author}{K.~{Ahn}}, \bibinfo{journal}{\mnras}  \bibinfo{volume}{423}
  (\bibinfo{year}{2012}) \bibinfo{pages}{2222--2253}.
\bibitem[{{Volonteri} and {Gnedin}(2009)}]{volonteri2009}
\bibinfo{author}{M.~{Volonteri}}, \bibinfo{author}{N.~Y. {Gnedin}},
  \bibinfo{journal}{\apj}  \bibinfo{volume}{703} (\bibinfo{year}{2009})
  \bibinfo{pages}{2113--2117}.
\bibitem[{{Madau} and {Haardt}(2015)}]{madau2015}
\bibinfo{author}{P.~{Madau}}, \bibinfo{author}{F.~{Haardt}},
  \bibinfo{journal}{\apjl}  \bibinfo{volume}{813} (\bibinfo{year}{2015})
  \bibinfo{pages}{L8}.
\bibitem[{{Mitra} et~al.(2018){Mitra}, {Choudhury}, and {Ferrara}}]{mitra2018}
\bibinfo{author}{S.~{Mitra}}, \bibinfo{author}{T.~R. {Choudhury}},
  \bibinfo{author}{A.~{Ferrara}}, \bibinfo{journal}{\mnras}
  \bibinfo{volume}{473} (\bibinfo{year}{2018}) \bibinfo{pages}{1416--1425}.
\bibitem[{{Giallongo} et~al.(2015){Giallongo}, {Grazian}, {Fiore}, and et.
  al.}]{giallongo2015}
\bibinfo{author}{E.~{Giallongo}}, \bibinfo{author}{A.~{Grazian}},
  \bibinfo{author}{F.~{Fiore}}, \bibinfo{author}{et. al.},
  \bibinfo{journal}{Astronomy and Astrophysics}  \bibinfo{volume}{578}
  (\bibinfo{year}{2015}) \bibinfo{pages}{A83}.
\bibitem[{{D'Aloisio} et~al.(2017){D'Aloisio}, {Upton Sanderbeck}, {McQuinn},
  {Trac}, and {Shapiro}}]{daloisio2017}
\bibinfo{author}{A.~{D'Aloisio}}, \bibinfo{author}{P.~R. {Upton Sanderbeck}},
  \bibinfo{author}{M.~{McQuinn}}, \bibinfo{author}{H.~{Trac}},
  \bibinfo{author}{P.~R. {Shapiro}}, \bibinfo{journal}{\mnras}
  \bibinfo{volume}{468} (\bibinfo{year}{2017}) \bibinfo{pages}{4691--4701}.
\bibitem[{{Kakiichi} et~al.(2017){Kakiichi}, {Graziani}, {Ciardi}, {Meiksin},
  {Compostella}, {Eide}, and {Zaroubi}}]{kakiichi2017}
\bibinfo{author}{K.~{Kakiichi}}, \bibinfo{author}{L.~{Graziani}},
  \bibinfo{author}{B.~{Ciardi}}, \bibinfo{author}{A.~{Meiksin}},
  \bibinfo{author}{M.~{Compostella}}, \bibinfo{author}{M.~B. {Eide}},
  \bibinfo{author}{S.~{Zaroubi}}, \bibinfo{journal}{\mnras}
  \bibinfo{volume}{468} (\bibinfo{year}{2017}) \bibinfo{pages}{3718--3736}.
\bibitem[{{Bolgar} et~al.(2018){Bolgar}, {Eames}, {Hottier}, and
  {Semelin}}]{bolgar2018}
\bibinfo{author}{F.~{Bolgar}}, \bibinfo{author}{E.~{Eames}},
  \bibinfo{author}{C.~{Hottier}}, \bibinfo{author}{B.~{Semelin}},
  \bibinfo{journal}{ArXiv e-prints}   (\bibinfo{year}{2018}).
\bibitem[{{Mapelli} et~al.(2006){Mapelli}, {Ferrara}, and
  {Pierpaoli}}]{mapelli2006}
\bibinfo{author}{M.~{Mapelli}}, \bibinfo{author}{A.~{Ferrara}},
  \bibinfo{author}{E.~{Pierpaoli}}, \bibinfo{journal}{\mnras}
  \bibinfo{volume}{369} (\bibinfo{year}{2006}) \bibinfo{pages}{1719--1724}.
\bibitem[{{Salvaterra} et~al.(2005){Salvaterra}, {Haardt}, and
  {Ferrara}}]{salvaterra2005}
\bibinfo{author}{R.~{Salvaterra}}, \bibinfo{author}{F.~{Haardt}},
  \bibinfo{author}{A.~{Ferrara}}, \bibinfo{journal}{\mnras}
  \bibinfo{volume}{362} (\bibinfo{year}{2005}) \bibinfo{pages}{L50--L54}.
\bibitem[{{McQuinn}(2012)}]{mcquinn2012}
\bibinfo{author}{M.~{McQuinn}}, \bibinfo{journal}{\mnras}
  \bibinfo{volume}{426} (\bibinfo{year}{2012}) \bibinfo{pages}{1349--1360}.
\bibitem[{{Evoli} et~al.(2014){Evoli}, {Mesinger}, and {Ferrara}}]{evoli2014}
\bibinfo{author}{C.~{Evoli}}, \bibinfo{author}{A.~{Mesinger}},
  \bibinfo{author}{A.~{Ferrara}}, \bibinfo{journal}{\jcap}
  \bibinfo{volume}{11} (\bibinfo{year}{2014}) \bibinfo{pages}{024}.
\bibitem[{{Furlanetto} et~al.(2006){Furlanetto}, {Oh}, and
  {Pierpaoli}}]{furlanetto2006b}
\bibinfo{author}{S.~R. {Furlanetto}}, \bibinfo{author}{S.~P. {Oh}},
  \bibinfo{author}{E.~{Pierpaoli}}, \bibinfo{journal}{\prd}
  \bibinfo{volume}{74} (\bibinfo{year}{2006}) \bibinfo{pages}{103502}.
\bibitem[{{Meier}(1976)}]{meier1976}
\bibinfo{author}{D.~L. {Meier}}, \bibinfo{journal}{Astrophysical Journal}
  \bibinfo{volume}{207} (\bibinfo{year}{1976}) \bibinfo{pages}{343--350}.
\bibitem[{{Gunn} and {Peterson}(1965)}]{gunn-peterson1965}
\bibinfo{author}{J.~E. {Gunn}}, \bibinfo{author}{B.~A. {Peterson}},
  \bibinfo{journal}{Astrophysical Journal}  \bibinfo{volume}{142}
  (\bibinfo{year}{1965}) \bibinfo{pages}{1633--1641}.
\bibitem[{{Becker} et~al.(2001){Becker}, {Fan}, {White}, and et.
  al.}]{becker2001}
\bibinfo{author}{R.~H. {Becker}}, \bibinfo{author}{X.~{Fan}},
  \bibinfo{author}{R.~L. {White}}, \bibinfo{author}{et. al.},
  \bibinfo{journal}{\aj}  \bibinfo{volume}{122} (\bibinfo{year}{2001})
  \bibinfo{pages}{2850--2857}.
\bibitem[{{Oesch} et~al.(2013){Oesch}, {Bouwens}, {Illingworth}, and et.
  al.}]{oesch2013}
\bibinfo{author}{P.~A. {Oesch}}, \bibinfo{author}{R.~J. {Bouwens}},
  \bibinfo{author}{G.~D. {Illingworth}}, \bibinfo{author}{et. al.},
  \bibinfo{journal}{Astrophysical Journal}  \bibinfo{volume}{773}
  (\bibinfo{year}{2013}) \bibinfo{pages}{75}.
\bibitem[{{Oesch} et~al.(2016){Oesch}, {Brammer}, {van Dokkum}, and et.
  al.}]{oesch2016}
\bibinfo{author}{P.~A. {Oesch}}, \bibinfo{author}{G.~{Brammer}},
  \bibinfo{author}{P.~G. {van Dokkum}}, \bibinfo{author}{et. al.},
  \bibinfo{journal}{Astrophysical Journal}  \bibinfo{volume}{819}
  (\bibinfo{year}{2016}) \bibinfo{pages}{129}.
\bibitem[{{Bouwens} et~al.(2007){Bouwens}, {Illingworth}, {Franx}, and
  {Ford}}]{bouwens2007}
\bibinfo{author}{R.~J. {Bouwens}}, \bibinfo{author}{G.~D. {Illingworth}},
  \bibinfo{author}{M.~{Franx}}, \bibinfo{author}{H.~{Ford}},
  \bibinfo{journal}{Astrophysical Journal}  \bibinfo{volume}{670}
  (\bibinfo{year}{2007}) \bibinfo{pages}{928--958}.
\bibitem[{{Bouwens} et~al.(2010){Bouwens}, {Illingworth}, {Gonz{\'a}lez}, and
  et. al.}]{bouwens2010}
\bibinfo{author}{R.~J. {Bouwens}}, \bibinfo{author}{G.~D. {Illingworth}},
  \bibinfo{author}{V.~{Gonz{\'a}lez}}, \bibinfo{author}{et. al.},
  \bibinfo{journal}{Astrophysical Journal}  \bibinfo{volume}{725}
  (\bibinfo{year}{2010}) \bibinfo{pages}{1587--1599}.
\bibitem[{{Bouwens} et~al.(2011{\natexlab{a}}){Bouwens}, {Illingworth},
  {Labbe}, and et. al.}]{bouwens2011a}
\bibinfo{author}{R.~J. {Bouwens}}, \bibinfo{author}{G.~D. {Illingworth}},
  \bibinfo{author}{I.~{Labbe}}, \bibinfo{author}{et. al.},
  \bibinfo{journal}{Nature}  \bibinfo{volume}{469}
  (\bibinfo{year}{2011}{\natexlab{a}}) \bibinfo{pages}{504--507}.
\bibitem[{{Bouwens} et~al.(2011{\natexlab{b}}){Bouwens}, {Illingworth},
  {Oesch}, and et. al.}]{bouwens2011b}
\bibinfo{author}{R.~J. {Bouwens}}, \bibinfo{author}{G.~D. {Illingworth}},
  \bibinfo{author}{P.~A. {Oesch}}, \bibinfo{author}{et. al.},
  \bibinfo{journal}{Astrophysical Journal}  \bibinfo{volume}{737}
  (\bibinfo{year}{2011}{\natexlab{b}}) \bibinfo{pages}{90}.
\bibitem[{{Bowler} et~al.(2014){Bowler}, {Dunlop}, {McLure}, and et.
  al.}]{bowler2014}
\bibinfo{author}{R.~A.~A. {Bowler}}, \bibinfo{author}{J.~S. {Dunlop}},
  \bibinfo{author}{R.~J. {McLure}}, \bibinfo{author}{et. al.},
  \bibinfo{journal}{Monthly Notices of the Royal Astronomical Society}
  \bibinfo{volume}{440} (\bibinfo{year}{2014}) \bibinfo{pages}{2810--2842}.
\bibitem[{{Bowler} et~al.(2015){Bowler}, {Dunlop}, {McLure}, and et.
  al.}]{bowler2015}
\bibinfo{author}{R.~A.~A. {Bowler}}, \bibinfo{author}{J.~S. {Dunlop}},
  \bibinfo{author}{R.~J. {McLure}}, \bibinfo{author}{et. al.},
  \bibinfo{journal}{Monthly Notices of the Royal Astronomical Society}
  \bibinfo{volume}{452} (\bibinfo{year}{2015}) \bibinfo{pages}{1817--1840}.
\bibitem[{{McLure} et~al.(2010){McLure}, {Dunlop}, and
  {Cirasuolo}}]{mclure2010}
\bibinfo{author}{R.~J. {McLure}}, \bibinfo{author}{J.~S. {Dunlop}},
  \bibinfo{author}{M.~e.~a. {Cirasuolo}}, \bibinfo{journal}{Monthly Notices of
  the Royal Astronomical Society}  \bibinfo{volume}{403} (\bibinfo{year}{2010})
  \bibinfo{pages}{960--983}.
\bibitem[{{McLure} et~al.(2011){McLure}, {Dunlop}, {de Ravel}, and et.
  al.}]{mclure2011}
\bibinfo{author}{R.~J. {McLure}}, \bibinfo{author}{J.~S. {Dunlop}},
  \bibinfo{author}{L.~{de Ravel}}, \bibinfo{author}{et. al.},
  \bibinfo{journal}{Monthly Notices of the Royal Astronomical Society}
  \bibinfo{volume}{418} (\bibinfo{year}{2011}) \bibinfo{pages}{2074--2105}.
\bibitem[{{McLure} et~al.(2013){McLure}, {Dunlop}, {Bowler}, and et.
  al.}]{mclure2013}
\bibinfo{author}{R.~J. {McLure}}, \bibinfo{author}{J.~S. {Dunlop}},
  \bibinfo{author}{R.~A.~A. {Bowler}}, \bibinfo{author}{et. al.},
  \bibinfo{journal}{Monthly Notices of the Royal Astronomical Society}
  \bibinfo{volume}{432} (\bibinfo{year}{2013}) \bibinfo{pages}{2696--2716}.
\bibitem[{{Bradley} et~al.(2012){Bradley}, {Trenti}, {Oesch}, and et.
  al.}]{bradley2012}
\bibinfo{author}{L.~D. {Bradley}}, \bibinfo{author}{M.~{Trenti}},
  \bibinfo{author}{P.~A. {Oesch}}, \bibinfo{author}{et. al.},
  \bibinfo{journal}{Astrophysical Journal}  \bibinfo{volume}{760}
  (\bibinfo{year}{2012}) \bibinfo{pages}{108}.
\bibitem[{{Bradley} et~al.(2014){Bradley}, {Zitrin}, {Coe}, and et.
  al.}]{bradley2014}
\bibinfo{author}{L.~D. {Bradley}}, \bibinfo{author}{A.~{Zitrin}},
  \bibinfo{author}{D.~{Coe}}, \bibinfo{author}{et. al.},
  \bibinfo{journal}{Astrophysical Journal}  \bibinfo{volume}{792}
  (\bibinfo{year}{2014}) \bibinfo{pages}{76}.
\bibitem[{{Castellano} et~al.(2010{\natexlab{a}}){Castellano}, {Fontana},
  {Paris}, and et. al.}]{castellano2010a}
\bibinfo{author}{M.~{Castellano}}, \bibinfo{author}{A.~{Fontana}},
  \bibinfo{author}{D.~{Paris}}, \bibinfo{author}{et. al.},
  \bibinfo{journal}{Astronomy and Astrophysics}  \bibinfo{volume}{524}
  (\bibinfo{year}{2010}{\natexlab{a}}) \bibinfo{pages}{A28}.
\bibitem[{{Castellano} et~al.(2010{\natexlab{b}}){Castellano}, {Fontana},
  {Boutsia}, and et. al.}]{castellano2010b}
\bibinfo{author}{M.~{Castellano}}, \bibinfo{author}{A.~{Fontana}},
  \bibinfo{author}{K.~{Boutsia}}, \bibinfo{author}{et. al.},
  \bibinfo{journal}{Astronomy and Astrophysics}  \bibinfo{volume}{511}
  (\bibinfo{year}{2010}{\natexlab{b}}) \bibinfo{pages}{A20}.
\bibitem[{{Osterbrock}(1989)}]{osterbrock1989}
\bibinfo{author}{D.~E. {Osterbrock}}, \bibinfo{title}{{Astrophysics of gaseous
  nebulae and active galactic nuclei}},  \bibinfo{year}{1989}.
\bibitem[{{Dey} et~al.(1998){Dey}, {Spinrad}, {Stern}, {Graham}, and
  {Chaffee}}]{dey1998}
\bibinfo{author}{A.~{Dey}}, \bibinfo{author}{H.~{Spinrad}},
  \bibinfo{author}{D.~{Stern}}, \bibinfo{author}{J.~R. {Graham}},
  \bibinfo{author}{F.~H. {Chaffee}}, \bibinfo{journal}{Astrophys. J. Letters}
  \bibinfo{volume}{498} (\bibinfo{year}{1998}) \bibinfo{pages}{L93--L97}.
\bibitem[{{Spinrad} et~al.(1998){Spinrad}, {Stern}, {Bunker}, {Dey},
  {Lanzetta}, {Yahil}, {Pascarelle}, and {Fern{\'a}ndez-Soto}}]{spinrad1998}
\bibinfo{author}{H.~{Spinrad}}, \bibinfo{author}{D.~{Stern}},
  \bibinfo{author}{A.~{Bunker}}, \bibinfo{author}{A.~{Dey}},
  \bibinfo{author}{K.~{Lanzetta}}, \bibinfo{author}{A.~{Yahil}},
  \bibinfo{author}{S.~{Pascarelle}}, \bibinfo{author}{A.~{Fern{\'a}ndez-Soto}},
  \bibinfo{journal}{Astrophysical Journal}  \bibinfo{volume}{116}
  (\bibinfo{year}{1998}) \bibinfo{pages}{2617--2623}.
\bibitem[{{Hu} et~al.(1998){Hu}, {Cowie}, and {McMahon}}]{hu1998}
\bibinfo{author}{E.~M. {Hu}}, \bibinfo{author}{L.~L. {Cowie}},
  \bibinfo{author}{R.~G. {McMahon}}, \bibinfo{journal}{\apjl}
  \bibinfo{volume}{502} (\bibinfo{year}{1998}) \bibinfo{pages}{L99--L103}.
\bibitem[{{Ouchi} et~al.(2005){Ouchi}, {Shimasaku}, {Akiyama}, and et.
  al.}]{ouchi2005}
\bibinfo{author}{M.~{Ouchi}}, \bibinfo{author}{K.~{Shimasaku}},
  \bibinfo{author}{M.~{Akiyama}}, \bibinfo{author}{et. al.},
  \bibinfo{journal}{Astrophysical Journal Letters}  \bibinfo{volume}{620}
  (\bibinfo{year}{2005}) \bibinfo{pages}{L1--L4}.
\bibitem[{{Taniguchi} et~al.(2005){Taniguchi}, {Ajiki}, {Nagao}, and et.
  al.}]{taniguchi2005}
\bibinfo{author}{Y.~{Taniguchi}}, \bibinfo{author}{M.~{Ajiki}},
  \bibinfo{author}{T.~{Nagao}}, \bibinfo{author}{et. al.},
  \bibinfo{journal}{Publications of the Astronomical Society of Japan}
  \bibinfo{volume}{57} (\bibinfo{year}{2005}) \bibinfo{pages}{165--182}.
\bibitem[{{Ouchi} et~al.(2008){Ouchi}, {Shimasaku}, {Akiyama}, and et.
  al.}]{ouchi2008}
\bibinfo{author}{M.~{Ouchi}}, \bibinfo{author}{K.~{Shimasaku}},
  \bibinfo{author}{M.~{Akiyama}}, \bibinfo{author}{et. al.},
  \bibinfo{journal}{Astrophysical Journal Supplementary Series}
  \bibinfo{volume}{176} (\bibinfo{year}{2008}) \bibinfo{pages}{301--330}.
\bibitem[{{Ouchi} et~al.(2010){Ouchi}, {Shimasaku}, {Furusawa}, and et.
  al.}]{ouchi2010}
\bibinfo{author}{M.~{Ouchi}}, \bibinfo{author}{K.~{Shimasaku}},
  \bibinfo{author}{H.~{Furusawa}}, \bibinfo{author}{et. al.},
  \bibinfo{journal}{astrophysical Journal}  \bibinfo{volume}{723}
  (\bibinfo{year}{2010}) \bibinfo{pages}{869--894}.
\bibitem[{{Iye} et~al.(2006){Iye}, {Ota}, {Kashikawa}, {Furusawa}, {Hashimoto},
  {Hattori}, {Matsuda}, {Morokuma}, {Ouchi}, and {Shimasaku}}]{iye2006}
\bibinfo{author}{M.~{Iye}}, \bibinfo{author}{K.~{Ota}},
  \bibinfo{author}{N.~{Kashikawa}}, \bibinfo{author}{H.~{Furusawa}},
  \bibinfo{author}{T.~{Hashimoto}}, \bibinfo{author}{T.~{Hattori}},
  \bibinfo{author}{Y.~{Matsuda}}, \bibinfo{author}{T.~{Morokuma}},
  \bibinfo{author}{M.~{Ouchi}}, \bibinfo{author}{K.~{Shimasaku}},
  \bibinfo{journal}{Nature}  \bibinfo{volume}{443} (\bibinfo{year}{2006})
  \bibinfo{pages}{186--188}.
\bibitem[{{Kashikawa} et~al.(2006){Kashikawa}, {Shimasaku}, {Malkan}, and et.
  al.}]{kashikawa2006}
\bibinfo{author}{N.~{Kashikawa}}, \bibinfo{author}{K.~{Shimasaku}},
  \bibinfo{author}{M.~A. {Malkan}}, \bibinfo{author}{et. al.},
  \bibinfo{journal}{Astrophysical Journal}  \bibinfo{volume}{648}
  (\bibinfo{year}{2006}) \bibinfo{pages}{7--22}.
\bibitem[{{Ota} et~al.(2008){Ota}, {Iye}, {Kashikawa}, and et. al.}]{ota2008}
\bibinfo{author}{K.~{Ota}}, \bibinfo{author}{M.~{Iye}},
  \bibinfo{author}{N.~{Kashikawa}}, \bibinfo{author}{et. al.},
  \bibinfo{journal}{Astrophysical Journal}  \bibinfo{volume}{677}
  (\bibinfo{year}{2008}) \bibinfo{pages}{12--26}.
\bibitem[{{Ota} et~al.(2010){Ota}, {Iye}, {Kashikawa}, and et. al.}]{ota2010}
\bibinfo{author}{K.~{Ota}}, \bibinfo{author}{M.~{Iye}},
  \bibinfo{author}{N.~{Kashikawa}}, \bibinfo{author}{et. al.},
  \bibinfo{journal}{astrophysica Journal}  \bibinfo{volume}{722}
  (\bibinfo{year}{2010}) \bibinfo{pages}{803--811}.
\bibitem[{{Taniguchi} et~al.(2009){Taniguchi}, {Murayama}, {Scoville}, and et.
  al.}]{taniguchi2009}
\bibinfo{author}{Y.~{Taniguchi}}, \bibinfo{author}{T.~{Murayama}},
  \bibinfo{author}{N.~Z. {Scoville}}, \bibinfo{author}{et. al.},
  \bibinfo{journal}{Astrophysical Journal}  \bibinfo{volume}{701}
  (\bibinfo{year}{2009}) \bibinfo{pages}{915--944}.
\bibitem[{{Kashikawa} et~al.(2011){Kashikawa}, {Shimasaku}, {Matsuda}, and et.
  al.}]{kashikawa2011}
\bibinfo{author}{N.~{Kashikawa}}, \bibinfo{author}{K.~{Shimasaku}},
  \bibinfo{author}{Y.~{Matsuda}}, \bibinfo{author}{et. al.},
  \bibinfo{journal}{Astrophysical Journal}  \bibinfo{volume}{734}
  (\bibinfo{year}{2011}) \bibinfo{pages}{119}.
\bibitem[{{Matthee} et~al.(2014){Matthee}, {Sobral}, {Swinbank}, and
  et~al.}]{matthee2014}
\bibinfo{author}{J.~J.~A. {Matthee}}, \bibinfo{author}{D.~{Sobral}},
  \bibinfo{author}{A.~M. {Swinbank}}, \bibinfo{author}{et~al.},
  \bibinfo{journal}{Monthly Notices of the Royal Astronomical Society}
  \bibinfo{volume}{440} (\bibinfo{year}{2014}) \bibinfo{pages}{2375--2387}.
\bibitem[{{Matthee} et~al.(2015){Matthee}, {Sobral}, {Santos},
  {R{\"o}ttgering}, {Darvish}, and {Mobasher}}]{matthee2015}
\bibinfo{author}{J.~{Matthee}}, \bibinfo{author}{D.~{Sobral}},
  \bibinfo{author}{S.~{Santos}}, \bibinfo{author}{H.~{R{\"o}ttgering}},
  \bibinfo{author}{B.~{Darvish}}, \bibinfo{author}{B.~{Mobasher}},
  \bibinfo{journal}{Monthly Notices of the Royal Astronomical Society}
  \bibinfo{volume}{451} (\bibinfo{year}{2015}) \bibinfo{pages}{400--417}.
\bibitem[{{Ouchi} et~al.(2018){Ouchi}, {Harikane}, {Shibuya}, and et.
  al.}]{ouchi2017}
\bibinfo{author}{M.~{Ouchi}}, \bibinfo{author}{Y.~{Harikane}},
  \bibinfo{author}{T.~{Shibuya}}, \bibinfo{author}{et. al.},
  \bibinfo{journal}{\pasj}  \bibinfo{volume}{70} (\bibinfo{year}{2018})
  \bibinfo{pages}{S13}.
\bibitem[{{Madau} and {Rees}(2000)}]{madau2000}
\bibinfo{author}{P.~{Madau}}, \bibinfo{author}{M.~J. {Rees}},
  \bibinfo{journal}{\apjl}  \bibinfo{volume}{542} (\bibinfo{year}{2000})
  \bibinfo{pages}{L69--L73}.
\bibitem[{{Hu} et~al.(2010){Hu}, {Cowie}, {Barger}, {Capak}, {Kakazu}, and
  {Trouille}}]{hu2010}
\bibinfo{author}{E.~M. {Hu}}, \bibinfo{author}{L.~L. {Cowie}},
  \bibinfo{author}{A.~J. {Barger}}, \bibinfo{author}{P.~{Capak}},
  \bibinfo{author}{Y.~{Kakazu}}, \bibinfo{author}{L.~{Trouille}},
  \bibinfo{journal}{\apj}  \bibinfo{volume}{725} (\bibinfo{year}{2010})
  \bibinfo{pages}{394--423}.
\bibitem[{{Santos} et~al.(2016){Santos}, {Sobral}, and {Matthee}}]{santos2016}
\bibinfo{author}{S.~{Santos}}, \bibinfo{author}{D.~{Sobral}},
  \bibinfo{author}{J.~{Matthee}}, \bibinfo{journal}{\mnras}
  \bibinfo{volume}{463} (\bibinfo{year}{2016}) \bibinfo{pages}{1678--1691}.
\bibitem[{{Ota} et~al.(2017){Ota}, {Iye}, {Kashikawa}, and et. al.}]{ota2017}
\bibinfo{author}{K.~{Ota}}, \bibinfo{author}{M.~{Iye}},
  \bibinfo{author}{N.~{Kashikawa}}, \bibinfo{author}{et. al.},
  \bibinfo{journal}{\apj}  \bibinfo{volume}{844} (\bibinfo{year}{2017})
  \bibinfo{pages}{85}.
\bibitem[{{Stark} et~al.(2015){Stark}, {Richard}, {Charlot}, and et.
  al.}]{stark2015}
\bibinfo{author}{D.~P. {Stark}}, \bibinfo{author}{J.~{Richard}},
  \bibinfo{author}{S.~{Charlot}}, \bibinfo{author}{et. al.},
  \bibinfo{journal}{\mnras}  \bibinfo{volume}{450} (\bibinfo{year}{2015})
  \bibinfo{pages}{1846--1855}.
\bibitem[{{Smit} et~al.(2015){Smit}, {Bouwens}, {Franx}, and et.
  al.}]{smit2015}
\bibinfo{author}{R.~{Smit}}, \bibinfo{author}{R.~J. {Bouwens}},
  \bibinfo{author}{M.~{Franx}}, \bibinfo{author}{et. al.},
  \bibinfo{journal}{\apj}  \bibinfo{volume}{801} (\bibinfo{year}{2015})
  \bibinfo{pages}{122}.
\bibitem[{{Roberts-Borsani} et~al.(2016){Roberts-Borsani}, {Bouwens}, {Oesch},
  and et. al.}]{roberts-borsani2016}
\bibinfo{author}{G.~W. {Roberts-Borsani}}, \bibinfo{author}{R.~J. {Bouwens}},
  \bibinfo{author}{P.~A. {Oesch}}, \bibinfo{author}{et. al.},
  \bibinfo{journal}{\apj}  \bibinfo{volume}{823} (\bibinfo{year}{2016})
  \bibinfo{pages}{143}.
\bibitem[{{Stark} et~al.(2015){Stark}, {Walth}, {Charlot}, and et.
  al.}]{stark2015b}
\bibinfo{author}{D.~P. {Stark}}, \bibinfo{author}{G.~{Walth}},
  \bibinfo{author}{S.~{Charlot}}, \bibinfo{author}{et. al.},
  \bibinfo{journal}{\mnras}  \bibinfo{volume}{454} (\bibinfo{year}{2015})
  \bibinfo{pages}{1393--1403}.
\bibitem[{{Schmidt} et~al.(2017){Schmidt}, {Huang}, {Treu}, and et.
  al.}]{schmidt2017}
\bibinfo{author}{K.~B. {Schmidt}}, \bibinfo{author}{K.-H. {Huang}},
  \bibinfo{author}{T.~{Treu}}, \bibinfo{author}{et. al.},
  \bibinfo{journal}{\apj}  \bibinfo{volume}{839} (\bibinfo{year}{2017})
  \bibinfo{pages}{17}.
\bibitem[{{Verhamme} et~al.(2008){Verhamme}, {Schaerer}, {Atek}, and
  {Tapken}}]{verhamme2008}
\bibinfo{author}{A.~{Verhamme}}, \bibinfo{author}{D.~{Schaerer}},
  \bibinfo{author}{H.~{Atek}}, \bibinfo{author}{C.~{Tapken}},
  \bibinfo{journal}{\aap}  \bibinfo{volume}{491} (\bibinfo{year}{2008})
  \bibinfo{pages}{89--111}.
\bibitem[{{Dayal} and {Ferrara}(2012)}]{dayal2012}
\bibinfo{author}{P.~{Dayal}}, \bibinfo{author}{A.~{Ferrara}},
  \bibinfo{journal}{\mnras}  \bibinfo{volume}{421} (\bibinfo{year}{2012})
  \bibinfo{pages}{2568--2579}.
\bibitem[{{Hutter} et~al.(2015){Hutter}, {Dayal}, and
  {M{\"u}ller}}]{hutter2015}
\bibinfo{author}{A.~{Hutter}}, \bibinfo{author}{P.~{Dayal}},
  \bibinfo{author}{V.~{M{\"u}ller}}, \bibinfo{journal}{\mnras}
  \bibinfo{volume}{450} (\bibinfo{year}{2015}) \bibinfo{pages}{4025--4034}.
\bibitem[{{Stark} et~al.(2011){Stark}, {Ellis}, and {Ouchi}}]{stark2011}
\bibinfo{author}{D.~P. {Stark}}, \bibinfo{author}{R.~S. {Ellis}},
  \bibinfo{author}{M.~{Ouchi}}, \bibinfo{journal}{\apjl}  \bibinfo{volume}{728}
  (\bibinfo{year}{2011}) \bibinfo{pages}{L2}.
\bibitem[{{Pentericci} et~al.(2014){Pentericci}, {Vanzella}, {Fontana}, and
  et~al.}]{pentericci2014}
\bibinfo{author}{L.~{Pentericci}}, \bibinfo{author}{E.~{Vanzella}},
  \bibinfo{author}{A.~{Fontana}}, \bibinfo{author}{et~al.},
  \bibinfo{journal}{\apj}  \bibinfo{volume}{793} (\bibinfo{year}{2014})
  \bibinfo{pages}{113}.
\bibitem[{{Hutter} et~al.(2014){Hutter}, {Dayal}, {Partl}, and
  {M{\"u}ller}}]{hutter2014}
\bibinfo{author}{A.~{Hutter}}, \bibinfo{author}{P.~{Dayal}},
  \bibinfo{author}{A.~M. {Partl}}, \bibinfo{author}{V.~{M{\"u}ller}},
  \bibinfo{journal}{\mnras}  \bibinfo{volume}{441} (\bibinfo{year}{2014})
  \bibinfo{pages}{2861--2877}.
\bibitem[{{Dunlop}(2013)}]{dunlop2013}
\bibinfo{author}{J.~S. {Dunlop}}, volume \bibinfo{volume}{396},  p.
  \bibinfo{pages}{223}.
\bibitem[{{Stark}(2016)}]{stark2016}
\bibinfo{author}{D.~P. {Stark}}, \bibinfo{journal}{\araa}  \bibinfo{volume}{54}
  (\bibinfo{year}{2016}) \bibinfo{pages}{761--803}.
\bibitem[{{Wiklind} et~al.(2008){Wiklind}, {Dickinson}, {Ferguson},
  {Giavalisco}, {Mobasher}, {Grogin}, and {Panagia}}]{wiklind2008}
\bibinfo{author}{T.~{Wiklind}}, \bibinfo{author}{M.~{Dickinson}},
  \bibinfo{author}{H.~C. {Ferguson}}, \bibinfo{author}{M.~{Giavalisco}},
  \bibinfo{author}{B.~{Mobasher}}, \bibinfo{author}{N.~A. {Grogin}},
  \bibinfo{author}{N.~{Panagia}}, \bibinfo{journal}{Astrophysical Journal}
  \bibinfo{volume}{676} (\bibinfo{year}{2008}) \bibinfo{pages}{781--806}.
\bibitem[{{Willott} et~al.(2010){Willott}, {Delorme}, {Reyl{\'e}}, and et.
  al.}]{willott2010}
\bibinfo{author}{C.~J. {Willott}}, \bibinfo{author}{P.~{Delorme}},
  \bibinfo{author}{C.~{Reyl{\'e}}}, \bibinfo{author}{et. al.},
  \bibinfo{journal}{Astrophysical Journal}  \bibinfo{volume}{139}
  (\bibinfo{year}{2010}) \bibinfo{pages}{906--918}.
\bibitem[{{Mortlock} et~al.(2010){Mortlock}, {Warren}, {Venemans}, and et.
  al.}]{mortlock2011}
\bibinfo{author}{D.~J. {Mortlock}}, \bibinfo{author}{S.~J. {Warren}},
  \bibinfo{author}{B.~P. {Venemans}}, \bibinfo{author}{et. al.},
  \bibinfo{journal}{\apj}  \bibinfo{volume}{725} (\bibinfo{year}{2010})
  \bibinfo{pages}{1011--1031}.
\bibitem[{{Willott} et~al.(2015){Willott}, {Bergeron}, and
  {Omont}}]{willott2015}
\bibinfo{author}{C.~J. {Willott}}, \bibinfo{author}{J.~{Bergeron}},
  \bibinfo{author}{A.~{Omont}}, \bibinfo{journal}{Astrophysical Journal}
  \bibinfo{volume}{801} (\bibinfo{year}{2015}) \bibinfo{pages}{123}.
\bibitem[{{Kashikawa} et~al.(2015){Kashikawa}, {Ishizaki}, {Willott}, and et.
  al.}]{kashikawa2015}
\bibinfo{author}{N.~{Kashikawa}}, \bibinfo{author}{Y.~{Ishizaki}},
  \bibinfo{author}{C.~J. {Willott}}, \bibinfo{author}{et. al.},
  \bibinfo{journal}{Astrophysical Journal}  \bibinfo{volume}{798}
  (\bibinfo{year}{2015}) \bibinfo{pages}{28}.
\bibitem[{{Jiang} et~al.(2016){Jiang}, {McGreer}, {Fan}, and et.
  al.}]{jiang2016}
\bibinfo{author}{L.~{Jiang}}, \bibinfo{author}{I.~D. {McGreer}},
  \bibinfo{author}{X.~{Fan}}, \bibinfo{author}{et. al.},
  \bibinfo{journal}{\apj}  \bibinfo{volume}{833} (\bibinfo{year}{2016})
  \bibinfo{pages}{222}.
\bibitem[{{Ono} et~al.(2017){Ono}, {Ouchi}, and {Harikane}}]{ono2017}
\bibinfo{author}{Y.~{Ono}}, \bibinfo{author}{M.~{Ouchi}},
  \bibinfo{author}{Y.~e.~a. {Harikane}}, \bibinfo{journal}{\pasj}
  (\bibinfo{year}{2017}).
\bibitem[{{Woosley} and {Bloom}(2006)}]{woosley2006}
\bibinfo{author}{S.~E. {Woosley}}, \bibinfo{author}{J.~S. {Bloom}},
  \bibinfo{journal}{Annual Reviews of Astronomy and Astrophysics}
  \bibinfo{volume}{44} (\bibinfo{year}{2006}) \bibinfo{pages}{507--556}.
\bibitem[{{Salvaterra} et~al.(2009){Salvaterra}, {Della Valle}, {Campana}, and
  et. al.}]{salvaterra2009}
\bibinfo{author}{R.~{Salvaterra}}, \bibinfo{author}{M.~{Della Valle}},
  \bibinfo{author}{S.~{Campana}}, \bibinfo{author}{et. al.},
  \bibinfo{journal}{Nature}  \bibinfo{volume}{461} (\bibinfo{year}{2009})
  \bibinfo{pages}{1258--1260}.
\bibitem[{{Tanvir} et~al.(2009){Tanvir}, {Fox}, {Levan}, and et.
  al.}]{tanvir2009}
\bibinfo{author}{N.~R. {Tanvir}}, \bibinfo{author}{D.~B. {Fox}},
  \bibinfo{author}{A.~J. {Levan}}, \bibinfo{author}{et. al.},
  \bibinfo{journal}{Nature}  \bibinfo{volume}{461} (\bibinfo{year}{2009})
  \bibinfo{pages}{1254--1257}.
\bibitem[{{Schechter}(1976)}]{schechter1976}
\bibinfo{author}{P.~{Schechter}}, \bibinfo{journal}{\apj}
  \bibinfo{volume}{203} (\bibinfo{year}{1976}) \bibinfo{pages}{297--306}.
\bibitem[{{Oesch} et~al.(2010){Oesch}, {Bouwens}, {Carollo}, and et.
  al.}]{oesch2010}
\bibinfo{author}{P.~A. {Oesch}}, \bibinfo{author}{R.~J. {Bouwens}},
  \bibinfo{author}{C.~M. {Carollo}}, \bibinfo{author}{et. al.},
  \bibinfo{journal}{\apjl}  \bibinfo{volume}{709} (\bibinfo{year}{2010})
  \bibinfo{pages}{L21--L25}.
\bibitem[{{Bouwens} et~al.(2017){Bouwens}, {Oesch}, {Illingworth}, {Ellis}, and
  {Stefanon}}]{bouwens2017b}
\bibinfo{author}{R.~J. {Bouwens}}, \bibinfo{author}{P.~A. {Oesch}},
  \bibinfo{author}{G.~D. {Illingworth}}, \bibinfo{author}{R.~S. {Ellis}},
  \bibinfo{author}{M.~{Stefanon}}, \bibinfo{journal}{\apj}
  \bibinfo{volume}{843} (\bibinfo{year}{2017}) \bibinfo{pages}{129}.
\bibitem[{{McLeod} et~al.(2016){McLeod}, {McLure}, and {Dunlop}}]{mcleod2016}
\bibinfo{author}{D.~J. {McLeod}}, \bibinfo{author}{R.~J. {McLure}},
  \bibinfo{author}{J.~S. {Dunlop}}, \bibinfo{journal}{\mnras}
  \bibinfo{volume}{459} (\bibinfo{year}{2016}) \bibinfo{pages}{3812--3824}.
\bibitem[{{Bouwens} et~al.(2016){Bouwens}, {Oesch}, {Labb{\'e}}, and et.
  al.}]{bouwens2016b}
\bibinfo{author}{R.~J. {Bouwens}}, \bibinfo{author}{P.~A. {Oesch}},
  \bibinfo{author}{I.~{Labb{\'e}}}, \bibinfo{author}{et. al.},
  \bibinfo{journal}{\apj}  \bibinfo{volume}{830} (\bibinfo{year}{2016})
  \bibinfo{pages}{67}.
\bibitem[{{Dayal} et~al.(2013){Dayal}, {Dunlop}, {Maio}, and
  {Ciardi}}]{dayal2013}
\bibinfo{author}{P.~{Dayal}}, \bibinfo{author}{J.~S. {Dunlop}},
  \bibinfo{author}{U.~{Maio}}, \bibinfo{author}{B.~{Ciardi}},
  \bibinfo{journal}{\mnras}  \bibinfo{volume}{434} (\bibinfo{year}{2013})
  \bibinfo{pages}{1486--1504}.
\bibitem[{{Grazian} et~al.(2012){Grazian}, {Castellano}, {Fontana}, and et.
  al.}]{grazian2012}
\bibinfo{author}{A.~{Grazian}}, \bibinfo{author}{M.~{Castellano}},
  \bibinfo{author}{A.~{Fontana}}, \bibinfo{author}{et. al.},
  \bibinfo{journal}{\aap}  \bibinfo{volume}{547} (\bibinfo{year}{2012})
  \bibinfo{pages}{A51}.
\bibitem[{{Bouwens} et~al.(2017){Bouwens}, {Illingworth}, {Oesch}, {Atek},
  {Lam}, and {Stefanon}}]{bouwens2017}
\bibinfo{author}{R.~J. {Bouwens}}, \bibinfo{author}{G.~D. {Illingworth}},
  \bibinfo{author}{P.~A. {Oesch}}, \bibinfo{author}{H.~{Atek}},
  \bibinfo{author}{D.~{Lam}}, \bibinfo{author}{M.~{Stefanon}},
  \bibinfo{journal}{\apj}  \bibinfo{volume}{843} (\bibinfo{year}{2017})
  \bibinfo{pages}{41}.
\bibitem[{{Oesch} et~al.(2018){Oesch}, {Bouwens}, {Illingworth}, {Labb{\'e}},
  and {Stefanon}}]{oesch2017}
\bibinfo{author}{P.~A. {Oesch}}, \bibinfo{author}{R.~J. {Bouwens}},
  \bibinfo{author}{G.~D. {Illingworth}}, \bibinfo{author}{I.~{Labb{\'e}}},
  \bibinfo{author}{M.~{Stefanon}}, \bibinfo{journal}{\apj}
  \bibinfo{volume}{855} (\bibinfo{year}{2018}) \bibinfo{pages}{105}.
\bibitem[{{Gnedin}(2016)}]{gnedin2016}
\bibinfo{author}{N.~Y. {Gnedin}}, \bibinfo{journal}{\apjl}
  \bibinfo{volume}{825} (\bibinfo{year}{2016}) \bibinfo{pages}{L17}.
\bibitem[{{Dijkstra} et~al.(2007){Dijkstra}, {Wyithe}, and
  {Haiman}}]{dijkstra2007}
\bibinfo{author}{M.~{Dijkstra}}, \bibinfo{author}{J.~S.~B. {Wyithe}},
  \bibinfo{author}{Z.~{Haiman}}, \bibinfo{journal}{\mnras}
  \bibinfo{volume}{379} (\bibinfo{year}{2007}) \bibinfo{pages}{253--259}.
\bibitem[{{Dayal} et~al.(2008){Dayal}, {Ferrara}, and {Gallerani}}]{dayal2008}
\bibinfo{author}{P.~{Dayal}}, \bibinfo{author}{A.~{Ferrara}},
  \bibinfo{author}{S.~{Gallerani}}, \bibinfo{journal}{\mnras}
  \bibinfo{volume}{389} (\bibinfo{year}{2008}) \bibinfo{pages}{1683--1696}.
\bibitem[{{Dayal} et~al.(2010){Dayal}, {Ferrara}, and {Saro}}]{dayal2010}
\bibinfo{author}{P.~{Dayal}}, \bibinfo{author}{A.~{Ferrara}},
  \bibinfo{author}{A.~{Saro}}, \bibinfo{journal}{\mnras}  \bibinfo{volume}{402}
  (\bibinfo{year}{2010}) \bibinfo{pages}{1449--1457}.
\bibitem[{{Bolton} and {Haehnelt}(2013)}]{bolton2013}
\bibinfo{author}{J.~S. {Bolton}}, \bibinfo{author}{M.~G. {Haehnelt}},
  \bibinfo{journal}{\mnras}  \bibinfo{volume}{429} (\bibinfo{year}{2013})
  \bibinfo{pages}{1695--1704}.
\bibitem[{{Weinberger} et~al.(2018){Weinberger}, {Kulkarni}, {Haehnelt},
  {Choudhury}, and {Puchwein}}]{weinberger2018}
\bibinfo{author}{L.~H. {Weinberger}}, \bibinfo{author}{G.~{Kulkarni}},
  \bibinfo{author}{M.~G. {Haehnelt}}, \bibinfo{author}{T.~R. {Choudhury}},
  \bibinfo{author}{E.~{Puchwein}}, \bibinfo{journal}{\mnras}
  \bibinfo{volume}{479} (\bibinfo{year}{2018}) \bibinfo{pages}{2564--2587}.
\bibitem[{{Dijkstra} and {Wyithe}(2012)}]{dijkstra2012}
\bibinfo{author}{M.~{Dijkstra}}, \bibinfo{author}{J.~S.~B. {Wyithe}},
  \bibinfo{journal}{\mnras}  \bibinfo{volume}{419} (\bibinfo{year}{2012})
  \bibinfo{pages}{3181--3193}.
\bibitem[{{Steidel} et~al.(2011){Steidel}, {Bogosavljevi{\'c}}, {Shapley},
  {Kollmeier}, {Reddy}, {Erb}, and {Pettini}}]{steidel2011}
\bibinfo{author}{C.~C. {Steidel}}, \bibinfo{author}{M.~{Bogosavljevi{\'c}}},
  \bibinfo{author}{A.~E. {Shapley}}, \bibinfo{author}{J.~A. {Kollmeier}},
  \bibinfo{author}{N.~A. {Reddy}}, \bibinfo{author}{D.~K. {Erb}},
  \bibinfo{author}{M.~{Pettini}}, \bibinfo{journal}{\apj}
  \bibinfo{volume}{736} (\bibinfo{year}{2011}) \bibinfo{pages}{160}.
\bibitem[{{Leclercq} et~al.(2017){Leclercq}, {Bacon}, {Wisotzki}, and et.
  al.}]{leclercq2017}
\bibinfo{author}{F.~{Leclercq}}, \bibinfo{author}{R.~{Bacon}},
  \bibinfo{author}{L.~{Wisotzki}}, \bibinfo{author}{et. al.},
  \bibinfo{journal}{\aap}  \bibinfo{volume}{608} (\bibinfo{year}{2017})
  \bibinfo{pages}{A8}.
\bibitem[{{De Barros} et~al.(2017){De Barros}, {Pentericci}, {Vanzella}, and
  et. al.}]{debarros2017}
\bibinfo{author}{S.~{De Barros}}, \bibinfo{author}{L.~{Pentericci}},
  \bibinfo{author}{E.~{Vanzella}}, \bibinfo{author}{et. al.},
  \bibinfo{journal}{\aap}  \bibinfo{volume}{608} (\bibinfo{year}{2017})
  \bibinfo{pages}{A123}.
\bibitem[{{Schenker} et~al.(2014){Schenker}, {Ellis}, {Konidaris}, and
  {Stark}}]{schenker2014}
\bibinfo{author}{M.~A. {Schenker}}, \bibinfo{author}{R.~S. {Ellis}},
  \bibinfo{author}{N.~P. {Konidaris}}, \bibinfo{author}{D.~P. {Stark}},
  \bibinfo{journal}{\apj}  \bibinfo{volume}{795} (\bibinfo{year}{2014})
  \bibinfo{pages}{20}.
\bibitem[{{Pentericci} et~al.(2011){Pentericci}, {Fontana}, {Vanzella}, and et.
  al.}]{pentericci2011}
\bibinfo{author}{L.~{Pentericci}}, \bibinfo{author}{A.~{Fontana}},
  \bibinfo{author}{E.~{Vanzella}}, \bibinfo{author}{et. al.},
  \bibinfo{journal}{\apj}  \bibinfo{volume}{743} (\bibinfo{year}{2011})
  \bibinfo{pages}{132}.
\bibitem[{{Ono} et~al.(2012){Ono}, {Ouchi}, {Mobasher}, and et. al.}]{ono2012}
\bibinfo{author}{Y.~{Ono}}, \bibinfo{author}{M.~{Ouchi}},
  \bibinfo{author}{B.~{Mobasher}}, \bibinfo{author}{et. al.},
  \bibinfo{journal}{\apj}  \bibinfo{volume}{744} (\bibinfo{year}{2012})
  \bibinfo{pages}{83}.
\bibitem[{{Tilvi} et~al.(2014){Tilvi}, {Papovich}, {Finkelstein}, and et.
  al.}]{tilvi2014}
\bibinfo{author}{V.~{Tilvi}}, \bibinfo{author}{C.~{Papovich}},
  \bibinfo{author}{S.~L. {Finkelstein}}, \bibinfo{author}{et. al.},
  \bibinfo{journal}{\apj}  \bibinfo{volume}{794} (\bibinfo{year}{2014})
  \bibinfo{pages}{5}.
\bibitem[{{Cassata} et~al.(2015){Cassata}, {Tasca}, {Le F{\`e}vre}, and et.
  al.}]{cassata2015}
\bibinfo{author}{P.~{Cassata}}, \bibinfo{author}{L.~A.~M. {Tasca}},
  \bibinfo{author}{O.~{Le F{\`e}vre}}, \bibinfo{author}{et. al.},
  \bibinfo{journal}{\aap}  \bibinfo{volume}{573} (\bibinfo{year}{2015})
  \bibinfo{pages}{A24}.
\bibitem[{{Curtis-Lake} et~al.(2012){Curtis-Lake}, {McLure}, {Pearce}, and et.
  al.}]{curtis-lake2012}
\bibinfo{author}{E.~{Curtis-Lake}}, \bibinfo{author}{R.~J. {McLure}},
  \bibinfo{author}{H.~J. {Pearce}}, \bibinfo{author}{et. al.},
  \bibinfo{journal}{\mnras}  \bibinfo{volume}{422} (\bibinfo{year}{2012})
  \bibinfo{pages}{1425--1435}.
\bibitem[{{Labb{\'e}} et~al.(2013){Labb{\'e}}, {Oesch}, {Bouwens}, and et.
  al.}]{labbe2013}
\bibinfo{author}{I.~{Labb{\'e}}}, \bibinfo{author}{P.~A. {Oesch}},
  \bibinfo{author}{R.~J. {Bouwens}}, \bibinfo{author}{et. al.},
  \bibinfo{journal}{\apjl}  \bibinfo{volume}{777} (\bibinfo{year}{2013})
  \bibinfo{pages}{L19}.
\bibitem[{{Gonz{\'a}lez} et~al.(2011){Gonz{\'a}lez}, {Labb{\'e}}, {Bouwens},
  {Illingworth}, {Franx}, and {Kriek}}]{gonzalez2011}
\bibinfo{author}{V.~{Gonz{\'a}lez}}, \bibinfo{author}{I.~{Labb{\'e}}},
  \bibinfo{author}{R.~J. {Bouwens}}, \bibinfo{author}{G.~{Illingworth}},
  \bibinfo{author}{M.~{Franx}}, \bibinfo{author}{M.~{Kriek}},
  \bibinfo{journal}{\apjl}  \bibinfo{volume}{735} (\bibinfo{year}{2011})
  \bibinfo{pages}{L34}.
\bibitem[{{Stark} et~al.(2013){Stark}, {Schenker}, {Ellis}, {Robertson},
  {McLure}, and {Dunlop}}]{stark2013}
\bibinfo{author}{D.~P. {Stark}}, \bibinfo{author}{M.~A. {Schenker}},
  \bibinfo{author}{R.~{Ellis}}, \bibinfo{author}{B.~{Robertson}},
  \bibinfo{author}{R.~{McLure}}, \bibinfo{author}{J.~{Dunlop}},
  \bibinfo{journal}{\apj}  \bibinfo{volume}{763} (\bibinfo{year}{2013})
  \bibinfo{pages}{129}.
\bibitem[{{Oesch} et~al.(2014){Oesch}, {Bouwens}, {Illingworth}, and et.
  al.}]{oesch2014}
\bibinfo{author}{P.~A. {Oesch}}, \bibinfo{author}{R.~J. {Bouwens}},
  \bibinfo{author}{G.~D. {Illingworth}}, \bibinfo{author}{et. al.},
  \bibinfo{journal}{\apj}  \bibinfo{volume}{786} (\bibinfo{year}{2014})
  \bibinfo{pages}{108}.
\bibitem[{{Duncan} et~al.(2014){Duncan}, {Conselice}, {Mortlock}, and et.
  al.}]{duncan2014}
\bibinfo{author}{K.~{Duncan}}, \bibinfo{author}{C.~J. {Conselice}},
  \bibinfo{author}{A.~{Mortlock}}, \bibinfo{author}{et. al.},
  \bibinfo{journal}{\mnras}  \bibinfo{volume}{444} (\bibinfo{year}{2014})
  \bibinfo{pages}{2960--2984}.
\bibitem[{{Grazian} et~al.(2015){Grazian}, {Fontana}, {Santini}, and et.
  al.}]{grazian2015}
\bibinfo{author}{A.~{Grazian}}, \bibinfo{author}{A.~{Fontana}},
  \bibinfo{author}{P.~{Santini}}, \bibinfo{author}{et. al.},
  \bibinfo{journal}{\aap}  \bibinfo{volume}{575} (\bibinfo{year}{2015})
  \bibinfo{pages}{A96}.
\bibitem[{{Song} et~al.(2016){Song}, {Finkelstein}, {Ashby}, and et.
  al.}]{song2016}
\bibinfo{author}{M.~{Song}}, \bibinfo{author}{S.~L. {Finkelstein}},
  \bibinfo{author}{M.~L.~N. {Ashby}}, \bibinfo{author}{et. al.},
  \bibinfo{journal}{\apj}  \bibinfo{volume}{825} (\bibinfo{year}{2016})
  \bibinfo{pages}{5}.
\bibitem[{{Ellis} et~al.(2013){Ellis}, {McLure}, {Dunlop}, and et.
  al.}]{ellis2013}
\bibinfo{author}{R.~S. {Ellis}}, \bibinfo{author}{R.~J. {McLure}},
  \bibinfo{author}{J.~S. {Dunlop}}, \bibinfo{author}{et. al.},
  \bibinfo{journal}{\apjl}  \bibinfo{volume}{763} (\bibinfo{year}{2013})
  \bibinfo{pages}{L7}.
\bibitem[{{Zheng} et~al.(2012){Zheng}, {Postman}, {Zitrin}, and et.
  al.}]{zheng2012}
\bibinfo{author}{W.~{Zheng}}, \bibinfo{author}{M.~{Postman}},
  \bibinfo{author}{A.~{Zitrin}}, \bibinfo{author}{et. al.},
  \bibinfo{journal}{\nat}  \bibinfo{volume}{489} (\bibinfo{year}{2012})
  \bibinfo{pages}{406--408}.
\bibitem[{{Coe} et~al.(2013){Coe}, {Zitrin}, {Carrasco}, and et. al.}]{coe2013}
\bibinfo{author}{D.~{Coe}}, \bibinfo{author}{A.~{Zitrin}},
  \bibinfo{author}{M.~{Carrasco}}, \bibinfo{author}{et. al.},
  \bibinfo{journal}{\apj}  \bibinfo{volume}{762} (\bibinfo{year}{2013})
  \bibinfo{pages}{32}.
\bibitem[{{Bouwens} et~al.(2014){Bouwens}, {Bradley}, {Zitrin}, and et.
  al.}]{bouwens2014b}
\bibinfo{author}{R.~J. {Bouwens}}, \bibinfo{author}{L.~{Bradley}},
  \bibinfo{author}{A.~{Zitrin}}, \bibinfo{author}{et. al.},
  \bibinfo{journal}{\apj}  \bibinfo{volume}{795} (\bibinfo{year}{2014})
  \bibinfo{pages}{126}.
\bibitem[{{Madau} et~al.(1998){Madau}, {Pozzetti}, and {Dickinson}}]{madau1998}
\bibinfo{author}{P.~{Madau}}, \bibinfo{author}{L.~{Pozzetti}},
  \bibinfo{author}{M.~{Dickinson}}, \bibinfo{journal}{\apj}
  \bibinfo{volume}{498} (\bibinfo{year}{1998}) \bibinfo{pages}{106--116}.
\bibitem[{{Bouwens} et~al.(2012){Bouwens}, {Illingworth}, {Oesch}, and et.
  al.}]{bouwens2012b}
\bibinfo{author}{R.~J. {Bouwens}}, \bibinfo{author}{G.~D. {Illingworth}},
  \bibinfo{author}{P.~A. {Oesch}}, \bibinfo{author}{et. al.},
  \bibinfo{journal}{\apj}  \bibinfo{volume}{754} (\bibinfo{year}{2012})
  \bibinfo{pages}{83}.
\bibitem[{{Cazaux} and {Tielens}(2004)}]{cazaux2004}
\bibinfo{author}{S.~{Cazaux}}, \bibinfo{author}{A.~G.~G.~M. {Tielens}},
  \bibinfo{journal}{\apj}  \bibinfo{volume}{604} (\bibinfo{year}{2004})
  \bibinfo{pages}{222--237}.
\bibitem[{{Hirashita} and {Ferrara}(2002)}]{hirashita2002}
\bibinfo{author}{H.~{Hirashita}}, \bibinfo{author}{A.~{Ferrara}},
  \bibinfo{journal}{\mnras}  \bibinfo{volume}{337} (\bibinfo{year}{2002})
  \bibinfo{pages}{921--937}.
\bibitem[{{Yamasawa} et~al.(2011){Yamasawa}, {Habe}, {Kozasa}, {Nozawa},
  {Hirashita}, {Umeda}, and {Nomoto}}]{yamasawa2011}
\bibinfo{author}{D.~{Yamasawa}}, \bibinfo{author}{A.~{Habe}},
  \bibinfo{author}{T.~{Kozasa}}, \bibinfo{author}{T.~{Nozawa}},
  \bibinfo{author}{H.~{Hirashita}}, \bibinfo{author}{H.~{Umeda}},
  \bibinfo{author}{K.~{Nomoto}}, \bibinfo{journal}{\apj}  \bibinfo{volume}{735}
  (\bibinfo{year}{2011}) \bibinfo{pages}{44}.
\bibitem[{{Omukai} et~al.(2005){Omukai}, {Tsuribe}, {Schneider}, and
  {Ferrara}}]{omukai2005}
\bibinfo{author}{K.~{Omukai}}, \bibinfo{author}{T.~{Tsuribe}},
  \bibinfo{author}{R.~{Schneider}}, \bibinfo{author}{A.~{Ferrara}},
  \bibinfo{journal}{\apj}  \bibinfo{volume}{626} (\bibinfo{year}{2005})
  \bibinfo{pages}{627--643}.
\bibitem[{{Dayal} et~al.(2010){Dayal}, {Hirashita}, and
  {Ferrara}}]{dayal2010dust}
\bibinfo{author}{P.~{Dayal}}, \bibinfo{author}{H.~{Hirashita}},
  \bibinfo{author}{A.~{Ferrara}}, \bibinfo{journal}{\mnras}
  \bibinfo{volume}{403} (\bibinfo{year}{2010}) \bibinfo{pages}{620--624}.
\bibitem[{{Matteucci} and {Greggio}(1986)}]{matteucci1986}
\bibinfo{author}{F.~{Matteucci}}, \bibinfo{author}{L.~{Greggio}},
  \bibinfo{journal}{\aap}  \bibinfo{volume}{154} (\bibinfo{year}{1986})
  \bibinfo{pages}{279--287}.
\bibitem[{{Feder} et~al.(1966){Feder}, {Russell}, {Lothe}, and
  {Pound}}]{feder1966}
\bibinfo{author}{J.~{Feder}}, \bibinfo{author}{K.~{Russell}},
  \bibinfo{author}{J.~{Lothe}}, \bibinfo{author}{G.~{Pound}},
  \bibinfo{journal}{Advances in Physics}  \bibinfo{volume}{15}
  (\bibinfo{year}{1966}) \bibinfo{pages}{111--178}.
\bibitem[{{Todini} and {Ferrara}(2001)}]{todini2001}
\bibinfo{author}{P.~{Todini}}, \bibinfo{author}{A.~{Ferrara}},
  \bibinfo{journal}{\mnras}  \bibinfo{volume}{325} (\bibinfo{year}{2001})
  \bibinfo{pages}{726--736}.
\bibitem[{{Wooden} et~al.(1993){Wooden}, {Rank}, {Bregman}, {Witteborn},
  {Tielens}, {Cohen}, {Pinto}, and {Axelrod}}]{wooden1993}
\bibinfo{author}{D.~H. {Wooden}}, \bibinfo{author}{D.~M. {Rank}},
  \bibinfo{author}{J.~D. {Bregman}}, \bibinfo{author}{F.~C. {Witteborn}},
  \bibinfo{author}{A.~G.~G.~M. {Tielens}}, \bibinfo{author}{M.~{Cohen}},
  \bibinfo{author}{P.~A. {Pinto}}, \bibinfo{author}{T.~S. {Axelrod}},
  \bibinfo{journal}{\apjs}  \bibinfo{volume}{88} (\bibinfo{year}{1993})
  \bibinfo{pages}{477--507}.
\bibitem[{{Elmhamdi} et~al.(2003){Elmhamdi}, {Danziger}, {Chugai}, and et.
  al.}]{elmhamdi2003}
\bibinfo{author}{A.~{Elmhamdi}}, \bibinfo{author}{I.~J. {Danziger}},
  \bibinfo{author}{N.~{Chugai}}, \bibinfo{author}{et. al.},
  \bibinfo{journal}{\mnras}  \bibinfo{volume}{338} (\bibinfo{year}{2003})
  \bibinfo{pages}{939--956}.
\bibitem[{{Kozasa} et~al.(2009){Kozasa}, {Nozawa}, {Tominaga}, {Umeda},
  {Maeda}, and {Nomoto}}]{kozasa2009}
\bibinfo{author}{T.~{Kozasa}}, \bibinfo{author}{T.~{Nozawa}},
  \bibinfo{author}{N.~{Tominaga}}, \bibinfo{author}{H.~{Umeda}},
  \bibinfo{author}{K.~{Maeda}}, \bibinfo{author}{K.~{Nomoto}}, in:
  \bibinfo{editor}{T.~{Henning}}, \bibinfo{editor}{E.~{Gr{\"u}n}},
  \bibinfo{editor}{J.~{Steinacker}} (Eds.), \bibinfo{booktitle}{Cosmic Dust -
  Near and Far}, volume \bibinfo{volume}{414} of
  \textit{\bibinfo{series}{Astronomical Society of the Pacific Conference
  Series}},  p.~\bibinfo{pages}{43}.
\bibitem[{{Gall} et~al.(2011){Gall}, {Hjorth}, and {Andersen}}]{gall2011}
\bibinfo{author}{C.~{Gall}}, \bibinfo{author}{J.~{Hjorth}},
  \bibinfo{author}{A.~C. {Andersen}}, \bibinfo{journal}{\aapr}
  \bibinfo{volume}{19} (\bibinfo{year}{2011}) \bibinfo{pages}{43}.
\bibitem[{{Rho} et~al.(2008){Rho}, {Kozasa}, {Reach}, {Smith}, {Rudnick},
  {DeLaney}, {Ennis}, {Gomez}, and {Tappe}}]{rho2008}
\bibinfo{author}{J.~{Rho}}, \bibinfo{author}{T.~{Kozasa}},
  \bibinfo{author}{W.~T. {Reach}}, \bibinfo{author}{J.~D. {Smith}},
  \bibinfo{author}{L.~{Rudnick}}, \bibinfo{author}{T.~{DeLaney}},
  \bibinfo{author}{J.~A. {Ennis}}, \bibinfo{author}{H.~{Gomez}},
  \bibinfo{author}{A.~{Tappe}}, \bibinfo{journal}{\apj}  \bibinfo{volume}{673}
  (\bibinfo{year}{2008}) \bibinfo{pages}{271--282}.
\bibitem[{{Rho} et~al.(2009){Rho}, {Reach}, {Tappe}, {Hwang}, {Slavin},
  {Kozasa}, and {Dunne}}]{rho2009}
\bibinfo{author}{J.~{Rho}}, \bibinfo{author}{W.~T. {Reach}},
  \bibinfo{author}{A.~{Tappe}}, \bibinfo{author}{U.~{Hwang}},
  \bibinfo{author}{J.~D. {Slavin}}, \bibinfo{author}{T.~{Kozasa}},
  \bibinfo{author}{L.~{Dunne}}, \bibinfo{journal}{\apj}  \bibinfo{volume}{700}
  (\bibinfo{year}{2009}) \bibinfo{pages}{579--596}.
\bibitem[{{Matsuura} et~al.(2011){Matsuura}, {Dwek}, {Meixner}, and et.
  al.}]{matsuura2011}
\bibinfo{author}{M.~{Matsuura}}, \bibinfo{author}{E.~{Dwek}},
  \bibinfo{author}{M.~{Meixner}}, \bibinfo{author}{et. al.},
  \bibinfo{journal}{Science}  \bibinfo{volume}{333} (\bibinfo{year}{2011})
  \bibinfo{pages}{1258}.
\bibitem[{{Indebetouw} et~al.(2014){Indebetouw}, {Matsuura}, {Dwek}, and et.
  al.}]{indebetouw2014}
\bibinfo{author}{R.~{Indebetouw}}, \bibinfo{author}{M.~{Matsuura}},
  \bibinfo{author}{E.~{Dwek}}, \bibinfo{author}{et. al.},
  \bibinfo{journal}{\apjl}  \bibinfo{volume}{782} (\bibinfo{year}{2014})
  \bibinfo{pages}{L2}.
\bibitem[{{Bianchi} and {Schneider}(2007)}]{bianchi2007}
\bibinfo{author}{S.~{Bianchi}}, \bibinfo{author}{R.~{Schneider}},
  \bibinfo{journal}{\mnras}  \bibinfo{volume}{378} (\bibinfo{year}{2007})
  \bibinfo{pages}{973--982}.
\bibitem[{{Zhukovska} et~al.(2008){Zhukovska}, {Gail}, and
  {Trieloff}}]{zhukovska2008}
\bibinfo{author}{S.~{Zhukovska}}, \bibinfo{author}{H.-P. {Gail}},
  \bibinfo{author}{M.~{Trieloff}}, \bibinfo{journal}{\aap}
  \bibinfo{volume}{479} (\bibinfo{year}{2008}) \bibinfo{pages}{453--480}.
\bibitem[{{Valiante} et~al.(2009){Valiante}, {Schneider}, {Bianchi}, and
  {Andersen}}]{valiante2009}
\bibinfo{author}{R.~{Valiante}}, \bibinfo{author}{R.~{Schneider}},
  \bibinfo{author}{S.~{Bianchi}}, \bibinfo{author}{A.~C. {Andersen}},
  \bibinfo{journal}{\mnras}  \bibinfo{volume}{397} (\bibinfo{year}{2009})
  \bibinfo{pages}{1661--1671}.
\bibitem[{{Dwek} et~al.(2007){Dwek}, {Galliano}, and {Jones}}]{dwek2007}
\bibinfo{author}{E.~{Dwek}}, \bibinfo{author}{F.~{Galliano}},
  \bibinfo{author}{A.~P. {Jones}}, \bibinfo{journal}{\apj}
  \bibinfo{volume}{662} (\bibinfo{year}{2007}) \bibinfo{pages}{927--939}.
\bibitem[{{Hirashita} et~al.(2014){Hirashita}, {Ferrara}, {Dayal}, and
  {Ouchi}}]{hirashita2014}
\bibinfo{author}{H.~{Hirashita}}, \bibinfo{author}{A.~{Ferrara}},
  \bibinfo{author}{P.~{Dayal}}, \bibinfo{author}{M.~{Ouchi}},
  \bibinfo{journal}{\mnras}  \bibinfo{volume}{443} (\bibinfo{year}{2014})
  \bibinfo{pages}{1704--1712}.
\bibitem[{{Hirashita} et~al.(2017){Hirashita}, {Burgarella}, and
  {Bouwens}}]{hirashita2017}
\bibinfo{author}{H.~{Hirashita}}, \bibinfo{author}{D.~{Burgarella}},
  \bibinfo{author}{R.~J. {Bouwens}}, \bibinfo{journal}{\mnras}
  \bibinfo{volume}{472} (\bibinfo{year}{2017}) \bibinfo{pages}{4587--4597}.
\bibitem[{{Nozawa} et~al.(2007){Nozawa}, {Kozasa}, {Habe}, {Dwek}, {Umeda},
  {Tominaga}, {Maeda}, and {Nomoto}}]{nozawa2007}
\bibinfo{author}{T.~{Nozawa}}, \bibinfo{author}{T.~{Kozasa}},
  \bibinfo{author}{A.~{Habe}}, \bibinfo{author}{E.~{Dwek}},
  \bibinfo{author}{H.~{Umeda}}, \bibinfo{author}{N.~{Tominaga}},
  \bibinfo{author}{K.~{Maeda}}, \bibinfo{author}{K.~{Nomoto}},
  \bibinfo{journal}{\apj}  \bibinfo{volume}{666} (\bibinfo{year}{2007})
  \bibinfo{pages}{955--966}.
\bibitem[{{Barlow} et~al.(2010){Barlow}, {Krause}, {Swinyard}, and et.
  al.}]{barlow2010}
\bibinfo{author}{M.~J. {Barlow}}, \bibinfo{author}{O.~{Krause}},
  \bibinfo{author}{B.~M. {Swinyard}}, \bibinfo{author}{et. al.},
  \bibinfo{journal}{\aap}  \bibinfo{volume}{518} (\bibinfo{year}{2010})
  \bibinfo{pages}{L138}.
\bibitem[{{Gomez} et~al.(2012){Gomez}, {Krause}, {Barlow}, and et.
  al.}]{gomez2012}
\bibinfo{author}{H.~L. {Gomez}}, \bibinfo{author}{O.~{Krause}},
  \bibinfo{author}{M.~J. {Barlow}}, \bibinfo{author}{et. al.},
  \bibinfo{journal}{\apj}  \bibinfo{volume}{760} (\bibinfo{year}{2012})
  \bibinfo{pages}{96}.
\bibitem[{{Otsuka} et~al.(2010){Otsuka}, {van Loon}, {Long}, and et.
  al.}]{otsuka2010}
\bibinfo{author}{M.~{Otsuka}}, \bibinfo{author}{J.~T. {van Loon}},
  \bibinfo{author}{K.~S. {Long}}, \bibinfo{author}{et. al.},
  \bibinfo{journal}{\aap}  \bibinfo{volume}{518} (\bibinfo{year}{2010})
  \bibinfo{pages}{L139}.
\bibitem[{{Draine}(2009)}]{draine2009}
\bibinfo{author}{B.~T. {Draine}}, in: \bibinfo{editor}{T.~{Henning}},
  \bibinfo{editor}{E.~{Gr{\"u}n}}, \bibinfo{editor}{J.~{Steinacker}} (Eds.),
  \bibinfo{booktitle}{Cosmic Dust - Near and Far}, volume \bibinfo{volume}{414}
  of \textit{\bibinfo{series}{Astronomical Society of the Pacific Conference
  Series}},  p. \bibinfo{pages}{453}.
\bibitem[{{Jones} and {Nuth}(2011)}]{jones2011}
\bibinfo{author}{A.~P. {Jones}}, \bibinfo{author}{J.~A. {Nuth}},
  \bibinfo{journal}{\aap}  \bibinfo{volume}{530} (\bibinfo{year}{2011})
  \bibinfo{pages}{A44}.
\bibitem[{{Asano} et~al.(2013){Asano}, {Takeuchi}, {Hirashita}, and
  {Inoue}}]{asano2013}
\bibinfo{author}{R.~S. {Asano}}, \bibinfo{author}{T.~T. {Takeuchi}},
  \bibinfo{author}{H.~{Hirashita}}, \bibinfo{author}{A.~K. {Inoue}},
  \bibinfo{journal}{Earth, Planets, and Space}  \bibinfo{volume}{65}
  (\bibinfo{year}{2013}) \bibinfo{pages}{213--222}.
\bibitem[{{Dwek}(1998)}]{dwek1998}
\bibinfo{author}{E.~{Dwek}}, \bibinfo{journal}{\apj}  \bibinfo{volume}{501}
  (\bibinfo{year}{1998}) \bibinfo{pages}{643}.
\bibitem[{{Inoue}(2011)}]{inoue2011}
\bibinfo{author}{A.~K. {Inoue}}, \bibinfo{journal}{Earth, Planets, and Space}
  \bibinfo{volume}{63} (\bibinfo{year}{2011}) \bibinfo{pages}{1027--1039}.
\bibitem[{{de Bennassuti} et~al.(2014){de Bennassuti}, {Schneider}, {Valiante},
  and {Salvadori}}]{debennassuti2014}
\bibinfo{author}{M.~{de Bennassuti}}, \bibinfo{author}{R.~{Schneider}},
  \bibinfo{author}{R.~{Valiante}}, \bibinfo{author}{S.~{Salvadori}},
  \bibinfo{journal}{\mnras}  \bibinfo{volume}{445} (\bibinfo{year}{2014})
  \bibinfo{pages}{3039--3054}.
\bibitem[{{Mancini} et~al.(2015){Mancini}, {Schneider}, {Graziani}, {Valiante},
  {Dayal}, {Maio}, {Ciardi}, and {Hunt}}]{mancini2015}
\bibinfo{author}{M.~{Mancini}}, \bibinfo{author}{R.~{Schneider}},
  \bibinfo{author}{L.~{Graziani}}, \bibinfo{author}{R.~{Valiante}},
  \bibinfo{author}{P.~{Dayal}}, \bibinfo{author}{U.~{Maio}},
  \bibinfo{author}{B.~{Ciardi}}, \bibinfo{author}{L.~K. {Hunt}},
  \bibinfo{journal}{\mnras}  \bibinfo{volume}{451} (\bibinfo{year}{2015})
  \bibinfo{pages}{L70--L74}.
\bibitem[{{McKee}(1989)}]{mckee1989}
\bibinfo{author}{C.~F. {McKee}}, \bibinfo{journal}{\apj}  \bibinfo{volume}{345}
  (\bibinfo{year}{1989}) \bibinfo{pages}{782--801}.
\bibitem[{{Lisenfeld} and {Ferrara}(1998)}]{lisenfeld1998}
\bibinfo{author}{U.~{Lisenfeld}}, \bibinfo{author}{A.~{Ferrara}},
  \bibinfo{journal}{\apj}  \bibinfo{volume}{496} (\bibinfo{year}{1998})
  \bibinfo{pages}{145--154}.
\bibitem[{{Seab} and {Shull}(1983)}]{seab1983}
\bibinfo{author}{C.~G. {Seab}}, \bibinfo{author}{J.~M. {Shull}},
  \bibinfo{journal}{\apj}  \bibinfo{volume}{275} (\bibinfo{year}{1983})
  \bibinfo{pages}{652--660}.
\bibitem[{{Ferrara} et~al.(2016){Ferrara}, {Viti}, and
  {Ceccarelli}}]{ferrara2016b}
\bibinfo{author}{A.~{Ferrara}}, \bibinfo{author}{S.~{Viti}},
  \bibinfo{author}{C.~{Ceccarelli}}, \bibinfo{journal}{\mnras}
  \bibinfo{volume}{463} (\bibinfo{year}{2016}) \bibinfo{pages}{L112--L116}.
\bibitem[{{Schaerer} et~al.(2015){Schaerer}, {Boone}, {Zamojski}, {Staguhn},
  {Dessauges-Zavadsky}, {Finkelstein}, and {Combes}}]{schaerer2015}
\bibinfo{author}{D.~{Schaerer}}, \bibinfo{author}{F.~{Boone}},
  \bibinfo{author}{M.~{Zamojski}}, \bibinfo{author}{J.~{Staguhn}},
  \bibinfo{author}{M.~{Dessauges-Zavadsky}},
  \bibinfo{author}{S.~{Finkelstein}}, \bibinfo{author}{F.~{Combes}},
  \bibinfo{journal}{\aap}  \bibinfo{volume}{574} (\bibinfo{year}{2015})
  \bibinfo{pages}{A19}.
\bibitem[{{Kanekar} et~al.(2013){Kanekar}, {Wagg}, {Chary}, and
  {Carilli}}]{kanekar2013}
\bibinfo{author}{N.~{Kanekar}}, \bibinfo{author}{J.~{Wagg}},
  \bibinfo{author}{R.~R. {Chary}}, \bibinfo{author}{C.~L. {Carilli}},
  \bibinfo{journal}{\apjl}  \bibinfo{volume}{771} (\bibinfo{year}{2013})
  \bibinfo{pages}{L20}.
\bibitem[{{Ota} et~al.(2014){Ota}, {Walter}, {Ohta}, and et. al.}]{ota2014}
\bibinfo{author}{K.~{Ota}}, \bibinfo{author}{F.~{Walter}},
  \bibinfo{author}{K.~{Ohta}}, \bibinfo{author}{et. al.},
  \bibinfo{journal}{\apj}  \bibinfo{volume}{792} (\bibinfo{year}{2014})
  \bibinfo{pages}{34}.
\bibitem[{{Ouchi} et~al.(2013){Ouchi}, {Ellis}, {Ono}, and et. al.}]{ouchi2013}
\bibinfo{author}{M.~{Ouchi}}, \bibinfo{author}{R.~{Ellis}},
  \bibinfo{author}{Y.~{Ono}}, \bibinfo{author}{et. al.},
  \bibinfo{journal}{\apj}  \bibinfo{volume}{778} (\bibinfo{year}{2013})
  \bibinfo{pages}{102}.
\bibitem[{{Maiolino} et~al.(2015){Maiolino}, {Carniani}, {Fontana}, and et.
  al.}]{maiolino2015}
\bibinfo{author}{R.~{Maiolino}}, \bibinfo{author}{S.~{Carniani}},
  \bibinfo{author}{A.~{Fontana}}, \bibinfo{author}{et. al.},
  \bibinfo{journal}{\mnras}  \bibinfo{volume}{452} (\bibinfo{year}{2015})
  \bibinfo{pages}{54--68}.
\bibitem[{{Watson} et~al.(2015){Watson}, {Christensen}, {Knudsen}, {Richard},
  {Gallazzi}, and {Micha{\l}owski}}]{watson2015}
\bibinfo{author}{D.~{Watson}}, \bibinfo{author}{L.~{Christensen}},
  \bibinfo{author}{K.~K. {Knudsen}}, \bibinfo{author}{J.~{Richard}},
  \bibinfo{author}{A.~{Gallazzi}}, \bibinfo{author}{M.~J. {Micha{\l}owski}},
  \bibinfo{journal}{Nature}  \bibinfo{volume}{519} (\bibinfo{year}{2015})
  \bibinfo{pages}{327--330}.
\bibitem[{{Laporte} et~al.(2017){Laporte}, {Ellis}, {Boone}, {Bauer},
  {Qu{\'e}nard}, {Roberts-Borsani}, {Pell{\'o}}, {P{\'e}rez-Fournon}, and
  {Streblyanska}}]{laporte2017}
\bibinfo{author}{N.~{Laporte}}, \bibinfo{author}{R.~S. {Ellis}},
  \bibinfo{author}{F.~{Boone}}, \bibinfo{author}{F.~E. {Bauer}},
  \bibinfo{author}{D.~{Qu{\'e}nard}}, \bibinfo{author}{G.~W.
  {Roberts-Borsani}}, \bibinfo{author}{R.~{Pell{\'o}}},
  \bibinfo{author}{I.~{P{\'e}rez-Fournon}},
  \bibinfo{author}{A.~{Streblyanska}}, \bibinfo{journal}{\apjl}
  \bibinfo{volume}{837} (\bibinfo{year}{2017}) \bibinfo{pages}{L21}.
\bibitem[{{Meurer} et~al.(1999){Meurer}, {Heckman}, and
  {Calzetti}}]{meurer1999}
\bibinfo{author}{G.~R. {Meurer}}, \bibinfo{author}{T.~M. {Heckman}},
  \bibinfo{author}{D.~{Calzetti}}, \bibinfo{journal}{\apj}
  \bibinfo{volume}{521} (\bibinfo{year}{1999}) \bibinfo{pages}{64--80}.
\bibitem[{{Finkelstein} et~al.(2012){Finkelstein}, {Papovich}, {Salmon}, and
  et. al.}]{finkelstein2012b}
\bibinfo{author}{S.~L. {Finkelstein}}, \bibinfo{author}{C.~{Papovich}},
  \bibinfo{author}{B.~{Salmon}}, \bibinfo{author}{et. al.},
  \bibinfo{journal}{\apj}  \bibinfo{volume}{756} (\bibinfo{year}{2012})
  \bibinfo{pages}{164}.
\bibitem[{{Dunlop} et~al.(2013){Dunlop}, {Rogers}, {McLure}, and et.
  al.}]{dunlop2013b}
\bibinfo{author}{J.~S. {Dunlop}}, \bibinfo{author}{A.~B. {Rogers}},
  \bibinfo{author}{R.~J. {McLure}}, \bibinfo{author}{et. al.},
  \bibinfo{journal}{\mnras}  \bibinfo{volume}{432} (\bibinfo{year}{2013})
  \bibinfo{pages}{3520--3533}.
\bibitem[{{Bouwens} et~al.(2014){Bouwens}, {Illingworth}, {Oesch}, and et.
  al.}]{bouwens2014}
\bibinfo{author}{R.~J. {Bouwens}}, \bibinfo{author}{G.~D. {Illingworth}},
  \bibinfo{author}{P.~A. {Oesch}}, \bibinfo{author}{et. al.},
  \bibinfo{journal}{\apj}  \bibinfo{volume}{793} (\bibinfo{year}{2014})
  \bibinfo{pages}{115}.
\bibitem[{{Rogers} et~al.(2014){Rogers}, {McLure}, {Dunlop}, and et.
  al.}]{rogers2014}
\bibinfo{author}{A.~B. {Rogers}}, \bibinfo{author}{R.~J. {McLure}},
  \bibinfo{author}{J.~S. {Dunlop}}, \bibinfo{author}{et. al.},
  \bibinfo{journal}{\mnras}  \bibinfo{volume}{440} (\bibinfo{year}{2014})
  \bibinfo{pages}{3714--3725}.
\bibitem[{{Wilkins} et~al.(2013){Wilkins}, {Bunker}, {Coulton}, {Croft}, {di
  Matteo}, {Khandai}, and {Feng}}]{wilkins2013}
\bibinfo{author}{S.~M. {Wilkins}}, \bibinfo{author}{A.~{Bunker}},
  \bibinfo{author}{W.~{Coulton}}, \bibinfo{author}{R.~{Croft}},
  \bibinfo{author}{T.~{di Matteo}}, \bibinfo{author}{N.~{Khandai}},
  \bibinfo{author}{Y.~{Feng}}, \bibinfo{journal}{\mnras}  \bibinfo{volume}{430}
  (\bibinfo{year}{2013}) \bibinfo{pages}{2885--2890}.
\bibitem[{{Mancini} et~al.(2016){Mancini}, {Schneider}, {Graziani}, {Valiante},
  {Dayal}, {Maio}, and {Ciardi}}]{mancini2016}
\bibinfo{author}{M.~{Mancini}}, \bibinfo{author}{R.~{Schneider}},
  \bibinfo{author}{L.~{Graziani}}, \bibinfo{author}{R.~{Valiante}},
  \bibinfo{author}{P.~{Dayal}}, \bibinfo{author}{U.~{Maio}},
  \bibinfo{author}{B.~{Ciardi}}, \bibinfo{journal}{\mnras}
  \bibinfo{volume}{462} (\bibinfo{year}{2016}) \bibinfo{pages}{3130--3145}.
\bibitem[{{Finkelstein} et~al.(2015){Finkelstein}, {Ryan}, {Papovich}, and et.
  al.}]{finkelstein2015}
\bibinfo{author}{S.~L. {Finkelstein}}, \bibinfo{author}{R.~E. {Ryan}, Jr.},
  \bibinfo{author}{C.~{Papovich}}, \bibinfo{author}{et. al.},
  \bibinfo{journal}{\apj}  \bibinfo{volume}{810} (\bibinfo{year}{2015})
  \bibinfo{pages}{71}.
\bibitem[{{Khakhaleva-Li} and {Gnedin}(2016)}]{lignedin2016}
\bibinfo{author}{Z.~{Khakhaleva-Li}}, \bibinfo{author}{N.~Y. {Gnedin}},
  \bibinfo{journal}{\apj}  \bibinfo{volume}{820} (\bibinfo{year}{2016})
  \bibinfo{pages}{133}.
\bibitem[{{Wilkins} et~al.(2018){Wilkins}, {Feng}, {Di Matteo}, {Croft},
  {Lovell}, and {Thomas}}]{wilkins2018}
\bibinfo{author}{S.~M. {Wilkins}}, \bibinfo{author}{Y.~{Feng}},
  \bibinfo{author}{T.~{Di Matteo}}, \bibinfo{author}{R.~{Croft}},
  \bibinfo{author}{C.~C. {Lovell}}, \bibinfo{author}{P.~{Thomas}},
  \bibinfo{journal}{\mnras}  \bibinfo{volume}{473} (\bibinfo{year}{2018})
  \bibinfo{pages}{5363--5369}.
\bibitem[{{Capak} et~al.(2015){Capak}, {Carilli}, {Jones}, and et.
  al.}]{capak2015}
\bibinfo{author}{P.~L. {Capak}}, \bibinfo{author}{C.~{Carilli}},
  \bibinfo{author}{G.~{Jones}}, \bibinfo{author}{et. al.},
  \bibinfo{journal}{\nat}  \bibinfo{volume}{522} (\bibinfo{year}{2015})
  \bibinfo{pages}{455--458}.
\bibitem[{{Micha{\l}owski}(2015)}]{michalowski2015}
\bibinfo{author}{M.~J. {Micha{\l}owski}}, \bibinfo{journal}{\aap}
  \bibinfo{volume}{577} (\bibinfo{year}{2015}) \bibinfo{pages}{A80}.
\bibitem[{{Behrens} et~al.(2018){Behrens}, {Pallottini}, {Ferrara},
  {Gallerani}, and {Vallini}}]{behrens2018}
\bibinfo{author}{C.~{Behrens}}, \bibinfo{author}{A.~{Pallottini}},
  \bibinfo{author}{A.~{Ferrara}}, \bibinfo{author}{S.~{Gallerani}},
  \bibinfo{author}{L.~{Vallini}}, \bibinfo{journal}{\mnras}
  \bibinfo{volume}{477} (\bibinfo{year}{2018}) \bibinfo{pages}{552--565}.
\bibitem[{{Mo} et~al.(1998){Mo}, {Mao}, and {White}}]{mo1998}
\bibinfo{author}{H.~J. {Mo}}, \bibinfo{author}{S.~{Mao}},
  \bibinfo{author}{S.~D.~M. {White}}  \bibinfo{volume}{295}
  (\bibinfo{year}{1998}) \bibinfo{pages}{319--336}.
\bibitem[{{Peebles}(1969)}]{peebles1969}
\bibinfo{author}{P.~J.~E. {Peebles}}, \bibinfo{journal}{\apj}
  \bibinfo{volume}{155} (\bibinfo{year}{1969}) \bibinfo{pages}{393}.
\bibitem[{{Barnes} and {Efstathiou}(1987)}]{barnes1987}
\bibinfo{author}{J.~{Barnes}}, \bibinfo{author}{G.~{Efstathiou}},
  \bibinfo{journal}{\apj}  \bibinfo{volume}{319} (\bibinfo{year}{1987})
  \bibinfo{pages}{575--600}.
\bibitem[{{Doroshkevich}(1970)}]{doro1970}
\bibinfo{author}{A.~G. {Doroshkevich}}, \bibinfo{journal}{Astrophysics}
  \bibinfo{volume}{6} (\bibinfo{year}{1970}) \bibinfo{pages}{320--330}.
\bibitem[{{Bullock} et~al.(2001){Bullock}, {Dekel}, {Kolatt}, {Kravtsov},
  {Klypin}, {Porciani}, and {Primack}}]{bullock2001}
\bibinfo{author}{J.~S. {Bullock}}, \bibinfo{author}{A.~{Dekel}},
  \bibinfo{author}{T.~S. {Kolatt}}, \bibinfo{author}{A.~V. {Kravtsov}},
  \bibinfo{author}{A.~A. {Klypin}}, \bibinfo{author}{C.~{Porciani}},
  \bibinfo{author}{J.~R. {Primack}}, \bibinfo{journal}{\apj}
  \bibinfo{volume}{555} (\bibinfo{year}{2001}) \bibinfo{pages}{240--257}.
\bibitem[{{Vitvitska} et~al.(2002){Vitvitska}, {Klypin}, {Kravtsov},
  {Wechsler}, {Primack}, and {Bullock}}]{vitvitska2002}
\bibinfo{author}{M.~{Vitvitska}}, \bibinfo{author}{A.~A. {Klypin}},
  \bibinfo{author}{A.~V. {Kravtsov}}, \bibinfo{author}{R.~H. {Wechsler}},
  \bibinfo{author}{J.~R. {Primack}}, \bibinfo{author}{J.~S. {Bullock}},
  \bibinfo{journal}{\apj}  \bibinfo{volume}{581} (\bibinfo{year}{2002})
  \bibinfo{pages}{799--809}.
\bibitem[{{Davis} and {Natarajan}(2009)}]{davis2009}
\bibinfo{author}{A.~J. {Davis}}, \bibinfo{author}{P.~{Natarajan}},
  \bibinfo{journal}{\mnras}  \bibinfo{volume}{393} (\bibinfo{year}{2009})
  \bibinfo{pages}{1498--1502}.
\bibitem[{{Shibuya} et~al.(2016){Shibuya}, {Ouchi}, {Kubo}, and
  {Harikane}}]{shibuya2016}
\bibinfo{author}{T.~{Shibuya}}, \bibinfo{author}{M.~{Ouchi}},
  \bibinfo{author}{M.~{Kubo}}, \bibinfo{author}{Y.~{Harikane}},
  \bibinfo{journal}{\apj}  \bibinfo{volume}{821} (\bibinfo{year}{2016})
  \bibinfo{pages}{72}.
\bibitem[{{Madau} and {Dickinson}(2014)}]{madau2014}
\bibinfo{author}{P.~{Madau}}, \bibinfo{author}{M.~{Dickinson}},
  \bibinfo{journal}{\araa}  \bibinfo{volume}{52} (\bibinfo{year}{2014})
  \bibinfo{pages}{415--486}.
\bibitem[{{Lotz} et~al.(2011){Lotz}, {Jonsson}, {Cox}, {Croton}, {Primack},
  {Somerville}, and {Stewart}}]{lotz2011}
\bibinfo{author}{J.~M. {Lotz}}, \bibinfo{author}{P.~{Jonsson}},
  \bibinfo{author}{T.~J. {Cox}}, \bibinfo{author}{D.~{Croton}},
  \bibinfo{author}{J.~R. {Primack}}, \bibinfo{author}{R.~S. {Somerville}},
  \bibinfo{author}{K.~{Stewart}}, \bibinfo{journal}{\apj}
  \bibinfo{volume}{742} (\bibinfo{year}{2011}) \bibinfo{pages}{103}.
\bibitem[{{Ferguson} et~al.(2004){Ferguson}, {Dickinson}, {Giavalisco}, and et.
  al.}]{ferguson2004}
\bibinfo{author}{H.~C. {Ferguson}}, \bibinfo{author}{M.~{Dickinson}},
  \bibinfo{author}{M.~{Giavalisco}}, \bibinfo{author}{et. al.},
  \bibinfo{journal}{\apjl}  \bibinfo{volume}{600} (\bibinfo{year}{2004})
  \bibinfo{pages}{L107--L110}.
\bibitem[{{Ravindranath} et~al.(2006){Ravindranath}, {Giavalisco}, {Ferguson},
  and et. al.}]{ravindranath2006}
\bibinfo{author}{S.~{Ravindranath}}, \bibinfo{author}{M.~{Giavalisco}},
  \bibinfo{author}{H.~C. {Ferguson}}, \bibinfo{author}{et. al.},
  \bibinfo{journal}{\apj}  \bibinfo{volume}{652} (\bibinfo{year}{2006})
  \bibinfo{pages}{963--980}.
\bibitem[{{Conselice} and {Arnold}(2009)}]{conselice2009}
\bibinfo{author}{C.~J. {Conselice}}, \bibinfo{author}{J.~{Arnold}},
  \bibinfo{journal}{\mnras}  \bibinfo{volume}{397} (\bibinfo{year}{2009})
  \bibinfo{pages}{208--231}.
\bibitem[{{Jiang} et~al.(2013){Jiang}, {Egami}, {Fan}, and et. al.}]{jiang2013}
\bibinfo{author}{L.~{Jiang}}, \bibinfo{author}{E.~{Egami}},
  \bibinfo{author}{X.~{Fan}}, \bibinfo{author}{et. al.},
  \bibinfo{journal}{\apj}  \bibinfo{volume}{773} (\bibinfo{year}{2013})
  \bibinfo{pages}{153}.
\bibitem[{{Kawamata} et~al.(2015){Kawamata}, {Ishigaki}, {Shimasaku}, {Oguri},
  and {Ouchi}}]{kawamata2015}
\bibinfo{author}{R.~{Kawamata}}, \bibinfo{author}{M.~{Ishigaki}},
  \bibinfo{author}{K.~{Shimasaku}}, \bibinfo{author}{M.~{Oguri}},
  \bibinfo{author}{M.~{Ouchi}}, \bibinfo{journal}{\apj}  \bibinfo{volume}{804}
  (\bibinfo{year}{2015}) \bibinfo{pages}{103}.
\bibitem[{{Bond} et~al.(2009){Bond}, {Gawiser}, {Gronwall}, {Ciardullo},
  {Altmann}, and {Schawinski}}]{bond2009}
\bibinfo{author}{N.~A. {Bond}}, \bibinfo{author}{E.~{Gawiser}},
  \bibinfo{author}{C.~{Gronwall}}, \bibinfo{author}{R.~{Ciardullo}},
  \bibinfo{author}{M.~{Altmann}}, \bibinfo{author}{K.~{Schawinski}},
  \bibinfo{journal}{\apj}  \bibinfo{volume}{705} (\bibinfo{year}{2009})
  \bibinfo{pages}{639--649}.
\bibitem[{{Dekel} et~al.(2009){Dekel}, {Sari}, and {Ceverino}}]{dekel2009b}
\bibinfo{author}{A.~{Dekel}}, \bibinfo{author}{R.~{Sari}},
  \bibinfo{author}{D.~{Ceverino}}, \bibinfo{journal}{\apj}
  \bibinfo{volume}{703} (\bibinfo{year}{2009}) \bibinfo{pages}{785--801}.
\bibitem[{{Romano-D{\'{\i}}az} et~al.(2011){Romano-D{\'{\i}}az}, {Choi},
  {Shlosman}, and {Trenti}}]{romanodiaz2011}
\bibinfo{author}{E.~{Romano-D{\'{\i}}az}}, \bibinfo{author}{J.-H. {Choi}},
  \bibinfo{author}{I.~{Shlosman}}, \bibinfo{author}{M.~{Trenti}},
  \bibinfo{journal}{\apjl}  \bibinfo{volume}{738} (\bibinfo{year}{2011})
  \bibinfo{pages}{L19}.
\bibitem[{{Feng} et~al.(2015){Feng}, {Di Matteo}, {Croft}, {Tenneti}, {Bird},
  {Battaglia}, and {Wilkins}}]{feng2015}
\bibinfo{author}{Y.~{Feng}}, \bibinfo{author}{T.~{Di Matteo}},
  \bibinfo{author}{R.~{Croft}}, \bibinfo{author}{A.~{Tenneti}},
  \bibinfo{author}{S.~{Bird}}, \bibinfo{author}{N.~{Battaglia}},
  \bibinfo{author}{S.~{Wilkins}}, \bibinfo{journal}{\apjl}
  \bibinfo{volume}{808} (\bibinfo{year}{2015}) \bibinfo{pages}{L17}.
\bibitem[{{Bouwens} et~al.(2004){Bouwens}, {Illingworth}, {Blakeslee},
  {Broadhurst}, and {Franx}}]{bouwens2004}
\bibinfo{author}{R.~J. {Bouwens}}, \bibinfo{author}{G.~D. {Illingworth}},
  \bibinfo{author}{J.~P. {Blakeslee}}, \bibinfo{author}{T.~J. {Broadhurst}},
  \bibinfo{author}{M.~{Franx}}, \bibinfo{journal}{\apjl}  \bibinfo{volume}{611}
  (\bibinfo{year}{2004}) \bibinfo{pages}{L1--L4}.
\bibitem[{{Ono} et~al.(2013){Ono}, {Ouchi}, {Curtis-Lake}, and et.
  al.}]{ono2013}
\bibinfo{author}{Y.~{Ono}}, \bibinfo{author}{M.~{Ouchi}},
  \bibinfo{author}{E.~{Curtis-Lake}}, \bibinfo{author}{et. al.},
  \bibinfo{journal}{\apj}  \bibinfo{volume}{777} (\bibinfo{year}{2013})
  \bibinfo{pages}{155}.
\bibitem[{{Huang} et~al.(2013){Huang}, {Ferguson}, {Ravindranath}, and
  {Su}}]{huang2013}
\bibinfo{author}{K.-H. {Huang}}, \bibinfo{author}{H.~C. {Ferguson}},
  \bibinfo{author}{S.~{Ravindranath}}, \bibinfo{author}{J.~{Su}},
  \bibinfo{journal}{\apj}  \bibinfo{volume}{765} (\bibinfo{year}{2013})
  \bibinfo{pages}{68}.
\bibitem[{{Holwerda} et~al.(2015){Holwerda}, {Bouwens}, {Oesch}, {Smit},
  {Illingworth}, and {Labbe}}]{holwerda2015}
\bibinfo{author}{B.~W. {Holwerda}}, \bibinfo{author}{R.~{Bouwens}},
  \bibinfo{author}{P.~{Oesch}}, \bibinfo{author}{R.~{Smit}},
  \bibinfo{author}{G.~{Illingworth}}, \bibinfo{author}{I.~{Labbe}},
  \bibinfo{journal}{\apj}  \bibinfo{volume}{808} (\bibinfo{year}{2015})
  \bibinfo{pages}{6}.
\bibitem[{{Shibuya} et~al.(2015){Shibuya}, {Ouchi}, and
  {Harikane}}]{shibuya2015}
\bibinfo{author}{T.~{Shibuya}}, \bibinfo{author}{M.~{Ouchi}},
  \bibinfo{author}{Y.~{Harikane}}, \bibinfo{journal}{\apjs}
  \bibinfo{volume}{219} (\bibinfo{year}{2015}) \bibinfo{pages}{15}.
\bibitem[{{Curtis-Lake} et~al.(2016){Curtis-Lake}, {McLure}, {Dunlop}, and et.
  al.}]{curtis-lake2016}
\bibinfo{author}{E.~{Curtis-Lake}}, \bibinfo{author}{R.~J. {McLure}},
  \bibinfo{author}{J.~S. {Dunlop}}, \bibinfo{author}{et. al.},
  \bibinfo{journal}{\mnras}  \bibinfo{volume}{457} (\bibinfo{year}{2016})
  \bibinfo{pages}{440--464}.
\bibitem[{{Venemans} et~al.(2005){Venemans}, {R{\"o}ttgering}, {Miley}, and et.
  al.}]{venemans2005}
\bibinfo{author}{B.~P. {Venemans}}, \bibinfo{author}{H.~J.~A.
  {R{\"o}ttgering}}, \bibinfo{author}{G.~K. {Miley}}, \bibinfo{author}{et.
  al.}, \bibinfo{journal}{\aap}  \bibinfo{volume}{431} (\bibinfo{year}{2005})
  \bibinfo{pages}{793--812}.
\bibitem[{{Overzier} et~al.(2008){Overzier}, {Bouwens}, {Cross}, and et.
  al.}]{overzier2008}
\bibinfo{author}{R.~A. {Overzier}}, \bibinfo{author}{R.~J. {Bouwens}},
  \bibinfo{author}{N.~J.~G. {Cross}}, \bibinfo{author}{et. al.},
  \bibinfo{journal}{\apj}  \bibinfo{volume}{673} (\bibinfo{year}{2008})
  \bibinfo{pages}{143--162}.
\bibitem[{{Malhotra} et~al.(2012){Malhotra}, {Rhoads}, {Finkelstein}, {Hathi},
  {Nilsson}, {McLinden}, and {Pirzkal}}]{malhotra2012}
\bibinfo{author}{S.~{Malhotra}}, \bibinfo{author}{J.~E. {Rhoads}},
  \bibinfo{author}{S.~L. {Finkelstein}}, \bibinfo{author}{N.~{Hathi}},
  \bibinfo{author}{K.~{Nilsson}}, \bibinfo{author}{E.~{McLinden}},
  \bibinfo{author}{N.~{Pirzkal}}, \bibinfo{journal}{\apjl}
  \bibinfo{volume}{750} (\bibinfo{year}{2012}) \bibinfo{pages}{L36}.
\bibitem[{{Tully} and {Fisher}(1977)}]{tully-fisher1977}
\bibinfo{author}{R.~B. {Tully}}, \bibinfo{author}{J.~R. {Fisher}},
  \bibinfo{journal}{\aap}  \bibinfo{volume}{54} (\bibinfo{year}{1977})
  \bibinfo{pages}{661--673}.
\bibitem[{{Faber} and {Jackson}(1976)}]{faber-jackson1976}
\bibinfo{author}{S.~M. {Faber}}, \bibinfo{author}{R.~E. {Jackson}},
  \bibinfo{journal}{\apj}  \bibinfo{volume}{204} (\bibinfo{year}{1976})
  \bibinfo{pages}{668--683}.
\bibitem[{{Liu} et~al.(2017){Liu}, {Mutch}, {Poole}, {Angel}, {Duffy}, {Geil},
  {Mesinger}, and {Wyithe}}]{liu2017}
\bibinfo{author}{C.~{Liu}}, \bibinfo{author}{S.~J. {Mutch}},
  \bibinfo{author}{G.~B. {Poole}}, \bibinfo{author}{P.~W. {Angel}},
  \bibinfo{author}{A.~R. {Duffy}}, \bibinfo{author}{P.~M. {Geil}},
  \bibinfo{author}{A.~{Mesinger}}, \bibinfo{author}{J.~S.~B. {Wyithe}},
  \bibinfo{journal}{\mnras}  \bibinfo{volume}{465} (\bibinfo{year}{2017})
  \bibinfo{pages}{3134--3142}.
\bibitem[{{Hathi} et~al.(2008){Hathi}, {Jansen}, {Windhorst}, {Cohen}, {Keel},
  {Corbin}, and {Ryan}}]{hathi2008}
\bibinfo{author}{N.~P. {Hathi}}, \bibinfo{author}{R.~A. {Jansen}},
  \bibinfo{author}{R.~A. {Windhorst}}, \bibinfo{author}{S.~H. {Cohen}},
  \bibinfo{author}{W.~C. {Keel}}, \bibinfo{author}{M.~R. {Corbin}},
  \bibinfo{author}{R.~E. {Ryan}, Jr.}, \bibinfo{journal}{\aj}
  \bibinfo{volume}{135} (\bibinfo{year}{2008}) \bibinfo{pages}{156--166}.
\bibitem[{{Mosleh} et~al.(2012){Mosleh}, {Williams}, {Franx}, {Gonzalez},
  {Bouwens}, {Oesch}, {Labbe}, {Illingworth}, and {Trenti}}]{mosleh2012}
\bibinfo{author}{M.~{Mosleh}}, \bibinfo{author}{R.~J. {Williams}},
  \bibinfo{author}{M.~{Franx}}, \bibinfo{author}{V.~{Gonzalez}},
  \bibinfo{author}{R.~J. {Bouwens}}, \bibinfo{author}{P.~{Oesch}},
  \bibinfo{author}{I.~{Labbe}}, \bibinfo{author}{G.~D. {Illingworth}},
  \bibinfo{author}{M.~{Trenti}}, \bibinfo{journal}{\apjl}
  \bibinfo{volume}{756} (\bibinfo{year}{2012}) \bibinfo{pages}{L12}.
\bibitem[{{Allen} et~al.(2016){Allen}, {Kacprzak}, {Glazebrook}, and et.
  al.}]{allen2016}
\bibinfo{author}{R.~J. {Allen}}, \bibinfo{author}{G.~G. {Kacprzak}},
  \bibinfo{author}{K.~{Glazebrook}}, \bibinfo{author}{et. al.},
  \bibinfo{journal}{ArXiv e-prints}   (\bibinfo{year}{2016}).
\bibitem[{{Barden} et~al.(2005){Barden}, {Rix}, {Somerville}, and et.
  al.}]{barden2005}
\bibinfo{author}{M.~{Barden}}, \bibinfo{author}{H.-W. {Rix}},
  \bibinfo{author}{R.~S. {Somerville}}, \bibinfo{author}{et. al.},
  \bibinfo{journal}{\apj}  \bibinfo{volume}{635} (\bibinfo{year}{2005})
  \bibinfo{pages}{959--981}.
\bibitem[{{Graham} and {Worley}(2008)}]{graham2008}
\bibinfo{author}{A.~W. {Graham}}, \bibinfo{author}{C.~C. {Worley}},
  \bibinfo{journal}{\mnras}  \bibinfo{volume}{388} (\bibinfo{year}{2008})
  \bibinfo{pages}{1708--1728}.
\bibitem[{{Bouwens} et~al.(2016){Bouwens}, {Aravena}, {Decarli}, and et.
  al.}]{bouwens2016}
\bibinfo{author}{R.~{Bouwens}}, \bibinfo{author}{M.~{Aravena}},
  \bibinfo{author}{R.~{Decarli}}, \bibinfo{author}{et. al.},
  \bibinfo{journal}{ArXiv e-prints}   (\bibinfo{year}{2016}).
\bibitem[{{Dow-Hygelund} et~al.(2007){Dow-Hygelund}, {Holden}, {Bouwens}, and
  et. al.}]{dow2007}
\bibinfo{author}{C.~C. {Dow-Hygelund}}, \bibinfo{author}{B.~P. {Holden}},
  \bibinfo{author}{R.~J. {Bouwens}}, \bibinfo{author}{et. al.},
  \bibinfo{journal}{\apj}  \bibinfo{volume}{660} (\bibinfo{year}{2007})
  \bibinfo{pages}{47--61}.
\bibitem[{{Zick} et~al.(2018){Zick}, {Weisz}, and {Boylan-Kolchin}}]{zick2018}
\bibinfo{author}{T.~O. {Zick}}, \bibinfo{author}{D.~R. {Weisz}},
  \bibinfo{author}{M.~{Boylan-Kolchin}}, \bibinfo{journal}{\mnras}
  (\bibinfo{year}{2018}).
\bibitem[{{Vrbanec} et~al.(2016){Vrbanec}, {Ciardi}, {Jeli{\'c}}, and et.
  al.}]{vrbanec2016}
\bibinfo{author}{D.~{Vrbanec}}, \bibinfo{author}{B.~{Ciardi}},
  \bibinfo{author}{V.~{Jeli{\'c}}}, \bibinfo{author}{et. al.},
  \bibinfo{journal}{\mnras}  \bibinfo{volume}{457} (\bibinfo{year}{2016})
  \bibinfo{pages}{666--675}.
\bibitem[{{Sobacchi} et~al.(2016){Sobacchi}, {Mesinger}, and
  {Greig}}]{sobacchi2016}
\bibinfo{author}{E.~{Sobacchi}}, \bibinfo{author}{A.~{Mesinger}},
  \bibinfo{author}{B.~{Greig}}, \bibinfo{journal}{\mnras}
  \bibinfo{volume}{459} (\bibinfo{year}{2016}) \bibinfo{pages}{2741--2750}.
\bibitem[{{Hutter} et~al.(2017){Hutter}, {Dayal}, {M{\"u}ller}, and
  {Trott}}]{hutter2017}
\bibinfo{author}{A.~{Hutter}}, \bibinfo{author}{P.~{Dayal}},
  \bibinfo{author}{V.~{M{\"u}ller}}, \bibinfo{author}{C.~M. {Trott}},
  \bibinfo{journal}{\apj}  \bibinfo{volume}{836} (\bibinfo{year}{2017})
  \bibinfo{pages}{176}.
\bibitem[{{Hutter} et~al.(2018){Hutter}, {Trott}, and {Dayal}}]{hutter2018}
\bibinfo{author}{A.~{Hutter}}, \bibinfo{author}{C.~M. {Trott}},
  \bibinfo{author}{P.~{Dayal}}, \bibinfo{journal}{\mnras}
  \bibinfo{volume}{479} (\bibinfo{year}{2018}) \bibinfo{pages}{L129--L133}.
\bibitem[{{Kubota} et~al.(2018){Kubota}, {Yoshiura}, {Takahashi}, {Hasegawa},
  {Yajima}, {Ouchi}, {Pindor}, and {Webster}}]{kubota2018}
\bibinfo{author}{K.~{Kubota}}, \bibinfo{author}{S.~{Yoshiura}},
  \bibinfo{author}{K.~{Takahashi}}, \bibinfo{author}{K.~{Hasegawa}},
  \bibinfo{author}{H.~{Yajima}}, \bibinfo{author}{M.~{Ouchi}},
  \bibinfo{author}{B.~{Pindor}}, \bibinfo{author}{R.~L. {Webster}},
  \bibinfo{journal}{\mnras}  \bibinfo{volume}{479} (\bibinfo{year}{2018})
  \bibinfo{pages}{2754--2766}.

\end{thebibliography}

\end{document}